\documentclass[10pt]{article}
\pdfoutput=1
\usepackage{mathStyle}
\usepackage{amsmath,amssymb,amsthm,latexsym,bbm,calc,wasysym,mathtools,empheq, bbold, yfonts}
\usepackage[english]{babel}
\usepackage{relsize}
\usepackage{graphicx,color}
\usepackage[dvipsnames]{xcolor}
\usepackage[makeroom]{cancel}

\usepackage{tikz-cd}
\usepackage{tikz}
\usetikzlibrary{shapes.geometric}
\usetikzlibrary{arrows,matrix,calc,scopes,decorations.markings,intersections}
\usetikzlibrary{arrows.meta}
\usetikzlibrary{decorations.pathreplacing,angles,quotes}
\usetikzlibrary{external}

\usepackage{xspace}
\usepackage{pifont,dsfont}
\usepackage{marvosym}
\usepackage{slashed}
\usepackage{subfigure}
\usepackage{sidecap}
\usepackage{stmaryrd}
\usepackage{empheq}

\usepackage[export]{adjustbox}
\mathtoolsset{showonlyrefs=true}
\usepackage{placeins}
\usepackage{enumitem}
\usepackage{verbatim}

\usepackage{amsthm}
\newtheoremstyle{mydef}%
	{0.9em} 
	{0.7em}
	{\hangindent=2em}
	{1.8em}
	{\scshape}
	{.}
	{.5em}
	{}%
\theoremstyle{mydef}
\newtheorem{definition}{Definition}
\numberwithin{definition}{section}

\theoremstyle{mydef}

\numberwithin{lemma}{section}

\theoremstyle{mydef}

\numberwithin{theorem}{section}

\theoremstyle{mydef}
\newtheorem{example}{Example}
\numberwithin{example}{section}

\theoremstyle{mydef}
\newtheorem{convention}{Convention}
\numberwithin{convention}{section}

\theoremstyle{mydef}

\numberwithin{property}{section}

\newcommand{\la}{\langle}
\newcommand{\ra}{\rangle}
\newcommand{\q}{\quad}
\newcommand{\nn}{\nonumber}
\newcommand{\rU}{{\rm U}}
\newcommand{\sss}{\scriptstyle}
\newcommand{\bul}{{\sss \bullet}}

\newcommand{\zo}{\text{\small $0$}}
\newcommand{\snum}[1]{\text{\small $#1$}}

\newcommand{\cM}{\mathcal{M}}
\newcommand{\cA}{\mathcal{A}}

\newcommand{\cC}{\mathcal{C}}
\newcommand{\mc}[1]{\mathcal{#1}}
\newcommand{\ket}[1]{| #1 \rangle}
\newcommand{\bra}[1]{\langle #1 |}
\newcommand{\Z}{\mc{Z}^{G}_{\omega}}
\newcommand{\Zo}[1]{\mc{Z}^{G}_{\omega}[#1]}
\newcommand{\cob}[3]{{_{#1}#2_{#3}}}
\newcommand{\Hilb}[1]{\mc{H}^{G}_{\omega}[#1]}
\newcommand{\Ham}[1]{\mathbb{H}^{G}_{\omega}(#1)}
\newcommand{\gs}[1]{\mc{V}^{G}_{\omega}[#1]}
\newcommand{\tubem}[1]{\mathfrak{T}[#1]}
\newcommand{\tubealg}[1]{{\rm Tube}_{\omega}^{G}(#1)}
\newcommand{\I}{\mathbb{I}}
\newcommand{\looDSy}[2]{%
	\mathrel{\raisebox{-0.15em}{$\substack{#1\\ \includegraphics[scale=1,valign=c]{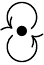} \\ #2}$}}%
}
\newcommand{\looSy}[1]{%
	\mathrel{\raisebox{0.36em}{$\substack{#1\\ \includegraphics[scale=1,valign=c]{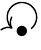}}$}}%
}

\makeatletter
\newcommand{\xRrightarrow}[2][]{\ext@arrow 0359\Rrightarrowfill@{#1}{#2}}
\newcommand{\Rrightarrowfill@}{\arrowfill@\equiv\equiv\Rrightarrow}
\newcommand{\xLleftarrow}[2][]{\ext@arrow 3095\Lleftarrowfill@{#1}{#2}}
\newcommand{\Lleftarrowfill@}{\arrowfill@\Lleftarrow\equiv\equiv}
\makeatother

\usepackage{soul}

\hypersetup{%
	bookmarks=true,         
	bookmarksopen=true,
	unicode=true,           
	pdftoolbar=true,        
	pdfmenubar=true,        
	pdffitwindow=false,     
	pdfstartview={FitH},    
	pdftitle={Tube algebras, excitations statistics and compactification in gauge models of  topological phases},
	pdfauthor={},
	pdfsubject={},          
	pdfcreator={},          
	pdfproducer={pdfLaTeX}, 
	pdfkeywords={string net} {topological phases} {category theory},
	pdfnewwindow=true,      
	colorlinks,
	citecolor=BrickRed,
	filecolor=BrickRed,
	linkcolor=BlueViolet,
	urlcolor=BrickRed,
	linktocpage=true
}

\tikzset{->1-/.style={decoration={
			markings,
			mark=at position #1 with {\arrow{Triangle[fill=black, width=3.5pt, length=3.5pt]}}},postaction={decorate}}}
\tikzset{->2-/.style={decoration={
					markings,
					mark=at position #1 with {\arrow{Triangle[fill=white, width=5pt, length=5pt]}}},postaction={decorate}}}

\tikzset{->-/.style={decoration={
			markings,
			mark=at position #1 with {\arrow{latex}}},postaction={decorate}}}

\tikzset{-<-/.style={decoration={
			markings,
			mark=at position #1 with {\arrow{latex reversed}}},postaction={decorate}}}

\tikzset{
	every picture/.append style={
		line join=round,
		line cap=round,
		line width = 0.8pt
	}
}
\tikzstyle{dot}=[dotted, line width=0.8pt]
\tikzstyle{op}=[opacity=0.4]
\tikzstyle{op6}=[opacity=0.6]
\tikzstyle{op5}=[opacity=0.5]
\tikzstyle{op4}=[opacity=0.4]
\tikzstyle{op3}=[opacity=0.3]
\tikzstyle{op2}=[opacity=0.2]
\tikzstyle{op1}=[opacity=0.1]
\tikzstyle{sha}=[shorten >= 0.3em]
\tikzstyle{shb}=[shorten <= 0.3em]
\tikzstyle{shab}=[shorten >= 0.3em, shorten <= 0.3em]

\newcommand{\interL}[5]{
	\path[name path=P#1#2] (#1) -- (#2);
	\path[name path=P#3#4] (#3) -- (#4);
	\path [name intersections={of=P#1#2 and P#3#4,by={sp}}];
	\draw[sha, opacity=#5] (#1) -- (sp);
	\draw[shb, opacity=#5] (sp) -- (#2);
}

\newcommand{\interLL}[7]{
	\path[name path=P#1#2] (#1) -- (#2);
	\path[name path=P#3#4] (#3) -- (#4);
	\path[name path=P#5#6] (#5) -- (#6);	
	\path [name intersections={of=P#1#2 and P#3#4,by={sp1}}];
	\path [name intersections={of=P#1#2 and P#5#6,by={sp2}}];	
	\draw[sha, opacity=#7] (#1) -- (sp1);
	\draw[shab, opacity=#7] (sp1) -- (sp2);
	\draw[shb, opacity=#7] (sp2) -- (#2);
}

\newcommand{\interLLp}[6]{
	\path[name path=P#1#2] (#1) -- (#2);	
	\path [name intersections={of=P#1#2 and P#3#4,by={sp1}}];
	\path [name intersections={of=P#1#2 and P#5#6,by={sp2}}];	
	\draw[sha] (#1) -- (sp1);
	\draw[shab] (sp1) -- (sp2);
	\draw[shb] (sp2) -- (#2);
}

\newcommand{\interLLL}[9]{
	\path[name path=P#1#2] (#1) -- (#2);
	\path[name path=P#3#4] (#3) -- (#4);
	\path[name path=P#5#6] (#5) -- (#6);
	\path[name path=P#7#8] (#7) -- (#8);		
	\path [name intersections={of=P#1#2 and P#3#4,by={sp1}}];
	\path [name intersections={of=P#1#2 and P#5#6,by={sp2}}];
	\path [name intersections={of=P#1#2 and P#7#8,by={sp3}}];		
	\draw[sha, opacity=#9] (#1) -- (sp1);
	\draw[shab, opacity=#9] (sp1) -- (sp2);
	\draw[shab, opacity=#9] (sp2) -- (sp3);
	\draw[shb, opacity=#9] (sp3) -- (#2);
}

\newcommand{\speL}[6]{
	\path[name path=P#1#2] (#1) -- (#2);
	\path [name intersections={of=P#1#2 and P#3#4,by={sp}}];
	\draw[#5] (#1) -- (sp);
	\draw[#6] (sp) -- (#2);
}

\newcommand{\speLL}[7]{
	\path[name path=P#1#2] (#1) -- (#2);
	\path [name intersections={of=P#1#2 and #3,by={sp1}}];
	\path [name intersections={of=P#1#2 and #4,by={sp2}}];
	\draw[#5] (#1) -- (sp1);
	\draw[#6] (sp1) -- (sp2);
	\draw[#7] (sp2) -- (#2);
}

\newcommand{\speC}[4]{
	\begin{scope}
		\path (#1) ++ (-0.5em,0) coordinate (sp1);
		\path (#2) ++ (0,-0.5em) coordinate (sp2);
		\clip (sp1) rectangle (sp2);
		\draw[#4] (#1) to[out=-85, in=265, looseness=0.5] (#3);
	\end{scope}
	\begin{scope}
		\path (#2) ++ (0,-5em) coordinate (sp2);
		\path (#3) ++ (0.5em,0) coordinate (sp3);
		\clip (sp2) rectangle (sp3);
		\draw[] (#1) to[out=-85, in=265, looseness=0.5] (#3);
	\end{scope}	
}

\newcommand{\speCC}[5]{
	\begin{scope}
		\path (#1) ++ (-0.5em,0) coordinate (sp1);
		\path (#2) ++ (0,-0.5em) coordinate (sp2);
		\clip (sp1) rectangle (sp2);
		\draw[#5] (#1) to[out=-85, in=265, looseness=0.5] (#4);
	\end{scope}
	\begin{scope}
		\path (#3) ++ (0,-5em) coordinate (sp2);
		\path (#4) ++ (0.5em,0) coordinate (sp3);
		\clip (sp2) rectangle (sp3);
		\draw[] (#1) to[out=-85, in=265, looseness=0.5] (#4);
	\end{scope}	
}

\newcommand{\speD}[6]{
	\begin{scope}
		\path [name intersections={of=P#1#2 and P#3#4,by={sp}}];
		\path (#1) ++ (-0.1em,0) coordinate (sp1);
		\path (sp) ++ (0.1em,0) coordinate (sp2);
		\clip (sp1) rectangle (sp2);
		\draw[#5] (#1) to[out=-90, in=135, looseness=1] (#2);
	\end{scope}
	\begin{scope}
		\path [name intersections={of=P#1#2 and P#3#4,by={sp}}];
		\path (sp) ++ (-0.1em,0.1em) coordinate (sp2);
		\path (#2) ++ (0.1em,-0.1em) coordinate (sp3);
		\clip (sp2) rectangle (sp3);
		\draw[#6] (#1) to[out=-90, in=135, looseness=1] (#2);
	\end{scope}	
}

\newcommand{\speE}[6]{
	\begin{scope}
		\path [name intersections={of=P#1#2 and P#3#4,by={sp}}];
		\path (#1) ++ (0.1em,0) coordinate (sp1);
		\path (sp) ++ (-0.1em,0) coordinate (sp2);
		\clip (sp1) rectangle (sp2);
		\draw[#5] (#1) to[out=-90, in=45, looseness=1] (#2);
	\end{scope}
	\begin{scope}
		\path [name intersections={of=P#1#2 and P#3#4,by={sp}}];
		\path (sp) ++ (0.1em,0.1em) coordinate (sp2);
		\path (#2) ++ (-0.1em,-0.1em) coordinate (sp3);
		\clip (sp2) rectangle (sp3);
		\draw[#6] (#1) to[out=-90, in=45, looseness=1] (#2);
	\end{scope}	
}

\tikzset{
	partial ellipse/.style args={#1:#2:#3}{
		insert path={+ (#1:#3) arc (#1:#2:#3)}
	}
}

\newcommand{\picBraid}[1]{
	\begin{tikzpicture}[scale=1.8, baseline=2.25em]
	\coordinate (0) at (0,0);
	\coordinate (1) at (0,1);
	\coordinate (2) at (0.7,0);
	\coordinate (3) at (0.7,1);
	\coordinate (4) at (1.4,0);
	\coordinate (5) at (1.4,1);	
	\coordinate (8) at (-0.8, 0.5);
	\coordinate (9) at (-0.5,0.5);
	\coordinate (10) at (0,0.5);
	\coordinate (11) at (0.7,0.5);
	\coordinate (12) at (1.4,0.5);
	\coordinate (13) at (1.9,0.5);
	\coordinate (14) at (2.2, 0.5);
	
	\ifthenelse{#1 =1}{
		\draw[name path=P01r] (0) to[out=0, in=-0, looseness =0.8] (1);	
		\draw[name path=P01l] (0) to[out =180, in=180, looseness =0.8] (1);
		\draw[name path=P23r] (2) to[out =0, in=0, looseness =0.8] (3);	
		\draw[name path=P23l] (2) to[out =180, in=180, looseness =0.8] (3);
		
		\draw[line width=0.6em, white] (9) -- (10);
		\draw[line width=0.6em, white] (0.3,0.5) -- (0.5,0.5);
		\draw[dotted] (8) -- (9);
		\speL{9}{11}{0}{1r}{sha}{shb};
		\coordinate (12) at (1.2,0.5);
		\coordinate (13) at (1.5,0.5);
		\speL{11}{12}{2}{3r}{sha}{shb};
		\draw[dotted] (12) -- (13);

		\node[below] at (0) {{$\sss{i}$}};
		\node[below] at (2) {{$\sss{i+1}$}};
		\node[above] at (9) {{$\sss{y}$}};
		\node[above] at (12) {{$\sss{y}$}};
	}{}
	\ifthenelse{#1 =2}{
		\draw[name path=P04, ->1-=0.3, ->1-=0.8] (0) to[out=15, in =165, looseness=1] (4);
		\draw[name path=P15, ->1-=0.3, ->1-=0.8] (1) to[out=-15, in=-165, looseness=1] (5);
		
		\draw[name path=P01r] (0) to[out=0, in=-0, looseness =0.8] (1);	
		\draw[name path=P01l] (0) to[out =180, in=180, looseness =0.8] (1);
		\draw[line width =0.6em,white] (2) to[out=0, in=0, looseness=0.8] (3); 
		\draw[name path=P23r] (2) to[out =0, in=0, looseness =0.8] (3);	
		\path[name path=P23l] (2) to[out =180, in=180, looseness =0.8] (3);
		\draw[name path=P45r] (4) to[out =0, in=0, looseness =0.8] (5);	
		\draw[name path=P45l, op4] (4) to[out =180, in=180, looseness =0.8] (5);

		\begin{scope}
		\path [name intersections={of=P15 and P23l,by={sp}}];
		\path (sp) ++ (-0.1em,0) coordinate (sp1);
		\path (3) ++ (0.1em,0.1em) coordinate (sp2);
		\clip (sp1) rectangle (sp2);
		\draw[] (2) to[out=175, in=185, looseness=0.8] (3);
		\end{scope}
		\begin{scope}
		\path [name intersections={of=P04 and P23l,by={sp}}];
		\path (sp) ++ (-0.1em,0) coordinate (sp1);
		\path (2) ++ (0.1em,-0.1em) coordinate (sp2);
		\clip (sp1) rectangle (sp2);
		\draw[] (2) to[out=175, in=185, looseness=0.8] (3);
		\end{scope}
		\begin{scope}
		\path [name intersections={of=P15 and P23l,by={sp1}}];
		\path [name intersections={of=P04 and P23l,by={sp2}}];
		\path (sp1) ++ (-0.5em,0) coordinate (sp3);
		\path (sp2) ++ (0.5em,0) coordinate (sp4);
		\clip (sp3) rectangle (sp4);
		\draw[op2] (2) to[out=175, in=185, looseness=0.8] (3);
		\end{scope}

		\draw[line width=0.6em, white] (9) -- (10);
		\draw[line width=0.6em, white] (0.35,0.5) -- (0.5,0.5);
		\draw[line width=0.6em, white] (1.1,0.5) -- (1.2,0.5);
		\draw[dotted] (8) -- (9);
		\draw[dotted] (13) -- (14);
		\speL{9}{11}{0}{1r}{sha}{shb, op4};
		\speL{11}{12}{2}{3r}{sha, op4}{shb, op4};
		\speL{12}{13}{4}{5r}{sha, op4}{shb};
		
		\node[below] at (0) {{$\sss{i}$}};
		\node[below] at (2) {{$\sss{i+1}$}};
		\node[below] at (4) {{$\sss{i}$}};
		\node[above] at (9) {{$\sss{y}$}};
		\node[above] at (13) {{$\sss{y}$}};
	}{}
	\ifthenelse{#1 =3}{
		\draw[name path=P01r] (0) to[out=0, in=-0, looseness =0.8] (1);	
		\draw[name path=P01l] (0) to[out =180, in=180, looseness =0.8] (1);
		\draw[name path=P23r] (2) to[out =0, in=0, looseness =0.8] (3);	
		\draw[name path=P23l] (2) to[out =180, in=180, looseness =0.8] (3);
		
		\draw[line width=0.6em, white] (9) -- (10);
		\draw[line width=0.6em, white] (0.3,0.5) -- (0.5,0.5);
		\draw[dotted] (8) -- (9);
		\speL{9}{11}{0}{1r}{sha}{shb};
		\coordinate (12) at (1.2,0.5);
		\coordinate (13) at (1.5,0.5);
		\speL{11}{12}{2}{3r}{sha}{shb};
		\draw[dotted] (12) -- (13);
		
		\node[below] at (0) {{$\sss{i+1}$}};
		\node[below] at (2) {{$\sss{i}$}};
		\node[above] at (9) {{$\sss{y}$}};
		\node[above] at (12) {{$\sss{y}$}};
	}{}
	\end{tikzpicture}
}

\newcommand{\pinchedinterval}[1]{
	\begin{tikzpicture}[scale=0.7,baseline={([yshift=-.5ex]current bounding box.center)}]
	\def\a{1.5}
	\coordinate (1) at (0,0);
	\coordinate (2) at (\a,0);
	\draw[] (1) to [bend right = 50]node[anchor=north]{$\overline{#1}$} (2);
	\draw[] (1) to [bend left =50]node[anchor=south]{$#1$} (2);
	\end{tikzpicture}
}

\newcommand{\interval}[1]{
	\begin{tikzpicture}[scale=0.7,baseline={([yshift=-.5ex]current bounding box.center)}]
	\def\a{1.5}
	\def\b{1.5}
	\coordinate (1) at (0,0);
	\coordinate (2) at (\a,0);
	\coordinate (3) at (0,\b);
	\coordinate (4) at (\a,\b);
	\draw[] (1) to node[anchor=north]{$\overline{#1}$} (2);
	\draw[] (3) to node[anchor=south]{$#1$} (4);
	\draw[] (1) to (3);
	\draw[] (2) to (4);
	\end{tikzpicture}
}

\newcommand{\clst}[1]{
	\begin{tikzpicture}[scale=0.5,baseline=-0.5em]
	\coordinate (0) at (-1,3);
	\coordinate (1) at (0.2,0); 
	\coordinate (2) at (-1,1);
	\coordinate (3) at (2.5,-1);
	\coordinate (4) at (2,1);
	\coordinate (5) at (-2,-1);
	\coordinate (6) at (0,-1.8);
	\coordinate (7) at (-2.5,1);
	\coordinate (8) at (0.3,1.8);
	\coordinate (9) at (-1.3,-2.2);
	\coordinate (10) at (-3,-1.5);
	
	\ifthenelse{#1 = 1}{
		\draw[] (1) -- (2);
		\draw[] (1) -- (3);
		\draw[] (1) -- (4);
		\draw[] (1) -- (5);
		\draw[] (1) -- (6);
		\draw[] (2) -- (4);	
		\draw[] (4) -- (3);
		\draw[] (3) -- (6);
		\draw[] (6) -- (5);
		\draw[] (5) -- (2);
		\draw[] (2) -- (7);
		\draw[] (7) -- (5);
		\draw[] (2) -- (8);
		\draw[] (8) -- (4);
		\draw[] (5) -- (9);
		\draw[] (9) -- (6);
		\draw[] (7) -- (8);	
		\draw[] (9) -- (10);
		\draw[] (5) -- (10);
		\draw[] (7) -- (10);	
	}{}	
	\ifthenelse{#1 = 2}{
		\draw[] (1) -- (2);
		\draw[] (1) -- (3);
		\draw[] (1) -- (4);
		\draw[] (1) -- (5);
		\draw[] (1) -- (6);
		\draw[] (2) -- (4);	
		\draw[] (4) -- (3);
		\draw[] (3) -- (6);
		\draw[] (6) -- (5);
		\draw[] (5) -- (2);
	}{}
	\ifthenelse{#1 = 3}{
		\draw[] (0) -- (1);
		\draw[] (0) -- (2);
		\draw[] (0) -- (3);
		\draw[] (0) -- (4);
		\draw[] (0) -- (5);
		\draw[] (0) -- (6);		
		\draw[] (1) -- (3);
		\draw[] (1) -- (6);
		\interL{1}{2}{0}{6}{1};
		\interL{1}{5}{0}{6}{1};
		\interL{1}{4}{0}{3}{1};	
		\interLLL{2}{4}{0}{6}{0}{1}{0}{3}{1};			
		\draw[] (4) -- (3);
		\draw[] (3) -- (6);
		\draw[] (6) -- (5);
		\draw[] (5) -- (2);
		\node[left] at (0) {${\scriptstyle 0'}$};
	}{}		
	\node[below, xshift=0.3em] at (1) {${\scriptstyle 0}$};
	
	\end{tikzpicture}	
}

\newcommand{\HamONE}[2]{
	\begin{tikzpicture}[scale=#1,baseline=-0.5em]
	\coordinate (0) at (0,1);
	\coordinate (1) at (1,-0.4);
	\coordinate (2) at (0.55,-1);
	\coordinate (3) at (-1.2,-0.55);
	\coordinate (4) at (0,-0.1);	
	
	\draw[] (0) -- (1);
	\draw[] (1) -- (2);
	\draw[] (2) -- (3);
	\draw[] (0) -- (3);
	\draw[] (0) -- (2);
	\draw[] (3) -- (4);
	\draw[] (0) -- (4);
	\draw[] (2) -- (4);
	\draw[shorten >= 0.3em] (1) -- (intersection of 1--4 and 0--2);
	\draw[shorten <= 0.3em] (intersection of 1--4 and 0--2) -- (4);
	\draw[shorten >= 0.3em] (1) -- (intersection of 1--3 and 0--2);
	\draw[shorten <= 0.3em, shorten >=0.3em] (intersection of 1--3 and 0--2) -- (intersection of 1--3 and 2--4);
	\draw[shorten <= 0.3em] (intersection of 1--3 and 2--4) -- (3);
	
	\ifthenelse{#2 = 1}{	
		\node[above] at (0) {{$\scriptstyle{0}$}};	
		\node[right] at (1) {{$\scriptstyle{1}$}};
		\node[below] at (2) {{$\scriptstyle{2}$}};
		\node[left] at (3) {{$\scriptstyle{3}$}};	
		\node[left, yshift=1.5] at (4) {{$\scriptstyle{4}$}};}{}
	
	\ifthenelse{#2 = 2}{	
		\node[above] at (0) {{$\scriptstyle{0}$}};	
		\node[right] at (1) {{$\scriptstyle{1}$}};
		\node[below] at (2) {{$\scriptstyle{2}$}};
		\node[left] at (3) {{$\scriptstyle{3}$}};	
		\node[left, yshift=1.5] at (4) {{$\scriptstyle{4'}$}};}{}
	\end{tikzpicture}
}

\newcommand{\HamTWO}[2]{
	\begin{tikzpicture}[scale=#1,baseline=-0.5em]
	\coordinate (0) at (0,1);
	\coordinate (1) at (1,-0.4);
	\coordinate (2) at (0.55,-1);
	\coordinate (3) at (-1.2,-0.55);
	\coordinate (4) at (0,-0.1);
	\coordinate (5) at (-1.5,1.1);	
	
	\draw[] (0) -- (1);
	\draw[] (1) -- (2);
	\draw[] (2) -- (3);
	\draw[shorten >= 0.3em] (0) -- (intersection of 0--3 and 2--5);
	\draw[shorten <= 0.3em] (intersection of 0--3 and 2--5) -- (3);
	\draw[] (0) -- (2);
	\draw[shorten >= 0.3em] (3) -- (intersection of 3--4 and 2--5);
	\draw[shorten <= 0.3em] (intersection of 3--4 and 2--5) -- (4);
	\draw[] (0) -- (4);
	\draw[] (2) -- (4);
	\draw[] (0) -- (5);
	\draw[] (3) -- (5);
	\draw[shorten >= 0.3em] (1) -- (intersection of 1--4 and 0--2);
	\draw[shorten <= 0.3em] (intersection of 1--4 and 0--2) -- (4);
	\draw[shorten >= 0.3em] (1) -- (intersection of 1--3 and 0--2);
	\draw[shorten <= 0.3em, shorten >=0.3em] (intersection of 1--3 and 0--2) -- (intersection of 1--3 and 2--4);
	\draw[shorten <= 0.3em, shorten >= 0.3em] (intersection of 1--3 and 2--4) -- (intersection of 1--3 and 2--5);
	\draw[shorten <= 0.3em] (intersection of 1--3 and 2--5) -- (3);
	\draw[shorten >= 0.3em] (4) -- (intersection of 0--3 and 4--5);
	\draw[shorten <= 0.3em] (intersection of 0--3 and 4--5) -- (5);
	\draw[] (2) -- (5);
	\draw[shorten >= 0.3em] (1) -- (intersection of 1--5 and 0--2);
	\draw[shorten >= 0.3em, shorten <= 0.3em] (intersection of 1--5 and 0--2) -- (intersection of 1--5 and 0--4);
	\draw[shorten >= 0.3em, shorten <= 0.3em] (intersection of 1--5 and 0--4) -- (intersection of 1--5 and 0--3);
	\draw[shorten <= 0.3em] (intersection of 1--5 and 0--3) -- (5);

	\ifthenelse{#2 = 1}{	
		\node[above] at (0) {{$\scriptstyle{0}$}};	
		\node[right] at (1) {{$\scriptstyle{1}$}};
		\node[below] at (2) {{$\scriptstyle{2}$}};
		\node[left] at (3) {{$\scriptstyle{3}$}};
		\node[above] at (5) {{$\scriptstyle{4'}$}};		
		\node[left, yshift=1.2, xshift=-0.05em] at (4) {{$\scriptstyle{4}$}};}{}
	\end{tikzpicture}
}

\newcommand{\PFourOne}[6]{
	\begin{tikzpicture}[scale=#1, baseline=#3 em, line width=#5 pt]
	\coordinate (0) at (-0.3,1);
	\coordinate (1) at (1,-0.4);
	\coordinate (2) at (0.2,-1);
	\coordinate (3) at (-1.2,-0.55);
	\coordinate (4) at (-0.3,-0.1);	
	
	\draw[] (0) -- (1);
	\draw[] (1) -- (2);
	\draw[] (2) -- (3);
	\draw[] (0) -- (3);
	\draw[] (0) -- (2);
	
	\ifthenelse{#2 = 1}{
		\ifthenelse{#6 = 1}{	
			\node[left, yshift=0.15em] at (4) {{$\scriptstyle{4}$}};}{}
		\draw[] (3) -- (4);
		\draw[] (0) -- (4);
		\draw[] (2) -- (4);
		\draw[shorten >= #4 em] (1) -- (intersection of 1--4 and 0--2);
		\draw[shorten <= #4 em] (intersection of 1--4 and 0--2) -- (4);
		\draw[shorten >= #4 em] (1) -- (intersection of 1--3 and 0--2);
		\draw[shorten <= #4 em, shorten >=#4 em] (intersection of 1--3 and 0--2) -- (intersection of 1--3 and 2--4);
		\draw[shorten <= #4 em] (intersection of 1--3 and 2--4) -- (3);
	}{}
	\ifthenelse{#2 = 2}{
		\draw[shorten >= #4 em] (1) -- (intersection of 1--3 and 0--2);
		\draw[shorten <= #4 em] (intersection of 1--3 and 0--2) -- (3); 
	}{}
	\ifthenelse{#6 = 1}{
		\node[above] at (0) {{$\scriptstyle{0}$}};	
		\node[right] at (1) {{$\scriptstyle{1}$}};
		\node[below] at (2) {{$\scriptstyle{2}$}};
		\node[left] at (3) {{$\scriptstyle{3}$}};
	}{}
	\end{tikzpicture}
}

\newcommand{\PFourOneneg}[6]{
	\begin{tikzpicture}[scale=#1, baseline=#3 em, line width=#5 pt]
	\begin{scope}[xscale=1,yscale=1]
	\coordinate (0) at (-0.3,1);
	\coordinate (1) at (1,-0.4);
	\coordinate (2) at (0.2,-1);
	\coordinate (3) at (-1.2,-0.55);
	\coordinate (4) at (-0.3,-0.1);	
	
	\draw[] (0) -- (1);
	\draw[] (1) -- (2);
	\draw[] (2) -- (3);
	\draw[] (0) -- (3);
	\draw[] (0) -- (2);
	
	\ifthenelse{#2 = 1}{
		\ifthenelse{#6 = 1}{	
			\node[left, yshift=0.15em] at (4) {{$\scriptstyle{}$}};}{}
		\draw[] (3) -- (4);
		\draw[] (0) -- (4);
		\draw[] (2) -- (4);
		\draw[shorten >= #4 em] (1) -- (intersection of 1--4 and 0--2);
		\draw[shorten <= #4 em] (intersection of 1--4 and 0--2) -- (4);
		\draw[shorten >= #4 em] (1) -- (intersection of 1--3 and 0--2);
		\draw[shorten <= #4 em, shorten >=#4 em] (intersection of 1--3 and 0--2) -- (intersection of 1--3 and 2--4);
		\draw[shorten <= #4 em] (intersection of 1--3 and 2--4) -- (3);
	}{}
	\ifthenelse{#2 = 2}{
		\draw[shorten >= #4 em] (1) -- (intersection of 1--3 and 0--2);
		\draw[shorten <= #4 em] (intersection of 1--3 and 0--2) -- (3); 
	}{}
	\ifthenelse{#6 = 1}{
		\node[above] at (0) {{$\scriptstyle{}$}};	
		\node[right] at (1) {{$\scriptstyle{}$}};
		\node[below] at (2) {{$\scriptstyle{}$}};
		\node[left] at (3) {{$\scriptstyle{}$}};
	}{}
	\end{scope}
	\end{tikzpicture}
}

\newcommand{\PTwoThree}[6]{
	\begin{tikzpicture}[scale=#1, baseline=#3 em, line width=#5 pt]
	\coordinate (0) at (-0.3,1);
	\coordinate (1) at (1,-0.4);
	\coordinate (2) at (0.2,-0.9);
	\coordinate (3) at (-1.2,-0.55);
	\coordinate (4) at (-0.2,-2);	
	
	\draw[] (0) -- (1);
	\draw[] (1) -- (2);
	\draw[] (2) -- (3);
	\draw[] (0) -- (3);
	\draw[] (0) -- (2);
	
	\ifthenelse{#2 = 1}{
		\ifthenelse{#6 = 1}{	
			\node[below] at (4) {{$\scriptstyle{4}$}};}{}
		\draw[] (3) -- (4);
		\draw[shorten >= #4 em] (0) -- (intersection of 0--4 and 2--3);
		\draw[shorten <= #4 em] (intersection of 0--4 and 2--3) -- (4);
		\draw[] (2) -- (4);
		\draw[] (1) -- (4);
		\draw[shorten >= #4 em] (1) -- (intersection of 1--3 and 0--2);
		\draw[shorten <= #4 em, shorten >=#4 em] (intersection of 1--3 and 0--2) -- (intersection of 1--3 and 0--4);
		\draw[shorten <= #4 em] (intersection of 1--3 and 0--4) -- (3);
	}{}
	\ifthenelse{#2 = 2}{
		\ifthenelse{#6 = 1}{	
	\node[below] at (4) {{$\scriptstyle{4}$}};}{}
	\draw[] (3) -- (4);
	\draw[] (2) -- (4);
	\draw[] (1) -- (4);
	\draw[shorten >= #4 em] (1) -- (intersection of 1--3 and 0--2);
	\draw[shorten <= #4 em] (intersection of 1--3 and 0--2) -- (3);
		}{}
	\ifthenelse{#6 = 1}{
		\node[above] at (0) {{$\scriptstyle{0}$}};	
		\node[right] at (1) {{$\scriptstyle{1}$}};
		\node[left, yshift=-0.25em] at (2) {{$\scriptstyle{2}$}};
		\node[left] at (3) {{$\scriptstyle{3}$}};
	}{}
	\end{tikzpicture}
}

\newcommand{\PTwoThreeneg}[6]{
	\begin{tikzpicture}[scale=#1, baseline=#3 em, line width=#5 pt]
	\coordinate (0) at (-0.3,1);
	\coordinate (1) at (1,-0.4);
	\coordinate (2) at (0.2,-0.9);
	\coordinate (3) at (-1.2,-0.55);
	\coordinate (4) at (-0.2,-2);	
	
	\draw[] (0) -- (1);
	\draw[] (1) -- (2);
	\draw[] (2) -- (3);
	\draw[] (0) -- (3);
	\draw[] (0) -- (2);
	
	\ifthenelse{#2 = 1}{
		\ifthenelse{#6 = 1}{	
			\node[below] at (4) {{$\scriptstyle{}$}};}{}
		\draw[] (3) -- (4);
		\draw[shorten >= #4 em] (0) -- (intersection of 0--4 and 2--3);
		\draw[shorten <= #4 em] (intersection of 0--4 and 2--3) -- (4);
		\draw[] (2) -- (4);
		\draw[] (1) -- (4);
		\draw[shorten >= #4 em] (1) -- (intersection of 1--3 and 0--2);
		\draw[shorten <= #4 em, shorten >=#4 em] (intersection of 1--3 and 0--2) -- (intersection of 1--3 and 0--4);
		\draw[shorten <= #4 em] (intersection of 1--3 and 0--4) -- (3);
	}{}
	\ifthenelse{#2 = 2}{
		\ifthenelse{#6 = 1}{	
			\node[below] at (4) {{$\scriptstyle{}$}};}{}
		\draw[] (3) -- (4);
		\draw[] (2) -- (4);
		\draw[] (1) -- (4);
		\draw[shorten >= #4 em] (1) -- (intersection of 1--3 and 0--2);
		\draw[shorten <= #4 em] (intersection of 1--3 and 0--2) -- (3);
	}{}
	\ifthenelse{#6 = 1}{
		\node[above] at (0) {{$\scriptstyle{}$}};	
		\node[right] at (1) {{$\scriptstyle{}$}};
		\node[left, yshift=-0.25em] at (2) {{$\scriptstyle{}$}};
		\node[left] at (3) {{$\scriptstyle{}$}};
	}{}
	\end{tikzpicture}
}

\newcommand{\preSlantProd}[2]{
	\begin{tikzpicture}[scale=#1, baseline=0.6em, line width=0.8 pt]
	\coordinate (0) at (0,0);
	\coordinate (1) at (2,0);
	\coordinate (2) at (1.2,0.5);
	\coordinate (3) at (-0.8,0.5);
	
	\ifthenelse{#2 = 1}{
		\draw[] (0) -- (1);
		\draw[] (1) -- (2);
		\draw[] (2) -- (3);
		\draw[] (0) -- (3);
		\draw[] (0) -- (2);
	}{}	

	\ifthenelse{#2 = 2}{
		\draw[] (0) -- (1);
		\draw[] (1) -- (2);
		\draw[] (2) -- (3);
		\draw[] (0) -- (3);
		\draw[] (1) -- (3);
	}{}	

	\node[below] at (0) {{$\scriptstyle{0}$}};	
	\node[below] at (1) {{$\scriptstyle{3}$}};
	\node[above] at (2) {{$\scriptstyle{2}$}};
	\node[above] at (3) {{$\scriptstyle{1}$}};
	\end{tikzpicture}
}

\newcommand{\slantProdA}[2]{
	\begin{tikzpicture}[scale=#1, baseline=2.5em, line width=0.8 pt]
	\coordinate (0) at (0,0);
	\coordinate (1) at (2,0);
	\coordinate (2) at (1.2,0.5);
	\coordinate (3) at (-0.8,0.5);
	\coordinate (4) at (0,2);
	\coordinate (5) at (2,2);
	\coordinate (6) at (1.2,2.5);
	\coordinate (7) at (-0.8,2.5);
	
	\ifthenelse{#2 = 1}{
		\draw[] (0) -- (1);
		\draw[] (1) -- (2);
		\draw[] (0) -- (3);
		\draw[] (4) -- (5);
		\draw[] (5) -- (6);
		\draw[] (6) -- (7);
		\draw[] (4) -- (7);
		\draw[] (0) -- (4);
		\draw[] (1) -- (5);
		\draw[] (3) -- (7);
		\draw[] (0) -- (7);
		\draw[] (0) -- (5);
		\draw[] (0) -- (2);
		\draw[] (4) -- (6);
		\draw[] (2) -- (5);
		\interL{0}{6}{4}{5}{1};
		\interLL{2}{6}{0}{5}{4}{5}{1};
		\interLLL{3}{6}{0}{7}{0}{4}{4}{5}{1};
		\draw[sha] (3) -- (intersection of 3--2 and 0--7);
		\draw[shab] (intersection of 2--3 and 0--4) -- (intersection of 2--3 and 0--6);
		\draw[shab] (intersection of 2--3 and 0--6) -- (intersection of 2--3 and 0--5);		
		\draw[shb] (intersection of 2--3 and 0--5) -- (2);
	}{}
	\ifthenelse{#2 = 2}{
		\draw[] (0) -- (1);
		\draw[] (1) -- (2);
		\draw[] (0) -- (3);
		\draw[] (4) -- (5);
		\draw[] (5) -- (6);
		\draw[] (6) -- (7);
		\draw[] (4) -- (7);
		\draw[] (0) -- (4);
		\draw[] (1) -- (5);
		\draw[] (3) -- (7);
		\draw[] (0) -- (7);
		\draw[] (0) -- (5);
		\draw[] (0) -- (2);
		\draw[] (2) -- (5);
		\draw[] (4) -- (6);
		\interL{5}{7}{4}{6}{1};
		\draw[sha] (2) -- (intersection of 2--6 and 0--5);
		\draw[shab] (intersection of 2--6 and 0--5) -- (intersection of 2--6 and 4--5);
		\draw[shb] (intersection of 2--6 and 5--7) -- (6);
		\draw[sha] (0) -- (intersection of 0--6 and 4--5);
		\draw[shb] (intersection of 0--6 and 5--7) -- (6);
		\draw[sha] (3) -- (intersection of 3--2 and 0--7);
		\draw[shab] (intersection of 2--3 and 0--4) -- (intersection of 2--3 and 0--6);
		\draw[shab] (intersection of 2--3 and 0--6) -- (intersection of 2--3 and 0--5);		
		\draw[shb] (intersection of 2--3 and 0--5) -- (2);
		\draw[sha] (3) -- (intersection of 3--6 and 0--7);
		\draw[shab] (intersection of 3--6 and 0--7) -- (intersection of 3--6 and 0--4);
		\draw[shab] (intersection of 3--6 and 0--4) -- (intersection of 3--6 and 4--5);
		\draw[shab] (intersection of 3--6 and 4--5) -- (intersection of 3--6 and 5--7);
		\draw[shb] (intersection of 3--6 and 5--7) -- (6);
	}{}
	\ifthenelse{#2 = 3}{
		\draw[] (0) -- (1);
		\draw[] (1) -- (2);
		\draw[] (0) -- (3);
		\draw[] (4) -- (5);
		\draw[] (5) -- (6);
		\draw[] (6) -- (7);
		\draw[] (4) -- (7);
		\draw[] (0) -- (4);
		\draw[] (1) -- (5);
		\draw[] (3) -- (7);
		\draw[] (0) -- (7);
		\draw[] (0) -- (5);
		\draw[] (0) -- (2);
		\draw[] (4) -- (6);
		\interL{5}{7}{4}{6}{1};
		\draw[] (2) -- (5);
		\draw[sha] (2) -- (intersection of 2--6 and 0--5);
		\draw[shab] (intersection of 2--6 and 0--5) -- (intersection of 2--6 and 3--5);
		\draw[shab] (intersection of 2--6 and 3--5) -- (intersection of 2--6 and 4--5);
		\draw[shb] (intersection of 2--6 and 5--7) -- (6);
		
		\interLL{3}{5}{0}{7}{0}{4}{1};
		
		\draw[sha] (3) -- (intersection of 3--2 and 0--7);
		\draw[shab] (intersection of 2--3 and 0--4) -- (intersection of 2--3 and 0--5);		
		\draw[shb] (intersection of 2--3 and 0--5) -- (2);
		
		\draw[sha] (3) -- (intersection of 3--6 and 0--7);
		\draw[shab] (intersection of 3--6 and 0--7) -- (intersection of 3--6 and 0--4);
		\draw[shab] (intersection of 3--6 and 0--4) -- (intersection of 3--6 and 4--5);
		\draw[shab] (intersection of 3--6 and 4--5) -- (intersection of 3--6 and 5--7);
		\draw[shb] (intersection of 3--6 and 5--7) -- (6);
	}{}
	\ifthenelse{#2 = 4}{
		\draw[] (0) -- (1);
		\draw[] (1) -- (2);
		\draw[] (0) -- (3);
		\draw[] (4) -- (5);
		\draw[] (5) -- (6);
		\draw[] (6) -- (7);
		\draw[] (4) -- (7);
		\draw[] (0) -- (4);
		\draw[] (1) -- (5);
		\draw[] (3) -- (7);
		\draw[] (0) -- (7);
		\draw[] (0) -- (5);
		\draw[] (4) -- (6);
		\interL{5}{7}{4}{6}{1};
		\draw[] (2) -- (5);
		\interL{0}{2}{1}{3}{1};
		\draw[sha] (2) -- (intersection of 2--6 and 0--5);
		\draw[shab] (intersection of 2--6 and 0--5) -- (intersection of 2--6 and 3--5);
		\draw[shab] (intersection of 2--6 and 3--5) -- (intersection of 2--6 and 4--5);
		\draw[shb] (intersection of 2--6 and 5--7) -- (6);
		
		\interLL{3}{5}{0}{7}{0}{4}{1};
		
		\draw[sha] (3) -- (intersection of 3--2 and 0--7);
		\draw[shab] (intersection of 2--3 and 0--4) -- (intersection of 2--3 and 0--5);		
		\draw[shb] (intersection of 2--3 and 0--5) -- (2);
		
		\draw[sha] (3) -- (intersection of 3--6 and 0--7);
		\draw[shab] (intersection of 3--6 and 0--7) -- (intersection of 3--6 and 0--4);
		\draw[shab] (intersection of 3--6 and 0--4) -- (intersection of 3--6 and 4--5);
		\draw[shab] (intersection of 3--6 and 4--5) -- (intersection of 3--6 and 5--7);
		\draw[shb] (intersection of 3--6 and 5--7) -- (6);
		
		\draw[sha] (3) -- (intersection of 3--1 and 0--7);
		\draw[shab] (intersection of 3--1 and 0--4) -- (intersection of 3--1 and 0--5);
		\draw[shb] (intersection of 3--1 and 0--5) -- (1);
		
	}{}
	\ifthenelse{#2 = 5}{
		\draw[] (0) -- (1);
		\draw[] (1) -- (2);
		\draw[] (0) -- (3);
		\draw[] (4) -- (5);
		\draw[] (5) -- (6);
		\draw[] (6) -- (7);
		\draw[] (4) -- (7);
		\draw[] (0) -- (4);
		\draw[] (1) -- (5);
		\draw[] (3) -- (7);
		\draw[] (0) -- (7);
		\draw[] (0) -- (5);
		\draw[] (5) -- (7);
		\draw[] (2) -- (5);
		\draw[sha] (2) -- (intersection of 2--6 and 0--5);
		\draw[shab] (intersection of 2--6 and 0--5) -- (intersection of 2--6 and 3--5);
		\draw[shab] (intersection of 2--6 and 3--5) -- (intersection of 2--6 and 4--5);
		\draw[shb] (intersection of 2--6 and 5--7) -- (6);
		
		\interLL{3}{5}{0}{7}{0}{4}{1};
		
		\draw[sha] (3) -- (intersection of 3--2 and 0--7);
		\draw[shab] (intersection of 2--3 and 0--4) -- (intersection of 2--3 and 0--5);		
		\draw[shb] (intersection of 2--3 and 0--5) -- (2);
		
		\draw[sha] (3) -- (intersection of 3--6 and 0--7);
		\draw[shab] (intersection of 3--6 and 0--7) -- (intersection of 3--6 and 0--4);
		\draw[shab] (intersection of 3--6 and 0--4) -- (intersection of 3--6 and 4--5);
		\draw[shab] (intersection of 3--6 and 4--5) -- (intersection of 3--6 and 5--7);
		\draw[shb] (intersection of 3--6 and 5--7) -- (6);
		
		\draw[sha] (3) -- (intersection of 3--1 and 0--7);
		\draw[shab] (intersection of 3--1 and 0--4) -- (intersection of 3--1 and 0--5);
		\draw[shb] (intersection of 3--1 and 0--5) -- (1);
	}{}

	\node[below] at (0) {{$\scriptstyle{0}$}};	
	\node[below] at (1) {{$\scriptstyle{3}$}};
	\node[below, yshift=0.15em] at (2) {{$\scriptstyle{2}$}};
	\node[left] at (3) {{$\scriptstyle{1}$}};
	\node[yshift=0.45em, xshift=0.1em] at (4) {{$\scriptstyle{0'}$}};	
	\node[above] at (5) {{$\scriptstyle{3'}$}};
	\node[above] at (6) {{$\scriptstyle{2'}$}};
	\node[left] at (7) {{$\scriptstyle{1'}$}};
	
	\end{tikzpicture}
}

\newcommand{\slantProdB}[2]{
	\begin{tikzpicture}[scale=#1, baseline=2.5em, line width=0.8 pt]
	\coordinate (0) at (0,0);
	\coordinate (1) at (2,0);
	\coordinate (2) at (1.2,0.5);
	\coordinate (3) at (-0.8,0.5);
	\coordinate (4) at (0,2);
	\coordinate (5) at (2,2);
	\coordinate (6) at (1.2,2.5);
	\coordinate (7) at (-0.8,2.5);
	
	\ifthenelse{#2 = 1}{
		\draw[] (0) -- (1);
		\draw[] (1) -- (2);
		\draw[sha] (2) -- (intersection of 2--3 and 0--6);
		\draw[shab] (intersection of 2--3 and 0--6) -- (intersection of 2--3 and 0--4);
		\draw[shb] (intersection of 2--3 and 0--7) -- (3);
		\draw[] (0) -- (3);
		\draw[] (4) -- (5);
		\draw[] (5) -- (6);
		\draw[] (6) -- (7);
		\draw[] (4) -- (7);
		\draw[] (0) -- (4);
		\draw[] (1) -- (5);
		\interLL{2}{6}{1}{4}{4}{5}{1};	
		\draw[] (3) -- (7);
		\draw[] (0) -- (2);
		\draw[] (4) -- (6);	
		\draw[] (1) -- (4);	
		\draw[] (0) -- (7);
		\interL{1}{6}{4}{5}{1};
		\interLL{2}{7}{0}{6}{0}{4}{1};									
		\interLL{0}{6}{1}{4}{4}{5}{1};
		\node[below] at (2) {{$\scriptstyle{1}$}};
		\node[above] at (4) {{$\scriptstyle{2'}$}};
	}{}
	\ifthenelse{#2 = 2}{
		\draw[] (0) -- (1);
		\draw[] (1) -- (2);
		\draw[sha] (2) -- (intersection of 2--3 and 0--4);
		\draw[shb] (intersection of 2--3 and 0--7) -- (3);
		\draw[] (0) -- (3);
		\draw[] (4) -- (5);
		\draw[] (5) -- (6);
		\draw[] (6) -- (7);
		\draw[] (4) -- (7);
		\draw[] (0) -- (4);
		\draw[] (1) -- (5);
		\draw[sha] (2) -- (intersection of 2--6 and 1--7);
		\draw[shab] (intersection of 2--6 and 1--4) -- (intersection of 2--6 and 4--5);
		\draw[shb] (intersection of 2--6 and 4--5) -- (6);
		\draw[] (3) -- (7);
		\draw[] (0) -- (2);
		\draw[] (4) -- (6);	
		\draw[] (1) -- (4);	
		\draw[] (0) -- (7);
		\interL{1}{6}{4}{5}{1};
		\interL{2}{7}{0}{4}{1};									
		\interL{1}{7}{0}{4}{1};
		\node[below] at (2) {{$\scriptstyle{1}$}};
		\node[above] at (4) {{$\scriptstyle{2'}$}};
	}{}
	\ifthenelse{#2 = 3}{
		\draw[] (0) -- (1);
		\draw[] (1) -- (2);
		\draw[sha] (2) -- (intersection of 2--3 and 0--4);
		\draw[shb] (intersection of 2--3 and 0--7) -- (3);
		\draw[] (0) -- (3);
		\draw[] (4) -- (5);
		\draw[] (5) -- (6);
		\draw[] (6) -- (7);
		\draw[] (4) -- (7);
		\draw[] (0) -- (4);
		\draw[] (1) -- (5);
		\draw[sha] (2) -- (intersection of 2--6 and 1--7);
		\draw[shab] (intersection of 2--6 and 1--4) -- (intersection of 2--6 and 4--5);
		\draw[shb] (intersection of 2--6 and 5--7) -- (6);
		\draw[] (3) -- (7);
		\draw[sha] (1) -- (intersection of 1--3 and 0--4);
		\draw[shb] (intersection of 1--3 and 0--7) -- (3);	
		\draw[] (5) -- (7);	
		\draw[] (1) -- (4);	
		\draw[] (0) -- (7);
		\draw[sha] (1) -- (intersection of 1--6 and 4--5);
		\draw[shb] (intersection of 1--6 and 5--7) -- (6);
		\interL{2}{7}{0}{4}{1};									
		\interL{1}{7}{0}{4}{1};
		\node[below, yshift=0.15em] at (2) {{$\scriptstyle{1}$}};
		\node[above, yshift=-0.27em, xshift=0.18em] at (4) {{$\scriptstyle{2'}$}};
	}{}

	\node[below] at (0) {{$\scriptstyle{2}$}};	
	\node[below] at (1) {{$\scriptstyle{3}$}};
	\node[left] at (3) {{$\scriptstyle{0}$}};	
	\node[above] at (5) {{$\scriptstyle{3'}$}};
	\node[above] at (6) {{$\scriptstyle{1'}$}};
	\node[left] at (7) {{$\scriptstyle{0'}$}};
	
	\end{tikzpicture}
}

\newcommand{\OneDCyl}[4]{
	\begin{tikzpicture}[scale=#1, baseline=-0.3em]
	\coordinate (1) at (0,0);
	\coordinate (2) at (2,0);
	
	\draw[] (1) --node[above]{${\sss #4}$} (2);
	\node[left] at (1) {{${\sss #2}$}};
	\node[right] at (2) {{${\sss #3}$}};
	\node[] at (1) {{$\bul$}};
	\node[] at (2) {{$\bul$}};
	\end{tikzpicture}
}

\newcommand{\OneDCylDouble}[6]{
	\begin{tikzpicture}[scale=#1, baseline=-0.3em]
	\coordinate (1) at (0,0);
	\coordinate (2) at (2,0);
	\coordinate (3) at (4,0);
	
	\draw[] (1) --node[above]{${\sss #5}$} (2);
	\draw[] (2) --node[above]{${\sss #6}$} (3);
	\node[left] at (1) {{${\sss #2}$}};
	\node[below] at (2) {{${\sss #3}$}};
	\node[right] at (3) {{${\sss #4}$}};
	\node[] at (1) {{$\bul$}};
	\node[] at (2) {{$\bul$}};
	\node[] at (3) {{$\bul$}};
	\end{tikzpicture}
}

\newcommand{\OneDCylZ}[8]{
	\begin{tikzpicture}[scale=#1, baseline=1em]
	\coordinate (1) at (0,0);
	\coordinate (2) at (2,0);
	\coordinate (3) at (4,0);
	\coordinate (4) at (2,2);
	
	\draw[] (1) --node[above]{${\sss #6}$} (2);
	\draw[] (2) --node[above]{${\sss #7}$} (3);
	\draw[] (1) -- (4);
	\draw[] (2) --node[right]{${\sss #8}$} (4);
	\draw[] (3) -- (4);
	\node[left] at (1) {{${\sss #2}$}};
	\node[above] at (4) {{${\sss #5}$}};
	\node[below] at (2) {{${\sss #3}$}};
	\node[right] at (3) {{${\sss #4}$}};
	\end{tikzpicture}
}

\newcommand{\OneDCylPinched}[6]{
	\begin{tikzpicture}[scale=#1, baseline=-2em]
	\coordinate (1) at (0,0);
	\coordinate (2) at (2,0);
	\coordinate (3) at (1,-1.5);
	
	\draw[] (1) --node[left]{${\sss #5}$} (3);
	\draw[] (2) --node[right]{${\sss #6}$} (3);
	\draw[] (1) -- (2);
	\node[left] at (1) {{${\sss #2}$}};
	\node[right] at (2) {{${\sss #3}$}};
	\node[below] at (3) {{${\sss #4}$}};
	\end{tikzpicture}
}

\newcommand{\GrCyl}[5]{
	\begin{tikzpicture}[scale=#1, baseline=-0.3em]
	\coordinate (1) at (0,0);
	\coordinate (2) at (2,0);
	
	\draw[] (1) --node[above]{${\sss #5}$} (2);
	\node[below] at (1) {{${\sss #2}$}};
	\node[below] at (2) {{${\sss #3}$}};
	\node[] at (1) {{$\bul$}};
	\node[above] at (1) {{${\sss #4}$}};
	\node[above] at (2) {{${\sss #4 ^{#5}}$}};
	\node[] at (2) {{$\bul$}};
	\end{tikzpicture}
}

\newcommand{\GrCylT}[6]{
	\begin{tikzpicture}[scale=#1, baseline=-0.3em]
	\coordinate (1) at (0,0);
	\coordinate (2) at (2,0);
	
	\draw[] (1) --node[above]{${\sss #6}$} (2);
	\node[below] at (1) {{${\sss #2}$}};
	\node[below] at (2) {{${\sss #3}$}};
	\node[] at (1) {{$\bul$}};
	\node[above] at (1) {{${\sss #4, #5}$}};
	\node[above] at (2) {{${\sss #4 ^{#6}, #5^{#6}}$}};
	\node[] at (2) {{$\bul$}};
	\end{tikzpicture}
}

\newcommand{\GrCylDouble}[7]{
	\begin{tikzpicture}[scale=#1, baseline=-0.3em]
	\coordinate (1) at (0,0);
	\coordinate (2) at (2,0);
	\coordinate (3) at (4,0);
	
	\draw[] (1) --node[above]{${\sss #6}$} (2);
	\draw[] (2) --node[above]{${\sss #7}$} (3);
	\node[below] at (1) {{${\sss #2}$}};
	\node[below] at (2) {{${\sss #3}$}};
	\node[below] at (3) {{${\sss #4}$}};
	\node[] at (1) {{$\bul$}};
	\node[] at (2) {{$\bul$}};
	\node[] at (3) {{$\bul$}};
	\node[above] at (1) {{${\sss #5}$}};
	\node[above] at (2) {{${\sss #5 ^{#6}}$}};
	\node[above] at (3) {{${\sss #5 ^{#6 #7}}$}};
	\end{tikzpicture}
}

\newcommand{\GrCylDoubleT}[8]{
	\begin{tikzpicture}[scale=#1, baseline=-0.3em]
	\coordinate (1) at (0,0);
	\coordinate (2) at (2,0);
	\coordinate (3) at (4,0);
	
	\draw[] (1) --node[above]{${\sss #7}$} (2);
	\draw[] (2) --node[above]{${\sss #8}$} (3);
	\node[below] at (1) {{${\sss #2}$}};
	\node[below] at (2) {{${\sss #3}$}};
	\node[below] at (3) {{${\sss #4}$}};
	\node[] at (1) {{$\bul$}};
	\node[] at (2) {{$\bul$}};
	\node[] at (3) {{$\bul$}};
	\node[above] at (1) {{${\sss #5, #6}$}};
	\node[above] at (2) {{${\sss #5 ^{#7}, #6 ^{#7}}$}};
	\node[above] at (3) {{${\sss #5 ^{#7 #8}, #6 ^{#7 #8}}$}};
	\end{tikzpicture}
}

\newcommand{\GrCylDoubleSpe}[8]{
	\begin{tikzpicture}[scale=#1, baseline=-0.3em]
	\coordinate (1) at (0,0);
	\coordinate (2) at (2,0);
	\coordinate (3) at (4,0);
	
	\draw[] (1) --node[above]{${\sss #6}$} (2);
	\draw[] (2) --node[above]{${\sss #7}$} (3);
	\node[below] at (1) {{${\sss #2}$}};
	\node[below] at (2) {{${\sss #3}$}};
	\node[below] at (3) {{${\sss #4}$}};
	\node[] at (1) {{$\bul$}};
	\node[] at (2) {{$\bul$}};
	\node[] at (3) {{$\bul$}};
	\node[above] at (1) {{${\sss #5}$}};
	\node[above] at (2) {{${\sss #5 ^{#6}}$}};
	\node[above] at (3) {{${\sss #8}$}};
	\end{tikzpicture}
}

\newcommand{\GrCylDoubleSpeT}[9]{
	\begin{tikzpicture}[scale=#1, baseline=-0.3em]
	\coordinate (1) at (0,0);
	\coordinate (2) at (2,0);
	\coordinate (3) at (4,0);
	
	\draw[] (1) --node[above]{${\sss #7 \;\;\;}$} (2);
	\draw[] (2) --node[above]{${\sss \;\;\; #8}$} (3);
	\node[below] at (1) {{${\sss #2}$}};
	\node[below] at (2) {{${\sss #3}$}};
	\node[below] at (3) {{${\sss #4}$}};
	\node[] at (1) {{$\bul$}};
	\node[] at (2) {{$\bul$}};
	\node[] at (3) {{$\bul$}};
	\node[above] at (1) {{${\sss #5,#6 \;\;\;\;\;\;}$}};
	\node[above] at (2) {{${\sss #5 ^{#7},#6 ^{#7}}$}};
	\node[above] at (3) {{${\sss \;\;\;\;\;\;\; #9}$}};
	\end{tikzpicture}
}

\newcommand{\GrCylZ}[8]{
	\begin{tikzpicture}[scale=#1, baseline=1em]
	\coordinate (1) at (0,0);
	\coordinate (2) at (2,0);
	\coordinate (3) at (4,0);
	\coordinate (4) at (2,2);
	
	\draw[] (1) --node[above]{${\sss #6}$} (2);
	\draw[] (2) --node[above]{${\sss #7}$} (3);
	\draw[] (1) -- (4);
	\draw[] (2) -- (4);
	\draw[] (3) -- (4);
	\node[below] at (1) {{${\sss #2}$}};
	\node[above, xshift=-0.2em] at (1) {{${\sss #8}$}};
	\node[above] at (4) {{${\sss #5}$}};
	\node[below] at (2) {{${\sss #3}$}};
	\node[below] at (3) {{${\sss #4}$}};
	\node[above, xshift=0.2em] at (3) {{${\sss #8 ^{#6 #7}}$}};
	\end{tikzpicture}
}

\newcommand{\GrCylZT}[9]{
	\begin{tikzpicture}[scale=#1, baseline=1em]
	\coordinate (1) at (0,0);
	\coordinate (2) at (2,0);
	\coordinate (3) at (4,0);
	\coordinate (4) at (2,2);
	
	\draw[] (1) --node[above]{${\sss #6}$} (2);
	\draw[] (2) --node[above]{${\sss #7}$} (3);
	\draw[] (1) -- (4);
	\draw[] (2) -- (4);
	\draw[] (3) -- (4);
	\node[below] at (1) {{${\sss #2}$}};
	\node[above, xshift=-0.2em] at (1) {{${\sss #8, #9}\;\;\,$}};
	\node[above] at (4) {{${\sss #5}$}};
	\node[below] at (2) {{${\sss #3}$}};
	\node[below] at (3) {{${\sss #4}$}};
	\node[above, xshift=0.2em] at (3) {{$\;\;\;\;\,{\sss #8 ^{#6 #7},  #9 ^{#6 #7}}$}};
	\end{tikzpicture}
}

\newcommand{\GrCylPinched}[6]{
	\begin{tikzpicture}[scale=#1, baseline=-2em]
	\coordinate (1) at (0,0);
	\coordinate (2) at (2,0);
	\coordinate (3) at (1,-1.5);
	
	\draw[] (1) --node[left]{${\sss #5}$} (3);
	\draw[] (2) --node[right]{${\sss #6}$} (3);
	\draw[] (1) -- (2);
	\node[left] at (1) {{${\sss #2}$}};
	\node[right] at (2) {{${\sss #3}$}};
	\node[below] at (3) {{${\sss #4}$}};
	\end{tikzpicture}
}

\newcommand{\TwoPachner}[2]{
	\begin{tikzpicture}[scale=#1, baseline=1.8em]
	\coordinate (1) at (0,0);
	\coordinate (2) at (2,0);
	\coordinate (3) at (0,2);
	\coordinate (4) at (2,2);
	
	\draw[] (1) -- (2);
	\draw[] (1) -- (3);
	\draw[] (2) -- (4);
	\draw[] (3) --(4);
	\ifthenelse{#2 = 1}{
		\draw[] (1) -- (4);
	}{
		\draw[] (2) -- (3);
	}		
	\end{tikzpicture}
}

\newcommand{\TwoDCyl}[5]{
	\begin{tikzpicture}[scale=#1, baseline=-2.5em,yscale=-1]
	\coordinate (1) at (0,0);
	\coordinate (2) at (2,0);
	\coordinate (3) at (0,2);
	\coordinate (4) at (2,2);
	
	\draw[] (1) --node[above]{${\sss #4}$} (2);
	\draw[] (1) --node[left]{${\sss #5}$} (3);
	\ifthenelse{\equal{#4}{}}{
		\draw[] (2) -- (4);
	}
	{
		\draw[] (2) --node[right]{${\sss #4^{-1}#5 #4}$} (4);
	}
	\draw[] (3) --node[below]{${\sss #4}$} (4);
	\draw[] (1) --node[above, right]{${\sss #5 #4}$} (4);		
	\node[above] at (1) {{${\sss #2}$}};
	\node[above] at (2) {{${\sss #3}$}};	
	\node[below] at (3) {{${\sss #2'}$}};
	\node[below] at (4) {{${\sss #3'}$}};
	\end{tikzpicture}
}

\newcommand{\TwoDCylImp}[5]{
	\begin{tikzpicture}[scale=#1, baseline=-2.5em,yscale=-1]
	\coordinate (1) at (0,0);
	\coordinate (2) at (2,0);
	\coordinate (3) at (0,2);
	\coordinate (4) at (2,2);
	
	\draw[] (1) --node[above]{${\sss #4}$} (2);
	\draw[] (1) --node[right]{${\sss #5}$} (3);
	\draw[] (2) -- (4);
	\draw[] (3) --node[below]{${\sss #4}$} (4);
	\draw[] (1) -- (4);		
	\node[above] at (1) {{${\sss #2}$}};
	\node[above] at (2) {{${\sss #3}$}};	
	\node[below] at (3) {{${\sss #2'}$}};
	\node[below] at (4) {{${\sss #3'}$}};
	\end{tikzpicture}
}

\newcommand{\TwoDCylDouble}[7]{
	\begin{tikzpicture}[scale=#1, baseline=-2.5em,yscale=-1]
	\coordinate (1) at (0,0);
	\coordinate (2) at (2,0);
	\coordinate (3) at (0,2);
	\coordinate (4) at (2,2);
	\coordinate (5) at (4,0);
	\coordinate (6) at (4,2);
	
	\draw[] (1) --node[above]{${\sss #5}$} (2);
	\draw[] (1) --node[right]{${\sss #7}$} (3);
	\draw[] (2) -- (4);
	\draw[] (3) --node[below]{${\sss #5}$} (4);
	\draw[] (1) -- (4);		
	\draw[] (2) --node[above]{${\sss #6}$} (5);
	\draw[] (4) --node[below]{${\sss #6}$} (6);
	\draw[] (5) -- (6);
	\draw[] (2) -- (6);
	
	\node[above] at (1) {{${\sss #2}$}};
	\node[above] at (2) {{${\sss #3}$}};	
	\node[below] at (3) {{${\sss #2'}$}};
	\node[below] at (4) {{${\sss #3'}$}};
	\node[above] at (5) {{${\sss #4}$}};
	\node[below] at (6) {{${\sss #4'}$}};
	\end{tikzpicture}
}

\newcommand{\TwoDCylPinched}[4]{
	\begin{tikzpicture}[scale=#1, baseline=-1em]
	\coordinate (1) at (0,0);
	\coordinate (2) at (2,0);
	\coordinate (3) at (1,-1.5);
	\coordinate (4) at (-2,1);
	\coordinate (5) at (0,1);
	\coordinate (6) at (-1,-0.5);
	
	\draw[] (1) -- (3);
	\draw[] (2) -- (3);
	\draw[] (1) -- (2);
	\draw[] (1) -- (4);
	\draw[] (4) -- (5);
	\interLLL{5}{6}{2}{4}{1}{4}{3}{4}{1};
	\draw[] (4) -- (6);
	\draw[] (3) -- (6);
	\draw[] (2) -- (5);
	\draw[] (2) -- (4); 
	\draw[] (3) -- (4);
	\interLL{2}{6}{1}{3}{3}{4}{1};
	
	\node[above] at (1) {{${\sss #2}$}};
	\node[above] at (2) {{${\sss #3}$}};
	\node[below] at (3) {{${\sss #4}$}};
	\node[above] at (4) {{${\sss #2'}$}};
	\node[above] at (5) {{${\sss #3'}$}};
	\node[below] at (6) {{${\sss #4'}$}};
	\end{tikzpicture}
}

\newcommand{\onecylinder}[1]{
	\begin{tikzpicture}[scale=#1, baseline=1.5em]
	\coordinate (1) at (0,0);
	\coordinate (2) at (0,1);
	\coordinate (3) at (1,0);
	\coordinate (4) at (1,1);

	\draw[] (1) -- (3);
	\draw[] (2) -- (4);
	\draw[] (1) to[out=180, in=180, looseness=0.8] (2);
	\draw[] (1) to[out=0, in=0, looseness=0.8] (2);
	\draw[op4] (3) to[out=180, in=180, looseness=0.8] (4);
	\draw[] (3) to[out=0, in=0, looseness=0.8] (4);
	\draw[line width =0.4pt, decoration={brace,mirror,raise=0.8em},decorate] (-0.2,0) -- (1.2,0);
	\node[] at (0.5,-0.5) {{${\mathbb S^1 \times \mathbb I}$}};
	\end{tikzpicture}
}

\newcommand{\onecylinderPure}[1]{
	\begin{tikzpicture}[scale=#1, baseline=1.5em]
	\coordinate (1) at (0,0);
	\coordinate (2) at (0,1);
	\coordinate (3) at (1,0);
	\coordinate (4) at (1,1);
	
	\draw[] (1) -- (3);
	\draw[] (2) -- (4);
	\draw[] (1) to[out=180, in=180, looseness=0.8] (2);
	\draw[] (1) to[out=0, in=0, looseness=0.8] (2);
	\draw[op4] (3) to[out=180, in=180, looseness=0.8] (4);
	\draw[] (3) to[out=0, in=0, looseness=0.8] (4);
	\end{tikzpicture}
}

\newcommand{\onecylinderb}[1]{
	\begin{tikzpicture}[scale=#1, baseline=1.5em]
	\coordinate (1) at (0,0);
	\coordinate (2) at (0,1);
	\coordinate (3) at (1,0);
	\coordinate (4) at (1,1);
	\coordinate (5) at (0.7,0);
	\coordinate (6) at (0.7,1);
	
	\draw[] (1) -- (5);
	\draw[] (2) -- (6);
	\draw[dotted] (5) -- (3);
	\draw[dotted] (6) -- (4);
	\draw[] (1) to[out=180, in=180, looseness=0.8] (2);
	\draw[] (1) to[out=0, in=0, looseness=0.8] (2);
	\draw[line width =0.4pt, decoration={brace,mirror,raise=0.8em},decorate] (-0.2,0) -- (1.2,0);
	\node[] at (0.5,-0.5) {{${\Sigma_{\rm 2d}^{\rm o}}$}};
	\end{tikzpicture}
}

\newcommand{\twocylinderb}[1]{
	\begin{tikzpicture}[scale=#1, baseline=1.5em]
	\coordinate (1) at (0,0);
	\coordinate (2) at (0,1);
	\coordinate (3) at (1,0);
	\coordinate (4) at (1,1);
	\coordinate (5) at (1.7,0);
	\coordinate (6) at (1.7,1);
	\coordinate (7) at (2,0);
	\coordinate (8) at (2,1);
	
	\draw[] (1) -- (3);
	\draw[] (2) -- (4);
	\draw[] (1) to[out=180, in=180, looseness=0.8] (2);
	\draw[] (1) to[out=0, in=0, looseness=0.8] (2);
	\draw[op4] (3) to[out=180, in=180, looseness=0.8] (4);
	\draw[] (3) to[out=0, in=0, looseness=0.8] (4);
	\draw[line width =0.4pt, decoration={brace,mirror,raise=0.8em},decorate] (-0.2,0) -- (2.2,0);
	\draw[] (3) -- (5);
	\draw[dotted] (5) -- (7);
	\draw[] (4) -- (6);
	\draw[dotted] (6) -- (8);
	\node[] at (1,-0.5) {{${\mathbb{S}^1\times \mathbb{I}\cup_{\mathbb{S}^1}\Sigma_{\rm 2d}^{\rm o}}$}};
	\end{tikzpicture}
}

\newcommand{\niceThreeCyl}[1]{
	\begin{tikzpicture}[scale=#1, baseline=2em]	
	\coordinate (1) at (0,0);
	\coordinate (2) at (0.8,0.1);
	\coordinate (3) at (2.2,0.1);
	\coordinate (4) at (3,0);
	
	\coordinate (5) at (0,1.5);
	\coordinate (6) at (0.8,1.6);
	\coordinate (7) at (2.2,1.6);
	\coordinate (8) at (3,1.5);
	
	\draw[] (1)  to[out=-90, in=270, looseness=0.8] (4);
	\draw[] (1) to[out=90, in=-270, looseness=0.8](4);
	\draw[] (2) to[out=-90, in=270, looseness=0.5, edge node={coordinate[pos=0.2] (sp1)}, edge node={coordinate[pos=0.8] (sp2)}](3);
	\draw[] (sp1) to[out=90, in=-270, looseness=0.5] (sp2);
	
	\draw[name path=P58d] (5)  to[out=-90, in=270, looseness=0.8] (8);
	\draw[] (5) to[out=90, in=-270, looseness=0.8](8);
	\draw[] (6) to[out=-90, in=270, looseness=0.5, edge node={coordinate[pos=0.2] (sp3)}, edge node={coordinate[pos=0.8] (sp4)}](7);
	\draw[] (sp3) to[out=90, in=-270, looseness=0.5] (sp4);
	
	\draw[dotted] (1) -- (5);
	\draw[dotted] (4) -- (8);
	
	\path[name path=Psp1sp3] (sp1) -- (sp3);
	\path [name intersections={of=P58d and Psp1sp3,by={sp5}}];
	\draw[dotted, op4] (sp3) -- (sp5);
	\draw[dotted] (sp5) -- (sp1);
	
	\path[name path=Psp2sp4] (sp2) -- (sp4);
	\path [name intersections={of=P58d and Psp2sp4,by={sp6}}];
	\draw[dotted, op4] (sp4) -- (sp6);
	\draw[dotted] (sp6) -- (sp2);
	\end{tikzpicture}
}

\newcommand{\threeCyl}[6]{
	\begin{tikzpicture}[scale=#1, baseline=2.5em, line width=0.8 pt]
	\coordinate (0) at (0,0);
	\coordinate (1) at (2,0);
	\coordinate (2) at (1.2,0.5);
	\coordinate (3) at (-0.8,0.5);
	\coordinate (4) at (0,2);
	\coordinate (5) at (2,2);
	\coordinate (6) at (1.2,2.5);
	\coordinate (7) at (-0.8,2.5);

	\draw[] (0) --node[below]{${\sss #6}$} (1);
	\draw[] (1) -- (2);
	\draw[] (0) --node[left, pos=0.3, yshift=-0.1em]{${\sss #4}$} (3);
	\draw[] (4) -- (5);
	\draw[] (5) -- (6);
	\draw[] (6) -- (7);
	\draw[] (4) -- (7);
	\draw[] (0) -- (4);
	\draw[] (1) -- (5);
	\draw[] (3) --node[left]{${\sss #5}$} (7);
	\interLL{1}{3}{0}{5}{0}{4}{1};
	\draw[] (5) -- (7);	
	\draw[] (0) -- (5);	
	\draw[] (3) -- (4);
	\draw[] (2) -- (5);						
	\interL{3}{5}{0}{4}{1};	
	\interLLL{3}{6}{0}{4}{4}{5}{5}{7}{1};		
	\interLL{2}{3}{0}{5}{0}{4}{1};	
	\draw[sha] (2) -- (intersection of 2--6 and 0--5);
	\draw[shab] (intersection of 2--6 and 0--5) -- (intersection of 2--6 and 3--5);
	\draw[shab] (intersection of 2--6 and 3--5) -- (intersection of 2--6 and 4--5);
	\draw[shb] (intersection of 2--6 and 5--7) -- (6);

	\node[below] at (0) {{$\scriptstyle{#2}'$}}; 
	\node[below] at (1) {{$\scriptstyle{#3'}$}}; 
	\node[right, yshift=0.1em] at (2) {{$\scriptstyle{#3}$}}; 
	\node[left] at (3) {{$\scriptstyle{#2}$}}; 
	\node[below, xshift=0.5em, yshift=0.2em] at (4) {{$\scriptstyle{\tilde{#2}'}$}};	
	\node[above] at (5) {{$\scriptstyle{\tilde{#3}'}$}}; 
	\node[above] at (6) {{$\scriptstyle{\tilde{#3}}$}}; 
	\node[above] at (7) {{$\scriptstyle{\tilde{#2}}$}}; 	
	\end{tikzpicture}
}

\newcommand{\threeCylDouble}[8]{
	\begin{tikzpicture}[scale=#1, baseline=2.5em, line width=0.8 pt]
	\coordinate (0) at (0,0);
	\coordinate (1) at (2,0);
	\coordinate (2) at (1.2,0.5);
	\coordinate (3) at (-0.8,0.5);
	\coordinate (4) at (0,2);
	\coordinate (5) at (2,2);
	\coordinate (6) at (1.2,2.5);
	\coordinate (7) at (-0.8,2.5);
	\coordinate (8) at (4,0);
	\coordinate (9) at (4,2);
	\coordinate (10) at (3.2,0.5);
	\coordinate (11) at (3.2,2.5);

	\draw[] (0) --node[below]{${\sss #7}$} (1);
	\draw[] (1) -- (2);
	\draw[] (0) --node[left, pos=0.3, yshift=-0.1em]{${\sss #5}$} (3);
	\draw[] (4) -- (5);
	\draw[] (5) -- (6);
	\draw[] (6) -- (7);
	\draw[] (4) -- (7);
	\draw[] (0) -- (4);
	\draw[] (1) -- (5);
	\draw[] (3) --node[left]{${\sss #6}$} (7);
	\interLL{1}{3}{0}{5}{0}{4}{1};
	\draw[] (5) -- (7);	
	\draw[] (0) -- (5);	
	\draw[] (3) -- (4);
	\draw[] (2) -- (5);						
	\interL{3}{5}{0}{4}{1};	
	\interLLL{3}{6}{0}{4}{4}{5}{5}{7}{1};		
	\interLL{2}{3}{0}{5}{0}{4}{1};	
	\draw[sha] (2) -- (intersection of 2--6 and 0--5);
	\draw[shab] (intersection of 2--6 and 0--5) -- (intersection of 2--6 and 3--5);
	\draw[shab] (intersection of 2--6 and 3--5) -- (intersection of 2--6 and 4--5);
	\draw[shb] (intersection of 2--6 and 5--7) -- (6);

	\draw[] (5) -- (9);
	\draw[] (9) -- (11);
	\draw[] (6) -- (11);
	\draw[] (9) -- (8);
	\draw[] (1) --node[below]{${\sss #8}$} (8);
	\draw[] (8) -- (10);
	\draw[] (1) -- (9);
	\draw[] (9) -- (10);
	\interLL{2}{10}{1}{5}{1}{9}{1};
	\interLL{2}{8}{1}{5}{1}{9}{1};
	\draw[] (6) -- (9);
	\interL{2}{9}{1}{5}{1};
	\interLLL{2}{11}{1}{5}{5}{9}{6}{9}{1};
	\draw[sha] (10) -- (intersection of 10--11 and 1--9);
	\draw[shab] (intersection of 10--11 and 1--9) -- (intersection of 10--11 and 2--9);
	\draw[shab] (intersection of 10--11 and 2--9) -- (intersection of 10--11 and 5--9);
	\draw[shb] (intersection of 10--11 and 6--9) -- (11);
	
	\node[below] at (0) {{$\scriptstyle{#2'}$}};	
	\node[below] at (1) {{$\scriptstyle{#3'}$}}; 
	\node[left, yshift=0.4em] at (2) {{$\scriptstyle{#3}$}}; 
	\node[left] at (3) {{$\scriptstyle{#2}$}}; 
	\node[right, yshift=0.1em] at (10) {{$\scriptstyle{#4}$}}; 
	\node[below] at (8) {{$\scriptstyle{#4'}$}}; 
	\node[below, xshift=0.5em, yshift=0.2em] at (4) {{$\scriptstyle{\tilde{#2}'}$}};	
	\node[below, xshift=0.5em, yshift=0.2em] at (5) {{$\scriptstyle{\tilde{#3}'}$}}; 
	\node[above] at (6) {{$\scriptstyle{\tilde{#3}}$}}; 
	\node[above] at (7) {{$\scriptstyle{\tilde{#2}}$}}; 
	\node[above] at (9) {{$\scriptstyle{\tilde{#4}'}$}}; 
	\node[above] at (11) {{$\scriptstyle{\tilde{#4}}$}}; 
	\end{tikzpicture}
}

\newcommand{\dualPTwoThree}[2]{
	\begin{tikzpicture}[scale=#1,baseline=3em]
	\coordinate (0) at (0, 0);
	\coordinate (1) at (2, 1);
	\coordinate (2) at (7, 1);
	\coordinate (3) at (9, 0.5);
	\coordinate (4) at (4, 2.5);
	\coordinate (5) at (3, 4);
	\coordinate (6) at (2.5, 6);
	\coordinate (7) at (6.5, 6);
	\coordinate (8) at (4.5, 7);
	\coordinate (9) at (0, 9);
	\coordinate (10) at (2, 10);
	\coordinate (11) at (7, 10);
	\coordinate (12) at (9, 9.5);
	\coordinate (13) at (4, 11.5);
	\coordinate (14) at (3, 13);
	\coordinate (15) at (4.5, 5);	
	\coordinate (16) at (4.5, 7);	
	
	\ifthenelse{#2 = 1}{
		\draw[] (0) -- (1);
		\draw[] (1) -- (2);
		\draw[] (2) -- (3);
		\draw[] (0) -- (9);
		\draw[] (3) -- (12);
		\draw[] (9) -- (10);
		\draw[] (10) -- (16);
		\draw[] (11) -- (16);
		\draw[] (1) -- (15);
		\draw[] (2) -- (15);
		\draw[name path=P1011] (10) -- (11);
		\draw[] (11) -- (12);
		\draw[op] (1) -- (4);
		\draw[op] (4) -- (2);
		\draw[name path=P1013] (10) -- (13);
		\draw[] (13) -- (11);
		\draw[] (13) -- (14);
		\speL{13}{16}{10}{11}{}{op4};
		\draw[] (15) -- (16);
		\draw[op] (15) -- (4);
		\draw[opacity=0.2] (4) -- (5);
		\speLL{14}{5}{P1013}{P1011}{}{op2}{op2};
		
		\node[opacity=0.6] at (5.2,9) {{$\scriptstyle{\bar a}$}};
		\node[] at (7,6) {{$\scriptstyle{b}$}};
		\node[] at (2,6) {{$\scriptstyle{\bar c}$}};
		\node[opacity=0.6] at (3.5,2.8) {{$\scriptstyle{d}$}};
	}{}

	\ifthenelse{#2 = 2}{
		\draw[] (0) -- (1);
		\draw[] (1) -- (2);
		\draw[] (2) -- (3);
		\draw[] (0) -- (9);
		\draw[] (1) -- (6);
		\draw[] (6) -- (7);
		\draw[] (2) -- (7);
		\draw[] (6) -- (10);
		\draw[] (7) -- (11);
		\draw[] (3) -- (12);
		\draw[] (9) -- (10);
		\draw[name path=P1011] (10) -- (11);
		\draw[] (11) -- (12);
		\draw[op] (1) -- (4);
		\draw[op] (4) -- (2);
		\draw[op] (6) -- (8);
		\draw[op] (7) -- (8);
		\draw[name path=P1013] (10) -- (13);
		\draw[] (13) -- (11);
		\draw[] (13) -- (14);
		\speL{13}{8}{10}{11}{}{op4};
		\draw[op4] (8) -- (4);
		\draw[opacity=0.2] (4) -- (5);
		\speLL{14}{5}{P1013}{P1011}{}{op2}{op2};
		
		\node[opacity=0.6] at (5.5,9) {{$\scriptstyle{\bar a}$}};
		\node[] at (7.7,6) {{$\scriptstyle{b}$}};
		\node[] at (1.3,6) {{$\scriptstyle{\bar c}$}};
		\node[opacity=0.6] at (3,2.8) {{$\scriptstyle{d}$}};
	}{}		
	\end{tikzpicture}
}

\newcommand{\specialPTwoThree}[2]{
	\begin{tikzpicture}[scale=#1,baseline=0em]
	\coordinate (0) at (0, 0);
	\coordinate (1) at (5, 0);
	\coordinate (3) at (-2, 3.5);
	\coordinate (4) at (7, 3.5);
	\coordinate (5) at (-2, -3.5);
	\coordinate (6) at (7, -3.5);

	\ifthenelse{#2 = 1}{
		\draw[] (3) to[out=90, in=-270, looseness=0.5] (4);
		\draw[name path=P34] (3) to[out=-90, in=270, looseness=0.5] (4);
		\draw[] (5) to[out=90, in=-270, looseness=0.5] (6);
		\draw[] (5) to[out=-90, in=270, looseness=0.5] (6);
		\coordinate (x1) at (-1.3,3.2);
		\coordinate (x2) at (-1.3,3.8);
		\coordinate (x3) at (-1.55,3.2);
		\coordinate (x4) at (-1.55,3.8);
		\draw[line width=0.5pt] (x1) to[bend left =30] (x2);
		\draw[line width=0.5pt] (x3) to[bend left =30] (x4);
		\coordinate (x5) at (-1.3,-3.8);
		\coordinate (x6) at (-1.3,-3.2);
		\coordinate (x7) at (-1.55,-3.8);
		\coordinate (x8) at (-1.55,-3.2);
		\draw[line width=0.5pt] (x5) to[bend left =30] (x6);
		\draw[line width=0.5pt] (x7) to[bend left =30] (x8);
		
		\node[] at (2.5,3.5) {{$\scriptstyle{\bar a}$}};
		\node[] at (2.5,-3.5) {{$\scriptstyle{\bar d}$}};	
	}{}
	
	\ifthenelse{#2 = 2}{
		\draw[op4 ] (0) to[out=90, in=-270, looseness=0.5, edge node={coordinate[pos=0.4] (2)}] (1);
		\draw[name path=P01] (0) to[out=-90, in=270, looseness=0.5] (1);
		\draw[] (3) to[out=90, in=-270, looseness=0.5, edge node={coordinate[pos=0.4] (10)}] (4);
		\draw[name path=P34] (3) to[out=-90, in=270, looseness=0.5] (4);
		\draw[op4] (5) to[out=90, in=-270, looseness=0.5, edge node={coordinate[pos=0.4] (11)}] (6);
		\draw[] (5) to[out=-90, in=270, looseness=0.5] (6);
		\speL{10}{2}{3}{4}{}{op4};
		\draw[] (0) -- (3);
		\draw[] (1) -- (4);
		\draw[] (0) -- (5);
		\draw[] (1) -- (6);	
		\speL{2}{11}{0}{1}{op2}{op4};
		
		\node[] at (4,3) {{$\scriptstyle{\bar a}$}};
		\node[op6] at (-0.3,-1.5) {{$\scriptstyle{d}$}};
		\node[opacity=0.6] at (2.5,0) {{$\scriptstyle{\overline{\! a+d \!}}$}};		
	}{}		
	\end{tikzpicture}	
}

\def\centerarc[#1](#2)(#3:#4:#5)
{ \draw[#1] ($(#2)+({#5*cos(#3)},{#5*sin(#3)})$) arc (#3:#4:#5); }

\newcommand{\specialSphere}[2]{
	\begin{tikzpicture}[scale=#1,baseline=1.3em]
	\coordinate (0) at (-0.85, 0);
	\coordinate (1) at (5.85, 0);
	\coordinate (2) at (2.5,2.7);
	\coordinate (3) at (-1,-1.6);
	\coordinate (4) at (9, -1.6);
	\coordinate (5) at (-4, 1.6);
	\coordinate (6) at (6, 1.6);
	
	\ifthenelse{#2 = 1}{
		\draw[] (3) -- (4);
		\draw[] (3) -- (5);
		\draw[] (4) -- (6);
		\centerarc[name path=P12](2.5,5)(0:360:4.3)
		\path[name path =P56] (5) -- (6);
		\path [name intersections={of=P56 and P12,by={sp1,sp2}}];
		\draw[] (5) -- (sp1);		
		\draw[] (sp2) -- (6);
		\draw[op2] (sp1) -- (sp2);
		
		\coordinate (x1) at (5.3,3);
		\coordinate (x2) at (5.8,4);
		\coordinate (x3) at (5.5,2.8);
		\coordinate (x4) at (6,3.8);
		\draw[line width=0.5pt] (x1) to[bend right =20] (x2);
		\draw[line width=0.5pt] (x3) to[bend right =20] (x4);
		\coordinate (x5) at (5.4,3.8);
		\coordinate (x6) at (6.1,3.2);
		\coordinate (x7) at (5.3,3.5);
		\coordinate (x8) at (6,2.9);		
		\draw[line width=0.5pt] (x5) -- (x6);
		\draw[line width=0.5pt] (x7) -- (x8);
		
		\node[] at (2.5,5) {{$\scriptstyle{a}$}};
		\node[] at (-2.2,0.9) {{$\scriptstyle{\bar d}$}};			
	}{}
	
	\ifthenelse{#2 = 2}{
		\draw[op4] (0) to[out=60, in=-240, looseness=0.5] (1);
		\draw[name path=P01] (0) to[out=-60, in=240, looseness=0.5] (1);
		\centerarc[name path=P12](2)(-42:222:4.3)
		\draw[] (3) -- (4);
		\draw[] (3) -- (5);
		\speL{4}{6}{1}{2}{}{op2};
		\speL{5}{6}{1}{2}{}{op2};
		
		\coordinate (x1) at (5.3,0.7);
		\coordinate (x2) at (5.8,1.7);
		\coordinate (x3) at (5.5,0.5);
		\coordinate (x4) at (6,1.5);
		\draw[line width=0.5pt] (x1) to[bend right =20] (x2);
		\draw[line width=0.5pt] (x3) to[bend right =20] (x4);
		\coordinate (x5) at (5.4,1.5);
		\coordinate (x6) at (6.1,0.9);
		\coordinate (x7) at (5.3,1.2);
		\coordinate (x8) at (6,0.6);		
		\draw[line width=0.5pt] (x5) -- (x6);
		\draw[line width=0.5pt] (x7) -- (x8);	
		
		\node[] at (2.5,2.7) {{$\scriptstyle{a}$}};
		\node[] at (-2.2,0.9) {{$\scriptstyle{\bar d}$}};
		\node[opacity=0.6] at (2.5,0) {{$\scriptstyle{\overline{\! a+d \!}}$}};		
	}{}		
	\end{tikzpicture}	
}

\newcommand{\dualPFourOne}[2]{
	\begin{tikzpicture}[scale=#1,baseline=1.5em]
	\coordinate (0) at (0.5, 1);
	\coordinate (1) at (0.5, 3);
	\coordinate (2) at (-2, 6);
	\coordinate (3) at (3, 6);
	\coordinate (4) at (0, 7.5);
	\coordinate (5) at (-4, 8);
	\coordinate (6) at (5, 8);
	\coordinate (7) at (0, 9.5);
	\coordinate (8) at (-1, 11);
	\coordinate (9) at (7, 7.5);
	\coordinate (10) at (-6, 7);
	\coordinate (11) at (-6,-2.25);
	\coordinate (12) at (-1,3.25);
	\coordinate (13) at (7,-0.625);

	\ifthenelse{#2 = 1}{
		\draw[name path = P01] (0) -- (1);
		\draw[name path = P12] (1) -- (2);
		\draw[name path = P13] (1) -- (3);
		\draw[name path = P23] (2) -- (3);
		\draw[name path = P24, op4] (2) -- (4);
		\draw[name path = P34, op4] (3) -- (4);
		\draw[name path = P56] (5) -- (6);
		\draw[name path = P57] (5) -- (7);
		\draw[name path = P67] (6) -- (7);
		\draw[name path = P25] (2) -- (5);
		\draw[name path = P36] (3) -- (6);
		\draw[name path = P69] (6) -- (9);
		\draw[name path = P913] (9) -- (13);
		\draw[name path = P013] (0) -- (13);
		\draw[name path = P510] (5) -- (10);
		\draw[name path = P011] (0) -- (11);
		\draw[name path = P1011] (10) -- (11);
		\draw[name path = P78] (7) -- (8);
		\speL{7}{4}{5}{6}{}{op4};
		\speL{4}{1}{2}{3}{op2}{op4};
		\speL{8}{12}{5}{7}{}{op2};
		\draw[op2] (0) -- (12);
		
		\node[opacity=0.6] at (2,7.3) {{$\scriptstyle{\bar a}$}};
		\node[] at (4,4) {{$\scriptstyle{b}$}};
		\node[] at (-3,4) {{$\scriptstyle{\bar c}$}};
		\node[opacity=0.6] at (-0.3,5) {{$\scriptstyle{d}$}};
	}{}

	\ifthenelse{#2 = 2}{
		\draw[name path = P01] (0) -- (1);
		\draw[] (1) -- (5);
		\draw[] (1) -- (6);
		\draw[name path = P56] (5) -- (6);
		\draw[name path = P57] (5) -- (7);
		\draw[name path = P67] (6) -- (7);
		\draw[name path = P69] (6) -- (9);
		\draw[name path = P913] (9) -- (13);
		\draw[name path = P013] (0) -- (13);
		\draw[name path = P510] (5) -- (10);
		\draw[name path = P011] (0) -- (11);
		\draw[name path = P1011] (10) -- (11);
		\draw[name path = P78] (7) -- (8);
		\speL{8}{12}{5}{7}{}{op2};
		\speL{7}{1}{5}{6}{}{op4};
		\draw[op2] (0) -- (12);
		
		\node[opacity=0.6] at (2,6.5) {{$\scriptstyle{\bar a}$}};
		\node[] at (4,4) {{$\scriptstyle{b}$}};
		\node[] at (-3,4) {{$\scriptstyle{\bar c}$}};
	}{}		
	\end{tikzpicture}
}

\newcommand{\convLabels}[2]{
	\begin{tikzpicture}[scale=#1,baseline=1.5em]
	\coordinate (0) at (0.5, 1);
	\coordinate (1) at (0.5, 3);
	\coordinate (2) at (-2, 6);
	\coordinate (3) at (3, 6);
	\coordinate (4) at (0, 7.5);
	\coordinate (5) at (-4, 8);
	\coordinate (6) at (5, 8);
	\coordinate (7) at (0, 9.5);
	\coordinate (8) at (-1, 11);
	\coordinate (9) at (7, 7.5);
	\coordinate (10) at (-6, 7);
	\coordinate (11) at (-6,-2.25);
	\coordinate (12) at (-1,3.25);
	\coordinate (13) at (7,-0.625);

	\draw[name path = P01] (0) -- (1);
	\draw[] (1) -- (5);
	\draw[] (1) -- (6);
	\draw[name path = P56] (5) -- (6);
	\draw[name path = P57] (5) -- (7);
	\draw[name path = P67] (6) -- (7);
	\draw[name path = P69] (6) -- (9);
	\draw[name path = P913] (9) -- (13);
	\draw[name path = P013] (0) -- (13);
	\draw[name path = P510] (5) -- (10);
	\draw[name path = P011] (0) -- (11);
	\draw[name path = P1011] (10) -- (11);
	\draw[name path = P78] (7) -- (8);
	\speL{8}{12}{5}{7}{}{op2};
	\speL{7}{1}{5}{6}{}{op4};
	\draw[op2] (0) -- (12);
	
	\ifthenelse{#2 = 1}{
		\node[op6] at (2,7) {{$\scriptstyle{b}$}};
		\node[op6] at (-0.2,2.8) {{$\scriptstyle{\bar{a}}$}};
		\node[] at (0.7,6) {{$\scriptstyle{c}$}};
	}{}
	\ifthenelse{#2 = 2}{
		\node[op6] at (2,7) {{$\scriptstyle{b}$}};
		\node[] at (4,4) {{$\scriptstyle{b+c}$}};
		\node[] at (-3.5,4) {{$\scriptstyle{a+b+c}$}};
		\node[op6] at (-0.2,2.8) {{$\scriptstyle{\bar{a}}$}};
		\node[op6] at (-2,7.4) {{$\scriptstyle{a+b}$}};
		\node[] at (0.7,6) {{$\scriptstyle{c}$}};
	}{}
	
	\end{tikzpicture}
}

\newcommand{\dualHamA}[2]{
	\begin{tikzpicture}[scale=#1,baseline=-2em]
	\coordinate (0) at (0, -7.5);
	\coordinate (1) at (0, -6);
	\coordinate (2) at (-6.5, 1.5);
	\coordinate (3) at (6.5, 1.5);
	\coordinate (4) at (-1, 4.6);
	\coordinate (5) at (-8, 3);
	\coordinate (6) at (8, 3);
	\coordinate (7) at (-1.5, 6.5);
	\coordinate (8) at (-3, 7.5);
	\coordinate (9) at (9.5, 2.75);
	\coordinate (10) at (-9.5, 2.75);
	\coordinate (11) at (-9.5,-9);
	\coordinate (13) at (9.5,-9);
	\coordinate (12) at (-3,-5.5);
	
	\coordinate (14) at (-6.5,-3.5);
	\coordinate (15) at (6.5,-3.5);
	\coordinate (16) at (0,3.5);
	\coordinate (17) at (-7,3.5);
	\coordinate (18) at (7,3.5);
	
	\ifthenelse{#2 = 1}{		
		\draw[name path = P01] (0) -- (1);
		\draw[name path = P12] (1) -- (2);
		\draw[name path = P13] (1) -- (3);
		\draw[name path = P23] (2) -- (3);
		\draw[name path = P56] (5) -- (6);
		\draw[name path = P57] (5) -- (7);
		\draw[name path = P67] (6) -- (7);
		\draw[name path = P25] (2) -- (5);
		\draw[name path = P36] (3) -- (6);
		\draw[name path = P69] (6) -- (9);
		\draw[name path = P913] (9) -- (13);
		\draw[name path = P013] (0) -- (13);
		\draw[name path = P510] (5) -- (10);
		\draw[name path = P011] (0) -- (11);
		\draw[name path = P1011] (10) -- (11);
		\draw[name path = P78] (7) -- (8);
		\draw[] (4) -- (7);
		\draw[op2] (0) -- (12);
		\speL{2}{4}{5}{6}{op4}{};
		\speL{3}{4}{5}{6}{op4}{}; 
		\path[name path=P812] (8) -- (12);
		\path [name intersections={of=P812 and P24,by={sp1}}];
		\speL{8}{sp1}{5}{7}{}{op2};	
		\speL{sp1}{12}{1}{2}{op1}{op2};	
		\path[name path=P14] (1) -- (4);
		\speLL{1}{4}{P23}{P56}{op4}{op2}{op4};
		
		\node[] at (2,4.2) {{$\scriptstyle{{\bar a}}$}};
		\node[] at (6,-3) {{$\scriptstyle{{b}}$}};
		\node[] at (-6,-3) {{$\scriptstyle{{\bar c}}$}};
		\node[op6] at (-4,0.2) {{$\scriptstyle{{d}}$}};		
	}{}

	\ifthenelse{#2 = 2}{
		\begin{scope}
			\clip (2) rectangle (6);
			\centerarc[name path=P00, op2](0,0)(0:360:3.5)
		\end{scope}
		\begin{scope}
			\clip (2) rectangle (13);
			\centerarc[op4](0,0)(0:360:3.5)
		\end{scope}	
		\begin{scope}
			\clip (6) rectangle (8);
			\centerarc[op4](0,0)(0:360:3.5)
		\end{scope}		
		\coordinate (x1) at (2,-2);
		\coordinate (x2) at (2.5,-1.2);
		\coordinate (x3) at (2.2,-2.2);
		\coordinate (x4) at (2.7,-1.4);
		\draw[line width=0.5pt, op4] (x1) to[bend right =20] (x2);
		\draw[line width=0.5pt, op4] (x3) to[bend right =20] (x4);
		\coordinate (x5) at (2.2,-1.3);
		\coordinate (x6) at (2.7,-1.8);
		\coordinate (x7) at (2,-1.5);
		\coordinate (x8) at (2.6,-2.1);		
		\draw[line width=0.5pt, op4] (x5) -- (x6);
		\draw[line width=0.5pt, op4] (x7) -- (x8);
		
		\draw[name path = P01] (0) -- (1);
		\draw[name path = P12] (1) -- (2);
		\draw[name path = P13] (1) -- (3);
		\draw[name path = P23] (2) -- (3);
		\draw[name path = P56] (5) -- (6);
		\draw[name path = P57] (5) -- (7);
		\draw[name path = P67] (6) -- (7);
		\draw[name path = P25] (2) -- (5);
		\draw[name path = P36] (3) -- (6);
		\draw[name path = P69] (6) -- (9);
		\draw[name path = P913] (9) -- (13);
		\draw[name path = P013] (0) -- (13);
		\draw[name path = P510] (5) -- (10);
		\draw[name path = P011] (0) -- (11);
		\draw[name path = P1011] (10) -- (11);
		\draw[name path = P78] (7) -- (8);
		\draw[] (4) -- (7);
		\draw[op2] (0) -- (12);
		\speL{2}{4}{5}{6}{op4}{};
		\speL{3}{4}{5}{6}{op4}{};
		\path[name path=P812] (8) -- (12);
		\path [name intersections={of=P812 and P24,by={sp1}}];
		\speL{8}{sp1}{5}{7}{}{op2};	
		\speL{sp1}{12}{1}{2}{op1}{op2};	
		\path[name path=P14] (1) -- (4);
		\path [name intersections={of=P14 and P00,by={sp2,sp3}}];
		\draw[op4] (1) -- (sp3);
		\speLL{sp3}{sp2}{P23}{P56}{op2}{op1}{op2};
		\draw[op4] (sp2) -- (4);
		
		\node[op6] at (0,0) {{$\scriptstyle{{\bar k}}$}};
		\node[] at (2,4.2) {{$\scriptstyle{{\bar a}}$}};
		\node[] at (6,-3) {{$\scriptstyle{{b}}$}};
		\node[] at (-6,-3) {{$\scriptstyle{{\bar c}}$}};
		\node[op6] at (-4,0.2) {{$\scriptstyle{{d}}$}};				
	}{}

	\ifthenelse{#2 = 3}{		
		\draw[name path = P01] (0) -- (1);
		\draw[name path = P12] (1) -- (2);
		\draw[name path = P13] (1) -- (3);
		\draw[name path = P23] (2) -- (3);
		\draw[name path = P56] (5) -- (6);
		\draw[name path = P57] (5) -- (7);
		\draw[name path = P67] (6) -- (7);
		\draw[name path = P25] (2) -- (5);
		\draw[name path = P36] (3) -- (6);
		\draw[name path = P69] (6) -- (9);
		\draw[name path = P913] (9) -- (13);
		\draw[name path = P013] (0) -- (13);
		\draw[name path = P510] (5) -- (10);
		\draw[name path = P011] (0) -- (11);
		\draw[name path = P1011] (10) -- (11);
		\draw[name path = P78] (7) -- (8);
		\draw[] (4) -- (7);
		\draw[op2] (0) -- (12);
		\speL{2}{4}{5}{6}{op4}{};
		\speL{3}{4}{5}{6}{op4}{};
		\path[name path=P812] (8) -- (12);
		\path [name intersections={of=P812 and P24,by={sp1}}];
		\speL{8}{sp1}{5}{7}{}{op2};	
		\speL{sp1}{12}{1}{2}{op1}{op2};	
		\path[name path=P14] (1) -- (4);
		\speLL{1}{4}{P23}{P56}{op4}{op2}{op4};
		
		\path[name path = P1415] (14) -- (15);
		\path[name path = P1416] (14) -- (16);
		\path[name path = P1516] (15) -- (16);
		\path [name intersections={of=P1415 and P12,by={sp3}}];
		\path [name intersections={of=P1415 and P13,by={sp4}}];		
		\path [name intersections={of=P1416 and P12,by={sp5}}];
		\path [name intersections={of=P1416 and P23,by={sp6}}];		
		\path [name intersections={of=P1516 and P13,by={sp7}}];
		\path [name intersections={of=P1516 and P23,by={sp8}}];
		
		\path[name path=Psp617] (sp6) -- (17);
		\path[name path=Psp818] (sp8) -- (18);
		\path[name path=P1718] (17) -- (18);
		\path [name intersections={of=Psp617 and P24,by={sp9}}];
		\path [name intersections={of=Psp818 and P34,by={sp10}}];
		\path [name intersections={of=P1718 and P24,by={sp11}}];
		\path [name intersections={of=P1718 and P34,by={sp12}}];
		
		\path[] (1) to[edge node={coordinate[pos=0.38] (sp13)}, edge node={coordinate[pos=0.6] (sp14)}] (4);
		\draw[] (sp3) -- (sp4);
		\draw[] (sp5) -- (sp6);	
		\draw[] (sp7) -- (sp8);
		\draw[op4] (sp6) -- (sp9);	
		\draw[op4] (sp8) -- (sp10);	
		\draw[] (sp11) -- (sp12);
		\draw[op4] (sp3) -- (sp13);
		\draw[op4] (sp4) -- (sp13);	
		\speL{sp5}{sp9}{2}{3}{op4}{op2};
		\speL{sp7}{sp10}{2}{3}{op4}{op2};	
		\speLL{sp11}{sp14}{P56}{P23}{op4}{op2}{op4};
		\speLL{sp12}{sp14}{P56}{P23}{op4}{op2}{op4};
		
		\node[op6] at (3.5,0.8) {{$\scriptstyle{{\bar k}}$}};
		\node[op6] at (0.1,-2.9) {{$\scriptstyle{{k}}$}};
		\node[op6] at (-3.5,0.8) {{$\scriptstyle{{\bar k}}$}};
		\node[op6] at (0,2.3) {{$\scriptstyle{{k}}$}};
		\node[] at (2,4.2) {{$\scriptstyle{{\bar a}}$}};
		\node[] at (6,-3) {{$\scriptstyle{{b}}$}};
		\node[] at (-6,-3) {{$\scriptstyle{{\bar c}}$}};	
		\node[op6] at (-5,0.8) {{${\sss d}$}};
		\node[op4] at (-1.3,2.3)	{{${\sss d}$}};
		\node[op6] at (-0.9,-4.1) {{${\sss d}$}};				
	}{}

	\ifthenelse{#2 = 4}{		
		\draw[name path = P01] (0) -- (1);
		\draw[name path = P12] (1) -- (2);
		\draw[name path = P13] (1) -- (3);
		\draw[name path = P23] (2) -- (3);
		\draw[name path = P56] (5) -- (6);
		\draw[name path = P57] (5) -- (7);
		\draw[name path = P67] (6) -- (7);
		\draw[name path = P25] (2) -- (5);
		\draw[name path = P36] (3) -- (6);
		\draw[name path = P69] (6) -- (9);
		\draw[name path = P913] (9) -- (13);
		\draw[name path = P013] (0) -- (13);
		\draw[name path = P510] (5) -- (10);
		\draw[name path = P011] (0) -- (11);
		\draw[name path = P1011] (10) -- (11);
		\draw[name path = P78] (7) -- (8);
		\draw[] (4) -- (7);
		\draw[op2] (0) -- (12);
		\speL{2}{4}{5}{6}{op4}{};
		\speL{3}{4}{5}{6}{op4}{}; 
		\path[name path=P812] (8) -- (12);
		\path [name intersections={of=P812 and P24,by={sp1}}];
		\speL{8}{sp1}{5}{7}{}{op2};	
		\speL{sp1}{12}{1}{2}{op1}{op2};	
		\path[name path=P14] (1) -- (4);
		\speLL{1}{4}{P23}{P56}{op4}{op2}{op4};
		
		\node[] at (2,4.2) {{$\scriptstyle{{\bar a}}$}};
		\node[] at (6,-3) {{$\scriptstyle{{b}}$}};
		\node[] at (-6,-3) {{$\scriptstyle{{\bar c}}$}};
		\node[op6] at (-3.6,0.2) {{$\scriptstyle{{d+k}}$}};		
	}{}
	
	\end{tikzpicture}
}

\newcommand{\dualHamB}[2]{
	\begin{tikzpicture}[scale=#1,baseline=1em]
	\coordinate (0) at (0, 0);
	\coordinate (1) at (0, 7);
	\coordinate (2) at (-2, 9);
	\coordinate (3) at (-2, 2);
	\coordinate (4) at (4, -1);
	\coordinate (5) at (4, 6);
	\coordinate (6) at (-3.5,-1.5);
	\coordinate (7) at (-3.5, 5.5);

	\draw[] (0) -- (1);	
	\draw[op4] (0) -- (3);
	\draw[] (1) -- (2);
	\draw[] (0) -- (4);
	\draw[] (1) -- (5);
	\draw[] (4) -- (5);
	\draw[] (6) -- (7);
	\draw[] (0) -- (6);
	\draw[name path = P17] (1) -- (7);
	\speL{2}{3}{1}{7}{}{op4};
	
	\ifthenelse{#2=1}{
		\node[] at (2,0.2) {{$\scriptstyle{{a}}$}};
		\node[] at (-2.2,0.2) {{$\scriptstyle{a+b}$}};
		\node[opacity=0.6] at (-1,2) {{$\scriptstyle{\bar{b}}$}};
	}{}
	\ifthenelse{#2=2}{
		\node[] at (2,3) {{$\scriptstyle{{a}}$}};
		\node[] at (-2.2,1) {{$\scriptstyle{\bar b}$}};
		\node[opacity=0.6] at (-1,4) {{$\scriptstyle{c}$}};
	}{}
	\ifthenelse{#2=3}{
		\node[] at (2,0.2) {{$\scriptstyle{{a}}$}};
		\node[] at (-2.2,0.2) {{$\scriptstyle{a+b}$}};
		\node[opacity=0.6] at (-1,2) {{$\scriptstyle{{b}}$}};
		\coordinate (x) at (-2.7,4);
		\coordinate (y) at (2.5,4.8);
		\coordinate (z) at (0.6,2.7);
		\draw[dotted, sha] (x) -- (intersection of x--y and 0--1);
		\draw[dotted, shb, ->1-=0.7] (intersection of x--y and 0--1) -- (y);
		\draw[->1-=0.8, dotted] (y) -- (z);
		\draw[->1-=0.6, dotted] (x) -- (z);
		
	}{}
	\end{tikzpicture}
}

\newcommand{\PPPONE}[1]{
	\begin{tikzpicture}[scale=1,baseline=0em, line width=0.8pt]
	\coordinate (0) at (0,1);
	\coordinate (1) at ({sin(60)} ,{cos(60)} );
	\coordinate (2) at ({sin(60)} ,{-cos(60)} );
	\coordinate (3) at (0,-1);
	\coordinate (4) at ({-sin(60)} ,{-cos(60)} );
	\coordinate (5) at ({-sin(60)} ,{cos(60)} );
	
	\draw[] (0) to (1);
	\draw[] (1) to (2);
	\draw[] (2) to (3);
	\draw[] (3) to (4);
	\draw[] (4) to (5);
	\draw[] (5) to (0);
	
	\draw[] (1) to (3);
	\draw[] (0) to (3);
	\draw[] (5) to (3);
	\coordinate (a) at (intersection of 3--1 and 0--2);
	\draw[shorten >= 0.3em] (0) to (a);
	\draw[shorten <= 0.3em, shorten >= 0em] (a) to (2);
	\coordinate (b) at (intersection of 5--3 and 4--2);
	\coordinate (c) at (intersection of 0--3 and 4--2);
	\coordinate (d) at (intersection of 3--1 and 4--2);
	\draw[shorten >= 0.3em] (4) to (b);
	\draw[shorten <= 0.3em, shorten >= 0.3em] (b) to (c);
	\draw[shorten <= 0.3em, shorten >= 0.3em] (c) to (d);
	\draw[shorten <= 0.3em] (d) to (2);
	\coordinate (e) at (intersection of 0--3 and 5--2);
	\coordinate (f) at (intersection of 3--1 and 5--2);
	\draw[shorten >= 0.3em] (5) to (e);
	\draw[shorten <= 0.3em, shorten >= 0.3em] (e) to (f);
	\draw[shorten <= 0.3em] (f) to (2);
	
	\ifthenelse{#1 = 1}{
		\node[above] at (0) {{$\scriptstyle{0}$}};	
		\node[right] at (1) {{$\scriptstyle{1}$}};
		\node[right] at (2) {{$\scriptstyle{2}$}};
		\node[below] at (3) {{$\scriptstyle{3}$}};
		\node[left] at (4) {{$\scriptstyle{4}$}};
		\node[left] at (5) {{$\scriptstyle{5}$}};}{}
	\end{tikzpicture}
}

\newcommand{\PPPTWO}[1]{
	\begin{tikzpicture}[scale=1,baseline=0em, line width=0.8pt]
	\coordinate (0) at (0,1);
	\coordinate (1) at ({sin(60)} ,{cos(60)} );
	\coordinate (2) at ({sin(60)} ,{-cos(60)} );
	\coordinate (3) at (0,-1);
	\coordinate (4) at ({-sin(60)} ,{-cos(60)} );
	\coordinate (5) at ({-sin(60)} ,{cos(60)} );
	
	\draw[] (0) to (1);
	\draw[] (1) to (2);
	\draw[] (2) to (3);
	\draw[] (3) to (4);
	\draw[] (4) to (5);
	\draw[] (5) to (0);
	
	\draw[] (1) to (3);
	\draw[] (0) to (3);
	\draw[] (5) to (3);
	\coordinate (a) at (intersection of 3--1 and 0--2);
	\draw[shorten >= 0.3em] (0) to (a);
	\draw[shorten <= 0.3em, shorten >= 0em] (a) to (2);
	\coordinate (b) at (intersection of 5--3 and 4--2);
	\coordinate (c) at (intersection of 0--3 and 4--2);
	\coordinate (d) at (intersection of 3--1 and 4--2);
	\draw[shorten >= 0.3em] (4) to (b);
	\draw[shorten <= 0.3em, shorten >= 0.3em] (b) to (c);
	\draw[shorten <= 0.3em, shorten >= 0.3em] (c) to (d);
	\draw[shorten <= 0.3em] (d) to (2);
	\coordinate (h) at (intersection of 0--4 and 5--2);
	\coordinate (e) at (intersection of 0--3 and 5--2);
	\coordinate (f) at (intersection of 3--1 and 5--2);
	\draw[shorten >= 0.3em] (5) to (h);
	\draw[shorten >= 0.3em, shorten <= 0.3em] (h) to (e);
	\draw[shorten <= 0.3em, shorten >= 0.3em] (e) to (f);
	\draw[shorten <= 0.3em] (f) to (2);
	\coordinate (g) at (intersection of 0--4 and 5--3);
	\draw[shorten >= 0.3em] (0) to (g);
	\draw[shorten <= 0.3em, shorten >= 0.3em] (g) to (4);
	
	\ifthenelse{#1 = 1}{
		\node[above] at (0) {{$\scriptstyle{0}$}};	
		\node[right] at (1) {{$\scriptstyle{1}$}};
		\node[right] at (2) {{$\scriptstyle{2}$}};
		\node[below] at (3) {{$\scriptstyle{3}$}};
		\node[left] at (4) {{$\scriptstyle{4}$}};
		\node[left] at (5) {{$\scriptstyle{5}$}};}{}
	\end{tikzpicture}
}

\newcommand{\PPPTHREE}[1]{
	\begin{tikzpicture}[scale=1,baseline=0em, line width=0.8pt]
	\coordinate (0) at (0,1);
	\coordinate (1) at ({sin(60)} ,{cos(60)} );
	\coordinate (2) at ({sin(60)} ,{-cos(60)} );
	\coordinate (3) at (0,-1);
	\coordinate (4) at ({-sin(60)} ,{-cos(60)} );
	\coordinate (5) at ({-sin(60)} ,{cos(60)} );
	
	\draw[] (0) to (1);
	\draw[] (1) to (2);
	\draw[] (2) to (3);
	\draw[] (3) to (4);
	\draw[] (4) to (5);
	\draw[] (5) to (0);
	
	\draw[] (1) to (3);
	\draw[] (0) to (3);
	\draw[] (5) to (3);
	\coordinate (a) at (intersection of 3--1 and 0--2);
	\coordinate (k) at (intersection of 1--4 and 0--2);
	\draw[shorten >= 0.3em] (0) to (k);
	\draw[shorten <= 0.3em, shorten >= 0.3em] (k) to (a);
	\draw[shorten <= 0.3em] (a) to (2);
	\coordinate (b) at (intersection of 5--3 and 4--2);
	\coordinate (c) at (intersection of 0--3 and 4--2);
	\coordinate (d) at (intersection of 3--1 and 4--2);
	\draw[shorten >= 0.3em] (4) to (b);
	\draw[shorten <= 0.3em, shorten >= 0.3em] (b) to (c);
	\draw[shorten <= 0.3em, shorten >= 0.3em] (c) to (d);
	\draw[shorten <= 0.3em] (d) to (2);
	\coordinate (h) at (intersection of 0--4 and 5--2);
	\coordinate (e) at (intersection of 0--3 and 5--2);
	\coordinate (f) at (intersection of 3--1 and 5--2);
	\draw[shorten >= 0.3em] (5) to (h);
	\draw[shorten >= 0.3em, shorten <= 0.3em] (h) to (e);
	\draw[shorten <= 0.3em, shorten >= 0.3em] (e) to (f);
	\draw[shorten <= 0.3em] (f) to (2);
	\coordinate (g) at (intersection of 0--4 and 5--3);
	\draw[shorten >= 0.3em] (0) to (g);
	\draw[shorten <= 0.3em, shorten >= 0.3em] (g) to (4);
	\coordinate (i) at (intersection of 0--3 and 1--4);
	\coordinate (j) at (intersection of 5--3 and 1--4);
	\draw[shorten >= 0.3em] (1) to (i);
	\draw[shorten <= 0.3em, shorten >= 0.3em] (i) to (j);
	\draw[shorten <= 0.3em] (j) to (4);
	
	\ifthenelse{#1 = 1}{	
		\node[above] at (0) {{$\scriptstyle{0}$}};	
		\node[right] at (1) {{$\scriptstyle{1}$}};
		\node[right] at (2) {{$\scriptstyle{2}$}};
		\node[below] at (3) {{$\scriptstyle{3}$}};
		\node[left] at (4) {{$\scriptstyle{4}$}};
		\node[left] at (5) {{$\scriptstyle{5}$}};}{}
	\end{tikzpicture}
}

\newcommand{\PPPFOUR}[1]{
	\begin{tikzpicture}[scale=1,baseline=0em, line width=0.8pt]
	\coordinate (0) at (0,1);
	\coordinate (1) at ({sin(60)} ,{cos(60)} );
	\coordinate (2) at ({sin(60)} ,{-cos(60)} );
	\coordinate (3) at (0,-1);
	\coordinate (4) at ({-sin(60)} ,{-cos(60)} );
	\coordinate (5) at ({-sin(60)} ,{cos(60)} );
	
	\draw[] (0) to (1);
	\draw[] (1) to (2);
	\draw[] (2) to (3);
	\draw[] (3) to (4);
	\draw[] (4) to (5);
	\draw[] (5) to (0);
	
	\draw[] (1) to (3);
	\draw[] (0) to (3);
	\draw[] (5) to (3);
	\coordinate (a) at (intersection of 3--1 and 0--2);
	\coordinate (k) at (intersection of 1--4 and 0--2);
	\coordinate (m) at (intersection of 1--5 and 0--2);
	\draw[shorten >= 0.3em] (0) to (m);
	\draw[shorten <= 0.3em, shorten >= 0.3em] (m) to (k);
	\draw[shorten <= 0.3em, shorten >= 0.3em] (k) to (a);
	\draw[shorten <= 0.3em] (a) to (2);
	\coordinate (b) at (intersection of 5--3 and 4--2);
	\coordinate (c) at (intersection of 0--3 and 4--2);
	\coordinate (d) at (intersection of 3--1 and 4--2);
	\draw[shorten >= 0.3em] (4) to (b);
	\draw[shorten <= 0.3em, shorten >= 0.3em] (b) to (c);
	\draw[shorten <= 0.3em, shorten >= 0.3em] (c) to (d);
	\draw[shorten <= 0.3em] (d) to (2);
	\coordinate (h) at (intersection of 0--4 and 5--2);
	\coordinate (e) at (intersection of 0--3 and 5--2);
	\coordinate (f) at (intersection of 3--1 and 5--2);
	\draw[shorten >= 0.3em] (5) to (h);
	\draw[shorten >= 0.3em, shorten <= 0.3em] (h) to (e);
	\draw[shorten <= 0.3em, shorten >= 0.3em] (e) to (f);
	\draw[shorten <= 0.3em] (f) to (2);
	\coordinate (g) at (intersection of 0--4 and 5--3);
	\draw[shorten >= 0.3em] (0) to (g);
	\draw[shorten <= 0.3em, shorten >= 0.3em] (g) to (4);
	\coordinate (i) at (intersection of 0--3 and 1--4);
	\coordinate (j) at (intersection of 5--3 and 1--4);
	\draw[shorten >= 0.3em] (1) to (i);
	\draw[shorten <= 0.3em, shorten >= 0.3em] (i) to (j);
	\draw[shorten <= 0.3em] (j) to (4);
	\coordinate (l) at (intersection of 0--4 and 5--1);
	\coordinate (m) at (intersection of 0--3 and 5--1);
	\draw[shorten >= 0.3em] (5) to (l);
	\draw[shorten <= 0.3em, shorten >= 0.3em] (l) to (m);
	\draw[shorten <= 0.3em] (m) to (1);
	
	\ifthenelse{#1 = 1}{	
		\node[above] at (0) {{$\scriptstyle{0}$}};	
		\node[right] at (1) {{$\scriptstyle{1}$}};
		\node[right] at (2) {{$\scriptstyle{2}$}};
		\node[below] at (3) {{$\scriptstyle{3}$}};
		\node[left] at (4) {{$\scriptstyle{4}$}};
		\node[left] at (5) {{$\scriptstyle{5}$}};}{}
	\end{tikzpicture}
}

\newcommand{\PPPFIVE}[1]{
	\begin{tikzpicture}[scale=1,baseline=0em, line width=0.8pt]
	\coordinate (0) at (0,1);
	\coordinate (1) at ({sin(60)} ,{cos(60)} );
	\coordinate (2) at ({sin(60)} ,{-cos(60)} );
	\coordinate (3) at (0,-1);
	\coordinate (4) at ({-sin(60)} ,{-cos(60)} );
	\coordinate (5) at ({-sin(60)} ,{cos(60)} );
	
	\draw[] (0) to (1);
	\draw[] (1) to (2);
	\draw[] (2) to (3);
	\draw[] (3) to (4);
	\draw[] (4) to (5);
	\draw[] (5) to (0);
	
	\draw[] (1) to (3);
	\draw[] (0) to (3);
	\draw[] (5) to (3);
	\coordinate (a) at (intersection of 3--1 and 0--2);
	\coordinate (k) at (intersection of 1--4 and 0--2);
	\coordinate (m) at (intersection of 1--5 and 0--2);
	\draw[shorten >= 0.3em] (0) to (m);
	\draw[shorten <= 0.3em, shorten >= 0.3em] (m) to (k);
	\draw[shorten <= 0.3em, shorten >= 0.3em] (k) to (a);
	\draw[shorten <= 0.3em] (a) to (2);
	\coordinate (b) at (intersection of 5--3 and 4--2);
	\coordinate (c) at (intersection of 0--3 and 4--2);
	\coordinate (d) at (intersection of 3--1 and 4--2);
	\draw[shorten >= 0.3em] (4) to (b);
	\draw[shorten <= 0.3em, shorten >= 0.3em] (b) to (c);
	\draw[shorten <= 0.3em, shorten >= 0.3em] (c) to (d);
	\draw[shorten <= 0.3em] (d) to (2);
	\coordinate (e) at (intersection of 0--3 and 5--2);
	\coordinate (f) at (intersection of 3--1 and 5--2);
	\draw[shorten >= 0.3em] (5) to (e);
	\draw[shorten <= 0.3em, shorten >= 0.3em] (e) to (f);
	\draw[shorten <= 0.3em] (f) to (2);
	\coordinate (m) at (intersection of 0--3 and 5--1);
	\draw[shorten >= 0.3em] (5) to (m);
	\draw[shorten <= 0.3em] (m) to (1);
	\coordinate (i) at (intersection of 0--3 and 1--4);
	\coordinate (j) at (intersection of 5--3 and 1--4);
	\draw[shorten >= 0.3em] (1) to (i);
	\draw[shorten <= 0.3em, shorten >= 0.3em] (i) to (j);
	\draw[shorten <= 0.3em] (j) to (4);
	
	\ifthenelse{#1 = 1}{	
		\node[above] at (0) {{$\scriptstyle{0}$}};	
		\node[right] at (1) {{$\scriptstyle{1}$}};
		\node[right] at (2) {{$\scriptstyle{2}$}};
		\node[below] at (3) {{$\scriptstyle{3}$}};
		\node[left] at (4) {{$\scriptstyle{4}$}};
		\node[left] at (5) {{$\scriptstyle{5}$}};}{}
	\end{tikzpicture}
}

\newcommand{\PPPSIX}[1]{
	\begin{tikzpicture}[scale=1,baseline=0em, line width=0.8pt]
	\coordinate (0) at (0,1);
	\coordinate (1) at ({sin(60)} ,{cos(60)} );
	\coordinate (2) at ({sin(60)} ,{-cos(60)} );
	\coordinate (3) at (0,-1);
	\coordinate (4) at ({-sin(60)} ,{-cos(60)} );
	\coordinate (5) at ({-sin(60)} ,{cos(60)} );
	
	\draw[] (0) to (1);
	\draw[] (1) to (2);
	\draw[] (2) to (3);
	\draw[] (3) to (4);
	\draw[] (4) to (5);
	\draw[] (5) to (0);
	
	\draw[] (1) to (3);
	\draw[] (0) to (3);
	\draw[] (5) to (3);
	\coordinate (a) at (intersection of 3--1 and 0--2);
	\coordinate (m) at (intersection of 1--5 and 0--2);
	\draw[shorten >= 0.3em] (0) to (m);
	\draw[shorten <= 0.3em, shorten >= 0.3em] (m) to (a);
	\draw[shorten <= 0.3em] (a) to (2);
	\coordinate (b) at (intersection of 5--3 and 4--2);
	\coordinate (c) at (intersection of 0--3 and 4--2);
	\coordinate (d) at (intersection of 3--1 and 4--2);
	\draw[shorten >= 0.3em] (4) to (b);
	\draw[shorten <= 0.3em, shorten >= 0.3em] (b) to (c);
	\draw[shorten <= 0.3em, shorten >= 0.3em] (c) to (d);
	\draw[shorten <= 0.3em] (d) to (2);
	\coordinate (e) at (intersection of 0--3 and 5--2);
	\coordinate (f) at (intersection of 3--1 and 5--2);
	\draw[shorten >= 0.3em] (5) to (e);
	\draw[shorten <= 0.3em, shorten >= 0.3em] (e) to (f);
	\draw[shorten <= 0.3em] (f) to (2);
	\coordinate (m) at (intersection of 0--3 and 5--1);
	\draw[shorten >= 0.3em] (5) to (m);
	\draw[shorten <= 0.3em] (m) to (1);
	
	\ifthenelse{#1 = 1}{	
		\node[above] at (0) {{$\scriptstyle{0}$}};	
		\node[right] at (1) {{$\scriptstyle{1}$}};
		\node[right] at (2) {{$\scriptstyle{2}$}};
		\node[below] at (3) {{$\scriptstyle{3}$}};
		\node[left] at (4) {{$\scriptstyle{4}$}};
		\node[left] at (5) {{$\scriptstyle{5}$}};}{}
	\end{tikzpicture}
}

\newcommand{\assocTWO}[0]{
	\begin{tikzpicture}[scale=1,baseline=0em]
	\matrix[matrix of math nodes,row sep =-2.5em,column sep=3.5em, ampersand replacement=\&] (m) {
		{}  \&[-1em] \PPPTWO{1} \& \PPPTHREE{1} \&[-1em] {}
		\\
		\PPPONE{1} \& {} \& {} \& \PPPFOUR{1}
		\\
		{} \& \PPPSIX{1}  \& \PPPFIVE{1} \& {} \\
	};	
	\path
	(m-2-1) edge[->, shorten <= -0.3em, shorten >= -0.3em, line width=0.4pt] node[above, sloped] {${\sss \omega(g[02345])}$}  (m-1-2)
	edge[->, shorten <= -0.3em, shorten >= -0.3em, line width=0.4pt] node[above, sloped] {$\sss{\omega(g[01235])}$} (m-3-2)
	(m-1-2) edge[->, shorten <= 0em, shorten >= 0em, line width=0.4pt] node[above] {$\sss{\omega(g[01234])}$}(m-1-3)
	(m-3-2) edge[->, shorten <= 0em, shorten >= 0em, line width=0.4pt] node[above] {$\sss{\omega(g[12345])}$} (m-3-3)
	(m-2-4) edge[<-, shorten <= -0.3em, shorten >= -0.3em, line width=0.4pt]  node[above, sloped] {$\sss{\omega(g[01245])}$}(m-1-3)
	edge[<-, shorten <= -0.3em, shorten >= -0.3em, line width=0.4pt]  node[above, sloped] {$\sss{\omega(g[01345])}$} (m-3-3)
	;
	\end{tikzpicture}
}

\newcommand{\loopMoveA}[1]{
	\begin{tikzpicture}[scale=0.3,baseline=3
	em]
	\coordinate (1) at (0,0);
	\coordinate (2) at (9, 0);
	\coordinate (3) at (-1.5, 14);
	\coordinate (4) at (1.5, 14);
	\coordinate (5) at (7.5, 14);
	\coordinate (6) at (10.5, 14);
	\coordinate (7) at (0, -4);
	\coordinate (8) at (9, -4);		
	\coordinate (9) at (-1.5, 4);
	\coordinate (10) at (1.5, 4);
	\coordinate (11) at (7.5, 4);
	\coordinate (12) at (10.5, 4);
	\coordinate (13) at (0, 9);
	\coordinate (14) at (9, 9);
	\coordinate (30) at (4.5 ,14);
	\coordinate (31) at (4.5, -4);
	\coordinate (40) at (0, 5);	
	\coordinate (41) at (0, 11);	
	\coordinate (42) at (2.8, 5);	
	\coordinate (43) at (2.8, 11);
	\coordinate (44) at (-3, 12);
	\coordinate (45) at (-3, -6);				
	
	\def\pa{0.2};
	\def\pb{0.3};
	\def\pc{0.4};
	
	\ifthenelse{#1 =1}{
		\draw[name path=P36d] (3) to[out=-90, in=270, looseness=0.5, edge node={coordinate[pos=\pa] (15)}, edge node={coordinate[pos=\pb] (16)}] (6);	
		\draw[name path=P36u] (3) to[out=90, in=-270, looseness=0.5] (6);
		
		\draw[name path=P45d] (4) to[out=-90, in=270, looseness=0.5, edge node={coordinate[pos=\pb] (17)}, edge node={coordinate[pos=\pc] (18)}] (5);	
		\draw[name path=P36u] (4) to[out=90, in=-270, looseness=0.5] (5);
		
		\path[name path=P12d] (1) to[out=-90, in=270, looseness=0.5, edge node={coordinate[pos=\pa] (19)}, edge node={coordinate[pos=\pb] (20)}, edge node={coordinate[pos=\pc] (29)}] (2);	
		\draw[op1, name path=P12u] (1) to[out=90, in=-270, looseness=0.5] (2);
		\speC{1}{19}{2}{op4};
		
		\path[name path=P78d] (7) to[out=-90, in=270, looseness=0.5, edge node={coordinate[pos=\pa] (21)}, edge node={coordinate[pos=\pc] (22)}] (8);	
		\draw[op1, name path=P78u] (7) to[out=90, in=-270, looseness=0.5] (8);
		\speC{7}{21}{8}{op4};
		
		\path[name path=P912d] (9) to[out=-90, in=270, looseness=0.5, edge node={coordinate[pos=\pa] (23)}, edge node={coordinate[pos=\pb] (24)}] (12);	
		\draw[op1, name path=P912u] (9) to[out=90, in=-270, looseness=0.5] (12);
		\speC{9}{23}{12}{op4};
		
		\draw[name path=P1011d, op3] (10) to[out=-90, in=270, looseness=0.5, edge node={coordinate[pos=\pb] (25)}, edge node={coordinate[pos=\pc] (26)}] (11);
		\draw[op1, name path=P1011u] (10) to[out=90, in=-270, looseness=0.5] (11);

		\path[name path=P1544] (15) -- (44);
		\path[name path=P1617] (16) -- (17);
		\path (13) ++ (0.3em,0.3em) coordinate (sp13A);
		\path (13) ++ (-0.3em,-0.3em) coordinate (sp13B);
		\path[name path=P313] (3) to[out=-90, in=135, looseness=1] (13);
		\speD{3}{13}{15}{44}{}{op4};
		\path[name path=P4sp13A] (4) to[out=-90, in=45, looseness=1] (sp13A);
		\speE{4}{sp13A}{16}{17}{}{op2};
		\draw[name path=P913, op4] (9) to[out=90, in=-135, looseness=1] (sp13B);
		\draw[name path=P1013, op2] (10) to[out=90, in=-45, looseness=1] (13);

		\path (14) ++ (-0.3em,0.3em) coordinate (sp14A);
		\path (14) ++ (0.3em,-0.3em) coordinate (sp14B);
		\path[name path=P5sp14A] (5) to[out=-90, in=135, looseness=1] (sp14A);
		\speD{5}{sp14A}{3}{6d}{}{op3};
		\draw[name path=P614] (6) to[out=-90, in=45, looseness=1] (14);
		\draw[name path=P1114, op3] (11) to[out=90, in=-135, looseness=1] (14);
		\draw[name path=P1214] (12) to[out=90, in=-45, looseness=1] (sp14B);

		\path[] (sp13B) to[out=-90, in=270, looseness=0.5, edge node={coordinate[pos=0.26] (27B)}, edge node={coordinate[pos=0.30] (28B)}] (sp14B);
		\speCC{sp13B}{27B}{28B}{sp14B}{op4};	
		\path[] (sp13A) to[out=-90, in=270, looseness=0.5, edge node={coordinate[pos=0.25] (27A)}, edge node={coordinate[pos=0.29] (28A)}] (sp14A);
		\speCC{sp13A}{27A}{28A}{sp14A}{op2};
		
		\draw[line width=0.8pt, op1] (sp13B) to[out=90, in=-270, looseness=0.5] (sp14B);
		\draw[line width=0.8pt, op1] (sp13A) to[out=90, in=-270, looseness=0.5] (sp14A);

		\draw[name path=P2340] (23) -- (40);
		\draw[name path=P2440] (24) -- (40);
		\draw[name path=P1541] (15) -- (41);
		\draw[name path=P1641] (16) -- (41);
		
		\draw[name path=P2542, op3] (25) -- (42);
		\draw[name path=P2642, op3] (26) -- (42);
		\speL{17}{43}{3}{6d}{}{op3};
		\speL{18}{43}{3}{6d}{}{op3};	
		
		\draw[op4] (27B) to[out=-135, in=90, looseness=1] (40);
		\draw[] (28B) to[out=-135, in=90, looseness=1] (40);
		\draw[op2] (27A) to[out=45, in=-90, looseness=1] (43);
		\draw[op3] (28A) to[out=45, in=-90, looseness=1] (43);
		\draw[op4] (27A) to[out=135, in=-90, looseness=1] (41);
		\draw[] (28A) to[out=135, in=-90, looseness=1] (41);
		\draw[op2] (27B) to[out=-45, in=90, looseness=1] (42);
		\draw[op3] (28B) to[out=-45, in=90, looseness=1] (42);

		\speL{30}{31}{4}{5d}{}{op2};
		\draw[op4] (1) -- (7);
		\draw[op4] (1) -- (9);
		\draw[] (2) -- (8);
		\draw[] (2) -- (12);
		\draw[op2] (1) -- (10);
		\draw[op3] (2) -- (11);
		\draw[] (21) -- (19);
		\draw[] (19) -- (23);
		\draw[] (22) -- (29);
		\draw[] (18) -- (30);
		\draw[op2] (22) -- (31);
		\draw[op3] (20) -- (25);
		\draw[] (20) -- (24);
		\draw[op3] (29) -- (26);
		\draw[] (44) -- (45);
		\draw[] (44) -- (15);
		\draw[] (45) -- (21);
		\draw[] (16) -- (17);
		
		\node[draw,circle,scale=0.3,fill] at (19) {};
		\node[draw,circle,scale=0.3,fill] at (20) {};
		\node[draw,circle,scale=0.3,fill] at (29) {};
				
		\node[] at (10,12.5) {{$\sss{a}$}};
		\node[] at (7,12.9) {{$\sss{b}$}};
		\node[] at (10,4) {{$\sss{b}$}};
		\node[opacity=0.8] at (7,4) {{$\sss{a}$}};
		\node[] at (-2,3) {{$\sss{x}$}};
		\node[opacity=0.8] at (2,11) {{$\sss{x}$}};
		\node[opacity=0.8] at (2,6) {{$\sss{x}$}};
		\node[opacity=0.6] at (3.5,7) {{$\sss{x}$}};	
	}{}

	\ifthenelse{#1 = 2}{
		\draw[name path=P36d] (3) to[out=-90, in=270, looseness=0.5, edge node={coordinate[pos=\pa] (15)}, edge node={coordinate[pos=\pb] (16)}] (6);	
		\draw[name path=P36u] (3) to[out=90, in=-270, looseness=0.5] (6);
		
		\draw[name path=P45d] (4) to[out=-90, in=270, looseness=0.5, edge node={coordinate[pos=\pb] (17)}, edge node={coordinate[pos=\pc] (18)}] (5);	
		\draw[name path=P36u] (4) to[out=90, in=-270, looseness=0.5] (5);
		
		\path[name path=P12d] (1) to[out=-90, in=270, looseness=0.5, edge node={coordinate[pos=\pa] (19)}, edge node={coordinate[pos=\pb] (20)}, edge node={coordinate[pos=\pc] (29)}] (2);	
		\draw[op1, name path=P12u] (1) to[out=90, in=-270, looseness=0.5] (2);
		\speC{1}{19}{2}{op4};
		
		\path[name path=P78d] (7) to[out=-90, in=270, looseness=0.5, edge node={coordinate[pos=\pa] (21)}, edge node={coordinate[pos=\pc] (22)}] (8);	
		\draw[op1, name path=P78u] (7) to[out=90, in=-270, looseness=0.5] (8);
		\speC{7}{21}{8}{op4};
		
		\path[name path=P912d] (9) to[out=-90, in=270, looseness=0.5, edge node={coordinate[pos=\pa] (23)}, edge node={coordinate[pos=\pb] (24)}] (12);	
		\draw[op1, name path=P912u] (9) to[out=90, in=-270, looseness=0.5] (12);
		\speC{9}{23}{12}{op4};
		
		\draw[name path=P1011d, op3] (10) to[out=-90, in=270, looseness=0.5, edge node={coordinate[pos=\pb] (25)}, edge node={coordinate[pos=\pc] (26)}] (11);
		\draw[op1, name path=P1011u] (10) to[out=90, in=-270, looseness=0.5] (11);

		\path[name path=P1544] (15) -- (44);
		\path[name path=P1617] (16) -- (17);	
		
		\speL{30}{31}{4}{5d}{}{op2};
		\draw[op4] (1) -- (7);
		\draw[op4] (1) -- (9);
		\draw[] (2) -- (8);
		\draw[] (2) -- (12);
		\draw[op2] (1) -- (10);
		\draw[op3] (2) -- (11);
		\draw[] (21) -- (19);
		\draw[] (19) -- (23);
		\draw[] (22) -- (29);
		\draw[] (18) -- (30);
		\draw[op2] (22) -- (31);
		\draw[op3] (20) -- (25);
		\draw[] (20) -- (24);
		\draw[op3] (29) -- (26);
		\draw[] (44) -- (45);
		\draw[] (44) -- (15);
		\draw[] (45) -- (21);
		\draw[name path = P1617] (16) -- (17);
		
		\draw[] (6) -- (12);
		\speL{5}{11}{3}{6d}{}{op3};
		\speL{18}{26}{3}{6d}{}{op3};
		\speL{17}{25}{3}{6d}{}{op3};
		\draw[] (16) -- (24);
		\speL{4}{10}{16}{17}{}{op2};
		\draw[] (15) -- (23);
		\speL{3}{9}{15}{44}{}{op4};
		
		\node[draw,circle,scale=0.3,fill] at (19) {};
		\node[draw,circle,scale=0.3,fill] at (20) {};
		\node[draw,circle,scale=0.3,fill] at (29) {};
		
		\node[] at (10,12.5) {{$\sss{a}$}};
		\node[] at (7,12.9) {{$\sss{b}$}};
		\node[] at (10,4) {{$\sss{a}$}};
		\node[opacity=0.8] at (7,4) {{$\sss{b}$}};
		\node[] at (-2,3) {{$\sss{x}$}};
		\node[opacity=0.8] at (2,8) {{$\sss{x}$}};
		\node[opacity=0.6] at (4.1,8) {{$\sss{x}$}};
	}{}
	
	\end{tikzpicture}
}

\newcommand{\loopMoveAred}[0]{
	\begin{tikzpicture}[scale=0.3,baseline=7em]
	\coordinate (3) at (-1.5, 14);
	\coordinate (4) at (1.5, 14);
	\coordinate (5) at (7.5, 14);
	\coordinate (6) at (10.5, 14);		
	\coordinate (9) at (-1.5, 4);
	\coordinate (10) at (1.5, 4);
	\coordinate (11) at (7.5, 4);
	\coordinate (12) at (10.5, 4);
	\coordinate (13) at (0, 9);
	\coordinate (14) at (9, 9);
	\coordinate (30) at (4.5 ,14);
	\coordinate (31) at (4.5, 4);
	\coordinate (40) at (0, 5);	
	\coordinate (41) at (0, 11);	
	\coordinate (42) at (2.8, 5);	
	\coordinate (43) at (2.8, 11);
	\coordinate (44) at (-3, 12);
	\coordinate (45) at (-3, 2);				
	
	\def\pa{0.2};
	\def\pb{0.3};
	\def\pc{0.4};

	\draw[name path=P36d] (3) to[out=-90, in=270, looseness=0.5, edge node={coordinate[pos=\pa] (15)}, edge node={coordinate[pos=\pb] (16)}] (6);	
	\draw[name path=P36u] (3) to[out=90, in=-270, looseness=0.5] (6);
	
	\draw[name path=P45d] (4) to[out=-90, in=270, looseness=0.5, edge node={coordinate[pos=\pb] (17)}, edge node={coordinate[pos=\pc] (18)}] (5);	
	\draw[name path=P36u] (4) to[out=90, in=-270, looseness=0.5] (5);

	\path[name path=P912d] (9) to[out=-90, in=270, looseness=0.5, edge node={coordinate[pos=\pa] (23)}, edge node={coordinate[pos=\pb] (24)}] (12);	
	\draw[op1, name path=P912u] (9) to[out=90, in=-270, looseness=0.5] (12);
	\speC{9}{23}{12}{op4};
	
	\draw[name path=P1011d, op3] (10) to[out=-90, in=270, looseness=0.5, edge node={coordinate[pos=\pb] (25)}, edge node={coordinate[pos=\pc] (26)}] (11);
	\draw[op1, name path=P1011u] (10) to[out=90, in=-270, looseness=0.5] (11);

	\path[name path=P1544] (15) -- (44);
	\path[name path=P1617] (16) -- (17);
	\path (13) ++ (0.3em,0.3em) coordinate (sp13A);
	\path (13) ++ (-0.3em,-0.3em) coordinate (sp13B);
	\path[name path=P313] (3) to[out=-90, in=135, looseness=1] (13);
	\speD{3}{13}{15}{44}{}{op4};
	\path[name path=P4sp13A] (4) to[out=-90, in=45, looseness=1] (sp13A);
	\speE{4}{sp13A}{16}{17}{}{op2};
	\draw[name path=P913, op4] (9) to[out=90, in=-135, looseness=1] (sp13B);
	\draw[name path=P1013, op2] (10) to[out=90, in=-45, looseness=1] (13);

	\path (14) ++ (-0.3em,0.3em) coordinate (sp14A);
	\path (14) ++ (0.3em,-0.3em) coordinate (sp14B);
	\path[name path=P5sp14A] (5) to[out=-90, in=135, looseness=1] (sp14A);
	\speD{5}{sp14A}{3}{6d}{}{op3};
	\draw[name path=P614] (6) to[out=-90, in=45, looseness=1] (14);
	\draw[name path=P1114, op3] (11) to[out=90, in=-135, looseness=1] (14);
	\draw[name path=P1214] (12) to[out=90, in=-45, looseness=1] (sp14B);
	
	\path[] (sp13B) to[out=-90, in=270, looseness=0.5, edge node={coordinate[pos=0.26] (27B)}, edge node={coordinate[pos=0.30] (28B)}] (sp14B);
	\speCC{sp13B}{27B}{28B}{sp14B}{op4};	
	\path[] (sp13A) to[out=-90, in=270, looseness=0.5, edge node={coordinate[pos=0.25] (27A)}, edge node={coordinate[pos=0.29] (28A)}] (sp14A);
	\speCC{sp13A}{27A}{28A}{sp14A}{op2};
	
	\draw[line width=0.8pt, op1] (sp13B) to[out=90, in=-270, looseness=0.5] (sp14B);
	\draw[line width=0.8pt, op1] (sp13A) to[out=90, in=-270, looseness=0.5] (sp14A);
	
	\draw[name path=P2340] (23) -- (40);
	\draw[name path=P2440] (24) -- (40);
	\draw[name path=P1541] (15) -- (41);
	\draw[name path=P1641] (16) -- (41);
	
	\draw[name path=P2542, op3] (25) -- (42);
	\draw[name path=P2642, op3] (26) -- (42);
	\speL{17}{43}{3}{6d}{}{op3};
	\speL{18}{43}{3}{6d}{}{op3};	
	
	\draw[op4] (27B) to[out=-135, in=90, looseness=1] (40);
	\draw[] (28B) to[out=-135, in=90, looseness=1] (40);
	\draw[op2] (27A) to[out=45, in=-90, looseness=1] (43);
	\draw[op3] (28A) to[out=45, in=-90, looseness=1] (43);
	\draw[op4] (27A) to[out=135, in=-90, looseness=1] (41);
	\draw[] (28A) to[out=135, in=-90, looseness=1] (41);
	\draw[op2] (27B) to[out=-45, in=90, looseness=1] (42);
	\draw[op3] (28B) to[out=-45, in=90, looseness=1] (42);
	
	\speL{30}{31}{4}{5d}{}{op2};
	\draw[] (18) -- (30);
	\draw[] (44) -- (45);
	\draw[] (44) -- (15);
	\draw[] (16) -- (17);
	\draw[] (45) -- (23);
	\draw[op3] (24) -- (25);
	\draw[op2] (26) -- (31); 
	
	\node[] at (10,12.5) {{$\sss{a}$}};
	\node[] at (7,12.9) {{$\sss{b}$}};
	\node[] at (10,4) {{$\sss{b}$}};
	\node[opacity=0.8] at (7,4) {{$\sss{a}$}};
	\node[] at (-2,3) {{$\sss{x}$}};
	\node[opacity=0.8] at (2,11) {{$\sss{x}$}};
	\node[opacity=0.8] at (2,6) {{$\sss{x}$}};
	\node[opacity=0.6] at (3.5,7) {{$\sss{x}$}};		
	\end{tikzpicture}
}

\newcommand{\loopMovie}[0]{
	\begin{tikzpicture}[scale=0.3,baseline=7em]
	\coordinate (0) at (-3, 16);
	\coordinate (1) at (-1.5, 18);
	\coordinate (2) at (1.5, 18);
	\coordinate (3) at (7.5, 18);
	\coordinate (4) at (10.5, 18);
	\coordinate (5) at (4.5 ,18);
	
	\coordinate (10) at (-3,1.2);		
	\coordinate (11) at (-1.5,3.2);
	\coordinate (12) at (1.5, 3.2);
	\coordinate (13) at (7.5, 3.2);
	\coordinate (14) at (10.5,3.2);
	\coordinate (15) at (4.5, 3.2);
	
	\coordinate (20) at (-3, 12.3);
	\coordinate (21) at (-0.75, 14.3);
	\coordinate (22) at (0.75, 14.3);
	\coordinate (23) at (8.25, 14.3);
	\coordinate (24) at (9.75, 14.3);
	\coordinate (25) at (4.5 ,14.3);
	
	\coordinate (30) at (-3, 8.6);
	\coordinate (31) at (0, 11);
	\coordinate (32) at (0, 10.6);
	\coordinate (33) at (9, 10.6);
	\coordinate (34) at (9, 11);
	\coordinate (35) at (4.5 ,10.6);
	
	\coordinate (50) at (-3, 4.9);
	\coordinate (51) at (-0.75,6.9);
	\coordinate (52) at (0.75,6.9);
	\coordinate (53) at (8.25,6.9);
	\coordinate (54) at (9.75,6.9);
	\coordinate (55) at (4.5,6.9);	
	
	\coordinate (60) at (14,18);
	\coordinate (61) at (14,3.2);	
						
	\def\pa{0.2};
	\def\pb{0.3};
	\def\pc{0.4};
		
	\draw[name path=P14d] (1) to[out=-90, in=270, looseness=0.5, edge node={coordinate[pos=\pa] (6)}, edge node={coordinate[pos=\pb] (7)}] (4);	
	\draw[name path=P14u] (1) to[out=90, in=-270, looseness=0.5] (4);	
	\draw[name path=P23d] (2) to[out=-90, in=270, looseness=0.5, edge node={coordinate[pos=\pb] (8)}, edge node={coordinate[pos=\pc] (9)}] (3);	
	\draw[name path=P23] (2) to[out=90, in=-270, looseness=0.5] (3);
	\draw[] (0) -- (6);
	\draw[] (7) -- (8);
	\draw[] (9) -- (5);
	
	\draw[name path=P2124d] (21) to[out=-90, in=270, looseness=0.5] (24);	
	\draw[name path=P2124u] (21) to[out=90, in=-270, looseness=0.5] (24);	
	\draw[name path=P2223d] (22) to[out=-90, in=270, looseness=0.5] (23);	
	\draw[name path=P2223u] (22) to[out=90, in=-270, looseness=0.5] (23);
	\interLLp{20}{25}{21}{24d}{22}{23d};
	
	\draw[name path=P3233d] (32) to[out=-90, in=270, looseness=0.5] (33);	
	\draw[name path=P3233u] (32) to[out=90, in=-270, looseness=0.5] (33);	
	\draw[white, line width=0.35em] (31) to[out=-90, in=270, looseness=0.5] (34);
	\draw[white, line width=0.35em] (31) to[out=90, in=-270, looseness=0.5] (34);			
	\draw[name path=P3134u] (31) to[out=90, in=-270, looseness=0.5] (34);
	\draw[name path=P3134d] (31) to[out=-90, in=270, looseness=0.5] (34);
	\interLLp{30}{35}{32}{33d}{31}{34d};
		
	\draw[name path=P5154d] (51) to[out=-90, in=270, looseness=0.5] (54);	
	\draw[name path=P5154u] (51) to[out=90, in=-270, looseness=0.5] (54);	
	\draw[name path=P5253d] (52) to[out=-90, in=270, looseness=0.5] (53);	
	\draw[name path=P5253u] (52) to[out=90, in=-270, looseness=0.5] (53);
	\interLLp{50}{55}{51}{54d}{52}{53d};
		
	\draw[] (11) to[out=-90, in=270, looseness=0.5, edge node={coordinate[pos=\pa] (16)}, edge node={coordinate[pos=\pb] (17)}] (14);	
	\draw[] (11) to[out=90, in=-270, looseness=0.5] (14);
	\draw[] (12) to[out=-90, in=270, looseness=0.5, edge node={coordinate[pos=\pb] (18)}, edge node={coordinate[pos=\pc] (19)}] (13);
	\draw[] (12) to[out=90, in=-270, looseness=0.5] (13);
	\draw[] (10) -- (16);
	\draw[] (17) -- (18);
	\draw[] (19) -- (15);
	
	\draw[->, >=Triangle] (60) -- (61);
	
	\foreach \x in {0,10,20,30,50} 
		\node[above] at (\x)  {{$\sss{x}$}};
	\foreach \x in {5,15,25,35,55} 
		\node[right] at (\x)  {{$\sss{x}$}};
	\node[above] at (7)  {{$\sss{x}$}};	
	\node[above] at (17)  {{$\sss{x}$}};
	\node[right] at (3) {{$\sss{b}$}};
	\node[right] at (4) {{$\sss{a}$}};
	\node[right] at (13) {{$\sss{a}$}};
	\node[right] at (14) {{$\sss{b}$}};
	\node[right] at (23) {{$\sss{b}$}};
	\node[right] at (24) {{$\sss{a}$}};
	\node[right, yshift=0.1em] at (34) {{$\sss{a}$}};
	\node[right, yshift=-0.1em] at (33) {{$\sss{b}$}};
	\node[right] at (53) {{$\sss{a}$}};
	\node[right] at (54) {{$\sss{b}$}};

	\end{tikzpicture}
}

\newcommand{\slantTwoTwoA}[1]{
	\begin{tikzpicture}[scale=0.3,baseline=2em]
	\coordinate (1) at (0, 0);
	\coordinate (2) at (12, 0);
	\coordinate (3) at (-3, 14);
	\coordinate (4) at (0, 14);
	\coordinate (5) at (3, 14);
	\coordinate (6) at (9, 14);
	\coordinate (7) at (12, 14);
	\coordinate (8) at (15, 14);
	\coordinate (11) at (0, -4);
	\coordinate (12) at (12, -4);
	\coordinate (13) at (-4,12.3);
	\coordinate (14) at (-4,-6.2);
	\coordinate (30) at (6,14);
	\coordinate (31) at (6,-4);

	\ifthenelse{#1 =1}{	
		\coordinate (9) at (-1.5, 7);
		\coordinate (10) at (13.5, 7);
		\def\pa{0.15};
		\def\pb{0.2};
		\def\pc{0.25};
		\def\pd{0.34};
		
		\draw[name path=P38d] (3) to[out=-90, in=270, looseness=0.5, edge node={coordinate[pos=\pa] (17)}, edge node={coordinate[pos=\pb] (20)}] (8);
		
		\draw[name path=P38u] (3) to[out=90, in=-270, looseness=0.5] (8);
			
		\draw[name path=P47d] (4) to[out=-90, in=270, looseness=0.5, edge node={coordinate[pos=\pb] (26)}, edge node={coordinate[pos=\pc] (27)}] (7);
		
		\draw[name path=P47u] (4) to[out=90, in=-270, looseness=0.5] (7);
		
		\draw[name path=P56d] (5) to[out=-90, in=270, looseness=0.5, edge node={coordinate[pos=\pc] (28)}, edge node={coordinate[pos=\pd] (29)}] (6);
		
		\draw[name path=P56u] (5) to[out=90, in=-270, looseness=0.5] (6);
		
		\path[name path=P12d] (1) to[out=-90, in=270, looseness=0.5, edge node={coordinate[pos=\pa] (18)}, edge node={coordinate[pos=\pc] (23)}, edge node={coordinate[pos=\pd] (24)}] (2);
		\speC{1}{18}{2}{op4};
		
		\path[name path=P910d] (9) to[out=-90, in=270, looseness=0.5, edge node={coordinate[pos=\pb] (21)}, edge node={coordinate[pos=\pc] (22)}] (10);
		\draw[name path=P1718] (17) -- (18);
		\path [name intersections={of=P910d and P1718,by={15}}];
		\speC{9}{15}{10}{op4};
		
		\draw[op1,name path=P910u] (9) to[out=90, in=-270, looseness=0.5] (10);
		
		\draw[op1,name path=P12u] (1) to[out=90, in=-270, looseness=0.5] (2);	
		
		\path[name path=P1112d] (11) to[out=-90, in=270, looseness=0.5, edge node={coordinate[pos=\pa] (19)}, edge node={coordinate[pos=\pd] (25)}] (12);
		\speC{11}{19}{12}{op4};		
		\draw[op1,name path=P1112u] (11) to[out=90, in=-270, looseness=0.5] (12);
		\draw[name path=P1317] (13) -- (17);	
		
		\draw[name path=P28] (2) -- (8);
		\draw[name path=P212] (2) -- (12);
		\draw[name path=P1314] (13) -- (14);
		\draw[] (18) -- (19);
		\draw[] (20) -- (21);
		\draw[] (22) -- (23);
		\draw[] (24) -- (25);
		\draw[name path=P2026] (20) -- (26);
		\draw[name path=P2728] (27) -- (28);
		\draw[op4] (1) -- (11);
		\draw[op4] (1) -- (9);
		\draw[] (14) -- (19);
		\draw[] (29) -- (30);
		\draw[opacity=0.2] (25) -- (31);
		
		\speL{3}{9}{13}{17}{}{op4};
		\speL{7}{10}{3}{8d}{}{op4};
		\speL{6}{2}{4}{7d}{}{op3};
		\speL{4}{9}{20}{26}{}{op3};
		\speL{5}{1}{27}{28}{}{op2};
		\speL{29}{24}{4}{7d}{}{op3};	
		\speL{26}{21}{3}{8d}{}{op4};	
		\speL{27}{22}{3}{8d}{}{op4};
		\speL{28}{23}{4}{7d}{}{op3};
		\speL{30}{31}{5}{6d}{}{op2};	
		
		\node[draw,circle,scale=0.3,fill] at (15) {};
		\node[draw,circle,scale=0.3,fill] at (21) {};
		\node[draw,circle,scale=0.3,fill] at (22) {};
		\node[draw,circle,scale=0.3,fill] at (18) {};
		\node[draw,circle,scale=0.3,fill] at (23) {};
		\node[draw,circle,scale=0.3,fill] at (24) {};
		
		\node[] at (14,12) {{$\sss{a}$}};
		\node[] at (11.5,12.5) {{$\sss{b}$}};
		\node[] at (8.6,12.9) {{$\sss{c}$}};
		\node[] at (-2,3) {{$\sss{x}$}};
		\node[opacity=0.8] at (0,10) {{$\sss{x}$}};
		\node[opacity=0.6] at (2.7,10) {{$\sss{x}$}};	
		\node[opacity=0.4] at (5.2,10) {{$\sss{x}$}};	
	}{}

	\ifthenelse{#1 =2}{	
		\coordinate (9) at (1.5, 7);
		\coordinate (10) at (10.5, 7);
		\def\pa{0.15};
		\def\pb{0.22};
		\def\pc{0.28};
		\def\pd{0.34};
		
		\draw[name path=P38d] (3) to[out=-90, in=270, looseness=0.5, edge node={coordinate[pos=\pa] (17)}, edge node={coordinate[pos=\pb] (20)}] (8);
		
		\draw[name path=P38u] (3) to[out=90, in=-270, looseness=0.5] (8);
		
		\draw[name path=P47d] (4) to[out=-90, in=270, looseness=0.5, edge node={coordinate[pos=\pb] (26)}, edge node={coordinate[pos=\pc] (27)}] (7);
		
		\draw[name path=P47u] (4) to[out=90, in=-270, looseness=0.5] (7);
		
		\draw[name path=P56d] (5) to[out=-90, in=270, looseness=0.5, edge node={coordinate[pos=\pc] (28)}, edge node={coordinate[pos=\pd] (29)}] (6);
		
		\draw[name path=P56u] (5) to[out=90, in=-270, looseness=0.5] (6);
		
		\path[name path=P12d] (1) to[out=-90, in=270, looseness=0.5, edge node={coordinate[pos=\pa] (18)}, edge node={coordinate[pos=\pb] (23)}, edge node={coordinate[pos=\pd] (24)}] (2);
		\speC{1}{18}{2}{op4};
		
		\path[name path=P910d] (9) to[out=-90, in=270, looseness=0.5, edge node={coordinate[pos=\pb] (21)}, edge node={coordinate[pos=\pc] (22)}] (10);
		\draw[name path=P1718] (17) -- (18);
		\speC{9}{21}{10}{op3};
		\path[name path= P2429] (24) -- (29);
		\path [name intersections={of=P910d and P2429,by={15}}];
		
		\draw[op1,name path=P910u] (9) to[out=90, in=-270, looseness=0.5] (10);
		
		\draw[op1,name path=P12u] (1) to[out=90, in=-270, looseness=0.5] (2);	
		
		\path[name path=P1112d] (11) to[out=-90, in=270, looseness=0.5, edge node={coordinate[pos=\pa] (19)}, edge node={coordinate[pos=\pd] (25)}] (12);
		\speC{11}{19}{12}{op4};		
		\draw[op1,name path=P1112u] (11) to[out=90, in=-270, looseness=0.5] (12);
		\draw[name path=P1317] (13) -- (17);	
		
		\draw[name path=P28] (2) -- (8);
		\draw[name path=P212] (2) -- (12);
		\draw[name path=P1314] (13) -- (14);
		\draw[] (18) -- (19);
		\draw[name path = P2023] (20) -- (23);
		\draw[] (21) -- (23);
		\draw[] (24) -- (25);
		\draw[name path=P2026] (20) -- (26);
		\draw[name path=P2728] (27) -- (28);
		\draw[op4] (1) -- (11);
		\draw[] (14) -- (19);
		\draw[] (29) -- (30);
		\draw[opacity=0.2] (25) -- (31);
		
		\speL{3}{1}{13}{17}{}{op4};
		\speL{7}{10}{3}{8d}{}{op4};
		\speL{6}{2}{4}{7d}{}{op3};
		\speL{4}{9}{20}{26}{}{op3};
		\speL{5}{9}{27}{28}{}{op2};
		\speL{29}{24}{4}{7d}{}{op3};	
		\speL{26}{21}{3}{8d}{}{op4};	
		\speL{27}{22}{3}{8d}{}{op4};
		\speL{28}{22}{4}{7d}{}{op3};
		\speL{30}{31}{5}{6d}{}{op2};
		\speLL{1}{9}{P1718}{P2023}{op3}{op4}{op3};	
		
		\node[draw,circle,scale=0.3,fill] at (15) {};
		\node[draw,circle,scale=0.3,fill] at (21) {};
		\node[draw,circle,scale=0.3,fill] at (22) {};
		\node[draw,circle,scale=0.3,fill] at (18) {};
		\node[draw,circle,scale=0.3,fill] at (23) {};
		\node[draw,circle,scale=0.3,fill] at (24) {};
		
		\node[] at (14,12) {{$\sss{a}$}};
		\node[] at (11,12.3) {{$\sss{b}$}};
		\node[] at (8.6,12.9) {{$\sss{c}$}};
		\node[] at (-2,3) {{$\sss{x}$}};
		\node[opacity=0.8] at (0.45,10) {{$\sss{x}$}};
		\node[opacity=0.6] at (3.2,10) {{$\sss{x}$}};	
		\node[opacity=0.4] at (5.2,10) {{$\sss{x}$}};
	}{}

	\end{tikzpicture}
}

\newcommand{\Rmove}[2]{
	\begin{tikzpicture}[scale=#1,baseline=1.5em, line width=0.8pt]
	\coordinate (0) at(0, 0);
	\coordinate (1) at (-2, 4);
	\coordinate (2) at (0 , 4);
	\coordinate (3) at (2, 4);
	\coordinate (4) at (0, 2);
	
	\ifthenelse{#2 = 1}{
		\draw[] (0) -- (4);	
		\draw[] (4) -- (1) node[anchor=center, pos=1.25] {{$\sss{b}$}};
		\draw[] (4) -- (3) node[anchor=center, pos=1.25] {{$\sss{a}$}};
	}{}
	
	\coordinate (5) at (0, 3);
	\ifthenelse{#2 = 2}{
		\draw[] (0) -- (4);	
		\draw[shorten >= 0.3em] (4) edge [out=10, in=-30, looseness=1.5] (5);
		\draw[shorten <= 0.3em] (5) -- (1) node[anchor=center, pos=1.25, yshift=0.1em] {{$\sss{b}$}};
		\draw[] (4) edge [out=170, in=210, looseness=1.5] (5);
		\draw[] (5) -- (3) node[anchor=center, pos=1.25, yshift=0.1em] {{$\sss{a}$}};
	}{}
	\ifthenelse{#2 = 3}{
		\draw[] (0) -- (4);	
		\draw[] (4) edge [out=10, in=-30, looseness=1.5] (5);
		\draw[] (5) -- (1) node[anchor=center, pos=1.25, yshift=0.1em] {{$\sss{b}$}};
		\draw[shorten >= 0.3em] (4) edge [out=170, in=210, looseness=1.5] (5);
		\draw[shorten <= 0.3em] (5) -- (3) node[anchor=center, pos=1.25, yshift=0.1em] {{$\sss{a}$}};
	}{}
	\end{tikzpicture}
}

\newcommand{\hexagonMoveONE}[2]{
	\begin{tikzpicture}[scale=#1,baseline=1.5em, line width=0.8pt]
	\coordinate (0) at(0, 0);
	\coordinate (1) at (0, 2);
	\coordinate (2) at (-2, 4);
	\coordinate (3) at (-4, 6);
	\coordinate (4) at (0, 6);
	\coordinate (5) at (4, 6);
	
	\ifthenelse{#2 = 1}{
		\draw[] (0) -- (1);
		\draw[] (1) -- (2);
		\draw[] (2) -- (3) node[anchor=center, pos=1.25] {{$\sss{c}$}};
		\draw[] (2) -- (4) node[anchor=center, pos=1.25] {{$\sss{d}$}};			
		\draw[] (1) -- (2, 4);
		\draw[] (2, 4) -- (5) node[anchor=center, pos=1.25] {{$\sss{e}$}};
	}{}	
	
	\ifthenelse{#2 = 2}{
		\draw[] (0) -- (1);
		\draw[] (1) -- (2);
		\draw[] (2) -- (3) node[anchor=center, pos=1.25] {{$\sss{a}$}};
		\draw[] (2) -- (4) node[anchor=center, pos=1.25] {{$\sss{b}$}};			
		\draw[] (1) -- (2, 4);
		\draw[] (2, 4) -- (5) node[anchor=center, pos=1.25] {{$\sss{c}$}};
	}{}	
	
	\ifthenelse{#2 = 3}{
		\draw[] (0) -- (1);
		\draw[] (1) -- (2);
		\draw[] (2) -- (3) node[anchor=center, pos=1.25] {{$\sss{c}$}};
		\draw[] (2) -- (4) node[anchor=center, pos=1.25] {{$\sss{g}$}};			
		\draw[] (1) -- (2, 4);
		\draw[] (2, 4) -- (5) node[anchor=center, pos=1.25] {{$\sss{h}$}};
	}{}	
	\end{tikzpicture}
}

\newcommand{\hexagonMoveTWO}[2]{
	\begin{tikzpicture}[scale=#1,baseline=1.5em, line width=0.8pt]
	\coordinate (0) at(0, 0);
	\coordinate (1) at (0, 2);
	\coordinate (2) at (-2, 4);
	\coordinate (3) at (-4, 6);
	\coordinate (4) at (0, 6);
	\coordinate (5) at (4, 6);
	\coordinate (6) at (0, 4);
	\coordinate (7) at (1, 2.5);
	\coordinate (8) at (3, 3.5);
	\coordinate (9) at (0, 4);
	\coordinate (10) at (2, 5);
	
	\coordinate (11) at (intersection of 10--4 and 2--5);
	
	\draw[] (0) -- (1);
	\draw[] (1) -- (8);
	\draw[] (8) edge [out=35, in=-30, looseness=1.5] (10);
	\draw[shorten >= 0.3em] (10) -- (11);
	\draw[shorten <= 0.3em] (11) -- (4);
	\draw[] (1) -- (3);
	\draw[] (2)  -- (11);
	\draw[] (11) -- (5);
	
	\ifthenelse{#2 = 1}{
		\path (2, 4) -- (5) node[anchor=center, pos=1.25] {{$\sss{e}$}};
		\path (2) -- (3) node[anchor=center, pos=1.25] {{$\sss{c}$}};
		\path (2, 4) -- (4) node[anchor=center, pos=1.25] {{$\sss{d}$}};
	}{}
	
	\ifthenelse{#2 = 2}{
		\path (2, 4) -- (5) node[anchor=center, pos=1.25] {{$\sss{c}$}};
		\path (2) -- (3) node[anchor=center, pos=1.25] {{$\sss{a}$}};
		\path (2, 4) -- (4) node[anchor=center, pos=1.25] {{$\sss{b}$}};
	}{}
	
	\ifthenelse{#2 = 3}{
		\path (2, 4) -- (5) node[anchor=center, pos=1.25] {{$\sss{h}$}};
		\path (2) -- (3) node[anchor=center, pos=1.25] {{$\sss{c}$}};
		\path (2, 4) -- (4) node[anchor=center, pos=1.25] {{$\sss{g}$}};
	}{}
	\end{tikzpicture}
}

\newcommand{\hexagonMoveTHREE}[2]{
	\begin{tikzpicture}[scale=#1,baseline=1.5em, line width=0.8pt]
	\coordinate (0) at(0, 0);
	\coordinate (1) at (0, 2);
	\coordinate (2) at (-2, 4);
	\coordinate (3) at (-4, 6);
	\coordinate (4) at (0, 6);
	\coordinate (5) at (4, 6);
	\coordinate (6) at (0, 3);
	
	\ifthenelse{#2 = 1}{
		\draw[] (0) -- (1);
		\draw[] (2) -- (3) node[anchor=center, pos=1.25] {{$\sss{c}$}};
		\draw[] (2) -- (4) node[anchor=center, pos=1.25] {{$\sss{d}$}};				
		
		\draw[shorten >= 0.3em] (1) edge [out=10, in=-30, looseness=1.5] (6);
		\draw[shorten <= 0.3em] (6) -- (2);
		\draw[] (1) edge [out=170, in=210, looseness=1.5] (6);
		\draw[] (6) -- (5);
		
		\path (2, 4) -- (5) node[anchor=center, pos=1.25] {{$\sss{e}$}};
	}{}
	
	\ifthenelse{#2 = 2}{
		\draw[] (0) -- (1);
		\draw[] (2) -- (3) node[anchor=center, pos=1.25] {{$\sss{a}$}};
		\draw[] (2) -- (4) node[anchor=center, pos=1.25] {{$\sss{b}$}};				
		
		\draw[shorten >= 0.3em] (1) edge [out=10, in=-30, looseness=1.5] (6);
		\draw[shorten <= 0.3em] (6) -- (2);
		\draw[] (1) edge [out=170, in=210, looseness=1.5] (6);
		\draw[] (6) -- (5);
		
		\path (2, 4) -- (5) node[anchor=center, pos=1.25] {{$\sss{c}$}};
	}{}
	\end{tikzpicture}
}

\newcommand{\hexagonMoveFOUR}[2]{
	\begin{tikzpicture}[scale=#1,baseline=1.5em, line width=0.8pt]
	\coordinate (0) at(0, 0);
	\coordinate (1) at (0, 2);
	\coordinate (2) at (-2, 4);
	\coordinate (3) at (-4, 6);
	\coordinate (4) at (0, 6);
	\coordinate (5) at (4, 6);
	\coordinate (6) at (0, 4);
	\coordinate (7) at (-1, 3);
	\coordinate (8) at (3, 3.5);
	\coordinate (10) at (2, 5);
	
	\coordinate (11) at (intersection of 10--4 and 2--5);
	\coordinate (12) at (intersection of 3--6 and 2--5);
	
	\draw[] (0) -- (1);
	\draw[] (1) -- (8);
	\draw[] (8) edge [out=35, in=-30, looseness=1.5] (10);
	\draw[shorten >= 0.3em] (10) -- (11);
	\draw[shorten <= 0.3em] (11) -- (4);
	\draw[] (1) -- (7);
	\draw[] (7) edge [out=135, in=205, looseness=1.5] (12);
	\draw[shorten >= 0.3em] (7) edge [out=10, in=-30, looseness=1.5] (12);
	\draw[shorten <= 0.3em] (12) -- (3);
	\draw[] (12) -- (5);
	
	\ifthenelse{#2 = 1}{
		\path (2, 4) -- (5) node[anchor=center, pos=1.25] {{$\sss{e}$}};
		\path (2) -- (3) node[anchor=center, pos=1.25] {{$\sss{c}$}};
		\path (2, 4) -- (4) node[anchor=center, pos=1.25] {{$\sss{d}$}};
	}{}
	
	\ifthenelse{#2 = 2}{
		\path (2, 4) -- (5) node[anchor=center, pos=1.25] {{$\sss{c}$}};
		\path (2) -- (3) node[anchor=center, pos=1.25] {{$\sss{a}$}};
		\path (2, 4) -- (4) node[anchor=center, pos=1.25] {{$\sss{b}$}};
	}{}
	\end{tikzpicture}
}

\newcommand{\hexagonMoveFIVE}[2]{
	\begin{tikzpicture}[scale=#1,baseline=1.5em, line width=0.8pt]
	\coordinate (0) at(0, 0);
	\coordinate (1) at (0, 2);
	\coordinate (2) at (2, 4);
	\coordinate (3) at (-4, 6);
	\coordinate (4) at (0, 6);
	\coordinate (5) at (4, 6);
	
	\ifthenelse{#2 = 2}{
		\draw[] (0) -- (1);
		\draw[] (1) -- (3);
		\draw[] (2) -- (4) node[anchor=center, pos=1.25] {{$\sss{b}$}};			
		\draw[] (1) -- (2, 4);
		\draw[] (2, 4) -- (5) node[anchor=center, pos=1.25] {{$\sss{c}$}};
		
		\path (-2, 4) -- (3) node[anchor=center, pos=1.25] {{$\sss{a}$}};
	}{}	
	
	\ifthenelse{#2 = 1}{
		\draw[] (0) -- (1);
		\draw[] (1) -- (3);
		\draw[] (2) -- (4) node[anchor=center, pos=1.25] {{$\sss{g}$}};			
		\draw[] (1) -- (2, 4);
		\draw[] (2, 4) -- (5) node[anchor=center, pos=1.25] {{$\sss{h}$}};
		
		\path (-2, 4) -- (3) node[anchor=center, pos=1.25] {{$\sss{c}$}};
	}{}	
	\end{tikzpicture}
}

\newcommand{\hexagonMoveSIX}[2]{
	\begin{tikzpicture}[scale=#1,baseline=1.5em, line width=0.8pt]
	\coordinate (0) at(0, 0);
	\coordinate (1) at (0, 2);
	\coordinate (2) at (2, 4);
	\coordinate (3) at (-4, 6);
	\coordinate (4) at (0, 6);
	\coordinate (5) at (4, 6);
	\coordinate (7) at (2, 5);
	
	\draw[] (0) -- (1);
	\draw[] (1) -- (3);
	\draw[] (2) edge [out=170, in=210, looseness=1.5] (7);	
	\draw[shorten >= 0.3em] (2) edge [out=45, in=-30, looseness=1.5] (7);
	\draw[shorten <= 0.3em] (7) -- (4);
	\draw[] (7) -- (5);	
	\draw[] (1) -- (2);
	
	\ifthenelse{#2 = 2}{	
		\path (-2, 4) -- (3) node[anchor=center, pos=1.25] {{$\sss{a}$}};
		\path (2) -- (4) node[anchor=center, pos=1.25] {{$\sss{b}$}};
		\path (2) -- (5) node[anchor=center, pos=1.25] {{$\sss{c}$}};
	}{}
	
	\ifthenelse{#2 = 1}{	
		\path (-2, 4) -- (3) node[anchor=center, pos=1.25] {{$\sss{c}$}};
		\path (2) -- (4) node[anchor=center, pos=1.25] {{$\sss{g}$}};
		\path (2) -- (5) node[anchor=center, pos=1.25] {{$\sss{h}$}};
	}{}
	\end{tikzpicture}
}

\newcommand{\Smove}[2]{
	\begin{tikzpicture}[scale=#1,baseline=0em, line width=0.8pt]
	\coordinate (0) at(0, 0);
	\coordinate (1) at (0, 2);
	\coordinate (2) at (2 , 0);
	\coordinate (3) at (0, -2);
	\coordinate (4) at (-2, 0);
	
	\ifthenelse{#2 = 1}{
		\draw[] (0) -- (1) node[anchor=center, pos=1.25] {{$\sss{a}$}};
		\draw[] (0) -- (2) node[anchor=center, pos=1.5] {{$\sss{b+c}$}};	
		\draw[] (0) -- (3) node[anchor=center, pos=1.25] {{$\sss{c}$}};	
		\draw[] (0) -- (4) node[anchor=center, pos=1.5] {{$\sss{a+b}$}};								
		
	}{}
	
	\coordinate (5) at (2, -2);
	\coordinate (6) at (4, 0);
	\ifthenelse{#2 = 2}{
		\draw[] (0) -- (1) node[anchor=center, pos=1.25] {{$\sss{a}$}};	
		\draw[] (0) -- (4) node[anchor=center, pos=1.5] {{$\sss{a+b}$}};
		\draw[] (0) -- (2) node[above, pos=0.5, yshift=-0.28em] {{$\sss{b}$}};	
		\draw[] (2) -- (5) node[anchor=center, pos=1.25] {{$\sss{c}$}};
		\draw[] (2) -- (6) node[anchor=center, pos=1.5] {{$\sss{b+c}$}};	
	}{}
	\end{tikzpicture}
}

\newcommand{\pent}[4]{
	\begin{tikzpicture}[scale=#3,baseline=-0.3em]
	\coordinate (0) at(0,1);
	\coordinate (1) at ({sin(72)} ,{cos(72)} );
	\coordinate (2) at ({sin(144)} ,{-cos(36)} );
	\coordinate (3) at ({-sin(144)} ,{-cos(36)} );
	\coordinate (4) at ({-sin(72)} ,{cos(72)} );
	
	\draw[] (0) to (1);
	\draw[] (1) to (2);
	\draw[] (2) to (3);
	\draw[] (3) to (4);
	\draw[] (4) to (0);
	
	\foreach \x in {#1, #2}
	{
		\ifthenelse{\x = 20 \OR \x = 02}{\draw[] (0) to (2);}{}
		\ifthenelse{\x = 30 \OR \x = 03}{\draw[] (0) to (3);}{}
		\ifthenelse{\x = 13 \OR \x = 31}{\draw[] (1) to (3);}{}
		\ifthenelse{\x = 14 \OR \x = 41}{\draw[] (1) to (4);}{}
		\ifthenelse{\x = 24 \OR \x = 42}{\draw[] (2) to (4);}{}
	}
	
	\ifthenelse{#4 = 1}{
		\node[above] at (0) {{$\scriptstyle{2}$}};	
		\node[right] at (1) {{$\scriptstyle{3}$}};
		\node[below] at (2) {{$\scriptstyle{4}$}};
		\node[below] at (3) {{$\scriptstyle{0}$}};
		\node[left] at (4) {{$\scriptstyle{1}$}};}{}	
	\end{tikzpicture}
}

\newcommand{\PrePTwoTwoCat}[0]{
	\begin{tikzpicture}[scale=1,baseline=-0.25 em, line width=0.4pt]
	\matrix[matrix of math nodes,row sep =3em,column sep=3em, ampersand replacement=\&] (m) {
		{\bul} \& {\bul} \\
		{\bul} \& {\bul}
		\\
	};	
	\path	
	(m-1-1) edge[->] node[pos=0.5, above] {${\sss b}$} (m-1-2)
	(m-2-1) edge[->] node[pos=0.5, left] {${\sss a}$} (m-1-1)
	(m-2-1) edge[->, out=40, in=140, looseness=1] coordinate[pos=0.5] (sp1) node[pos=0.5, above, yshift=-0.3em, xshift=0.7em] {${\sss (a \cdot b) \cdot c}$} (m-2-2)
	(m-2-1) edge[->, out=-40, in=-140, looseness=1] coordinate[pos=0.5] (sp2) node[pos=0.5, below] {${\sss a \cdot (b \cdot c)}$} (m-2-2)
	(m-1-2) edge[->] node[pos=0.5, right] {${\sss c}$} (m-2-2)
	(m-2-1) edge[->] node[pos=0.5, above, sloped] {${\sss a \cdot b}$} (m-1-2)
	(sp1) edge[double, double equal sign distance, -implies, line cap=butt, shab] node[pos=0.5, right] {${\sss \alpha}$} (sp2)	
	;
	\end{tikzpicture}
	\q = \q
	\begin{tikzpicture}[scale=1,baseline=-0.25 em, line width=0.4pt]
	\matrix[matrix of math nodes,row sep =3em,column sep=3em, ampersand replacement=\&] (m) {
		{\bul} \& {\bul} \\
		{\bul} \& {\bul}
		\\
	};	
	\path	
	(m-1-1) edge[->] node[pos=0.5, above] {${\sss b}$} (m-1-2)
	(m-2-1) edge[->] node[pos=0.5, left] {${\sss a}$} (m-1-1)
	(m-2-1) edge[->] node[pos=0.5, below] {${\sss a \cdot (b \cdot c)}$} (m-2-2)
	(m-1-2) edge[->] node[pos=0.5, right] {${\sss c}$} (m-2-2)
	(m-1-1) edge[->] node[pos=0.5, above, sloped] {${\sss b \cdot c}$} (m-2-2)	
	;
	\end{tikzpicture}
}

\newcommand{\PTwoTwoCat}[0]{
	\begin{tikzpicture}[scale=1,baseline=-0.25 em, line width=0.4pt]
		\matrix[matrix of math nodes,row sep =3em,column sep=3em, ampersand replacement=\&] (m) {
			{\bul} \& {\bul} \\
			{\bul} \& {\bul}
			\\
		};	
		\path	
		(m-1-1) edge[->] node[pos=0.5, above] {${\sss b}$} (m-1-2)
		(m-2-1) edge[->] node[pos=0.5, left] {${\sss a}$} (m-1-1)
		(m-2-1) edge[->] node[pos=0.5, below] {${\sss a \cdot b \cdot c}$} (m-2-2)
		(m-1-2) edge[->] node[pos=0.5, right] {${\sss c}$} (m-2-2)
		(m-2-1) edge[->] node[pos=0.5, above, sloped] {${\sss a \cdot b}$} (m-1-2)	
		;
	\end{tikzpicture}
	\q\xRightarrow{\alpha_{a,b,c}}\q
	\begin{tikzpicture}[scale=1,baseline=-0.25 em, line width=0.4pt]
		\matrix[matrix of math nodes,row sep =3em,column sep=3em, ampersand replacement=\&] (m) {
			{\bul} \& {\bul} \\
			{\bul} \& {\bul}
			\\
		};	
		\path	
		(m-1-1) edge[->] node[pos=0.5, above] {${\sss b}$} (m-1-2)
		(m-2-1) edge[->] node[pos=0.5, left] {${\sss a}$} (m-1-1)
		(m-2-1) edge[->] node[pos=0.5, below] {${\sss a \cdot b \cdot c}$} (m-2-2)
		(m-1-2) edge[->] node[pos=0.5, right] {${\sss c}$} (m-2-2)
		(m-1-1) edge[->] node[pos=0.5, above, sloped] {${\sss b \cdot c}$} (m-2-2)	
		;
	\end{tikzpicture}
}

\newcommand{\PTwoThreeCat}[1]{
	\begin{tikzpicture}[scale=1,baseline=0em, line width=0.4pt]
	\matrix[matrix of math nodes,row sep =0.7em,column sep=0.7em, ampersand replacement=\&] (m) {
		{} \& {} \& {} \& {\bul} \&[-1em] {} \& {} \& {} \& {}
		\\
		{} \& {} \& {} \& {} \& {} \& {} \& {} \& {}
		\\
		{} \& {} \& {} \& {} \& {} \& {} \& {} \& {}
		\\			
		{} \& {} \& {} \& {} \& {} \& {} \& {} \& {}
		\\
		{} \& {} \& {} \& {} \& {} \& {} \& {} \& {\bul}
		\\[-0.5em]		
		{\bul} \& {} \& {} \& {} \& {} \& {} \& {} \& {}
		\\[-0.3em]			
		{} \& {} \& {} \& {} \& {} \& {\bul} \& {}	\& {}
		\\
		{} \& {} \& {} \& {} \& {} \& {} \& {} \& {}
		\\
		{} \& {} \& {} \& {} \& {} \& {} \& {} \& {}
		\\
		{} \& {} \& {} \& {} \& {} \& {} \& {} \& {}
		\\
		{} \& {} \& {} \& {\bul} \& {} \& {} \& {} \& {}
		\\
	};	
	\ifthenelse{#1 =1}{
		\path
		(m-1-4) to[edge node={coordinate[pos=0.5] (sp1)}] (m-5-8)
		(sp1) edge[->] (m-5-8)
		(m-1-4) edge[<-] node[pos=0.35, above, sloped] {${\sss a}$} (m-6-1)
		(m-5-8) edge[<-, shorten <= 0.2em] node[pos=1, below, sloped] {${\sss c \cdot d}$} (m-7-6)
		(m-7-6) edge[->, shorten >= -0.1em] node[pos=0.4, above, sloped] {${\sss c}$} (m-11-4)
		(m-6-1) edge[->] node[pos=0.5, below, sloped] {${\sss a \cdot b \cdot c}$} (m-11-4)
		(m-5-8) edge[<-, sha] node[pos=0.7, below, sloped] {${\sss  d}$} (m-11-4)
		(m-1-4) edge[sha]  (intersection of m-1-4--m-11-4 and m-6-1--m-7-6)
		(intersection of m-1-4--m-11-4 and m-6-1--m-7-6) edge[->, shb, sha] node[pos=0.25, below, sloped] {${\sss b \cdot c}$} (m-11-4)
		(m-5-8) edge[sha, <-] node[pos=0.6, below, sloped] {${\sss a \cdot b \cdot c \cdot d}$} (intersection of m-5-8--m-6-1 and m-1-4--m-7-6)
		(intersection of m-5-8--m-6-1 and m-1-4--m-7-6) edge[shab] (intersection of m-5-8--m-6-1 and m-1-4--m-11-4) 
		(intersection of m-5-8--m-6-1 and m-1-4--m-11-4) edge[shab] (m-6-1)
		
		(m-9-3) edge[double, double equal sign distance, -implies, line cap=butt, out=25, in=-150, looseness=1] node[pos=0.3, left, xshift=0.27em] {${\sss \alpha^0_{a,b,c}}$} (m-4-3)
		(m-4-3) edge[double, double equal sign distance, -implies, line cap=butt, out=-45, in=-135, looseness=0.8] node[pos=0.4, below, yshift=0.2em, xshift=-0.1em] {${\sss \alpha^0_{a,b \cdot c,d}}$} (m-4-6)
		(m-4-6) edge[double, double equal sign distance, -implies, line cap=butt, out=100, in=10, looseness=5] node[pos=0.5, right] {${\sss \alpha^0_{b,c,d}}$} (m-5-6)
		
		(m-7-6) edge[<-] node[pos=0.7, below, sloped] {${\sss a \cdot b}$} (m-6-1)
		(m-1-4) edge[->] node[pos=0.4, above, sloped] {${\sss  b}$} (m-7-6)
		(m-1-4) edge[] node[pos=0.4, above, sloped] {${\sss b \cdot c \cdot d}$} (sp1)
	;}{}
	\ifthenelse{#1 =2}{
		\path
		(m-1-4) edge[->] node[pos=0.2, above, sloped] {${\sss b \cdot c \cdot d}$} (m-5-8)
		edge[<-] node[pos=0.35, above, sloped] {${\sss a}$} (m-6-1)
		(m-5-8) edge[<-, shorten <= 0.2em] node[pos=1, below, sloped] {${\sss c \cdot d}$} (m-7-6)
		(m-7-6) edge[->, shorten >= -0.1em] node[pos=0.4, above, sloped] {${\sss c}$} (m-11-4)
		(m-6-1) edge[->] node[pos=0.5, below, sloped] {${\sss a \cdot b \cdot c}$} (m-11-4)
		(m-5-8) edge[<-, sha] node[pos=0.7, below, sloped] {${\sss  d}$} (m-11-4)
		(m-5-8) edge[sha, <-]  (intersection of m-5-8--m-6-1 and m-1-4--m-7-6)
		(intersection of m-5-8--m-6-1 and m-1-4--m-7-6) edge[shab] node[pos=0.6, above, sloped] {${\sss a \cdot b \cdot c \cdot d}$} (m-6-1)
		
		(m-9-3) edge[double, double equal sign distance, -implies, line cap=butt, out=25, in=-90, looseness=1, shorten >=-0.1em] node[pos=0.6, left, xshift=0.1em] {${\sss \alpha^0_{a \cdot b,c,d}}$} (m-6-4)
		(m-6-4) edge[double, double equal sign distance, -implies, line cap=butt, out=90, in=160, looseness=1, shorten >=-1em] node[pos=0.7, left, yshift=0.2em] {${\sss \alpha^0_{a,b,c \cdot d}}$} (m-4-6)
		
		(m-7-6) edge[<-] node[pos=0.7, below, sloped] {${\sss a \cdot b}$} (m-6-1)
		(m-1-4) edge[->] node[pos=0.4, above, sloped] {${\sss  b}$} (m-7-6)
	;}{}
	\end{tikzpicture}
}

\newcommand{\assocLoop}[1]{
	\begin{tikzpicture}[scale=1,baseline=-0.5em, line width=0.4pt]
	
	\ifthenelse{#1 = 1}{
		\matrix[matrix of math nodes,row sep =3em,column sep=3em, ampersand replacement=\&] (m) {
			(\looSy{x^a}) \& (\looSy{x^{ab}}) \\
			(\looSy{x}) \& (\looSy{x^{abc}})
			\\
		};
		\path
		(m-1-1) edge[->, transform canvas={yshift=-0.5em}] node[pos=0.5, above] {${\sss b}$} (m-1-2)
		(m-2-1) edge[->] node[pos=0.5, left] {${\sss a}$} (m-1-1)
		(m-2-1) edge[->, transform canvas={yshift=-0.5em}] node[pos=0.5, below] {${\sss a \cdot b \cdot c}$} (m-2-2)
		(m-1-2) edge[->] node[pos=0.5, right] {${\sss c}$} (m-2-2)
		(m-2-1) edge[->, shorten >= -0.5em, transform canvas={yshift=-0.5em}] node[pos=0.5, above, sloped] {${\sss a \cdot b}$} (m-1-2)
		;
	}{}
	
	\ifthenelse{#1 = 2}{
		\matrix[matrix of math nodes,row sep =3em,column sep=3em, ampersand replacement=\&] (m) {
			(\looSy{x^a}) \& (\looSy{x^{ab}}) \\
			(\looSy{x}) \& (\looSy{x^{abc}})
			\\
		};
		\path
		(m-1-1) edge[->, transform canvas={yshift=-0.5em}] node[pos=0.5, above] {${\sss b}$} (m-1-2)
		(m-2-1) edge[->] node[pos=0.5, left] {${\sss a}$} (m-1-1)
		(m-2-1) edge[->, transform canvas={yshift=-0.5em}] node[pos=0.5, below] {${\sss a \cdot b \cdot c}$} (m-2-2)
		(m-1-2) edge[->] node[pos=0.5, right] {${\sss c}$} (m-2-2)
		(m-1-1) edge[->, shorten >=-0.5em, transform canvas={yshift=-0.5em}] node[pos=0.5, above, sloped] {${\sss b \cdot c}$} (m-2-2)
		;
	}{}
	\end{tikzpicture}
}

\newcommand{\Qalgebra}[1]{
	\begin{tikzpicture}[scale=0.3,baseline=1.2em]
	\coordinate (0) at (-3, 0);
	\coordinate (1) at (-1.5, 2);
	\coordinate (2) at (1.5, 2);
	\coordinate (3) at (7.5, 2);
	\coordinate (4) at (10.5, 2);
	\coordinate (5) at (4.5, 2);

	\def\pa{0.2};
	\def\pb{0.3};
	\def\pc{0.4};
	
	\ifthenelse{#1 =1}{
		\path[name path=P14d] (1) to[out=-90, in=270, looseness=0.5, edge node={coordinate[pos=\pa] (6)}, edge node={coordinate[pos=\pb] (7)}] (4);	
		\path[name path=P14u] (1) to[out=90, in=-270, looseness=0.5] (4);	
		\draw[name path=P23d] (2) to[out=-90, in=270, looseness=0.5, edge node={coordinate[pos=\pb] (8)}, edge node={coordinate[pos=\pc] (9)}] (3);	
		\draw[name path=P23] (2) to[out=90, in=-270, looseness=0.5] (3);
		\draw[] (7) -- (8);
		\draw[] (9) -- (5);
			
		\node[right] at (5)  {{$\sss{x}$}};
		\node[above] at (7)  {{$\sss{x}$}};	
		\node[right] at (3) {{$\sss{a}$}};
	}{}

	\ifthenelse{#1 =2}{
		\path[name path=P14d] (1) to[out=-90, in=270, looseness=0.5, edge node={coordinate[pos=\pa] (6)}, edge node={coordinate[pos=\pb] (7)}] (4);	
		\path[name path=P14u] (1) to[out=90, in=-270, looseness=0.5] (4);	
		\draw[name path=P23d] (2) to[out=-90, in=270, looseness=0.5, edge node={coordinate[pos=\pb] (8)}, edge node={coordinate[pos=\pc] (9)}] (3);	
		\draw[name path=P23] (2) to[out=90, in=-270, looseness=0.5] (3);
		\draw[] (7) -- (8);
		\draw[] (9) -- (5);
		
		\node[right] at (5)  {{$\sss{x}$}};
		\node[above] at (7)  {{$\sss{x}$}};	
		\node[right] at (3) {{$\sss{b}$}};
	}{}

	\ifthenelse{#1 =3}{
		\draw[name path=P14d] (1) to[out=-90, in=270, looseness=0.5, edge node={coordinate[pos=\pa] (6)}, edge node={coordinate[pos=\pb] (7)}] (4);	
		\draw[name path=P14u] (1) to[out=90, in=-270, looseness=0.5] (4);	
		\draw[name path=P23d] (2) to[out=-90, in=270, looseness=0.5, edge node={coordinate[pos=\pb] (8)}, edge node={coordinate[pos=\pc] (9)}] (3);	
		\draw[name path=P23] (2) to[out=90, in=-270, looseness=0.5] (3);
		\draw[] (0) -- (6);
		\draw[] (7) -- (8);
		\draw[] (9) -- (5);

		\node[above] at (0)  {{$\sss{x}$}};
		\node[right] at (5)  {{$\sss{x}$}};
		\node[above] at (7)  {{$\sss{x}$}};	
		\node[right] at (3) {{$\sss{b}$}};
		\node[right] at (4) {{$\sss{a}$}};
	}{}

	\ifthenelse{#1 =4}{
		\path[name path=P14d] (1) to[out=-90, in=270, looseness=0.5, edge node={coordinate[pos=\pa] (6)}, edge node={coordinate[pos=\pb] (7)}] (4);	
		\path[name path=P14u] (1) to[out=90, in=-270, looseness=0.5] (4);	
		\draw[name path=P23d] (2) to[out=-90, in=270, looseness=0.5, edge node={coordinate[pos=\pb] (8)}, edge node={coordinate[pos=\pc] (9)}] (3);	
		\draw[name path=P23] (2) to[out=90, in=-270, looseness=0.5] (3);
		\draw[] (7) -- (8);
		\draw[] (9) -- (5);
		
		\node[right] at (5)  {{$\sss{x}$}};
		\node[above] at (7)  {{$\sss{x}$}};	
		\node[right] at (3) {{$\sss{a+b}$}};
	}{}

	\end{tikzpicture}
}

\newcommand{\prism}[1]{
	\begin{tikzpicture}[scale=#1, baseline=2.5em, line width=0.8 pt]
	\coordinate (0) at (0,0);
	\coordinate (1) at (2,0);
	\coordinate (3) at (1,0.5);
	\coordinate (4) at (0,2);
	\coordinate (5) at (2,2);
	\coordinate (7) at (1,2.5);
	
	\draw[] (0) -- (1);
	\draw[] (0) -- (3);
	\draw[] (4) -- (5);
	\draw[] (4) -- (7);
	\draw[] (0) -- (4);
	\draw[] (1) -- (5);
	\draw[] (5) -- (7);	
	\draw[] (0) -- (5);	
	\draw[] (1) -- (3);
	\interLL{3}{7}{0}{5}{4}{5}{1};
	\interL{0}{7}{4}{5}{1};
	\interLL{1}{7}{0}{5}{4}{5}{1};

	\node[below] at (0) {{$\scriptstyle{0}$}};	
	\node[below] at (1) {{$\scriptstyle{1}$}};
	\node[below] at (3) {{$\scriptstyle{2}$}};
	\node[above] at (4) {{$\scriptstyle{0'}$}};	
	\node[above] at (5) {{$\scriptstyle{1'}$}};
	\node[above] at (7) {{$\scriptstyle{2'}$}};
	
	\end{tikzpicture}
}

\newcommand{\simplex}[1]{
	\begin{tikzpicture}[scale=#1, baseline=0.5em, line width=0.8 pt]
	\coordinate (0) at (0,0);
	\coordinate (1) at (2,0);
	\coordinate (3) at (1,0.5);

	\draw[] (0) -- (1);
	\draw[] (0) -- (3);
	\draw[] (1) -- (3);
		
	\node[below] at (0) {{$\scriptstyle{0}$}};	
	\node[below] at (1) {{$\scriptstyle{1}$}};
	\node[above] at (3) {{$\scriptstyle{2}$}};
	\end{tikzpicture}
}

\newcommand{\niceTorus}[1]{
	\begin{tikzpicture}[scale=#1, baseline=0em]	
	\coordinate (1) at (0,0);
	\coordinate (2) at (0.8,0.1);
	\coordinate (3) at (2.2,0.1);
	\coordinate (4) at (3,0);
	\coordinate (5) at (0.4,0.1);
	\coordinate (6) at (2.6,0.1);
	
	\draw[] (1)  to[out=-90, in=270, looseness=0.8, edge node={coordinate[pos=0.535] (sp5)}] (4);
	\draw[] (1) to[out=90, in=-270, looseness=0.8](4);
	\draw[] (2) to[out=-90, in=270, looseness=0.5, edge node={coordinate[pos=0.2] (sp1)}, edge node={coordinate[pos=0.8] (sp2)}, edge node={coordinate[pos=0.55] (sp6)}](3);
	\draw[] (sp1) to[out=90, in=-270, looseness=0.5] (sp2);
	
	\draw[line width=0.4pt] (5) to[out=-90, in= 270, looseness=0.7] (6);
	\draw[line width=0.4pt, ->1-=0.3] (5) to[out=90, in =-270, looseness=0.7] (6);
	\draw[line width=0.4pt, ->1-=0.75] (sp6) to[out=-10, in=30, looseness=0.6] (sp5);
	\draw[line width=0.4pt, dotted] (sp6) to[out=190, in=2000, looseness=0.7] (sp5);	
	
	\node[] at (1.4,-0.55) {{$\sss{x}$}};
	\node[] at (0.6,-0.35) {{$\sss{y}$}};
	\end{tikzpicture}
}

\newcommand{\niceTorusThreaded}[1]{
	\begin{tikzpicture}[scale=#1, baseline=0em]	
	\coordinate (1) at (0,0);
	\coordinate (2) at (0.8,0.1);
	\coordinate (3) at (2.2,0.1);
	\coordinate (4) at (3,0);
	\coordinate (5) at (0.4,0.1);
	\coordinate (6) at (2.6,0.1);
	\coordinate (7) at (1.1, 1.6);
	\coordinate (8) at (1.9, 1.6);
	\coordinate (9) at (1.1,-1.6);
	\coordinate (10) at (1.9,-1.6);
	
	\draw[op6, name path=P14d] (1)  to[out=-90, in=270, looseness=0.8, edge node={coordinate[pos=0.535] (sp5)}] (4);
	\draw[op6, name path= P23d] (2) to[out=-90, in=270, looseness=0.5, edge node={coordinate[pos=0.2] (sp1)}, edge node={coordinate[pos=0.8] (sp2)}, edge node={coordinate[pos=0.55] (sp6)}](3);

	\draw[op6, line width=0.4pt] (5) to[out=-90, in= 270, looseness=0.7] (6);
	\draw[op6, line width=0.4pt, ->1-=0.75] (sp6) to[out=-10, in=30, looseness=0.6] (sp5);
	\draw[op6, line width=0.4pt, dotted] (sp6) to[out=190, in=2000, looseness=0.7] (sp5);	
	
	\node[op6] at (1.4,-0.55) {{$\sss{x}$}};
	\node[op6] at (0.6,-0.35) {{$\sss{y}$}};

	\centerarc[name path=Pc2](1.5,0)(0:360:2)
	\draw[] (7) to[out=-90, in= 270, looseness=0.8] (8);
	\draw[] (7) to[out=90, in= -270, looseness=0.6] (8);
	\draw[op6] (9) to[out=-90, in= 270, looseness=0.6] (10);
	\draw[op3] (9) to[out=90, in= -270, looseness=0.8] (10);
	
	\begin{scope}
		\clip (-2,2) rectangle (9);
		\draw[op6, line width=0.4pt, ->1-=0.3] (5) to[out=90, in =-270, looseness=0.7] (6);
		\draw[op6] (sp1) to[out=90, in=-270, looseness=0.5] (sp2);
		\draw[op6] (1) to[out=90, in=-270, looseness=0.8](4);
	\end{scope}	
	
	\begin{scope}
		\clip (4,2) rectangle (10);
		\draw[op6, line width=0.4pt, ->1-=0.3] (5) to[out=90, in =-270, looseness=0.7] (6);
		\draw[op6] (sp1) to[out=90, in=-270, looseness=0.5] (sp2);
		\draw[op6] (1) to[out=90, in=-270, looseness=0.8](4);
	\end{scope}	
	
	\begin{scope}
		\clip (7) rectangle (10);
		\draw[op3, line width=0.4pt, ->1-=0.3] (5) to[out=90, in =-270, looseness=0.7] (6);
		\draw[op3] (sp1) to[out=90, in=-270, looseness=0.5] (sp2);
		\draw[op3] (1) to[out=90, in=-270, looseness=0.8](4);
	\end{scope}

	\path[name path=P79] (7) -- (9);
	\path[name path=P810] (8) -- (10);
	\path [name intersections={of=P79 and P23d,by={spp1}}];
	\path [name intersections={of=P79 and P14d,by={spp2}}];
	\path [name intersections={of=P810 and P23d,by={spp3}}];
	\path [name intersections={of=P810 and P14d,by={spp4}}];
	\draw[op6] (7) -- (spp1);
	\draw[op3] (spp1) -- (spp2);
	\draw[op6] (spp2) -- (9); 
	\draw[op6] (8) -- (spp3);
	\draw[op3] (spp3) -- (spp4);
	\draw[op6] (spp4) -- (10); 	
	\end{tikzpicture}
}

\newcommand{\picDelta}[4]{
	\begin{tikzpicture}[scale=1.8, baseline=2.25em]
	\coordinate (0) at (0,0);
	\coordinate (1) at (0,1);
	\coordinate (2) at (0.7,0);
	\coordinate (3) at (0.7,1);
	\coordinate (4) at (1.4,0);
	\coordinate (5) at (1.4,1);	
	\coordinate (8) at (-0.8, 0.5);
	\coordinate (9) at (-0.5,0.5);
	\coordinate (10) at (0,0.5);
	\coordinate (11) at (0.7,0.5);
	\coordinate (12) at (1.4,0.5);
	\coordinate (13) at (1.9,0.5);
	\coordinate (14) at (2.2, 0.5);
	
	\ifthenelse{#1 =1}{
		\draw[name path=P01r] (0) to[out=0, in=-0, looseness =0.8] (1);	
		\draw[name path=P01l] (0) to[out =180, in=180, looseness =0.8] (1);
		\draw[name path=P23r] (2) to[out =0, in=0, looseness =0.8] (3);	
		\draw[name path=P23l] (2) to[out =180, in=180, looseness =0.8] (3);
		
		\draw[line width=0.6em, white] (9) -- (10);
		\draw[line width=0.6em, white] (0.3,0.5) -- (0.5,0.5);
		\draw[dotted] (8) -- (9);
		\speL{9}{11}{0}{1r}{sha}{shb};
		\coordinate (12) at (1.2,0.5);
		\coordinate (13) at (1.5,0.5);
		\speL{11}{12}{2}{3r}{sha}{shb};
		\draw[dotted] (12) -- (13);
		
		\node[below] at (0) {{$\sss{#3}$}};
		\node[below] at (2) {{$\sss{#4}$}};
		\node[above] at (9) {{$\sss{#2}$}};
		\node[above] at (12) {{$\sss{#2}$}};
	}{}
	\ifthenelse{#1 =2}{
		\draw[name path=P01r] (0) to[out=0, in=-0, looseness =0.8] (1);	
		\draw[name path=P01l] (0) to[out =180, in=180, looseness =0.8] (1);
		
		\draw[line width=0.6em, white] (9) -- (10);
		\draw[line width=0.6em, white] (0.3,0.5) -- (0.5,0.5);
		\draw[dotted] (8) -- (9);
		\speL{9}{11}{0}{1r}{sha}{shb};
		\draw[dotted] (11) -- (1,0.5);
		
		\node[below] at (0) {{$\sss{#3}$}};
		\node[above] at (9) {{$\sss{#2}$}};
		\node[above] at (11) {{$\sss{#2}$}};
	}{}
	\end{tikzpicture}
}

\title{\boldmath Tube algebras, excitations statistics and compactification in gauge models \\of  topological phases}

\author[\Square]{Alex Bullivant,}
\author[\pentagon, \hexagon]{Clement Delcamp}

\affiliation[\Square]{Department of Pure Mathematics, University of Leeds, Leeds, LS2 9JT, UK}

\affiliation[\pentagon]{Max-Planck-Institut f{\"u}r Quantenoptik \\ Hans-Kopfermann-Str. 1, 85748 Garching, Germany}
\affiliation[\hexagon]{Munich Center for Quantum Science and Technology (MCQST)\\ Schellingstr. 4, D-80799 M{\"u}nchen}

\emailAdd{a.l.bullivant@leeds.ac.uk}
\emailAdd{clement.delcamp@mpq.mpg.de}

\abstract{\\~\\ We consider lattice Hamiltonian realizations of ($d$+1)-dimensional Dijkgraaf-Witten theory. In (2+1)d, it is well-known that the Hamiltonian yields point-like excitations classified by irreducible representations of the twisted quantum double. This can be confirmed using a tube algebra approach. In this paper, we propose a generalisation of this strategy that is valid in any dimensions. We then apply this generalisation to derive the algebraic structure of loop-like excitations in (3+1)d, namely the twisted quantum triple. The irreducible representations of the twisted quantum triple algebra correspond to the simple loop-like excitations of the model. Similarly to its (2+1)d counterpart, the twisted quantum triple comes equipped with a compatible comultiplication map and an $R$-matrix that encode the fusion and the braiding statistics of the loop-like excitations, respectively. Moreover, we explain using the language of loop-groupoids how a model defined on a manifold that is $n$-times compactified can be expressed in terms of another model in $n$-lower dimensions. This can in turn be used to recast higher-dimensional tube algebras in terms of lower dimensional analogues.}

\begin{document} 
	\vspace*{-2em}
	\maketitle
	\flushbottom
	\newpage

\section{Introduction}
Succinctly, a phase of matter is defined by an equivalence class of physical systems sharing certain common features of interest.
The subtlety in this definition is then to describe equivalence relations which capture physically insightful properties. Our best description of quantum many-body systems is manifest through the language of quantum field theory. In this way a \emph{quantum phases of matter} may be defined by an equivalence classes of quantum field theories whereby each quantum model constitutes a concrete realisation of the given phase. For gapped quantum field theories, the infra-red limit of the theory admits an effective field theory description in terms of a \emph{topological quantum field theory} (TQFT)\cite{Atiyah:1989vu, Baez:1995xq, Lurie:2009keu, Freed:2012hx}. From this observation, an important class of quantum phases of matter is given by so-called \emph{topological phases of matter}, which are typically defined by homotopy classes of gapped quantum models whose low energy effective field theories realise given TQFTs.
A consequence of this definition is that two quantum states are described by the same TQFT if, and only if, they can be related by an adiabatic evolution which does not close the energy gap. In practice, this signifies that ground states of a given gapped system must remain in the same phase under \emph{local unitary transformations}. In the discrete setting, local unitary transformations can be performed in order to implement a wave function renormalisation group flow. Equivalence classes of wave functions under such transformations can therefore be interpreted as so-called \emph{fixed point wave functions}. These fixed point wave functions admit a TQFT description and are thus expected to capture the defining \emph{long-range entanglement} pattern signifying \emph{topological order} \cite{Chen:2010gda}.

In this paper, we are interested in physical realisations of topological phases that have a \emph{gauge theory} interpretation. Such models are referred to as \emph{gauge models of topological phases}. The low energy limit of the corresponding phases are described by a particularly manageable class of \emph{fully-extended} topological quantum field theories known as \emph{Dijkgraaf-Witten} theories \cite{dijkgraaf1990topological}. Given a closed ($d$+1)-manifold, the input data of a Dijkgraaf-Witten theory is a finite group $G$ and a cohomology class $[\omega] \in H^{d+1}(BG, \rU(1))$. The corresponding partition function can be conveniently defined by summing over homotopy classes of maps from the ($d$+1)-manifold to the classifying space  $BG$ of $G$.
Given a triangulated manifold, the partition function can be recast as a lattice gauge theory so that the sum is now performed over $G$-colourings, i.e. $G$-labelings of the one-skeleton of the triangulation that are subject to local constraints. This latter formulation turns out to be also valid in the case of manifolds with boundary. The definition of the partition function for a special family of \emph{cobordisms} can then be utilised to define lattice Hamiltonian realisations of the theory \cite{Hu:2012wx, Wan:2014woa}. These constitute the exactly solvable models of interest for the present manuscript.

In (2+1)d, it is well-known that the Dijkgraaf-Witten lattice Hamiltonian yields point-like bulk excitations that are classified by the irreducible representations of the \emph{twisted Drinfel'd quantum double} \cite{Drinfeld:1989st, Dijkgraaf1991, Dijkgraaf:1990ne}. These bulk excitations come in three types, namely \emph{electric charges}, \emph{magnetic fluxes} and \emph{dyons}, i.e. electric charge-magnetic flux composites. The quantum double, which is an example of \emph{quasi-triangular quasi-Hopf algebra}, not only provides the classification of these (anyonic) excitations but also their \emph{fusion} and \emph{braiding} statistics. More precisely, as a Hopf algebra, the quantum double comes equipped with a comultiplication rule from which the tensor product of irreducible representations can be defined, while as a quasi-triangular Hopf algebra it comes equipped with a so-called $R$-matrix from which a braid group representation on the irreducible modules can be derived. In the case where the input cocycle is chosen to be trivial, the model reduces to the so-called \emph{Kitaev's quantum double} model \cite{Kitaev1997}. 

There exist several strategies to uncover that the algebraic structure underlying the (2+1)d bulk excitations is indeed the twisted quantum double. One fruitful approach consists of defining explicit operators from the algebra of local symmetries that generate and measure the excitations \cite{Kitaev1997}. Alternatively, we can consider a generalization of Ocneanu's tube algebra \cite{ocneanu1994chirality, ocneanu2001operator, Lan:2013wia, DDR1, Aasen:2017ubm}. This approach relies on the crucial remark that the physical properties of a given excitation localized within a subregion are encoded into the boundary conditions of the open manifold obtained after removing this subregion. This is the approach we follow in this manuscript. More specifically, the tube algebra approach utilises the length scale invariance of the renormalisation flow fixed point in order to define an algebra defined by gluing states of the twice-punctured sphere along the boundary, which in turn reproduces the multiplication rule of the twisted quantum double.

It turns out that the tube algebra approach can be generalized to all dimensions \cite{Delcamp:2017pcw, Delcamp:2018efi, AB}. For instance, in (3+1)d the relevant manifold is the one obtained by cutting open the three-torus along one direction. We show in detail that this tube algebra yields a generalization of the twisted quantum double referred to as the$  $ \emph{twisted quantum triple} whose irreducible modules classify the simple bulk loop-like excitations \cite{Delcamp:2017pcw, AB}. More precisely, the irreducible modules can be labeled by three components, namely two magnetic fluxes and one electric charge quantum numbers, so that one of the flux quantum numbers, referred to as the \emph{threading flux}, constraints the remaining magnetic flux and electric charge quantum numbers of the loop excitation.  Similarly to its (2+1)d counterpart, the twisted quantum triple comes equipped with a coalgebraic- and quasi-triangular-like structure that allows a description of the fusion and the braiding of the bulk excitations. Specifically, we show these correspond to the fusion and the braiding processes of two loop-like excitations labeled by a magnetic flux and an electric charge, sharing the same threading flux.
More generally, the braiding structure of loop-like objects in the 3-disk is governed by the so-called \emph{necklace groups} \cite{bellingeri2016braid,Bullivant:2018djw} when the threading flux is non-trivial and the \emph{loop-braid group} \cite{lin2005motion,Baez:2006un,Bullivant:2018pju} when the threading flux is trivial. It can be shown that the loop-like excitations of the twisted quantum triple naturally define representations of such motion groups \cite{ABrowell2}. 
In this way, the twisted quantum triple algebra provides a rigorous framework to describe the processes of interest in the condensed matter literature.

\bigskip \noindent
In the literature, fusion and braiding processes of loop-like excitations have often been described through \emph{dimensional reduction} arguments \cite{Levin:2012yb, Wang:2014oya,Wang:2014xba, Tiwari:2016zru}. This approach relies on the idea that upon \emph{compactification} of one of the spatial directions, a given topological model can be expressed in terms of another model in one lower dimension. In such context, the statistics of loop-like excitations in (3+1)d can be expressed in terms of the statistics of point-like excitations in (2+1)d.

This approach has most notably been applied in (3+1)d by considering the ground state subspace of the Dijkgraaf-Witten model for the 3-torus $\mathbb{T}^{3}$ \cite{Jiang:2014ksa,Wan:2014woa,Wang:2014oya,moradi2015universal}. In the (2+1)d Dijkgraaf-Witten model the topological spin, fusion and braiding statistics of anyons can be understood from the ground state subspace of the torus $\mathbb{T}^{2}$ by considering the action of the mapping class group of the torus ${\rm SL}(2,\mathbb{Z})$ on the corresponding states. Most investigations of the (3+1)d Dijkgraaf-Witten model have then utilised dimensional reduction techniques to consider the action of the ${\rm SL}(2,\mathbb{Z})$ subgroup of the mapping class group of the 3-torus ${\rm SL}(3,\mathbb{Z})$ on the ground state subspace of the 3-torus so as to infer the spin and braiding in analogy with the (2+1)d model. Although the notion of dimensional reduction and the statistics of loop-like excitations are indeed related, we explain that it is not necessary to use the former to describe the latter. In order to emphasize this point, we make the mechanisms at play precise by constructing explicitly the equivalent lower-dimensional model using the technology of \emph{loop-groupoids} \cite{willerton2008twisted, bartlett2009unitary}. We refer throughout the manuscript to such models as \emph{lifted} models. With this approach, we can not only clarify the nature of the input data of the lifted model, namely a loop-groupoid cocycle, but also construct explicitly the relevant Hilbert spaces in terms of loop-groupoid coloured graph-states.

This notion of lifted models in terms of loop groupoid is valid in any dimensions. Furthermore it can be iterated. This means that given a manifold that is $n$-times compactified, it is possible to express the original model in terms of another model in $n$-lower dimensions. In particular, we use this result in order to recast the higher-dimensional tube algebras in terms of the (1+1)d one, hence allowing for a particularly compact derivation and definition of the twisted quantum double and twisted quantum triple algebras.

In (2+1)d, the fusion and the braiding of point-like excitations can be made rather intuitive by means of a graphical calculus. In some cases, it may also make tedious computations a lot easier to perform. In this manuscript, we make a first step towards defining a graphical calculus for the statistics of loop-like excitations. In some ways, this provides a more physical description of the processes compared to the more rigorous and mathematical treatment provided by the twisted quantum triple. To do so, we propose a definition of the (3+1)d Hamiltonian realisation of the Dijkgraaf-Witten theory in terms of membrane-nets condensate and exploit the notion of lifted models. This alternative picture suggests a way to derive graphical identities that correspond to the algebraic definitions. For simplicity and in order to focus on the specificity of dealing with loop-like objects, we do so in the abelian case.

\begin{center}
{\bf Organization of the paper}
\end{center}
In sec.~\ref{sec:DW}, we review the definition of the Dijkgraaf-Witten theory both as a sigma model and as a lattice gauge theory. We pay particular attention to the definition of the partition function in the case where the manifold has a boundary. This is subsequently used to define the lattice Hamiltonian realisation of Dijkgraaf-Witten theory. In sec.~\ref{sec:tube}, a general framework in terms of tube algebras is presented to study excitations yielded by the Hamiltonian model in general dimensions. Three examples are studied in detail, namely the (1+1)d, the (2+1)d and the loop (3+1)d tube algebras. The notion of lifted models is also introduced in this section. The irreducible representations of these tube algebras that classify the simple excitations of the corresponding models are introduced in sec.~\ref{sec:simple}. Furthermore, the compatible comultiplication and $R$-matrices are defined, which in turn determine the fusion and the braiding of the excitations. In particular, we explain how the $R$-matrix of the twisted quantum triple algebra gives rise to the loop-braiding statistics as usually studied in the condensed matter literature. Sec.~\ref{sec:cat} reviews among other things the concept of loop groupoids. Apart from clarifying the meaning of the 2-cochains twisting the multiplication and comultiplication rules of the tube algebras, it makes more precise the notion of lifted models introduced in sec.~\ref{sec:tube}. Specifically, we show that given a manifold with one direction compactified, there exist a lower-dimensional model defined in terms of loop-groupoid colourings that is equivalent. This is finally used to express the $n$-dimensional tube algebra as an $n$-times lifted version of the (1+1)d one. In app.~\ref{sec:membrane} we introduce in the abelian case an alternative formulation of the (3+1)d model in terms of membrane-net condensation.

\newpage
\section{Dijkgraaf-Witten model\label{sec:DW}}
In this section, we review the construction of the Dijkgraaf-Witten partition function, as well as the definition and the main properties of its lattice Hamiltonian realisation.
\subsection{Partition function for closed manifolds}

In \cite{dijkgraaf1990topological}, Dijkgraaf and Witten introduced a topological gauge theory for any finite group $G$ in spacetime dimension $d$+1. They further showed that different $G$-gauge models are classified by cohomology classes $[\omega] \in H^{d+1}(BG, \rU(1))$ where $BG$ is the \emph{classifying space} of the group $G$, that is the topological space whose only non-vanishing homotopy group is the \emph{fundamental group} and it equals the group itself, i.e. $\pi_1(BG)=G$ \cite{eilenberg1953groups}. Given a finite group $G$ and a closed oriented ($d$+1)-manifold $\cM$, the partition function is performed over homotopy classes of maps $[\gamma]:\cM \to BG$, while the topological action is provided by the canonical pairing $\langle  \gamma^\star \omega, [\cM]\rangle$ between the pull-back of the cocycle $\omega \in Z^{d+1}(BG,\mathbb R / \mathbb Z)$ onto $\cM$ and the fundamental class $[\cM] \in H_{d+1}(\cM, \mathbb Z)$ of $\cM$. Putting everything together, we obtain a \emph{sigma model} with target space the classifying space $BG$ and the partition function explicitly reads
\begin{equation}
	\label{eq:DW_Z1}
	\mathcal{Z}_\omega^G[\cM] = \frac{1}{|G|^{b_0}}\sum_{[\gamma]: \cM \to BG} \langle  \gamma^\star \omega, [\cM]\rangle
\end{equation}
where the 0-th Betti number $b_0$ counts the number of connected components of $\cM$. Since the fundamental group is the only non-vanishing homotopy
group of $BG$, homotopy classes of maps $\cM \to BG$ can be expressed as \emph{homomorphisms} from the fundamental group $\pi_{1}(\cM)$ into $G$, up to simultaneous conjugation. We notate the set of such maps via ${\rm Hom}(\pi_1(\cM),G)/G$. This statement is merely the fact that the topology can be detected by holonomies along non-contractible closed curves only. Utilising this relation the partition function can be rewritten as:
\begin{equation}
	\label{eq:DW_Z2}
	\mathcal{Z}_\omega^G[\cM] = \frac{1}{|G|^{b_0}}\sum_{\gamma \in {\rm Hom}(\pi_1(\cM),G)/G} \langle  \gamma^\star \omega, [\cM]\rangle \; .
\end{equation}
Note that the expression above is only valid in the case where the manifold is closed. Indeed, if the manifold has a boundary, the fundamental class $[\cM]$ of the manifold cannot be defined and therefore the topological action cannot be written as $\langle  \gamma^\star \omega, [\cM]\rangle$. We will now derive another expression for the partition function that can be extended to open manifolds, ie. compact manifolds with boundary.

We begin by endowing the oriented ($d$+1)-manifold $\cM$ with a triangulation $\cM_{\triangle}$. More specifically, throughout this manuscript we will consider triangulations $\cM_{\triangle}$ of a manifold $\cM$ as a $\triangle$-complex whose geometric realisation is homeomorphic to $\cM$. Furthermore, we will require that all triangulations are equipped with a complete ordering of the vertex set $v_0 < v_1 < \ldots < v_{|\cM_{\triangle}^{0}|}$ where $|\cM_{\triangle}^{0}|$ is the total number of vertices. The ordering of the vertex set has the important feature that it naturally equips the edges (1-simplices) of $\cM$ with the structure of a directed graph, where we choose the convention that each edge is directed from the lowest to highest order vertex. Given a ($d$+1)-simplex $\triangle^{(d+1)}\in\cM_{\triangle}$, there are two possible configurations for the vertices which determines an orientation we notate via $\epsilon(\triangle^{(d+1)}) = \pm 1$. Since in the following we mainly focus on (3+1)d Dijkgraaf-Witten theory, we only provide below the explicit orientation conventions for 3- and 4-simplices:
\begin{convention}[\emph{Orientation of a 3-simplex}\label{conv:orThree}]
	Consider a 3-simplex $\triangle^{(3)} \equiv (v_0v_1v_2v_3)$ such that $v_0 < v_1 < v_2 <v_3$. Pick the 2-simplex $(v_0v_1v_2)$ and look at the remaining vertex $(v_3)$ through the 2-simplex. If the vertices $v_0$, $v_1$ and $v_2$ are organized in an clockwise fashion, the orientation of the 3-simplex $\epsilon(\triangle^{(3)})$ is $+1$, and $-1$ otherwise. 
\end{convention}

\begin{convention}[\emph{Orientation of a 4-simplex}\label{conv:orFour}]
	Consider a 4-simplex $\triangle^{(4)}$, pick one of the 3-simplices $\triangle^{(3)} \equiv (v_0v_1v_2v_3) \subset \triangle^{(4)}$ such that $v_0<v_1<v_2<v_3$ and determine its orientation according to conv.~\ref{conv:orThree}. The remaining vertex is denoted by $v$. If it takes an even number of permutations to bring the list $\{v,v_0,v_1,v_2,v_3\}$ to the ascending ordered one, then $\epsilon(v_0v_1v_2v_3v) = \epsilon(v_0v_1v_2v_3)$, and $\epsilon(v_0v_1v_2v_3v) = -\epsilon(v_0v_1v_2v_3)$ otherwise. 
\end{convention}
\noindent
The fundamental class of $\cM$ can now be expressed as
\begin{equation}
	[\cM] = \sum_{\triangle^{(d+1)}\subset \cM_{\triangle}} \epsilon(\triangle^{(d+1)}) \triangle^{(d+1)}
\end{equation}
so that the topological action in \eqref{eq:DW_Z2} can be decomposed as
\begin{equation}
	\langle \gamma^\star\omega, [\cM] \rangle = \prod_{\triangle^{(d+1)}\subset \cM_{\triangle}} \langle \omega, \triangle^{(d+1)} \rangle  \equiv \prod_{\triangle^{(d+1)}\subset \cM_{\triangle}} \omega(\triangle^{(d+1)})^{\epsilon(\triangle^{(d+1)})} \; .
\end{equation}
This last expression is also valid in the case where the manifold has a boundary. It remains to find an explicit expression for $\omega(\triangle^{(d+1)})$.  

\bigskip \noindent
Due to the path-connectedness of the classifying space $BG$, one may smoothly deform maps $\gamma\in {\rm Hom}(\pi_{1}(\cM),G)/G$ such that every $0$-simplex in $\cM$ is mapped to the same point in $BG$ and such that the space of paths in $BG$, which is $G$ up to homotopy, is mapped to the $1$-simplices of $\cM_{\triangle}$. Contractible paths are thus mapped to the identity group element. In practice, this means that every directed 1-simplex $(v_0v_1) \subset \cM_{\triangle}$ is assigned a group element $g_{v_0v_1}$ such that for every 2-simplex $(v_0v_1v_2)$ whose boundary is associated with a contractible path, the 1-cocycle condition (or flatness constraint) $ g_{v_0v_1} \cdot g_{v_1v_2} \cdot g_{v_0v_2}^{-1} = \mathbb{1}$ is imposed. Such a labeling of the 1-simplices defines a local description of a $G$-flat connection and is referred to as a $G$-colouring. We denote the set of $G$-colourings by ${\rm Col}(\cM_{\triangle},G)$.

This provides an algebraic expression for the topological action $\omega(\triangle^{(d+1)})$.  Given a ($d$+1)-simplex $\triangle^{(d+1)} \equiv (v_0v_1 \ldots v_{d+1}) \in \cM_{\triangle}$ such that $v_0<v_1<\ldots<v_{d+1}$ and $g\in {\rm Col}(\cM_{\triangle},G)$, we notate by $g[\triangle^{d+1}]$ the restriction of the colouring $g$ to the edges of $\triangle^{(d+1)}$ which is specified by the $d$+1 independent gauge fields $g_{v_0v_1}, g_{v_1v_2}, \dots, g_{v_{d}v_{d+1}}$. Using these conventions, we define the evaluation of the cocycle $\omega$ on the $G$-coloured ($d$+1)-simplex $ (v_0v_1 \ldots v_{d+1})$ as
\begin{equation}
	\omega(g[v_0v_1\ldots v_{d+1}]) \equiv \omega(g_{v_0v_1},g_{v_1v_2},\ldots,g_{v_dv_{d+1}}) \; .
\end{equation}
It is implicit in this construction that the cohomology $H^{d+1}(BG, \rU(1))$ of simplicial cocycles of $BG$ is equal to the group cohomology $H^{d+1}(G,\rU(1))$ of algebraic cocycles on $G$ whose definition is briefly recalled below:
\begin{definition}[\emph{Group cohomology}]
	Let $G$ be a finite group and $\cA$ a $G$-module whose action is denoted by $\triangleright$. We define an $n$-cochain on $G$ as a function $\omega_n: G^n \to \cA$.\footnote{When no confusion is possible, we will often drop the subscript $n$ in $\omega_n$.} The space of $n$-cochains on $G$ is denoted by $C^{n}(G,\cA)$. A coboundary operator $d^{(n)}: C^{n}(G,\cA) \to C^{n+1}(G,\cA)$ can be defined on the space of $n$-cochains via
	\begin{align}
		\label{eq:coBdry}
		&d^{(n)}\omega_n(g_1, \ldots, g_{n+1}) \\ \nn
		& \q = g_1 \triangleright \omega_n(g_2, \ldots, g_{n+1})\, \omega_n(g_1, \ldots g_n)^{(-1)^{n+1}} \prod_{i=1}^n \omega_n(g_1, \ldots,g_{i-1},g_{i}\cdot g_{i+1}, g_{i+2}, \ldots, g_{n+1})^{(-1)^i} \; .
	\end{align}
	It follows from the definition that $d^{(n+1)} \circ \, d^{(n)} = 0$.
	An $n$-cochain satisfying the equation $d^{(n)}\omega_n = 1 $ is referred to as an $n$-cocycle and the space of $n$-cocycles is denoted by $Z^{n}(G,\cA)$. We define an $n$-coboundary as an $n$-cocycle of the form $\omega_n = d^{(n-1)}\omega_{n-1}$. The subgroup of $n$-coboundaries is denoted by $B^n(G,\cA)$ and finally the $n$-th cohomology group of algebraic cocycles reads
	\begin{equation}
		H^n(G,\cA) = \frac{Z^n(G,\cA)}{B^n(G,\cA)} = \frac{{\rm Ker}\, d^{(n)}}{{\rm Im}\, d^{(n-1)}} \; .
	\end{equation}
\end{definition}
\noindent 
Throughout this manuscript, we take the $G$-module $\cA$ to be the abelian group $\rU(1)$ and the group action $\triangleright$ to be trivial. The partition function of a closed manifold $\cM$ for the discrete version of the Dijkgraaf-Witten partition function finally reads:
\begin{equation}
	\label{eq:DW_Z3}
	\mathcal{Z}_\omega^G[\cM] = \frac{1}{|G|^{|\cM_\triangle^{0}|}}\sum_{g \in {\rm Col}(\cM_{\triangle},G)} \prod_{\triangle^{(d+1)}\subset \cM_{\triangle}}  \omega(g[\triangle^{(d+1)}])^{\epsilon(\triangle^{(d+1)})} \; .
\end{equation}
The evaluation $\mathcal{Z}_\omega^G[\cM]\in \mathbb{C}$ is independent of the choice of triangulation $\cM_{\triangle}$ of $\cM$. In particular, this means that the partition function in invariant under so-called Pachner moves.\footnote{Given a piecewise linear manifold $\cM$ endowed with a triangulation $\cM_\triangle$, a Pachner move replaces $\cM_\triangle$ by another triangulation $\cM_\triangle'$ homeomorphic to $\cM$. In other words, given two triangulations of the same manifold, it is always possible to obtain one from the other via a finite sequence of Pachner moves.} This follows from the cocycle condition $d^{(d+1)} \omega = 1$ where $d^{(d+1)}$ is the group coboundary operator as defined in \eqref{eq:coBdry}. 

\subsection{Partition function for open manifolds\label{sec:DWopen}}
As announced earlier, formula \eqref{eq:DW_Z3} can be extended to the case of open manifolds.
A particularly important class of open manifolds in the following is provided by so-called \emph{cobordisms}.
Given a pair of oriented \emph{closed} $d$-dimensional manifolds $\mc{C}_{0}$ and $\mc{C}_{1}$, a ($d$+1)-dimensional cobordism from $\mc{C}_{0}$ to $\mc{C}_{1}$ is a compact oriented ($d$+1)-manifold $\mc{C}$ with boundary $\partial \mc{C}=\overline{\mc{C}_{0}}\sqcup \mc{C}_{1}$ where $\overline{\mc{C}_{0}}$ is the manifold $\mc{C}_{0}$ with orientation reversed. Given a triangulation $\mc{C}_{\triangle}$ of $\mc{C}$ with boundary triangulation $\partial \mc{C}_{\triangle}=\overline{\mc{C}_{\triangle,0}}\sqcup \mc{C}_{\triangle,1}$, the partition function defines a linear operator
\begin{align}
	&\mc{Z}^{G}_{\omega}[\mc{C}_{\triangle}]:\Hilb{\mc{C}_{\triangle,0}}\rightarrow \Hilb{\mc{C}_{\triangle,1}} \q \text{with} \q  \Hilb{\mc{C}_{\triangle,*}}\equiv\bigotimes_{\triangle^{(1)}\subset \mc{C}_{\triangle,*}}\mathbb{C}[G] \; ,
\end{align}
where $\mathbb{C}[G]$ is the Hilbert space spanned by complex linear combinations of the orthonormal basis elements $\{\ket{g}\}_{\forall g\in G}$. More explicitly, one has 
\begin{align*}\label{eq:DW_Zclosed}
	\mc{Z}^{G}_{\omega}[\mc{C}_{\triangle}]
	=
	\frac{1}{|G|^{|\mc{C}_{\triangle}^{0}|-\frac{1}{2}| \partial \mc{C}_{\triangle}^{0} |  } }
	&\sum_{g\in{\rm Col}(\mc{C}_{\triangle},G)}\prod_{\triangle^{(d+1)}\subset \mc{C}_{\triangle}}
	\!\!\! \omega(g[\triangle^{(d+1)}])^{\epsilon(\triangle^{(d+1)})}
	\!\!\! \bigotimes_{\triangle^{(1)}\subset\mc{C}_{\triangle,1} }
	\!\!\!
	\ket{g[\triangle^{(1)}]}
	\!\!\! \bigotimes_{\triangle^{(1)}\subset\mc{C}_{\triangle,0} }
	\!\!\!
	\bra{g[\triangle^{(1)}]} \; .
\end{align*}
The operator $\mc{Z}^{G}_{\omega}[\mc{C}_{\triangle}]$ is boundary relative triangulation independent, i.e. it is independent of the choice of triangulation of ${\rm int}(\mc{C}_{\triangle}):=\mc{C}_{\triangle}\backslash \partial \mc{C}_{\triangle}$ but does depend on the choice of boundary triangulation.

Given a pair of triangulated cobordisms $\mc{C}$ and $\mc{C'}$ with boundaries $\partial \mc{C}_{\triangle}=\overline{\mc{C}_{\triangle,0}}\sqcup \mc{C}_{\triangle,1}$ and $\partial \mc{C'}_{\triangle}=\overline{\mc{C}_{\triangle,1}}\sqcup \mc{C}_{\triangle,2}$, we can consider a new triangulated cobordism $\mc{C}_{\triangle}\cup_{\mc{C}_{\triangle,1}}\mc{C}'_{\triangle}$ obtained by gluing $\cC_\triangle$ and $\cC'_\triangle$ along their common boundary component $\mc{C}_{\triangle,1}$ such that
\begin{align}
	\Z[\mc{C'}_{\triangle}]\Z[\mc{C}_{\triangle}]=\Z[\mc{C}_{\triangle}\cup_{\mc{C}_{\triangle,1}}\mc{C}'_{\triangle}] \; .
\end{align}
Additionally, operators of this form satisfy the unitarity condition
\begin{align}
	\Zo{\overline{\mc{C}_{\triangle}}}=\Zo{\mc{C}_{\triangle}}^{\dagger} \; .
\end{align}

\medskip \noindent
Let us now focus on a special kind of cobordisms. Let $\Sigma$ be a $d$-dimensional surface, $\Sigma^{\mathbb I}\equiv \Sigma\times[0,1] \equiv \Sigma \times \mathbb I$ defines a cobordism with triangulation $\Sigma^{\mathbb I}_{\triangle}$ such that $\partial\Sigma^{\mathbb I}_{\triangle}=\overline{\Sigma_{\triangle}}\sqcup \Sigma_{\triangle}$. As a consequence of the boundary relative triangulation independence of $\mc{Z}^{G}_{\omega}$, we find the relations
\begin{align}
	\Zo{\Sigma^{\mathbb I}_{\triangle}}\Zo{\Sigma^{\mathbb I}_{\triangle}} &=\Zo{\Sigma^{\mathbb I}_{\triangle}}
	\nonumber\\
	\Z[\Sigma^{\mathbb I}_{\triangle}]&=\Z[\Sigma^{\mathbb I}_{\triangle}]^{\dagger} \; ,
\end{align}
so that $\Z[\Sigma^{\mathbb I}_{\triangle}]$ defines an \emph{Hermitian} projector. In this way, we define
\begin{align}\label{eq:gs1}
	{\rm Im}\Z[\Sigma^{\mathbb I}_{\triangle}]\equiv\gs{\Sigma_{\triangle}}\subseteq \mc{H}^G_\omega[\Sigma_{\triangle}]
\end{align}
to be the \emph{physical state space} of $\mc{Z}^{G}_{\omega}$ associated to the triangulation $\Sigma_{\triangle}$. A consequence of this definition is that for all states $\ket{\psi}\in \gs{\Sigma_{\triangle}}$
\begin{align}
	\Z[\Sigma^{\I}_{\triangle}]\triangleright \ket{\psi}=\ket{\psi} \; .
\end{align}
In sec.~\ref{sec:DWHam}, we will define an exactly solvable model that is the lattice Hamiltonian realisation of the Dijkgraaf-Witten model given by a sum of local mutually commuting projection operators. The Hamiltonian is defined in such a way that the ground state subspace for a closed triangulated $d$-manifold $\Sigma_{\triangle}$ is naturally identified with the physical state space $\mc{V}^G_\omega[\Sigma_{\triangle}]$, and the ground state projector is identified with  $\Z[\Sigma^{\mathbb I}_{\triangle}]$.

Note that in order to recover equation \eqref{eq:DW_Z3} as the limiting case of equation \eqref{eq:DW_Zclosed} for $\partial\mc{C}_{\triangle}=\varnothing$, we choose the convention that the empty set $\varnothing$ can be thought of as a closed oriented $d$-manifold such that $\overline{\varnothing}=\varnothing$. Thus, it follows that any closed  ($d$+1)-manifold $\mc{M}$ can be seen as a cobordism with boundary $\partial\mc{M}=\overline{\varnothing}\sqcup \varnothing$. Accordingly, we choose the conventions  $\Hilb{\varnothing}:=\mathbb{C}$ and $g[\varnothing]=\mathbb{1}$, for all $G$-colourings $g\in{\rm Col}(\mc{M},G)$. Putting everything together, this ensures that \eqref{eq:DW_Zclosed} does reduce to \eqref{eq:DW_Z3} when $\mc{C}_{\triangle}$ is a closed manifold.

\bigskip \noindent
Let us finally introduce yet another special class of open manifolds which will be particularly useful in the subsequent discussion, namely \emph{pinched intervals}:
\begin{definition}[\emph{Pinched interval}\label{def:pinched}]
	Let $\Xi$ be an oriented $d$-manifold with possibly non-empty boundary, the pinched interval $\Xi\times_{\rm p}\I$ of $\Xi$ is the quotient manifold
	\begin{align}
	\Xi\times_{\rm p}\I\equiv\Xi\times \I \; /\sim
	\end{align}
	where the equivalence relation $\sim$ is defined such that $ (a,i)\sim (a,i')$, for all $(a,i),(a,i')\in \partial\Xi\times \I $.
\end{definition}
\noindent
An immediate consequence of def.~\ref{def:pinched} is that $\partial (\Xi\times_{\rm p}\I)=\overline{\Xi}\cup_{\partial\Xi} \Xi$ and $\overline{\Xi}\cap \Xi=\partial\Xi$. By comparison $\partial (\Xi\times \I)=\overline{\Xi}\cup \Xi\cup (\partial\Xi\times \I)$. In order to illustrate this property, let us consider the following simple examples:
\begin{equation}
	[0,1]\times_{\rm p}[0,1]=
	\pinchedinterval{}
	\q , \q
	[0,1]\times [0,1]=
	\interval{} \; .
\end{equation}
Additionally, if $\partial \Xi=\varnothing$, then we can directly identify $\Xi\times_{\rm p}\I=\Xi\times \I$.

We now define the partition function $\Z$ for pinched interval cobordisms.
Let $\Xi_{\triangle},\Xi_{\triangle'}$ be a pair of triangulations of $\Xi$ such that $\partial\Xi_{\triangle}=\partial\Xi_{\triangle'}$ and
$\cob{\triangle}{\Xi}{\triangle'}$
a triangulation of $\Xi\times_{\rm p} \I$ such that $\partial (\cob{\triangle}{\Xi}{\triangle'})=\overline{\Xi_{\triangle}}\cup_{\partial\Xi_{\triangle'}}\Xi_{\triangle'}$, then
\begin{equation}
	\mc{Z}^{G}_{\omega}[\cob{\triangle}{\Xi}{\triangle'}]
	=
	\frac{1}{|G|^{ \#(\cob{\triangle}{\Xi}{\triangle'})} }
	\sum_{g\in{\rm Col}(\cob{\triangle}{\Xi}{\triangle'},G)}
	\prod_{\triangle^{(d+1)}\subset\cob{\triangle}{\Xi}{\triangle'}}
	\!\!\! \omega(g[\triangle^{(d+1)}])^{\epsilon(\triangle^{(d+1)})}
	\!\!\!
	\bigotimes_{\triangle^{(1)}\subset\Xi_{\triangle'} }
	\!\!\! \ket{g[\triangle^{(1)}]}
	\!\!\!
	\bigotimes_{\triangle^{(1)}\subset\Xi_{\triangle} }
	\!\!\! \bra{g[\triangle^{(1)}]} 
\end{equation}
where $\#(\cob{\triangle}{\Xi}{\triangle'}) := |\cob{\triangle}{\Xi}{\triangle'}^{0}|-\frac{1}{2}| \partial \cob{\triangle}{\Xi}{\triangle'}^{0} |-\frac{1}{2}|\partial \Xi_{\triangle}^{0}|$.
Given an oriented $d$-dimensional manifold equipped with triangulation $\Sigma_{\triangle}$ and $\Xi_{\triangle}\subseteq \Sigma_{\triangle}$ a subcomplex, there is a natural action of $\Zo{\cob{\triangle}{\Xi}{\triangle'}}$ that defines a linear map
\begin{align}
	\Zo{\cob{\triangle}{\Xi}{\triangle'}}
	:\Hilb{\Sigma_{\triangle}}\rightarrow \Hilb{\Sigma_{\triangle'}} \; ,
\end{align}
which in turn descends to a unitary isomorphism
\begin{align}\label{eq:gsiso}
	\Zo{\cob{\triangle}{\Xi}{\triangle'}}:\gs{\Sigma_{\triangle}}\xrightarrow{\sim} \gs{\Sigma_{\triangle'}}
\end{align}
where $\Sigma_{\triangle'}$ is a triangulation of $\Sigma$ for which the subcomplex $\Xi_{\triangle}$ replaced by $\Xi_{\triangle'}$. The fact $\Zo{\cob{\triangle}{\Xi}{\triangle'}}$ is a unitary isomorphism on the physical state space follows from the relations
\begin{align}
	\Zo{\cob{\triangle}{\Xi}{\triangle'}}^{\dagger}
	\Zo{\cob{\triangle}{\Xi}{\triangle'}}
	=
	\Zo{\cob{\triangle}{\Xi}{\triangle}}
\end{align}
and
\begin{align}
	\Zo{\cob{\triangle}{\Xi}{\triangle}}
	\Zo{\Sigma^{\I}_{\triangle}}
	=
	\Zo{\Sigma^{\I}_{\triangle}}
	=
	\Zo{\Sigma^{\I}_{\triangle}}
	\Zo{\cob{\triangle}{\Xi}{\triangle}} \;
\end{align}
which are a consequence of the boundary relative triangulation independence of $\Z$. Most importantly, this isomorphism implies that for a closed oriented $d$-manifold $\Sigma$, any two triangulations $\Sigma_{\triangle},\Sigma_{\triangle'}$ give rise to \emph{isomorphic} state spaces $\gs{\Sigma_{\triangle}}\simeq \gs{\Sigma_{\triangle'}}$, and hence the dimension of the state space is a triangulation \emph{independent} quantity.

\subsection{Lattice Hamiltonian realisation of Dijkgraaf-Witten theory}\label{sec:DWHam}

In this section, we present the lattice Hamiltonian realisation of the partition function \eqref{eq:DW_Z3} \cite{Hu:2012wx,Wan:2014woa} whose ground state subspace corresponds to the \emph{physical} Hilbert space defined in equation \eqref{eq:gs1}. Let $\Sigma$ be a closed oriented $d$-manifold equipped with a triangulation $\Sigma_{\triangle}$. The input for the model is given by a pair $(G,\omega)$, where $G$ is a finite group and $\omega$ is a representative normalised\footnote{A normalised ($d$+1)-cocycle $\omega$ is cocycle which gives the identity when any of the input group elements are the group identity.} ($d$+1)-cocycle in a cohomology class $[\omega]\in H^{d+1}(G,\rU(1))$. The microscopic Hilbert space of the model is given by
\begin{align}
	\Hilb{\Sigma_{\triangle}}\equiv \bigotimes_{\triangle^{(1)}\subset \Sigma_{\triangle}}\mathbb{C}[G]\;.
\end{align}
Letting $g$ be a $G$-labeling of $\Sigma_{\triangle}$, we call the state $\ket{g}\in \Hilb{\Sigma_{\triangle}}$ a \emph{graph-state}. Given a graph-state $\ket{g}$, we use the notation $g_{v_0v_1}\equiv g[v_{0}v_{1}]\in G$ to define the group element associated to the oriented edge $(v_0v_1) \equiv \triangle^{(1)} \subset \Sigma_\triangle$.

The Hamiltonian is defined in terms of two classes of operators, namely $\mathbb{B}_{\triangle^{(2)}}$ which act on the 2-simplices of $\Sigma_{\triangle}$, and $\mathbb{A}_{\triangle^{(0)}}$ which act on a local neighbourhood of the vertices of $\Sigma_{\triangle}$, such that
\begin{align}\label{eq:Ham}
	\Ham{\Sigma_{\triangle}}
	=-\sum_{\triangle^{(2)}\subset\Sigma_{\triangle}}\mathbb{B}_{\triangle^{(2)}}
	-\sum_{\triangle^{(0)}\subset\Sigma_{\triangle}}\mathbb{A}_{\triangle^{(0)}}\;.
\end{align}
The operator $\mathbb B_{(v_0v_1v_2)}$ for $(v_0v_1v_2)\subset \Sigma_{\triangle}$ is defined by its action on a graph-state $\ket{g}\in\Hilb{\Sigma_{\triangle}}$ as follows:
\begin{equation}
\mathbb B_{(v_0v_1v_2)} \triangleright | g \rangle = \delta_{g_{v_0v_1} g_{v_1v_2}\, , \,g_{v_0v_2}} | g \rangle
\end{equation}
which can be extended linearly to an operator on any state $\ket{\psi}\in\Hilb{\Sigma_{\triangle}}$.
The $\mathbb B$-operators provide an energy penalty for non-flat $G$-connections of $\Sigma_{\triangle}$.

For every vertex $\triangle^{(0)} \equiv (v_0)$ of $\Sigma_{\triangle}$, the operator $\mathbb A_{(v_0)}$ acts on the subcomplex $\Xi_{v_{0}}:={\rm cl} \circ {\rm st}(v_0)\subset \Sigma_{\triangle}$. Here ${\rm cl}(*)$ is the \emph{closure} operation and ${\rm st}(*)$ is the \emph{star} operation such that $\Xi_{v_0}$ is the smallest subcomplex of $\Sigma_{\triangle}$ that contains all the simplices which share $v_0$ as a subsimplex \cite{Williamson:2016evv}. We define $\mathbb A_{(v_0)}$ in terms of the triangulated pinched interval $\cob{\Xi_{v_{0}}\!}{\Xi}{\, \Xi_{v_{0}}}$ of which we choose a triangulation via
\begin{align}
	\cob{\Xi_{v_{0}}\!}{\Xi}{\, \Xi_{v_{0}}}:=
	(\snum{v_0'}) \sqcup_{\rm j} {\rm cl} \circ {\rm st}(v_0)\;.
\end{align}
Here $ {*}\sqcup_{\rm j} {*}$ is the \emph{join} operation, where the join of two simplices $ \triangle^{(p)} \equiv (v_0 v_1 \ldots v_p) $ and $\triangle^{(q)} \equiv (v_{p+1}v_{p+2} \ldots v_{p+q+1})$ is a new simplex $\triangle^{(p)} \sqcup_{\rm j} \triangle^{(q)} \equiv (v_0v_1 \ldots v_{p+q+1})$ and $v_{0}<v_{0'}<v_1$ is an auxiliary vertex which respects the ordering of $v_{0}$ with respect to all other vertices in $\Sigma_{\triangle}$. Let us illustrate these different definitions with a two-dimensional example:
\begin{equation}
	\Sigma_{{\rm 2d}, \triangle} = \clst{1} 
\end{equation}
so that
\begin{equation}
	\snum{(0')} \sqcup_{\rm j} {\rm cl} \circ {\rm st} (0) = 
	\snum{(0')} \sqcup_{\rm j} \; \clst{2} \; =  \;	\clst{3} \; .
\end{equation}
In this notation, for a given $\ket{\psi}\in\Hilb{\Sigma_{\triangle}}$, we finally define the action of $\mathbb{A}_{(v_{0})}$ via
\begin{align}
	\label{eq:opA}
	\mathbb{A}_{v_{0}}\triangleright \ket{\psi}=\Zo{(\snum{v_0'}) \sqcup_{\rm j} {\rm cl} \circ {\rm st}(v_0)}\ket{\psi} \; .
\end{align}
For instance, in (3+1)d the action of the operator $\mathbb{A}_{(4)}$ on a vertex $\snum{(4)}$ shared by four 3-simplices explicitly reads
\begin{align}
	\mathbb{A}_{(4)} \, \triangleright \Bigg| \HamONE{1}{1} \Bigg\rangle 
	&= \mathcal{Z}_\pi^G\Bigg[\HamTWO{1}{1}\Bigg]  \, \Bigg|\HamONE{1}{1}\Bigg\rangle \\[-1.5em] \nn
	& = \frac{1}{|G|}\sum_{k \in G} 
	\frac{\pi(ab,c,d,k) \, \pi(a,b,cd,k)}{\pi(b,c,d,k) \, \pi(a,bc,d,k) }
	\Bigg|\HamONE{1}{2}\Bigg\rangle \; .
\end{align}
\noindent
where $\pi \in Z^4(G,\rU(1))$. Note that the $G$-colouring was left implicit and we made use of the shorthand notation $g_{01}\equiv a$, $g_{12} \equiv b$, $g_{23}\equiv c$, $g_{34}\equiv d$ and $g_{44'}\equiv k$.

It follows directly from the definitions that the $\mathbb B$-operators are mutually commuting projection operators and that any $\mathbb A$-operator commutes with any $\mathbb B$-operator, and vice versa. The only non-trivial commutation relation corresponds to the situation where two $\mathbb A$-operators act on two vertices that are shared by the same 1-simplex. Let us consider for instance a 1-simplex $(v_0v_1)$ and let us compare the action of $\mathbb A_{(v_0)} \circ \mathbb A_{(v_1)}$ and $\mathbb A_{(v_1)} \circ \mathbb A_{(v_0)}$. It follows from the definition that the amplitudes of $\mathbb{A}_{(v_0)}$ and $\mathbb A_{(v_1)}$ are $\mathcal Z^G_\omega[(v_0') \sqcup_{\rm j} {\rm cl} \circ {\rm st}(v_0)]$ and $\mathcal Z^G_\omega[(v_1') \sqcup_{\rm j} {\rm cl} \circ {\rm st}(v_1)]$, respectively. However, the results of ${\rm cl} \circ {\rm st}(v_1)$ and ${\rm cl} \circ {\rm st}(v_0)$ depends on the order in which we act with the operators. Indeed, the action of $\mathbb A_{(v_0)} \circ \mathbb A_{(v_1)}$ is such that $(v_0) \subset {\rm cl}\circ {\rm st}(v_1)$ and $(v_1') \subset {\rm cl}\circ {\rm st}(v_0)$, while the action of $\mathbb A_{(v_1)} \circ \mathbb A_{(v_0)}$ is such that $(v_1) \subset {\rm cl}\circ {\rm st}(v_0)$ and $(v_0') \subset {\rm cl}\circ {\rm st}(v_1)$. In both cases, the overall amplitude is provided by $\mathcal{Z}_\omega^G[((v_0')\sqcup_{\rm j} {\rm cl} \circ {\rm st}(v_0)) \cup ((v_1') \sqcup_{\rm j} {\rm cl}\circ {\rm st}(v_1)) ]$. However, in the former case, the simplicial complex $((v_0')\sqcup_{\rm j} {\rm cl} \circ {\rm st}(v_0)) \cup ((v_1') \sqcup_{\rm j} {\rm cl}\circ {\rm st}(v_1))$ contains the 1-simplex $(v_0v_1')$ while in the latter case it contains $(v_0'v_1)$. These two complexes share the same topology and boundary so that they can be related by a finite sequence of Pachner moves. Topological invariance of the partition function then guarantees that the amplitudes are the same, and therefore that the operators $\mathbb A_{(v_0)}$ and $\mathbb A_{(v_1)}$ commute. Furthermore, that $\mathbb A_{(v_0)}$ is a \emph{projection} operator follows directly from the definition in terms of $\Z$ applied to a triangulated \emph{pinched} interval. Consequently, the Hamiltonian is a sum of \emph{mutually commuting projection operators} such that the model is exactly solvable. 

\bigskip \noindent
Let us conclude the definition of the Hamiltonian realisation by elucidating the relation between the ground state subspace of the Hamiltonian and the corresponding physical state space of the Dijkgraaf-Witten model. By definition, a ground state $\ket{\psi}$ of $\Ham{\Sigma_{\triangle}}$ is given by a linear superposition of graph-states such that the conditions $\mathbb{A}_{\triangle^{(0)}} \triangleright \ket{\psi}=\ket{\psi}$ and $\mathbb{B}_{\triangle^{(2)}} \triangleright\ket{\psi}=\ket{\psi}$ are satisfied for all $\triangle^{(0)},\triangle^{(2)}\subset \Sigma_{\triangle}$. Noting that the set of operators $\{\mathbb{A}_{\triangle^{(0)}},\mathbb{B}_{\triangle^{(2)}}\}_{\forall \triangle^{(0)},\triangle^{(2)}\subset \Sigma_{\triangle}}$ are all mutually commuting projection operators, we define the \emph{ground state projector} $\mathbb{P}_{\Sigma_{\triangle}}$ via:
\begin{align}
	\label{eq:projGS}
	\mathbb{P}_{\Sigma_{\triangle}}:= \prod_{\triangle^{(0)}\subset \Sigma_{\triangle}}\mathbb A_{\triangle^{(0)}} \prod_{\triangle^{(2)}\subset \Sigma_{\triangle}}\mathbb B_{\triangle^{(2)}} \;
	= \prod_{\triangle^{(0)}\subset \Sigma_{\triangle}}\mathbb A_{\triangle^{(0)}} \; .
\end{align}
The second equality in the above follows from the fact that the operator $\mathbb{A}_{(v_{0})}$ naturally enforces the flatness condition in ${\rm cl} \circ {\rm st}(v_0)\subseteq \Sigma_{\triangle}$, but $\cup_{(v_{0})\subset\Sigma_{\triangle}}{\rm cl} \circ {\rm st}(v_0)=\Sigma_{\triangle}$, and thus the term $\prod_{\triangle^{(2)}\subset \Sigma_{\triangle}}\mathbb B_{\triangle^{(2)}}$ is superfluous in the definition of the ground state projector $\mathbb{P}_{\Sigma_{\triangle}}$. Utilising the definition $\mathbb{A}_{(v_{0})}=\Zo{(\snum{v_0'}) \sqcup_{\rm j} {\rm cl} \circ {\rm st}(v_0)}$, we can naturally make the identification:
\begin{align}
	\prod_{\triangle^{(0)} \subset \Sigma_{\triangle}}\mathbb A_{\triangle^{(0)}}
	=
	\Zo{\Sigma^{\I}_{\triangle}}
\end{align}
where each ordering of the product of $\mathbb A_{\triangle^{(0)}}$ defines a different boundary relative triangulation of $\Sigma^{\I}_{\triangle}$. But the operator $\Z$ is invariant under such choices, hence the equality. In this way we can identify the ground state subspace of $\Ham{\Sigma_{\triangle}}$ with the physical state space $\gs{\Sigma_{\triangle}}$ of the Dijkgraaf-Witten model where
\begin{align}
	\label{eq:idSub}
	{\rm Im} \, \mathbb{P}_{\Sigma_{\triangle}}={\rm Im} \, \Zo{\Sigma^{\I}_{\triangle}}\equiv\gs{\Sigma_{\triangle}}
\end{align}
following from equation \eqref{eq:gs1}.

\subsection{Fixed point wave functions\label{sec:fixedpoint}}

\begin{figure}[]
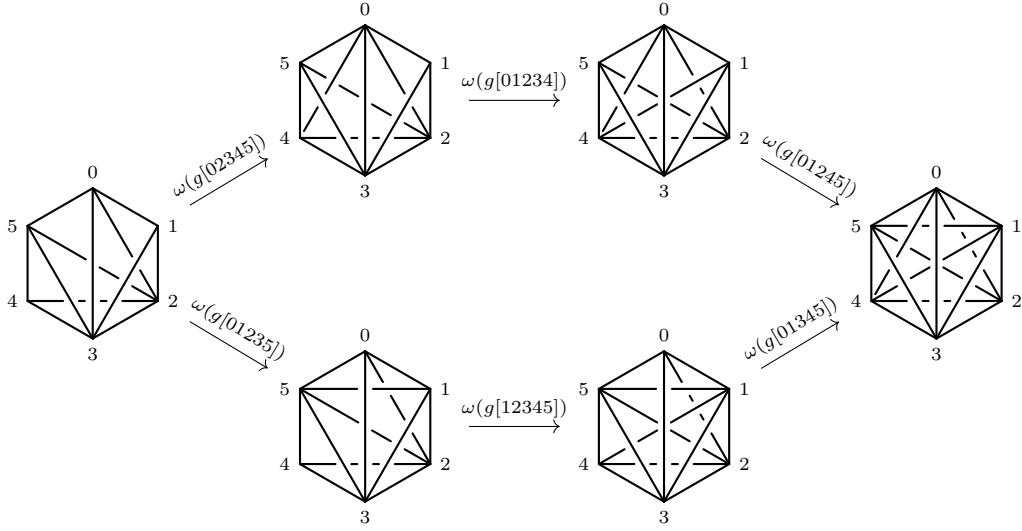

	\center
	\assocTWO{}
	\caption{Coherence relation of the $2 \leftrightharpoons 3$ Pachner moves. Given a complex obtained as the gluing of three 3-simplices, two different sequences of $2 \leftrightharpoons 3$ Pachner moves result in the same complex obtained as the gluing of six 3-simplices. Each arrow is decorated by the amplitude of the corresponding move obtained as the evaluation of $\omega$ on the 4-simplex whose boundary provides the 3-simplices involved in the corresponding $2 \leftrightharpoons 3$ move. This coherence relation is satisfied if $\omega$ is a 4-cocycle.}
	\label{fig:pentagonator}
\end{figure}

In equation \eqref{eq:gsiso}, it was shown using the language of pinched interval operators that given a pair of triangulations $\Sigma_{\triangle}$ and $\Sigma_{\triangle'}$ of a closed oriented three-manifold $\Sigma$, the corresponding state spaces were isomorphic, i.e. $\gs{\Sigma_{\triangle}}\simeq \gs{\Sigma_{\triangle'}}$. In light of the identification \eqref{eq:idSub} between the physical state space of the topological theory and the Hamiltonian ground state subspace, this informs us that under a local change of triangulation, ground states remain in the same gapped phase. In other words, to a change of triangulation corresponds a local unitary transformation that performs an adiabatic evolution of the system that preserves the gap. These local transformations can in turn be used in order to generate a renormaliation group flow so that gapped ground states can be interpreted as \emph{fixed point wave functions} \cite{Chen:2010gda}.

An important class of pinched interval operators for the following discussion are the so-called \emph{Pachner operators}. For a given compact $\Sigma$, any two triangulations of $\Sigma$ can be mutated between each other by a finite set of Pachner moves \cite{PACHNER1991129}. For instance, in three dimensions, we distinguish two sets of invertible Pachner moves given by the $1\leftrightharpoons 4$ and $2\leftrightharpoons 3$ moves:
\begin{equation}
	\PFourOne{1}{1}{-0.8}{0.3}{0.8}{0}
	\;\; \xrightleftharpoons[1-4]{4-1} \;\;
	\PFourOne{1}{2}{-0.8}{0.3}{0.8}{0}
	\q\q  , \q\q
	\PTwoThree{1}{1}{-1.5}{0.3}{0.8}{0}
	\;\; \xrightleftharpoons[2-3]{3-2} \;\;
	\PTwoThree{1}{2}{-1.5}{0.3}{0.8}{0} \; .
\end{equation}
Each $d$-dimensional Pachner move is derived from a ($d$+1)-simplex $\triangle^{(4)}$, which defines a triangulation of the ($d$+1)-disk, and such that the boundary $\partial \triangle^{(d+1)}$ defines a triangulation of the $d$-sphere $\mathbb S^{d}$. But a hemispherical decomposition of $\partial \triangle^{(d+1)}=\overline{{\textsf N}}\cup_{\partial \textsf{S}=\partial \overline{\textsf N}} \textsf{S}$ into two subcomplexes $\textsf{N}$ and $\textsf{S}$ (the north and south hemispheres) is such that the ($d$+1)-simplex defines a pinched interval which changes a subcomplex given by $\textsf{N}$ of a triangulated three-manifold to have triangulation $\textsf{S}$. Using the Dijkgraaf-Witten partition function $\Z$, such a pinched interval can be lifted to a unitary isomorphism on $\gs{\Sigma_{\triangle}}$. Furthermore, boundary relative triangulation independence of the partition function guarantees that the isomorphism is independent of the order or choices of Pachner operators between two triangulated three-manifolds as illustrated in fig.~\ref{fig:pentagonator} for the three-dimensional Pachner moves. For the sake of concreteness, we consider below two examples of three-dimensional  Pachner operators:
\begin{align}
	\label{eq:P41}
	\mathbb{P}_{4 \rightharpoonup 1}
	&:=\frac{1}{|G|^\frac{1}{2}}\sum_{g\in {\rm Col}((01234),G)} \!\!\!\! \omega(g[\snum{01234}])^{\epsilon(01234)}
	\Bigg|g\bigg[ \PFourOneneg{1}{2}{-0.8}{0.3}{0.8}{1}\bigg] \Bigg\rangle
	\Bigg\langle g\bigg[ \PFourOneneg{1}{1}{-0.8}{0.3}{0.8}{1}\bigg]\Bigg| \;\;\;\;\;\;\;
	\\
	\label{eq:P23}
	\mathbb{P}_{3 \rightharpoonup 2}
	&:=\sum_{g\in {\rm Col}((01234),G)}
	\!\!\! \omega(g[\snum{01234}])^{\epsilon(01234)}
	\Bigg|g\bigg[ \PTwoThreeneg{1}{2}{-1.5}{0.3}{0.8}{1}\bigg] \Bigg\rangle
	\Bigg\langle g\bigg[ \PTwoThreeneg{1}{1}{-1.5}{0.3}{0.8}{1}\bigg]\Bigg|
\end{align}
where the vertex enumeration is left implicit, since such enumeration defines an orientation for the corresponding 4-simplex according to conv.~\ref{conv:orFour}, and instead insert the orientation dependency in the amplitude. Finally, we define $\mathbb{P}_{1 \rightharpoonup 4}=\mathbb{P}^{\dagger}_{4 \rightharpoonup  1}$ and $\mathbb{P}_{2 \rightharpoonup  3}=\mathbb{P}^{\dagger}_{3 \rightharpoonup  2}$.

\newpage

\section{Tube algebras and excitations\label{sec:tube}}

In this section, we present an approach to study and classify excitations in topological models. This approach consists in revealing the algebraic structure underlying the excitations yielded by the lattice Hamiltonian. First, we present the general framework, then we provide some lower-dimensional examples, and finally we reveal and study in detail the algebraic structure relevant for the (3+1)d model. 

\subsection{General framework}
Given a closed oriented $d$-manifold $\Sigma$ equipped with a choice of triangulation $\Sigma_{\triangle}$, we introduced in equation \eqref{eq:Ham} the lattice Hamiltonian $\Ham{\Sigma_{\triangle}}$ whose ground state subspace $\gs{\Sigma_{\triangle}}$ is spanned by linear combinations of graph-states on $\Sigma_{\triangle}$ that satisfy the stabiliser conditions $\mathbb{A}_{\triangle^{(0)}}\triangleright |\psi\rangle=|\psi\rangle$ and $\mathbb{B}_{\triangle^{(2)}}\triangleright |\psi\rangle=|\psi\rangle$ for all $\triangle^{(0)},\triangle^{(2)}\subset \Sigma_{\triangle}$. This ground state subspace corresponds to a translational invariant state with constant energy density for all local neighbourhoods of $\Sigma_{\triangle}$.

We define an \emph{excitation} in the model to be a connected subcomplex of $\Sigma_{\triangle}$ with an energy density higher than that of the ground state. In terms of the Hamiltonian, an excitation is obtained by violating the stabiliser constraints $\mathbb{A}_{\triangle^{(0)}}\triangleright |\psi\rangle=|\psi\rangle$ and $\mathbb{B}_{\triangle^{(2)}}\triangleright |\psi\rangle=|\psi\rangle$ at 0- and 2-simplices of the subcomplex. Recalling that the Hamiltonian constraints $\mathbb{A}_{\triangle^{(0)}}$ and $\mathbb{B}_{\triangle^{(2)}}$ enforce twisted gauge invariance and flatness, respectively, we call a state $|\psi\rangle$ whereby $\mathbb{A}_{\triangle^{(0)}}\triangleright|\psi\rangle=0$ for one vertex $\triangle^{(0)}$ an \emph{electric charge excitation}, and a state for which $\mathbb{B}_{\triangle^{(2)}}\triangleright|\psi\rangle=0$ for one 2-simplex $\triangle^{(2)}$ a \emph{magnetic flux excitation}. There are numerous equivalent approaches to classifying excitations in the theory, such as the construction of string and membrane operators from the algebra of local symmetries. In this section, we instead utilise a generalisation of the Ocneanu's tube algebra \cite{ocneanu1994chirality, ocneanu2001operator}. The cornerstone of this approach is that the physical properties of any excitation associated with a given subcomplex of $\Sigma_{\triangle}$ are encoded as boundary conditions of the triangulation obtained by removing this subcomplex from $\Sigma_{\triangle}$. As such, it is possible to classify excitations by classifying boundary conditions.

Given an open manifold $\Sigma^{\rm o}$ with triangulation $\Sigma^{\rm o}_{\triangle}$ we define the Hamiltonian of equation \eqref{eq:Ham} as follows:
\begin{equation}
	\label{eq:Hamopen}
	\Ham{\Sigma^{\rm o}_{\triangle}} = 
	- \sum_{\triangle^{(2)}\subset {\rm int}(\Sigma^{\rm o}_{\triangle})}\mathbb B_{\triangle^{(2)}}
	- \sum_{\triangle^{(0)}\subset {\rm int}(\Sigma^{\rm o}_{\triangle})}\mathbb A_{\triangle^{(0)}}  \; ,
\end{equation}
where ${\rm int}(\Sigma^{\rm o}_{\triangle}):=\Sigma^{\rm o}_{\triangle}\backslash \partial \Sigma^{\rm o}_{\triangle}$. Since the Hamiltonian does not mix graph-states with different $G$-connections on $\partial\Sigma^{\rm o}_{\triangle}$, we say $\Ham{\Sigma^{\rm o}_{\triangle}}$ has \emph{open boundary conditions}. In the presence of such open boundary conditions, the ground state subspace naturally admits a decomposition via
\begin{align}
	\label{eq:grading}
	\gs{\Sigma^{\rm o}_{\triangle}}=\bigoplus_{a\in {\rm Col}(\partial \Sigma^{\rm o}_\triangle, G)}\gs{\Sigma^{\rm o}_{\triangle}}_{a}
\end{align}
where ${\rm Col}(\partial \Sigma^{\rm o}_\triangle, G)$ notates the set of $G$-colourings (or flat $G$-connections) of $\partial \Sigma^{\rm o}_{\triangle}$, and for $a\in {\rm Col}(\partial \Sigma^{\rm o}_\triangle, G)$, $\gs{\Sigma^{\rm o}_{\triangle}}_{a}$ is the ground state subspace consisting of linear superpositions of graph-states with boundary colouring $a$. Generically, a boundary condition $a\in {\rm Col}(\partial \Sigma^{\rm o}_\triangle, G)$ defines a set of excitations which are a linear superposition of magnetic flux and electric charge excitations. In order to find states with well-defined electric charge and magnetic flux, we will instead find an alternative basis for $\gs{\Sigma^{\rm o}_{\triangle}}$, namely the so-called \emph{fusion basis} \cite{DDR1, Delcamp:2017pcw, AB}.

\bigskip \noindent
We demonstrated previously that for closed spatial manifolds, there exists unitary isomorphisms between the ground state subspace associated with different choices of triangulation. Moreover, the corresponding equivalence classes can be interpreted as fixed point wave functions and are in one-to-one correspondence with the ground state wave functions of the lattice Hamiltonian. For open spatial manifolds this triangulation independence is not manifest on the triangulation choice of the boundary. We will now introduce a related `symmetry' of the ground state subspace for open manifolds with respect to the gluing of spatial \emph{tubes}.\footnote{We use the terminology `symmetry' here loosely as the symmetry is given by an algebra rather than a group as in the usual context.} 

Let us begin with a simple observation: Given a manifold $\Sigma^{\rm o}$ with non-empty boundary, we can always glue a copy of $\partial\Sigma^{\rm o}\times \I$ to $\Sigma^{\rm o}$ without modifying the topology, i.e. there exists an orientation preserving diffeomorphism such that $\partial\Sigma^{\rm o}\times \I\cup_{\partial\Sigma^{\rm o}}\Sigma^{\rm o}\simeq \Sigma^{\rm o}$. Let us illustrate this property with a lower-dimensional example: Let $\Sigma_{\rm 2d}^{\rm o}$ be a 2-manifold with boundary $\mathbb{S}^1$, we can glue a copy of the \emph{cylinder} $\mathbb{S}^1\times \I$ along $\partial\Sigma_{\rm 2d}^{\rm o}=\mathbb{S}^1$ in order to form $\mathbb{S}^1\times \I\cup_{\mathbb{S}^1}\Sigma^{\rm o}_{\rm 2d}\simeq \Sigma^{\rm o}_{\rm 2d}$. This gluing operation can be depicted as follows:
\begin{equation}
	\onecylinder{1.3}
	\onecylinderb{1.3}
	\xrightarrow{\text{gluing}}
	\twocylinderb{1.3}
	\simeq
	\onecylinderb{1.3} \; .
\end{equation}
In general spatial dimensions, making use of the operators that perform triangulation changes, this gluing process can be extended to a generalised symmetry of the ground state subspace. Let $\Sigma^{\rm o}_{\triangle}$ be a $d$-dimensional open manifold and $\Xi_{\triangle}= \partial\Sigma^{\rm o}_{\triangle}$ its boundary, we define $\tubem{\Xi_{\triangle}}$ to be a triangulation of $\Xi_{\triangle}\times \I$ with $\partial\tubem{\Xi_{\triangle}}=\overline{\Xi_{\triangle}}\sqcup \Xi_{\triangle}$. Henceforth, we call $\tubem{\Xi_{\triangle}}$ the \emph{tube} of $\Xi_{\triangle}$. Following from \eqref{eq:grading}, we decompose $\gs{\tubem{\Xi_{\triangle}}}$ via
\begin{align}
	\label{eq:gradingTube}
	\gs{\tubem{\Xi_{\triangle}}}
	\equiv\bigoplus_{a \in {\rm Col}(\Xi_{\triangle} \times \{0\},G) \atop b\in {\rm Col}(\Xi_{\triangle} \times \{1\},G)}
	\!\!\! \gs{\tubem{\Xi_{\triangle}}}_{a,b} \; .
\end{align}
We then define the injective map
\begin{align}
	\mathfrak{G} \,: \, \gs{\Sigma^{\rm o}_{\triangle}}\otimes \gs{\tubem{\Xi_{\triangle}}}
	\xrightarrow{\q}
	\!\!\! \bigoplus_{a \in  {\rm Col}(\partial \Sigma^{\rm o}_\triangle,G) \atop {a' \in  {\rm Col}(\Xi_{\triangle} \times \{0\},G) \atop b \in {\rm Col}(\Xi_{\triangle} \times \{1\},G)}}
	\!\!\!
	\gs{\Sigma^{\rm o}_{\triangle}}_{a}\otimes \gs{\tubem{\Xi_{\triangle}}}_{a',b}\; \subseteq \;
	\Hilb{\Sigma^{\rm o}_{\triangle}\cup_{\Xi_{\triangle} }\tubem{\Xi_{\triangle}}}
\end{align}
which acts on states $\ket{\psi_{a}}\in\gs{\Sigma^{\rm o}_{\triangle}}_{a}$ and $\ket{\varphi_{a',b}}\in \gs{\tubem{\Xi_{\triangle}}}_{a',b}$ by identifying boundary conditions on the gluing interface, i.e.
\begin{equation}
	\mathfrak{G} \,: \,\ket{\psi_{a}}\otimes \ket{\varphi_{a',b}}\mapsto \delta_{a,a'} \, \ket{\psi_{a}}\otimes \ket{\varphi_{a',b}} \; .
\end{equation}
This can be linearly extended to states with mixed grading. Most importantly, the image of the map $\mathfrak{G}$ is a subspace of $	\Hilb{\Sigma^{\rm o}_{\triangle}\cup_{\Xi_{\triangle} }\tubem{\Xi_{\triangle}}}$ which differs from the ground state subspace $\gs{\Sigma^{\rm o}_{\triangle}\cup_{\Xi_{\triangle} }\tubem{\Xi_{\triangle}}}$ because the Hamiltonian operators may be violated on the gluing interface. Letting
$\gs{\Sigma^{\rm o}_{\triangle}\cup_{\Xi_{\triangle} }\tubem{\Xi_{\triangle}}}
\xrightarrow{\sim}
\gs{\Sigma^{\rm o}_{\triangle}}$
define the triangulation changing unitary isomorphism between the two ground state subspaces, we define the operator $\star$ as the following composition of maps:
\begin{align*}
	\star \,: \,
	\gs{\Sigma^{\rm o}_{\triangle}}\otimes \gs{\tubem{\Xi_{\triangle}}}
	\xrightarrow{\mathfrak{G}}
	\Hilb{\Sigma^{\rm o}_{\triangle}\cup_{\Xi_{\triangle} }\tubem{\Xi_{\triangle}}}
	\xrightarrow{\mathbb{P}_{\Sigma^{\rm o}_{\triangle}\cup_{\Xi_{\triangle} }\tubem{\Xi_{\triangle}}}}
	\gs{\Sigma^{\rm o}_{\triangle}\cup_{\Xi_{\triangle} }\tubem{\Xi_{\triangle}}}
	\xrightarrow{\sim}
	\gs{\Sigma^{\rm o}_{\triangle}} \; ,
\end{align*}
where $\mathbb{P}$ is the projection map defined in \eqref{eq:projGS}. In particular, this implies a map 
\begin{align}
	\star:\gs{\tubem{\Xi_{\triangle}}}\otimes \gs{\tubem{\Xi_{\triangle}}}\rightarrow \gs{\tubem{\Xi_{\triangle}}}
\end{align}
that enriches the Hilbert space $\gs{\tubem{\Xi_{\triangle}}}$ with the structure of a finite dimensional algebra denoted by $\tubealg{\Xi_{\triangle}}$. Similarly, we can interpret $\gs{\Sigma^{\rm o}_{\triangle}}$ as defining a module over $\tubealg{\Xi_{\triangle}}$. It was shown in \cite{AB} that the algebra $\tubealg{\Xi_{\triangle}}$ is an \emph{associative semi-simple $*$-algebra} for any choice of triangulated boundary $\Xi_{\triangle}$, so that the ground state subspace $\gs{\Sigma^{\rm o}_{\triangle}}$ can be decomposed in terms of simple modules under the action on $\tubealg{\Xi_{\triangle}}$ by
\begin{align}
	\gs{\Sigma^{\rm o}_{\triangle}}=\bigoplus_{m\in \mathfrak{M}}\gs{\Sigma^{\rm o}_{\triangle}}_{m} \;,
\end{align}
where $\gs{\Sigma^{\rm o}_{\triangle}}_{m}$ is a simple $\tubealg{\Xi_{\triangle}}$ module and $\mathfrak{M}$ denotes the set of simple $\tubealg{\Xi_{\triangle}}$ modules up to isomorphism.

\bigskip \noindent
The approach described above can be used to classify the excitations of the theory. First, let us make a simple observation: Given an open manifold $\Sigma^{\rm o}$, it is always possible to find a collar neighbourhood of $\partial \Sigma^{\rm o}$  that is diffeomorphic to $\partial \Sigma^{\rm o} \times \I$. More specifically, given a triangulated manifold $\Sigma^{\rm o}_{\triangle}$ with a connected boundary component $\Xi_{\triangle}$, using triangulation changes we can always pick a representative ground state Hilbert $\gs{\Sigma^{\rm o}_{\triangle'}}$ isomorphic to $\gs{\Sigma^{\rm o}_{\triangle}}$ whereby $\Sigma^{\rm o}_{\triangle'}$ is a triangulation equivalent to $\Sigma_{\triangle}$ such that a local neighbourhood of the boundary $\Xi_{\triangle}$ is of the form $\tubem{\Xi_{\triangle}}$. Using this isomorphism, we can localise the above gluing map to act only on degrees of freedom restricted to the local neighbourhood of the boundary given by $\tubem{\Xi_{\triangle}}$. This means that we can classify boundary conditions for $\Xi_{\triangle}$, and hence excitations contained in a subcomplex bounded by $\Xi_{\triangle}$, in terms of the simple modules of the regular module of $\tubem{\Xi_{\triangle}}$, i.e. $\tubem{\Xi_{\triangle}}$ considered as a $\tubem{\Xi_{\triangle}}$-module.

A consequence of the semi-simplicity of $\tubealg{\Xi_{\triangle}}$ is that there are only \emph{finitely} many irreducible excitations in the theory and all other excitations are a linear superposition of such excitations.
In practice there are two non-canonical choices in defining the tube algebra. First a triangulation of the boundary $\Xi_{\triangle}$ and then a triangulation of $\tubem{\Xi_{\triangle}}$. For a given choice of boundary triangulation $\Xi_{\triangle}$ it was shown in \cite{AB} that any two triangulations of $\tubem{\Xi_{\triangle}}$ define isomorphic algebras. Independence of the choice of boundary triangulation is slightly more subtle. Although the tube algebra, and a fortiori the simple modules, depends on the choice of triangulation of the boundary, any two triangulations of the same boundary manifold yields \emph{Morita} equivalent tube algebras. Morita equivalence is a weaker relation between two algebras than isomorphism in the sense that two algebras can have different dimension while maintaining Morita equivalence. Instead, Morita equivalence states that the simple modules of two algebras are in one-one correspondence.\footnote{Morita equivalence is defined as follows: Let $\mathcal{A}_1$ and $\mathcal{A}_2$ be two associative algebras. Then $\mathcal{A}_1$ is Morita equivalent to $\mathcal{A}_2$ if and only if there exists a pair $(_{\mathcal{A}_1}\mathcal{P}_{\mathcal{A}_2}, _{\mathcal{A}_2}\mathcal{Q}_{\mathcal{A}_1})$ of $\mathcal{A}_1$--$\mathcal{A}_2$- and $\mathcal{A}_2$--$\mathcal{A}_1$-bimodules, respectively, such that ${_{\mathcal{A}_1}\mathcal{P}_{\mathcal{A}_2}} \otimes_{\mathcal{A}_2}  {_{\mathcal{A}_2}\mathcal{Q}_{\mathcal{A}_1}} \simeq \mathcal{A}_1$ and ${_{\mathcal{A}_2}\mathcal{P}_{\mathcal{A}_1}} \otimes_{\mathcal{A}_1}  {_{\mathcal{A}_1}\mathcal{Q}_{\mathcal{A}_2}} \simeq \mathcal{A}_2$. Morita equivalence is an important concept in the study of tube algebras as the dimension of the tube algebra has a strict dependency on the triangulation choice of the boundary. However, any two choices of boundary triangulations define Morita equivalent algebras such that the simple excitations of the two algebras are in one-one correspondence and thus the Morita equivalence class of simple modules is a triangulation independent quantity.}

\subsection{Tube algebra in (1+1)d}\label{sec:1+1dtube}
Ultimately we are interested in the excitations yielded by the (3+1)d lattice Hamiltonian model described in sec.~\ref{sec:DWHam}, but it is instructive to first consider some lower-dimensional examples. Here we present the simplest non-trivial example of the tube algebra approach for the lattice Hamiltonian realisation of (1+1)d Dijkgraaf-Witten theory. 

Let $\Sigma_{\rm 1d}$ be a 1d surface equipped with a triangulation $\Sigma_{{\rm 1d}, \triangle}$. The input for the model is given by a pair $(G,\beta)$, where $G$ is a finite group and $\beta$ is a representative normalised 2-cocycle in a cohomology class $[\beta]\in H^{2}(G,\rU(1))$. As explained earlier in the general case, the models assigns to every oriented edge $(v_0v_1) \equiv \triangle^{(1)} \subset \Sigma_{{\rm 1d },\triangle}$ a group element $g[v_0v_1]$ such that the microscopic Hilbert space of the model is given by
\begin{align}
	\mathcal{H}^G_\beta[\Sigma_{{\rm 1d}, \triangle}]
	\equiv \bigotimes_{\triangle^{(1)}\subset \Sigma_{{\rm 1d}, \triangle}}\mathbb{C}[G]\;.
\end{align}
In (1+1)d there is a unique choice of boundary, namely the 0-dimensional point $\mathbb{o}$. Taking the triangulation of the point to be  $0$-simplex, we can define its tube $\tubem{\mathbb{o}}=\triangle^{(1)}$ as a 1-simplex. Any other triangulation would give rise to an isomorphic vector space and thus isomorphic algebra, so that we are free to choose the simplest triangulation without loss of generality. Graphically, we depict this 1d tube as
\begin{align}
	\tubem{\mc{\mathbb{o}}}:=
	\OneDCyl{0.8}{0}{1}{}
\end{align}
so that the corresponding Hilbert space $\gs{\tubem{\mathbb{o}}}$ reads
\begin{align}
	\gs{\tubem{\mathbb{o}}}
	={\rm Span}_{\mathbb{C}}
	\big\{
	\, \big| g[\OneDCyl{0.8}{0}{1}{}] \big\rangle \, 
	\big\}_{\forall g\in {\rm Col}(\mathfrak{T}[\mathbb{o}],G)} 
	\equiv 
	{\rm Span}_{\mathbb{C}}
	\big\{
	\, \big| \OneDCyl{0.8}{0}{1}{a} \big\rangle \, 
	\big\}_{\forall a\in G}\; ,
\end{align}
which is equipped with the canonical inner product
\begin{align}
	\big\langle \, \OneDCyl{0.8}{0}{1}{a} \, \big| \, \OneDCyl{0.8}{0}{1}{b} \, \big\rangle
	=\delta_{a,b} \; .
\end{align}
Note that since there is a unique configuration of the point $\mathbb{o}$, the grading \eqref{eq:gradingTube} is in this case trivial. Utilising the previous discussion, we can now define the algebra product on $\gs{\tubem{\mathbb{o}}}$ as follows:
\begin{align}
	\nn\big|
	\OneDCyl{0.8}{0}{1}{a}
	\big\rangle
	\star 
	\big|
	\OneDCyl{0.8}{1}{2}{b}
	\big\rangle
	&= \mathbb{P}_{\tubem{\mathbb{o}} \cup_{\mathbb{o}} \tubem{\mathbb{o}}} \circ \mathfrak{G} \triangleright
	\big( \, \big| \OneDCyl{0.8}{0}{1}{a} \big\rangle
	\otimes \big| \OneDCyl{0.8}{1}{2}{b} \big\rangle \, \big) \\
	&=
	\mathbb{P}_{\tubem{\mathbb{o}} \cup_{\mathbb{o}} \tubem{\mathbb{o}}} \triangleright \big( \, \big|
	\OneDCylDouble{0.8}{0}{1}{2}{a}{b}
	\big\rangle \, \big) \; .
\end{align}
Applying definition \eqref{eq:opA}, the action of the operator $\mathbb{P}$ is expressed in terms of the partition function $\mathcal{Z}^G_\beta$ as follows:
\begin{align}
	\nn
	&\mathbb{P}_{\tubem{\mathbb{o}} \cup_{\mathbb{o}} \tubem{\mathbb{o}}} \triangleright \big( \, \big|
	\OneDCylDouble{0.8}{0}{1}{2}{a}{b}
	\big\rangle \, \big) \\[-0.5em]
	& \q = \mathcal{Z}^G_\beta \Bigg[\OneDCylZ{0.8}{0}{1}{2}{1'}{a}{b}{} \Bigg]
	\big|
	\OneDCylDouble{0.8}{0}{1}{2}{a}{b}
	\big\rangle
	\\
	& \q =
	\nn
	\frac{1}{|G|}\sum_{k}\frac{\beta(a,k)}{\beta(k,k^{-1}b)}
	\big|
	\OneDCylDouble{0.8}{0}{1'}{2}{ak}{k^{-1}b}
	\big\rangle \; .
\end{align}
It now remains to apply the triangulation changing isomorphism between ground states subspaces so as to recover the initial triangulation. Following sec.~\ref{sec:DWopen} and \ref{sec:fixedpoint}, this isomorphism is expressed as the 2d partition function for the pinched interval cobordism given by the 2-simplex $\snum{(012)}$. Explicitly, the triangulation changing operator reads
\begin{align}
	\mathcal{Z}^G_\beta \Bigg[
	\OneDCylPinched{0.8}{0}{2}{1'}{}{}
	\Bigg]
	=\frac{1}{|G|^\frac{1}{2}}\sum_{c,d\in G}\beta(c,d)
		\big| \, \OneDCyl{0.8}{0}{2}{cd} \, \big\rangle \big\langle \, \OneDCylDouble{0.8}{0}{1'}{2}{c}{d} \, \big|
\end{align}
so that
\begin{equation}
	\big| \, \OneDCylDouble{0.8}{0}{1'}{2}{ak}{k^{-1}b} \, \big\rangle
	\simeq \frac{1}{|G|^\frac{1}{2}}\, \beta(ak,k^{-1}b) 
	\big| \,  \OneDCyl{0.8}{0}{2}{ab} \, \big\rangle \; .
\end{equation}
Putting everything together, the algebra product of ${\rm Tube}^G_\beta(\mathbb o)$ is given by
\begin{equation}
	\label{eq:1+1dalg}
	\big|
	\OneDCyl{0.8}{0}{1}{a}
	\big\rangle
	\star 
	\big|
	\OneDCyl{0.8}{1}{2}{b}
	\big\rangle
	=
	\frac{1}{|G|^\frac{1}{2}}\beta(a,b)
	\big|
	\OneDCyl{0.8}{0}{2}{ab}
	\big\rangle \; ,
\end{equation}
where we made use of the 2-cocycle condition $d^{(2)}\beta(a,k,k^{-1}b) =1$.

\subsection{Tube algebra in (2+1)d: Twisted quantum double}
We continue our discussion with a second example of tube algebra for the lattice Hamiltonian realisation of (2+1)d Dijkgraaf-Witten model. Given a 2d surface $\Sigma_{\rm 2d}$ equipped with a triangulation $\Sigma_{{\rm 2d},\triangle}$, the input data of the model is a pair $(G,\alpha)$, where $G$ is a finite group and $\alpha$ is a representative normalised 3-cocycle in a cohomology class $[\alpha] \in H^3(G,\rU(1))$. As before, the models assigns to every oriented edge $(v_0v_1) \equiv \triangle^{(1)} \subset \Sigma_{{\rm 2d },\triangle}$ a group element $g[v_0v_1]$ such that the microscopic Hilbert space of the model is given by
\begin{align}
	\mathcal{H}^G_\alpha[\Sigma_{{\rm 2d}, \triangle}]
	\equiv \bigotimes_{\triangle^{(1)}\subset \Sigma_{{\rm 2d}, \triangle}}\mathbb{C}[G]\;.
\end{align}
As explained earlier in the general case, given a compact two-dimensional surface, any two triangulations can be mutated between each other by a finite set of Pachner moves. In two dimensions, we distinguish two sets of invertible Pachner moves given by the $1\leftrightharpoons 3$ and $2\leftrightharpoons 2$ moves. Each Pachner move is now derived from a $3$-simplex $\triangle^{(3)}$ which defines a pinched interval. Using the 3d Dijkgraaf-Witten partition function, such a pinched interval can in turn be lifted to a unitary isomorphism on the corresponding ground state subspace. For instance the Pachner operator associated with the $2 \rightharpoonup 2$ move explicitly reads
\begin{align}
	\label{eq:P22}
	\mathbb{P}_{2 \rightharpoonup 2}
	&:=\frac{1}{|G|^\frac{1}{2}}\sum_{g\in {\rm Col}((0123),G)} \!\!\!\! \alpha(g[\snum{0123}])^{\epsilon(0123)}
	\Bigg|g\bigg[ \; \TwoPachner{0.8}{1} \; \bigg] \Bigg\rangle
	\Bigg\langle g\bigg[\; \TwoPachner{0.8}{2} \; \bigg]\Bigg|
\end{align}
where the vertex enumeration is left implicit.

In (2+1)d there is a unique choice of closed boundary manifold, namely the circle (or 1-sphere) $\mathbb{S}^{1}$. Up to Morita equivalence we can choose the circle to be triangulated by a single 1-simplex with both vertices identified, and we refer to this triangulation as $\mathbb{S}^{1}_{\triangle}$. It follows that the corresponding tube $\tubem{\mathbb{S}^{1}_{\triangle}}$ of $\mathbb{S}^1_\triangle$ is provided by a triangulated quadrilateral with two opposite edges identified, i.e.
\begin{equation}
	\tubem{\mathbb{S}^{1}_{\triangle}}:=
	\TwoDCyl{0.8}{0}{1}{}{} \equiv	\onecylinderPure{1.3}
\end{equation}
with the identification of vertices $\snum{(0)} \equiv \snum{(0')}$, $\snum{(1)} \equiv \snum{(1')}$ and edge $\snum{(01)} \equiv \snum{(0'1')}$.
The Hilbert space $\mathcal{V}^G_\alpha[\mathfrak{T}[\mathbb{S}^{1}_{\triangle}]]$ is thus spanned by $G$-coloured graph-states of the form
\begin{align}
	\nn
	\mathcal{V}^G_\alpha[\mathfrak{T}[\mathbb{S}^{1}_{\triangle}]]
	={\rm Span}_{\mathbb{C}}
	\Bigg\{
	\, \Bigg| g\bigg[ \TwoDCyl{0.8}{0}{1}{}{} \bigg] \Bigg\rangle \, 
	\Bigg\}_{\forall g\in {\rm Col}(\mathfrak{T}[\mathbb{S}^1_\triangle],G)} 
	&\equiv 
	{\rm Span}_{\mathbb{C}}
	\Bigg\{
	\, \Bigg| \TwoDCyl{0.8}{0}{1}{a}{x} \Bigg\rangle \, 
	\Bigg\}_{\forall a,x\in G} \\
	\label{eq:Hilb2D}
	&=:{\rm Span}_{\mathbb C}
	\big\{ \,
	\ket{(\looSy{x})\xrightarrow{a}}
	\, \big\}_{\forall a,x \in G} \; ,
\end{align}
where the shorthand notation introduced in the last line will be justified in sec.~\ref{sec:cat}. Since we can distinguish several boundary configurations, this Hilbert space admits a non-trivial grading:
\begin{align}
	\mathcal{V}^G_\alpha[\mathfrak{T}[\mathbb{S}^{1}_{\triangle}]]
	\equiv\bigoplus_{a,x \in G}
	\!\!\! \mathcal{V}^G_\alpha[\mathfrak{T}[\mathbb{S}^{1}_{\triangle}]]_{x,a^{-1}xa} \; .
\end{align}
Following exactly the same steps as before, we define the algebra product on $	\mathcal{V}^G_\alpha[\mathfrak{T}[\mathbb{S}^{1}_{\triangle}]]$ as follows:
\begin{align*}
	\nn\Bigg|
	\TwoDCylImp{0.8}{0}{1}{a}{x_1}
	\Bigg\rangle
	\star 
	\Bigg|
	\TwoDCylImp{0.8}{1}{2}{b}{x_2}
	\Bigg\rangle
	&= \mathbb{P}_{\tubem{\mathbb{S}^1_\triangle} \cup_{\mathbb{S}^1_\triangle} \tubem{\mathbb{S}^1_\triangle}} \circ \mathfrak{G} \triangleright
	\Bigg( \, \Bigg| \TwoDCylImp{0.8}{0}{1}{a}{x_1} \Bigg\rangle
	\otimes \Bigg| \TwoDCylImp{0.8}{1}{2}{b}{x_2} \Bigg\rangle  \Bigg) \\
	&=
	\delta_{x_2,a^{-1}x_1a} \,\mathbb{P}_{\tubem{\mathbb{S}^1_\triangle} \cup_{\mathbb{S}^1_\triangle} \tubem{\mathbb{S}^1_\triangle}} \triangleright \Bigg( \, \Bigg|
	\TwoDCylDouble{0.8}{0}{1}{2}{a}{b}{x_{1}}
	\Bigg\rangle  \Bigg) \; ,
\end{align*}
where some of the $G$-labels are left implicit since they can be deduced from the flatness constraints. Applying definition \eqref{eq:opA}, the action of the operator $\mathbb{P}$ then reads
\begin{align*}
	\mathbb{P}\triangleright 
	\Bigg|
	\TwoDCylDouble{0.8}{0}{1}{2}{a}{b}{x_{1}}
	\Bigg\rangle  \Bigg) 
	 =
	\frac{1}{|G|}\sum_{k}
	\frac{\tau_{x_{1}}(\alpha)(a,k)}{\tau_{a^{-1}x_{1}a}(\alpha)(k,k^{-1}b)}
	\Bigg|
	\TwoDCylDouble{0.8}{0}{1}{2}{ak}{k^{-1}b}{x_{1}} \Bigg\rangle
\end{align*}
where we defined
\begin{align}
	\label{eq:S1trans3}
	\tau_{x}(\alpha)(a,b) := \frac{\alpha(x,a,b) \, \alpha(a,b,(ab)^{-1}xab)}{\alpha(a,a^{-1}xa,b)} \; .
\end{align}
Henceforth, we refer to the function $\tau(\alpha)$ as the $\mathbb S^1$-transgression of $\alpha$, for reasons that will be clarified in sec.~\ref{sec:cat}.\footnote{Note that if the group $G$ is abelian, then $\tau(\alpha)$ defines a normalised group 2-cocycle.} Repeated application of the 3-cocycle condition $d^{(3)}\alpha=1$ implies the following properties:
\begin{equation}
	\label{eq:twisted2Coc}
	\frac{\tau_{a^{-1}xa}(\alpha)(b,c) \, \tau_{x}(\alpha)(a,bc)}{\tau_{x}(\alpha)(ab,c) \, \tau_{x}(\alpha)(a,b)}=1 \; ,
\end{equation}
\begin{equation}
\label{eq:twisted2Cocc}
	\tau_{\mathbb 1_G}(\alpha)(a,b)=\tau_{x}(\alpha)(\mathbb 1_G,b)=\tau_{x}(\alpha)(a,\mathbb 1_G) = 1
	\q \text{and}  \q 
	\tau_{x}(\alpha)(a,a^{-1})=\tau_{a^{-1}xa}(\alpha)(a^{-1},a) \; ,
\end{equation}
for all $x,a,b,c \in G$.
It now remains to apply the triangulation changing isomorphism between ground states subspaces so as to recover the initial triangulation. Following sec.~\ref{sec:DWopen} and \ref{sec:fixedpoint}, this isomorphism is expressed as the 3d partition function for the pinched interval cobordism whose triangulation is provided by  the cartesian product $\snum{(012)^+} \times \mathbb{S}^1$:
\begin{equation}
	\snum{(012)}^{+} \times \mathbb{S}^{1} :=
	\TwoDCylPinched{0.8}{2}{0}{1}{}{}
	\equiv \snum{(00'1'2')}^+ \cup \snum{(011'2')}^{-} \cup \snum{(0122')}^{+} \; .
\end{equation}
The corresponding triangulation changing operator reads
\begin{align*}
	\mathcal{Z}^G_\alpha [\snum{(012)^+ \times \mathbb{S}^1}]
	=\frac{1}{|G|^\frac{1}{2}}\sum_{y,c,d\in G}\tau_y(\alpha)(c,d)
	\Bigg| \, \TwoDCylImp{0.8}{0}{1}{cd}{y} \, \Bigg\rangle 
	\Bigg\langle \, \TwoDCylDouble{0.8}{0}{1}{2}{c}{d}{y} \, \Bigg|
\end{align*}
so that
\begin{equation}
	\Bigg| \, \TwoDCylDouble{0.8}{0}{1}{2}{ak}{k^{-1}b}{x_{1}} \, \Bigg\rangle
	\simeq \frac{1}{|G|^\frac{1}{2}}\, \tau_{x_{1}}(\alpha)(ak,k^{-1}b) 
	\Bigg| \, \TwoDCylImp{0.8}{0}{2}{ab}{x_{1}} \, \Bigg\rangle \; .
\end{equation}
Putting everything together, and using the notation introduced in eq.~\eqref{eq:Hilb2D}, the algebra product of ${\rm Tube}^G_\alpha(\mathbb S^1_\triangle)$ is given by
\begin{equation}\label{eq:2dalgprod}
	|(\looSy{x_1})\xrightarrow{a} \rangle
	\star
	|(\looSy{x_2})\xrightarrow{b} \rangle
	=
	\delta_{x_2,a^{-1}x_1a} \, \frac{1}{|G|^\frac{1}{2}}\, \tau_{x_1}(\alpha)(a,b) \,
	|(\looSy{x_1})\xrightarrow{ab} \rangle
\end{equation}
where we made use of the condition
\begin{align}
\frac{\tau_{x_{1}}(\alpha)(a,k)\tau_{x_{1}}(\alpha)(ak,k^{-1}b)}{\tau_{a^{-1}xa}(\alpha)(k,k^{-1}b)  }=\tau_{x_{1}}(\alpha)(a,b)
\end{align}
which follows from \eqref{eq:twisted2Coc}.
This algebra was first defined by Roche, Dijkgraaf et al and is often referred to as the twisted quantum double of a finite group \cite{Dijkgraaf1991}.
Interestingly, we note there are strong similarities between the (1+1)d and the (2+1)d cases. It turns out that these similarities will persist in the (3+1)d case. As a matter of fact, we will explain in sec.~\ref{sec:cat} how these can be exploited in order to define higher-dimensional tube algebras in terms of the (1+1)d one.

\subsection{Loop tube algebra in (3+1)d: Twisted quantum triple\label{sec:twistedTriple}}

In this section we consider an example of the (3+1)d tube algebra.
Let $\Sigma_{\rm 3d}$ be a 3d surface equipped with a triangulation $\Sigma_{{\rm 3d}, \triangle}$. The input for the model is given by a pair $(G,\pi)$, where $G$ is a finite group and $\pi$ is a representative normalised 4-cocycle in a cohomology class $[\pi]\in H^{4}(G,\rU(1))$. As before, the models assigns to every oriented edge $(v_0v_1) \equiv \triangle^{(1)} \subset \Sigma_{{\rm 1d },\triangle}$ a group element $g[v_0v_1]$ such that the microscopic Hilbert space is the same as earlier.

Generally in (3+1)d, a boundary can be defined by each closed surface which in turn can be classified by its \emph{genus}. Here we derive the tube algebra for the lattice Hamiltonian realisation of (3+1)d Dijkgraaf-Witten theory associated to a triangulation of the torus $\mathbb{T}^{2}$. The representation theory of this algebra will then be derived and utilised in order to classify the loop-like excitations of the model, where by loop-like excitation we mean an excitation with the topology of the circle $\mathbb{S}^1$. Given a spatial 3-manifold $\Sigma$, a \emph{regular neighbourhood} of a loop is given by the solid torus $\mathbb D^{2}\times \mathbb{S}^1$, and as such loop-like excitations are classified by the boundary conditions of the torus $\mathbb{T}^{2}=\partial (\mathbb{D}^{2}\times \mathbb{S}^1)$. This means that we need to derive the tube algebra associated with the manifold $\mathbb{T}^2 \times \I$. We define a triangulation of $\mathbb{T}^{2}\times \I$ as follows
\begin{align}
	\tubem{\mathbb{T}^{2}_{\triangle}}:=\threeCyl{0.8}{0}{1}{}{}{} \equiv \;
	\niceThreeCyl{1}
\end{align}
where we make the identifications $\snum{(0)}\equiv\snum{(0')}\equiv\snum{(\tilde 0)}\equiv\snum{(\tilde{0}')}$, $\snum{(1)}\equiv\snum{(1')}\equiv\snum{(\tilde 1)}\equiv\snum{(\tilde{1}')}$, $\snum{(01)}\equiv\snum{(0'1')}\equiv \snum{(\tilde 0 \tilde 1)} \equiv \snum{(\tilde{0}' \tilde{1}')} $, $\snum{(00') }\equiv \snum{(\tilde{0} \tilde{0}')}$, $\snum{(0\tilde 0) }\equiv \snum{(0' \tilde{0}')}$, $\snum{(11') }\equiv \snum{(\tilde{1} \tilde{1}')}$ and $\snum{(1 \tilde 1) }\equiv \snum{(1' \tilde{1}')}$. 
The ground state subspace $\gs{\tubem{\mathbb{T}^{2}_{\triangle}}}$ is then provided by all superpositions of $G$-coloured graph-states of the form 
\begin{equation}
	\label{eq:Hilb3D}
	\mathcal{V}^G_\pi[\mathfrak{T}[\mathbb{T}^{2}_{\triangle}]]
	\equiv 
	{\rm Span}_{\mathbb{C}}
	\Bigg\{
	\, \Bigg| \threeCyl{0.8}{0}{1}{x}{y}{a}  \Bigg\rangle \, 
	\Bigg\}_{\substack{\forall a,y\in G \\ \forall x \in Z_y}}=:{\rm Span}_{\mathbb C}
	\big\{ \,
	\ket{(\looDSy{x}{y})\xrightarrow{a}}
	\, \big\}_{\substack{\forall a,y \in G \\ \forall x \in Z_y}} \; ,
\end{equation}
where $Z_y = \{ x \in G \, | \, xy = yx\}$ so that the $G$-colouring is specified by $a:=g[\snum{01}]$, $x:=g[\snum{00'}]$ and $y:=g[\snum{0' \tilde{0}'}]$ with $xy=yx$. As before the $G$-colouring of the remaining edges is left implicit since it follows from the different identifications and flatness constraints. Once again, the specific choice of shorthand notation we make will be justified in sec.~\ref{sec:cat}.

Let us now define the algebra product on $	\mathcal{V}^G_\pi[\mathfrak{T}[\mathbb{T}^{2}_{\triangle}]]$. Since the derivation follows exactly the same steps as for the two previous examples, we will present it in a more succinct way. 
In terms of the basis states provided above, we have
\begin{align}
	&\Bigg| \threeCyl{0.8}{0}{1}{x_1}{y_1}{a} \Bigg\rangle \star \Bigg| \threeCyl{0.8}{1}{2}{x_2}{y_2}{b} \Bigg\rangle 
	\\[-2em]
	& \q \q =\delta_{x_2,a^{-1}x_1a} \, \delta_{y_2,a^{-1}y_1a} \, \mathbb{P}_{\mathfrak{T}[\mathbb{T}^2_\triangle] \cup_{\mathbb{T}^2_\triangle} \mathfrak{T}[\mathbb{T}^2_\triangle]} \triangleright
	\Bigg| \threeCylDouble{0.8}{0}{1}{2}{x_1}{y_1}{a}{b} \Bigg\rangle \; .
\end{align}
The action of the projection operator $\mathbb{P}$ then reads
\begin{align*}
	\mathbb{P}\triangleright 
	\Bigg|
	\threeCylDouble{0.8}{0}{1}{2}{x_1}{y_1}{a}{b}
	\Bigg\rangle 
	=
	\frac{1}{|G|}\sum_{k}
	\frac{\tau_{x_1,y_1}(\pi)(a,k)}{\tau_{x_1^a,y_1^a}(\pi)(k,k^{-1}b)}
	\Bigg|
	\threeCylDouble{0.8}{0}{1}{2}{x_1}{y_1}{ak}{k^{-1}b}
	\Bigg\rangle 
\end{align*}
where we introduced the notation $x^a := a^{-1}xa$ and defined
\begin{align}
	\tau_{x,y}^2(\pi)(a,b) := \frac{\tau_y(\pi)(x,a,b) \, \tau_y(\pi)(a,b,x^{ab})}{\tau_y(\pi)(a,x^a,b)} 
\end{align}
as the $\mathbb S^1$-transgression of $\tau(\pi)$ that is itself defined according to
\begin{equation}
	\label{eq:S1trans4}
	\tau_y(\pi)(a,b,c) = \frac{\pi(a,y^a,b,c) \, \pi(a,b,c,y^{abc})}{\pi(y,a,b,c) \, \pi(a,b,y^{ab},c)} \; .
\end{equation}
We also refer to $\tau(\pi)$ as the $\mathbb S^1$-transgression of $\pi$ so that $\tau^2(\pi)$ is the $\mathbb S^1$-transgression of the $\mathbb S ^1$-transgression of $\pi$. Henceforth, we will therefore refer to the function $\tau^2(\pi)$ as the $\mathbb T^2$-transgression of $\pi$, for reasons that will be clarified in sec.~\ref{sec:cat}.  Repeated application of the 4-cocycle condition $d^{(4)}\pi=1$ implies the following properties:
\begin{equation}
	\label{eq:twisted3Coc}
	\frac{\tau_{a^{-1}ya}(\pi)(b,c,d) \, \tau_{y}(\pi)(a,bc,d) \, \tau_y(\pi)(a,b,c)}{\tau_{y}(\pi)(ab,c,d) \, \tau_{y}(\pi)(a,b,cd)}=1 \; ,
\end{equation}
\begin{equation}
	\tau_{\mathbb 1_G}(\pi)(a,b,c)=\tau_{y}(\pi)(\mathbb 1_G,b,c) = \tau_{y}(\pi)(a,\mathbb 1_G,c)=\tau_{y}(\pi)(a,b,\mathbb 1_G) = 1 \; , 
\end{equation}
\begin{equation}
	\tau_{y}(\pi)(a,a^{-1},a)=\tau_{a^{-1}ya}(\pi)^{-1}(a^{-1},a,a^{-1})
\end{equation}
for all $y,a,b,c \in G$.
Repeated application of \eqref{eq:twisted3Coc} in turn implies the subsequent properties:
\begin{equation}
	\label{eq:twisted2Coc2}
	\frac{\tau_{a^{-1}xa,a^{-1}ya}^2(\pi)(b,c) \, \tau_{x,y}^2(\pi)(a,bc)}{\tau_{x,y}^2(\pi)(ab,c) \, \tau^2_{x,y}(\pi)(a,b)}=1 \; ,
\end{equation}
\begin{equation}
\label{eq:twisted2Cocc2}
	\tau_{\mathbb 1_G,y}^2(\pi)(a,b)=\tau_{x,\mathbb 1_G}^2(\pi)(a,b) = \tau_{x,y}^2(\pi)(\mathbb 1_G,b)=\tau_{x,y}^2(\pi)(a,\mathbb 1_G) = 1 \; , 
\end{equation}
\begin{equation}
	\tau_{x,y}^2(\pi)(a,a^{-1})=\tau_{a^{-1}xa,a^{-1}ya}^2(\pi)(a^{-1},a)	
	 \; ,
\end{equation}
for all $y,a,b,c \in G$ and $x \in Z_y$. It now remains to apply the triangulation changing isomorphism between ground states subspaces so as to recover the initial triangulation. Following sec.~\ref{sec:DWopen} and \ref{sec:fixedpoint}, this isomorphism is expressed as the 4d partition function for the pinched interval cobordism whose triangulation is provided by  the cartesian product $\snum{(012)^+} \times \mathbb T^2$:
\begin{alignat}{5}
	&\snum{(012)}^{+} \times\mathbb{T}^{2}
	& \,= \,& \,
	\big( \snum{(0122')}^{+} &\, \cup \,&
	\snum{(011'2')}^{-}&\, \cup \,&
	\snum{(00'1'2')}^{+} \big) &\, \times \,& \mathbb{S}^{1}
	\nn\\
	& {} &\, = \,& \,
	\snum{(0122'\tilde{2}')}^{+} &\, \cup \,& \snum{(012\tilde{2}\tilde{2}')}^{-} &\, \cup \,& \snum{(01\tilde{1}\tilde{2}\tilde{2}')}^{+} &\, \cup \,& \snum{(0\tilde{0}\tilde{1}\tilde{2}\tilde{2}')}^{-}
	\nn\\
	&{} &\, \cup \,& \,
	\snum{(011'2'\tilde{2}')}^{-} &\, \cup \,& \snum{(011'\tilde{1}'\tilde{2}')}^{+}&\, \cup \,& \snum{(01\tilde{1}\tilde{1}'\tilde{2}')}^{-}&\, \cup \,& \snum{(0\tilde{0}\tilde{1}\tilde{1}'\tilde{2}')}^{+}
	\nn\\
	& {} &\, \cup \,& \,
	\snum{(00'1'2'\tilde{2}')}^{+}&\, \cup \,& \snum{(00'1'\tilde{1}'\tilde{2}')}^{-}&\, \cup \,& \snum{(00'\tilde{0}'\tilde{1}'\tilde{2}')}^{+}&\, \cup \,& \snum{(0\tilde{0}\tilde{0}'\tilde{1}'\tilde{2}')}^{-} \; .
\end{alignat}
The corresponding triangulation changing operator is $\mathcal{Z}^G_\pi [\snum{(012)^+ \times \mathbb{T}^2}]$ such that
\begin{align}
	\label{eq:isoTube3D}
	\Bigg | \threeCylDouble{0.8}{0}{1}{2}{x_1}{y_1}{ak}{k^{-1}b} \Bigg\rangle
	\simeq
	\tau_{x,y}^2(\pi)(ak,k^{-1}b)
	\Bigg| \threeCyl{0.8}{0}{2}{x_1}{y_1}{ab} \Bigg\rangle \; .
\end{align}
Putting everything together, and using the notation introduced in eq.~\eqref{eq:Hilb3D}, the algebra product of ${\rm Tube}^G_\pi(\mathbb T^2_\triangle)$ is given by
\begin{equation}\label{eq:tqtalgprod}
	| (\looDSy{x_1}{y_1} ) \xrightarrow{a} \rangle
	\star 
	| (\looDSy{x_1}{y_2}) \xrightarrow{b} \rangle
	= 
	\delta_{x_2,a^{-1}x_1a} \, \delta_{y_2,a^{-1}y_1a} \,
	\frac{1}{|G|^\frac{1}{2}} \, \tau_{x_1,y_{1}}^2(\pi)(a,b) \,
	| (\looDSy{x_1}{y_1}) \xrightarrow{ab} \rangle
\end{equation}
where we made use of the condition \eqref{eq:twisted2Coc2}.

\subsection{Compactification and lifted models\label{sec:Lifting1}}

Before studying in more detail these tube algebras and their representation theory, we would like to make several comments regarding compactification. 
In the previous part, we presented in detail the tube algebras for the (1+1)d, (2+1)d and (3+1)d Hamiltonian realisations of Dijkgraaf-Witten theory, the input data for these models being a finite group $G$ and a group 2-, 3- and 4-cocycle, respectively. Regardless of the spacetime dimension, the multiplication of the tube algebra is always `twisted' by a 2-cochain. In (1+1)d, this 2-cochain happens to be the input group 2-cocycle, and in higher dimensions it is provided by iterated $\mathbb S^1$-transgression of the input cocycle. We have only treated the (2+1)d and (3+1)d cases, but this result persists for any tube algebra associated with a manifold of the form $\mathbb S^1 \times \cdots \times \mathbb S^1 \times \I$. This suggests a way to express a given ($d$+1)-dimensional tube algebra in terms of lower-dimensional ones. Relatedly, upon compactification of one of the spatial directions, a given model can be decomposed into a `sum' of lower-dimensional topological models labeled by a group variable corresponding to the holonomy along the compactified direction, hence an effective \emph{dimensional reduction}. We sketch this compactification mechanism here and postpone its rigorous treatment to sec.~\ref{sec:cat} after the necessary tools have been introduced.

\bigskip \noindent
We consider the ($d$+1)-dimensional Dijkgraaf-Witten model applied to ($d$+1)-cobordisms of the form $\mc{C}\times \mathbb{S}^{1}$, where $\mc{C}$ is a $d$-dimensional cobordism. Henceforth, we call such cobordisms \emph{lifted}. Recall that the discrete partition function of the ($d$+1)-dimensional Dijkgraaf-Witten model is obtained by summing over $G$-colourings of the triangulation, while the topological action is provided by a ($d$+1)-cocycle. We would like to show that this partition function for a lifted cobordism $\mc{C}\times \mathbb{S}^{1}$ can be expressed in terms of a partition function for $\mathcal{C}$.  Let us begin by defining a convention for \emph{lifting} an oriented and compact triangulated $d$-manifold $\mathcal{M}_{\triangle}$ to a triangulation of $\mathcal{M}_{\triangle}\times\mathbb{S}^{1}$:
\begin{convention}[\emph{Lifting of a $d$-dimensional triangulation}\label{conv:dlifting}] 
	Let $\mc{M}$ be an oriented $d$-dimensional manifold endowed with a homogeneous triangulation $\mc{M}_{\triangle}$. We construct an oriented triangulation $\mc{M}^{\mathbb{S}^{1}}_{\triangle}$ for $\mc{M}^{\mathbb{S}^{1}}:=\mc{M}\times \mathbb{S}^{1}$ as follows:
	For each $d$-simplex $\triangle^{(d)} \equiv (v_0v_1v_2\ldots v_{d}) \subset \mc{M}_{\triangle}$ with orientation $\epsilon(\triangle^{(d)})=\pm 1$ we define a $d$-prism to be
	\begin{align*}
		(v_0v'_{0}v'_1\ldots v'_{d})^{(-1)^{d}\epsilon(\triangle^{(d)})}
		\cup
		(v_0v_{1}v'_1v'_2\ldots v'_{d})^{(-1)^{d-1}\epsilon(\triangle^{(d)})}
		\cup \cdots\cup
		(v_0v_1\ldots v_{d}v'_{d})^{\epsilon(\triangle^{(d)})}
	\end{align*}
	with vertex ordering $v_{0}< \cdots < v_{d} <v_{0'}< \cdots < v_{d'}$. The union of all such prisms forms a triangulation of $\mc{M}_{\triangle}\times \I$. In order to obtain the oriented triangulation $\mc{M}_\triangle^{\mathbb S^1}$, it remains to compactify the triangulation by identifying $(v_{0}\ldots v_{d})$ and $(v'_{0}\ldots v'_{d})$ for each $d$-simplex. For instance, the lifted triangulation  $\triangle^{(2)}\times\mathbb{S}^{1}$ of the 2-simplex $\triangle^{(2)}\equiv \snum{(012)}^+$ reads
	\begin{align}
		\label{eq:prismcol}
		\simplex{0.8} \times\mathbb{S}^{1}:=\prism{0.8} 
	\end{align}
	such that $\snum{(0)}<\snum{(1)}<\snum{(2)}<\snum{(0')}<\snum{(1')}<\snum{(2')}$, $\snum{(0)}\equiv  \snum{(0')}$, $\snum{(1)} \equiv \snum{(1')}$, $\snum{(2)} \equiv \snum{(2')}$, $\snum{(01)} \equiv \snum{(0'1')}$, $\snum{(12)} \equiv \snum{(1'2')}$ and $\snum{(02)} \equiv \snum{(0'2')}$.
\end{convention}
\noindent
In order to construct the Dijkgraaf-Witten partition function of a lifted triangulated cobordism, we need to describe the set of $G$-colourings ${\rm Col}(\mc{C}_{\triangle}\times\mathbb{S}^{1},G)$ in terms of colourings of $\mc{C}_{\triangle}$. Recall that a $G$-colouring assigns to every directed 1-simplex $(v_0v_1) \subset \mathcal{C}_{\triangle}$ a group element $g_{v_0v_1} \equiv g[v_0v_1]$ such that for every 2-simplex $(v_0v_1v_2)$ whose boundary is associated with a contractible path, the 1-cocycle condition (or flatness constraint) $ g_{v_0v_1} \cdot g_{v_1v_2} \cdot g_{v_0v_2}^{-1} = \mathbb{1}$ is imposed. Note that no matter the dimension of $\mathcal{C}_\triangle$, the set ${\rm Col}(\mc{C}_{\triangle},G)$ of $G$-colourings on $\mathcal{C}_\triangle$ only depends on the 0-, 1- and 2-simplices of the triangulation. But all the 0-,1- and 2-simplices of $\mathcal{C}_\triangle \times \mathbb S_1$ that are not included in $\mathcal{C}_\triangle$  arises from the lifting of the 1-simplices in $\mathcal{C}_\triangle$ according to conv.~\ref{conv:dlifting}. This means that to determine a $G$-colouring of $\mathcal{C}_\triangle \times \mathbb S^1$, it is enough to consider the lifting of every  one-simplex in $\mathcal{C}_\triangle$.

Let $g\in{\rm Col}(\mc{C}_{\triangle},G)$ be a $G$-colouring such that $g[v_0v_1]=g_{v_0v_1}\in G$. In order to specify a $G$-colouring of $(v_{0}v_{1}) \times \mathbb{S}^{1}$ from $g\in{\rm Col}(\mc{C}_{\triangle},G)$, it is enough to specify a colouring $g_{v_0v_0'}\in G$ of $(v_{0}v'_{0})\subset (v_{0}v_{1}) \times \mathbb{S}^{1}$. Indeed, the $G$-colouring of the remaining edges can be deduced from the identifications and the flatness constraints holding at every 2-simplex, i.e.  $g_{v_0'v_1'}=g_{v_0v_1}$, $g_{v_0'v_1}=g_{v_0v_1}^{-1}g_{v_0v_0'}$ and $g_{v_1v_1'}=g_{v_0v_1}^{-1}g_{v_0v_0'}g_{v_0v_1} \equiv g^{g_{v_0v_1}}_{v_0v_0'}$. By proceeding this way, we can define a $G$-colouring $g$ of $\mc{C}_{\triangle}\times\mathbb{S}^{1}$ in terms of a colouring $\mathfrak{g}$ of $\mc{C}_{\triangle}$ which specifies a $G$-labeling for both 0- and 1-simplices. Given a 1-simplex $(v_0v_1)$, this colouring assigns $\mathfrak{g}[v_0v_1] = g[v_0v_1] = g_{v_0v_1}$ to the bulk of the 1-simplex as before, and $\mathfrak{g}[v_0] = g[v_0v_0'] = g_{v_0v_0'}$, $\mathfrak{g}[v_1]=g[v_1v_1'] =g_{v_0v_0'}^{g_{v_0v_1}}$ to its boundary 0-simplices. 

In order to provide the discrete version of the partition function $\mathcal{Z}^G_\omega[\mathcal{C}_\triangle \times \mathbb S ^1]$ in terms of a lower-dimensional partition function, it remains to provide an algebraic expression for the corresponding topological action. Recall that given a ($d$+1)-simplex in $\mathcal{C}_\triangle \times \mathbb S^1$, an algebraic expression for the topological action $\omega(\triangle^{(d+1)})$ is provided as follows:  Let $\triangle^{(d+1)} \equiv (v_0v_1 \ldots v_{d+1}) \subset \mc{C}_\triangle \times \mathbb S^1$ such that $v_0<v_1<\ldots<v_{d+1}$ and $g\in {\rm Col}(\mc{C}_{\triangle} \times \mathbb{S}^1,G)$, the restriction $g[\triangle^{(d+1)}]$ of the $G$-colouring $g$ to the edges of $\triangle^{(d+1)}$ is specified by the $d$+1 independent gauge fields $g_{v_0v_1}, g_{v_1v_2}, \dots, g_{v_{d}v_{d+1}}$. Using these conventions, the evaluation of the cocycle $\omega$ on the $G$-coloured ($d$+1)-simplex $ (v_0v_1 \ldots v_{d+1})$ reads $\omega(g[v_0v_1\ldots v_{d+1}]) := \omega(g_{v_0v_1},g_{v_1v_2},\ldots,g_{v_dv_{d+1}})$.

Let us now consider a positively oriented $d$-simplex $\triangle^{(d)} \equiv (v_{0}\ldots v_{d})^+$ which we lift to
$(v_{0} \ldots v_{d})\times \mathbb{S}^{1}$ according convention \ref{conv:dlifting}.
Given a $G$-colouring $g\in{\rm Col}(\mc{C}_\triangle \times \mathbb{S}^{1},G)$, the amplitude associated with this lifted $d 
$-simplex reads: 
\begin{align}
	\omega(g[v_0v'_{0}v'_1v'_2 \ldots v'_{d}])^{(-1)^{d}} \,
	\omega(g[v_0v_{1}v'_1v'_2 \ldots v'_{d}])^{(-1)^{d-1}}
	\cdots
	\omega(g[v_0v_1v_2\ldots v_{d}v'_{d}]) \; .
\end{align}
Choosing the notation $g[v_0v_0'] \equiv x$ and $g[v_{i-1}v_{i}] \equiv g_i$, we write this cocycle data $\tau_x(\omega)(g_1,\ldots,g_d)$. It turns out that in (2+1)d and (3+1)d, this corresponds exactly to equations \eqref{eq:S1trans3} and \eqref{eq:S1trans4} that defines the $\mathbb S^1$-transgression of a 3- and 4-cocycle, respectively. More generally, the amplitude associated with a lifted $d$-simplex provides the defining formula for the $\mathbb S^1$-transgression of a ($d$+1)-cocycle. Let us now consider the colouring $\mathfrak{g}$ of $\mathcal{C}_\triangle$ compatible with the $G$-colouring of $\mathcal{C}_\triangle \times \mathbb S^1$, i.e. such that $\mathfrak{g}[v_0] = x$ and $\mathfrak{g}[v_{i-1}v_i] = g_i$. The amplitude associated with the $G$-coloured lifted $d$-simplex $(v_0\ldots v_d) \times \mathbb S^1$ is then equal to the evaluation of the $\mathbb S^1$-transgression $\tau(\omega)$ of $\omega$ on the $d$-simplex coloured by $\mathfrak{g}$, i.e.
\begin{align}
	\tau(\omega)(\mathfrak{g}[v_{0}\ldots v_{d}]):=\tau_{\mathfrak{g}[v_0]}(\omega)(\mathfrak{g}[v_0v_1],\mathfrak{g}[v_1v_2],\ldots,\mathfrak{g}[v_{d-1}v_d]) = \tau_x(g_1,\ldots,g_d) \; .
\end{align}
Using these conventions,  we can write the ($d$+1)-dimensional Dijkgraaf-Witten partition for $\mc{C}_{\triangle}\times\mathbb{S}^{1}$ as a $d$-dimensional Dijkgraaf-Witten model where the sum is over colourings $\mathfrak{g}$ as defined above, while the topological action is provided by the $\mathbb S^1$-transgression of the original ($d$+1)-cocycle. 

\bigskip \noindent
Since the lattice Hamiltonian realisation of the theory is defined solely in terms of the partition function, it is now easy to define a \emph{lifted} $d$-dimensional model on a $d$-dimensional surface $\Sigma$ that is equivalent to a ($d$+1)-dimensional model on $\Sigma \times \mathbb S^1$. The resulting Hamiltonian model in turn provides yet another interpretation to the $\mathbb S^1$-transgression of a cocycle which we illustrate here for the (3+1)d case. Recall that in (2+1)d the 3-cocycle $\alpha$ provides the topological action of the 3d partition function and, relatedly,  it arises as the amplitude of the $2 \leftrightharpoons 2$ Pachner operators as defined in \eqref{eq:P22}. In light of the discussion above, we expect the  $\mathbb S^1$-transgression $\tau(\pi)$ of $\pi \in Z^4(G,\rU(1))$ to arise as the amplitude of the pinched interval operator related to a lifted version of the  $2 \leftrightharpoons 2$ Pachner move. More specifically, $\tau(\pi)$ appears as the amplitude of the move obtained by lifting the complexes appearing in the definition of the $2 \leftrightharpoons 2$ move according to conv.~\ref{conv:dlifting}. This  move can be heuristically depicted as
\begin{equation}
	\label{eq:compact22pachner}
	\preSlantProd{0.8}{1} \times \mathbb{S}^1 \; \xrightarrow{\q\q\q} \; \preSlantProd{0.8}{2} \times \mathbb{S}^1 
\end{equation}
where the 3-complexes on the l.h.s and the r.h.s are obtained by lifting the 2d triangulations to three dimensions so as to obtain cubes whose top and bottom faces are identified. Once we apply conv.~\ref{conv:dlifting} to both sides of \eqref{eq:compact22pachner}, we obtain two complexes made of six 3-simplices. It turns out that these two complexes are related via a sequence of $2 \leftrightharpoons 3$ moves. Graphically, this sequence of transformations reads
\begin{align}
	\label{eq:lifted22move}
	&\slantProdA{0.8}{1} \xrightarrow{\pi (g[00'1'2'3'])^{-1}} \slantProdA{0.8}{2} \\[-0.7em] 
	&\xrightarrow[\pi(g{[}011'2'3'{]})]{\pi (g[0122'3'])^{-1}} \slantProdA{0.8}{3} 
	\xrightarrow{\pi (g[01233'])} \slantProdA{0.8}{4} 
	\xrightarrow{\q 1 \q } \slantProdA{0.8}{5} 
\end{align}
where each arrow is decorated by the amplitude associated with the corresponding Pachner operator according to \eqref{eq:P23}. Note that during the second step as well as the last one, a trivial $2 \leftrightharpoons 3$  move (obtained by setting one of the edge colourings to the identity) is used, which does not contribute to the collective amplitude. Setting $g[\snum{0'0}] \equiv x$, $g[\snum{01}] \equiv g[\snum{0'1'}] = a$, $g[\snum{12}] \equiv g[\snum{1'2'}]\equiv b$ and $g[\snum{23}] = g[\snum{2'3'}] \equiv c$, the collective amplitude of the local transformations performed above reads
\begin{equation}
	\tau_x(\pi)(a,b,c) = \frac{\pi(a,a^{-1}xa,b,c) \, \pi(a,b,c,(abc)^{-1}xabc)}{\pi(x,a,b,c) \, \pi(a,b,(ab)^{-1}xab,c)} \; ,
\end{equation}
which is precisely the definition \eqref{eq:S1trans4} of the $\mathbb S^1$-transgression of $\pi$ as expected.
Henceforth, we refer to the  move defined above as the lifted ${2 \rightleftharpoons 2}$ move, and notate it $({2 \rightleftharpoons 2}) \times \mathbb S^1$. Note finally that it is now possible to express the isomorphism in eq.~\eqref{eq:isoTube3D} as a sequence of three lifted ${2 \rightleftharpoons 2}$ moves so that the collective amplitude is provided by the $\mathbb T^2$-transgression of $\pi$.

\bigskip \noindent
\emph{All the ideas presented in this part are made more precise in sec.~\ref{sec:cat} using the technology of loop groupoid. There we also explain in detail how higher tube algebras can be expressed in terms of the (1+1)d one as suggested earlier.}

\newpage
\section{Representation theory and simple excitations statistics\label{sec:simple}}
In this section, we construct explicitly the simple modules of the tube algebras introduced previously. These simple modules in turn classify the simple excitations of the corresponding model. Furthermore, we introduce the comultiplication maps and the $R$-matrices compatible with the tube algebras multiplications rules. These can then be used to describe the statistics of the simple excitations. 

\subsection{Simple representations of the (1+1)d tube algebra\label{sec:1+1dsimple}}

Let us first discuss the simple excitations in the (1+1)d Dijkgraaf-Witten Hamiltonian model using the language of projective group representations. This discussion will be further generalised in the proceeding sections to discuss the simple excitations in the (2+1)d and (3+1)d models.

Given a pair $G,\beta$ where $G$ is a finite group and $\beta\in Z^{2}(G,\rU(1))$ a normalised 2-cocycle, the $\beta$-twisted group algebra $\mathbb{C}^\beta[G]$ is the algebra defined by the vector space ${\rm Span}_{\mathbb{C}}\{ \ket{\xrightarrow{a}} \}_{\forall a\in G}$ with algebra product defined by
\begin{align}
	\ket{\xrightarrow{a}} \star \ket{\xrightarrow{b}}=\beta(a,b)\ket{ \xrightarrow{ab}} \; ,  \q \forall \; a,b \in G \; .
\end{align}
It is useful to note how the conditions satisfied by $\beta$ defining a normalised 2-cocycle manifest in the properties of ${\mathbb{C}^\beta[G]}$. Firstly, the normalisation of $\beta$, i.e. $\beta(\mathbb 1_{G},a)=\beta(a,\mathbb 1_{G})=1$ for all $a\in G$, ensures the relation
\begin{align}
	\ket{\xrightarrow{a}} \star \ket{\xrightarrow{\mathbb 1_{G}}}=\ket{\xrightarrow{a}}=\ket{\xrightarrow{\mathbb 1_{G}}} \star \ket{\xrightarrow{a}}\; ,  \q \forall \; a \in G \; .
\end{align}
Secondly, the 2-cocycle equation
\begin{align}
	d^{(2)}\beta(a,b,c)=
	\frac{
		\beta(b,c) \, \beta(a,bc)
	}
	{
		\beta(ab,c) \, \beta(a,b)
	}
	=
	1
\end{align}
for all $a,b,c\in G$, ensures that ${\mathbb{C}^\beta[G]}$ is an associative algebra, i.e.
\begin{align}
	(\ket{\xrightarrow{a}} \star \ket{\xrightarrow{b}}) \star \ket{\xrightarrow{c}}= \ket{\xrightarrow{a}} \star (\ket{\xrightarrow{b}} \star \ket{\xrightarrow{c}})
	\; ,  \q \forall \; a,b,c \in G \; .
\end{align}
Finally, similarly to the untwisted group algebra, each element admits an inverse that takes the form
\begin{align}
	\ket{\xrightarrow{a}}^{-1}=\frac{1}{\beta(a,a^{-1})}\ket{\xrightarrow{a^{-1}}} \; ,  \q \forall \; a \in G 
\end{align}
such that
\begin{align}
	\ket{\xrightarrow{a}}^{-1}\star \ket{\xrightarrow{a}}=\ket{\xrightarrow{\mathbb 1_{G}}}=\ket{\xrightarrow{a}} \star \ket{\xrightarrow{a}}^{-1} \; ,
\end{align}
as expected.
We note that in the limiting case where $\beta$ is a trivial 2-cocycle, i.e. $\beta(a,b)=1$ for all $a,b\in G$, the twisted group algebra reduces to the untwisted group algebra $\mathbb{C}[G]$.

Comparing with equation \eqref{eq:1+1dalg}, we realize that, up to the normalisation factor $|G|^{-\frac{1}{2}}$, the (1+1)d tube algebra corresponds to a $\beta$-twisted group algebra. It follows that we can classify \emph{simple excitations} in the (1+1)d Dijkgraaf-Witten model as simple representations of ${\mathbb{C}^\beta[G]}$. In the study of simple representations of $\mathbb{C}^\beta[G]$, many of the familiar results from the complex representation theory of the group algebra still apply. In particular, a representation of $\mathbb{C}^\beta[G]$ is provided by a pair $(\mathcal{D}^R,V_R)$ where $V_R$ is a complex vector space and $\mathcal{D}:\mathbb{C}^\beta[G]\rightarrow {\rm End}(V)$ is an algebra homomorphism. A representation $(\mathcal{D}^R,V_R)$ is called \emph{simple} if the only proper subspace $W\subset V_R$, whereby $\mathcal{D}^R(a) \triangleright w \in W$ for all $w\in W$ and $a\in G$, is the trivial vector space. Akin to the untwisted case, $\mathbb{C}^\beta[G]$ is a \emph{semi-simple} algebra such that all representations are isomorphic to a direct sum of simple representations.

Let $(\mathcal{D}^{R},V_{R})$ define a simple representation of $\mathbb{C}^\beta[G]$. In light of the correspondence with the (1+1)d tube algebra, we interpret the vector space $V_{R}$ with the internal Hilbert space of a point particle localised on the boundary of the interval. The homomorphism $\mathcal{D}^R$ then defines how the internal vector space of the point particle is acted upon by the linearised symmetry of gluing. We then interpret the label $R$ of the simple representation as the \emph{charge quantum number}, which is well defined since such a label is invariant under the action of both the Hamiltonian and the tube algebra.

For the following discussion, it is useful to collect some basic properties of the representations of $\mathbb{C}^\beta[G]$. Let $\{(\mathcal{D}^{R},V_{R})\}_R$ denote the set of simple representations of $\mathbb{C}^\beta[G]$ up to isomorphism, the corresponding matrix elements $\mathcal{D}^{R}_{mn}$ for $m,n\in\{1,\ldots,{\rm dim}(V_{R}) \}$ satisfy the following conditions:
\begin{alignat}{2}
	\label{eq:1dLinearity}
	&\sum^{{\rm dim}(V_{R})}_{n=1}\mathcal{D}^{R}_{mn}(|\xrightarrow{a}\rangle)\mathcal{D}^{R}_{no}(|\xrightarrow{b} \rangle)=\beta(a,b)\mathcal{D}^{R}_{mo}(|\xrightarrow{ab}\rangle) & \q & \text{(Linearity)}
	\\
	\label{eq:1dConjugate}
	&\overline{\mathcal{D}^{R}_{mn}(|\xrightarrow{a} \rangle)}=\frac{1}{\beta(a,a^{-1})}\mathcal{D}^{R}_{nm}(|\xrightarrow{a^{-1}}\rangle)
	& \q & \text{(Complex conjugation)}
	\\
	\label{eq:1dOrtho}
	&\frac{1}{|G|}\sum_{a\in G}\mathcal{D}^{R}_{mn}(|\xrightarrow{a}\rangle)\overline{\mathcal{D}^{R'}_{m'n'}(|\xrightarrow{a})\rangle}=\frac{\delta_{\mathcal{D}^{R},\mathcal{D}^{R'}}}{{\rm dim}(V_R) }\delta_{m,m'}\delta_{n,n'}
	& \q & \text{(Orthogonality)}
	\\
	\label{eq:1dComplete}
	&\frac{1}{|G|}\sum_{\{\mathcal{D}^{R}\}}\sum_{m,n}{\rm dim}(V_{R})\mathcal{D}^{R}_{mn}(|\xrightarrow{a}\rangle)\overline{\mathcal{D}^{R}_{mn}(|\xrightarrow{a'}\rangle)}=\delta_{a,a'}
	& \q & \text{(Completeness)}
\end{alignat}
for all $a,b\in G$.

\subsection{Simple representations of the (2+1)d tube algebra\label{sec:2+1dsimple}}

We now consider the simple excitations of the (2+1)d Hamiltonian model whose tube algebra was shown to be equivalent to the twisted quantum double algebra. In order to find the simple excitations, we first choose to decompose the algebra into a direct sum of simpler sub-algebras whose simple representations can be described in terms of simple twisted group representations as discussed above. To this end we begin with a simple observation: Let $C_1,C_2 \subset G$ be two disjoint conjugacy classes of $G$, then for any pair of $G$-coloured graph-states of the form 
\begin{equation*}
	{|(\looSy{x_1})\xrightarrow{a} \rangle} \, ,\, {|(\looSy{x_2})\xrightarrow{b} \rangle}
\end{equation*}
such that $x_{1}\in C_1$ and $x_{2}\in C_2$ it follows from the ${\rm Tube}^G_\alpha(\mathbb S^1_\triangle)$ algebra product defined in \eqref{eq:2dalgprod} that
\begin{align}
|(\looSy{x_1})\xrightarrow{a} \rangle
\star
|(\looSy{x_2})\xrightarrow{b} \rangle
=0 \; .
\end{align}
A consequence of this observation is that each conjugacy class $C\subset G$ naturally defines a sub-algebra ${\rm Tube}^G_\alpha(\mathbb S^1_\triangle)_{C}\subset{\rm Tube}^G_\alpha(\mathbb S^1_\triangle)$ given by
\begin{align}
	&{\rm Tube}^G_\alpha(\mathbb S^1_\triangle)_{C}:={\rm Span}_{\mathbb{C}}\big\{ \,
	\ket{(\looSy{x})\xrightarrow{a}}
	\, \big\}_{\substack{\forall x\in C \\ \forall a \in G}} \; .
\end{align}
Noting that the set of conjugacy classes forms a partition of $G$ it follows that
\begin{align}
	&{\rm Tube}^G_\alpha(\mathbb S^1_\triangle)=\bigoplus_{C} {\rm Tube}^G_\alpha(\mathbb S^1_\triangle)_{C} \; ,
\end{align}
where the direct sum is over the set of all conjugacy classes of $G$. Utilising this decomposition of ${\rm Tube}^G_\alpha(\mathbb S^1_\triangle)$, we can find the simple modules in terms of the simple modules of ${\rm Tube}^G_\alpha(\mathbb S^1_\triangle)_{C}$ for each $C\subset G$. At this point it is illustrative to note that when $C=\{\mathbb 1_{G}\}$ is the conjugacy class of the identity element of $G$, the corresponding sub-algebra is given by the (1+1)d tube algebra of sec.~$\ref{sec:1+1dtube}$ with $\beta$ given by the trivial 2-cocycle, i.e. $\beta(a,b)=1$ for all $a,b\in G$. The corresponding simple representations are then discussed in sec.~\ref{sec:1+1dsimple}.

Given a conjugacy class $C$, we now describe the simple modules of ${\rm Tube}^G_\alpha(\mathbb S^1_\triangle)_{C}$. To this end we first introduce some notation.
We begin by notating each element in $C$ by $c_{i}$ for $i\in\{1,\ldots, |C| \}$. In the following, we will call $c_{1}\in C$ the \emph{representative element of $C$}.
We next define the set $Q_{C}:=\{q_{1},\ldots, q_{|C|} \}$ such that each $q_{i}\in G$ is defined by a non-canonical choice of element in $G$ satisfying the conditions $c_{1}=q^{-1}_{i}c_{i}q_{i}$ and $q_{1}:=\mathbb 1_{G}$.
Finally, we define the \emph{stabiliser group} of $C$ by $Z_{C}:=\{a \in G \, | \, c_{1}=a^{-1}c_{1}a  \}$, i.e. the subgroup of $G$ consisting of elements of $G$ that commute with $c_{1}$. In the extreme case that we take the group $G$ to be abelian each group element forms a conjugacy class and the stabiliser subgroup is just the group itself vastly simplifying the previous construction.

Utilising the conventions outlined above, it follows from equations (\ref{eq:twisted2Coc},\ref{eq:twisted2Cocc}) that $\tau_{c_{1}}(\alpha)\in Z^{2}(Z_{C},\rU(1))$ defines a normalised 2-cocycle of $Z_{C}$ when $\alpha\in Z^{3}(G,\rU(1))$ is a normalised 3-cocycle of $G$. For each simple $\tau_{c_{1}}(\alpha)$-projective representation $(\mathcal{D}^R,V_R)$ of $Z_{C}$ we can then define a simple representation of the twisted quantum double by a homomorphism $\mc{D}^{C,R}:{\rm Tube}^G_\alpha(\mathbb S^1_\triangle)_{C}\rightarrow {\rm End}(V_{C,R})$ where 
\begin{align}\label{eq:VCR}
	V_{C,R}:=
	{\rm Span}_{\mathbb{C}}\{ \ket{c_{i},v_{m}}\; |\; \forall \,  i=1,\ldots,|C| \; {\rm and} \;  m=1,\ldots,{\rm dim}(V_{R}) \}
\end{align}
and for $i,j\in\{1,\ldots,|C| \}$, $m,n\in\{1,\ldots,{\rm dim}(V_R) \}$ we define
\begin{align}
	\label{eq:TQDrepelement}
	\mc{D}^{C,R}_{im,jn} \big(|(\looSy{x})\xrightarrow{a} \rangle \big)
	:=
	\delta_{x,c_{i}}\delta_{x^a,c_{j}}
	\frac{\tau_{c_{1}}(\alpha)(q^{-1}_{i},a)}
	{\tau_{c_{1}}(\alpha)(q^{-1}_{i}aq_{j},q^{-1}_{j})}
	\mc{D}^{R}_{m,n}(|\xrightarrow{q^{-1}_{i}aq_{j}}\rangle)
\end{align}
such that
\begin{align}
	\mc{D}^{C,R}\big(|(\looSy{x})\xrightarrow{a} \rangle \big)
	:=\frac{1}{|G|^{\frac{1}{2}}}
	\sum^{|C|}_{i,j=1}\sum^{{\rm dim}(V_{R})}_{m,n=1}
	\mc{D}^{C,R}_{im,jn}\big(|(\looSy{x})\xrightarrow{a} \rangle \big)
	\ket{c_{i},v_{m}}
	\bra{c_{j},v_{n}} \; .
\end{align}
If follows from the definition and the linearity condition \eqref{eq:1dLinearity} that these matrices indeed define an algebra homomorphism, i.e.
\begin{align}
	\label{eq:2dLinearity}
	\sum_{j,n}
	\mc{D}^{C,R}_{im,jn}\big(|(\looSy{x_1})\xrightarrow{a} \rangle \big)
	\mc{D}^{C,R}_{jn,ko}\big(|(\looSy{x_2})\xrightarrow{b} \rangle \big)
	=
	\delta_{x_2,x_1^{a}} \, \tau_{x_1}(\alpha)(a,b)
	\mc{D}^{C,R}_{im,ko}\big(|(\looSy{x})\xrightarrow{ab} \rangle \big)
	\; .
\end{align}
Furthermore, the matrices satisfy the following conjugation relation
\begin{align}
	\overline{\mc{D}^{C,R}_{im,jn}(|(\looSy{x})\xrightarrow{a} \rangle)}
	=
	\frac{1}{\tau_{x}(\alpha)(a,a^{-1})}
	\mc{D}^{C,R}_{jn,im}\big(|(\looSy{x^{a}})\xrightarrow{a^{-1}} \rangle\big) \; ,
\end{align}
as well as the following orthogonality and completeness conditions
\begin{align}
	&\frac{1}{|G|}\sum_{x,a\in G}
	\mc{D}^{C,R}_{im,jn}\big(|(\looSy{x})\xrightarrow{a} \rangle \big)
	\overline{\mc{D}^{C',R'}_{i'm',j'n'}\big(|(\looSy{x})\xrightarrow{a} \rangle \big)}
	=
	\frac{\delta_{C,C'} \delta_{\mathcal{D}^R, \mathcal{D}^{R'}}}{|C|{\rm dim}(V_R)}
	\delta_{i,i'}\delta_{j,j'}\delta_{m,m'}\delta_{n,n'}
	\label{groupoidorthog2}
	\\
	&\frac{1}{|G|}\sum_{\{C,\mc{D}^R\}}
	\sum_{\substack{
			i,m\\
			j,n
	}}
	|C|{\rm dim}(V_R)\mc{D}^{C,R}_{im,jn}\big(|(\looSy{x})\xrightarrow{a} \rangle\big)
	\overline{\mc{D}^{C,R}_{im,jn}\big(|(\looSy{x'})\xrightarrow{a'} \rangle \big)}
	=\delta_{x,x'}\delta_{a,a'}
	\label{groupoidorthog1}
\end{align}
generalising (\ref{eq:1dConjugate}) for projective group representations.

From the previous discussion follows that a simple representation of ${\rm Tube}^G_\alpha(\mathbb S^1_\triangle)$ can be specified by a pair $(C,R)$ where $C$ is a conjugacy class of $G$ and $R$ is a simple representation of  $\mathbb{C}^{\tau_{c_{1}}(\alpha)}[Z_{C}]$. The conjugacy class $C$ represents the set of boundary colourings that are related by the action of the tube algebra, whereas the representation $R$ decomposes the action of the tube algebra that leaves the boundary colouring invariant. In other words, $R$ describes the symmetries of the boundary under the action of the tube algebra. Therefore, we interpret the label $C$ as a \emph{magnetic flux quantum number} and $R$ as an \emph{electric charge quantum number}. The vector space $V_{C,R}$ defined in \eqref{eq:VCR} thus describes the internal Hilbert space of a point particle in the (2+1)d Dijkgraaf-Witten model so that an element $\ket{c_{i},v_{m}}\in V_{C,R}$ defines a particle with well defined flux $c_{i}\in C$ and charge $v_{m}\in V_{R}$. A general excitation is finally obtained as a superposition of simple excitations with the corresponding internal vector space given by the direct sum of simple representations. 

\subsection{Twisted quantum double comultiplication and fusion of point-like excitations\label{sec:particlefusion}}

In this section we expand on the properties of the twisted quantum double in relation to the fusion of point particles in (2+1)d Dijkgraaf-Witten model. This structure will be generalised in sec.~\ref{sec:loopfusion} to the case of loop-like excitations in (3+1)d Dijkgraaf-Witten theory.

Given a pair of simple point particles with internal Hilbert spaces $V_{C_1,R_1}$ and $V_{C_2,R_2}$, respectively, we can consider their joint Hilbert in the absence of any external constraints as the space given by the tensor product $V_{C_1,R_1}\otimes V_{C_2,R_2}$. In order to understand how the twisted quantum double algebra acts on the corresponding two-particle Hilbert space $V_{C_1,R_1}\otimes V_{C_2,R_2}$, we introduce the \emph{comultiplication map} $\Delta:{\rm Tube}^G_\alpha(\mathbb S^1_\triangle)\rightarrow {\rm Tube}^G_\alpha(\mathbb S^1_\triangle)\otimes {\rm Tube}^G_\alpha(\mathbb S^1_\triangle)$ defined as
\begin{align}
	\Delta\big(|(\looSy{x})\xrightarrow{a} \rangle \big)
	:=\sum_{x_{1},x_{2}}\, \delta_{x_{1}x_{2},x}\gamma_{a}(\alpha)(x_{1},x_{2}) \,
	|(\looSy{x_{1}})\xrightarrow{g} \rangle
	\otimes
	|(\looSy{x_{2}})\xrightarrow{g} \rangle
\end{align}
where
\begin{align}
	\gamma_{a}(\alpha)(x_{1},x_{2}):=\frac{\alpha(x_{1},x_{2},a)\alpha(a,x^{a}_{1},x^{a}_{2})}{\alpha(x_{1},a,x^{a}_{2})} \; .
\end{align}
Most importantly, $\Delta$ defines an algebra homomorphism, i.e.
\begin{align}
	\Delta \big(|(\looSy{x_1})\xrightarrow{a} \rangle\star |(\looSy{x_2})\xrightarrow{b} \rangle \big)
	=
	\Delta \big(|(\looSy{x_1})\xrightarrow{a} \rangle \big)\star \Delta \big(|(\looSy{x_2})\xrightarrow{b} \rangle \big) \; ,
\end{align}
which follows from
\begin{align}
	\tau_{x_{1}}(\alpha)(a,b) \,
	\tau_{x_{2}}(\alpha)(a,b) \,
	\gamma_{a}(\alpha)(x_{1},x_{2}) \,
	\gamma_{b}(\alpha)(x_{1}^a,x_{2}^a)
	=\tau_{x_{1}x_{2}}(\alpha)(a,b) \,
	\gamma_{ab}(x_{1},x_{2}) \; .
\end{align}
This last relation descends from the definitions of $\tau(\alpha)$ and $\gamma(\alpha)$ together with the 3-cocycle condition satisfied by $\alpha$. In general, the comultiplication is not associative but instead satisfies the \emph{quasi-coassociativity} relation: 
\begin{align}
	(\Delta\otimes {\rm id})\circ \Delta\big(|(\looSy{x})\xrightarrow{a}\rangle \big)
	=
	\phi \, ({\rm id}\otimes\Delta)\circ \Delta\big(|(\looSy{x})\xrightarrow{a}\rangle\big)\phi^{-1}
\end{align}
where we introduced the \emph{twist} $\phi$ defined as
\begin{align}
	\phi=\sum_{x_{1},x_{2},x_{3}}\alpha^{-1}(x_{1},x_{2},x_{3}) \,
	|(\looSy{x_{1}})\xrightarrow{\mathbb 1_{G}}\rangle
	\otimes
	|(\looSy{x_{2}})\xrightarrow{\mathbb 1_{G}}\rangle
	\otimes
	|(\looSy{x_{3}})\xrightarrow{\mathbb 1_{G}}\rangle \; .
\end{align}
Given a three-particle vector space $(V \otimes W) \otimes Z$, the twist $\phi$ induces the module isomorphism
\begin{equation}
	\phi: (V \otimes W) \otimes Z \xrightarrow{
	\sim} V \otimes (W \otimes Z) \; .
\end{equation}
The quasi-coassociativity follows from the cocycle data relation
\begin{align}
	\gamma_{a}(\alpha)(x_{1},x_{2}) \,
	\gamma_{a}(\alpha)(x_{1}x_{2},x_{3}) \,
	\alpha(x_{1}^a,x_{2}^a,x_{3}^a)
	=
	\gamma_{a}(\alpha)(x_{2},x_{3}) \,
	\gamma_{a}(\alpha)(x_{1},x_{2}x_{3}) \,
	\alpha(x_{1},x_{2},x_{3}) \; .
\end{align}

\bigskip \noindent
Given a pair of representations $(\mc{D}^{C_1,R_1},V_{C_1,R_1})$ and $(\mc{D}^{C_2,R_2},V_{C_2,R_2})$ of the twisted quantum double, the comultiplication $\Delta$ allows us to define the tensor product representation $\mc{D}^{C_1,R_1}\otimes \mc{D}^{C_2,R_2}(\Delta):{\rm Tube}^G_\alpha(\mathbb S^1_\triangle)\otimes {\rm Tube}^G_\alpha(\mathbb S^1_\triangle)\rightarrow V_{C_1,R_1}\otimes V_{C_2,R_2}$ where
\begin{align}
	\nn
	\mc{D}^{C_1,R_1}\otimes \mc{D}^{C_2,R_2}\big(\Delta \big(|(\looSy{x})\xrightarrow{a}\rangle \big) \big)
	=\sum_{x_{1},x_{2}}\delta_{x_1x_2,x}\, \gamma_{a}(\alpha)(x_{1},x_{2}) \,
	\mc{D}^{C_1,R_1}\big(|(\looSy{x_{1}})\xrightarrow{a}\rangle \big)
	\otimes
	\mc{D}^{C_2,R_2}\big(|(\looSy{x_{2}})\xrightarrow{a}\rangle\big)
\end{align}
which is compatible with the algebra product since $\Delta$ defines an algebra homomorphism. The semi-simplicity of ${\rm Tube}^G_\alpha(\mathbb S^1_\triangle)$ implies that such tensor product representations are generically not simple and as such admit a decomposition into a direct sum of irreducible representations:
\begin{align}
	V_{C_1,R_1} \otimes V_{C_2,R_2}\simeq \bigoplus_{\{ (C_3,R_3) \}}N^{(C_3,R_3)}_{(C_1,R_1),(C_2,R_2)}V_{C_3,R_3} \; .
\end{align}
Here $N^{(C_3,R_3)}_{(C_1,R_1),(C_2,R_2)}\in \mathbb{Z}^{+}_{0}$ is a non-negative integer, called the \emph{fusion multiplicity}, that defines how many times each irreducible representation $(\mathcal{D}^{C_3,R_3},V_{C_3,R_3})$ occurs in the decomposition of ${(\mc{D}^{C_1,R_1} \otimes \mc{D}^{C_2,R_2}(\Delta),V_{C_1,R_1}\otimes V_{C_2,R_2})}$. Using the orthogonality relations of the representations, we find the following expression for the fusion multiplicities:
\begin{align}
	&N^{(C_3,R_3)}_{(C_1,R_1),(C_2,R_2)}
	:=
	\frac{1}{|G|}\sum_{x,a\in G}
	\text{tr}
	\Big(
	\mc{D}^{C_1,R_1}\otimes \mc{D}^{C_2,R_2}\big(\Delta\big(|(\looSy{x})\xrightarrow{a}\rangle \big) \big)
	\overline{\mc{D}^{C_3,R_3}\big(|(\looSy{x})\xrightarrow{a}\rangle \big)}
	\Big) \; .
\end{align}

\subsection{Twisted quantum double $R$-matrix and braiding of point-like excitations\label{sec:particleBraiding}}

Given a set of identical particles, the transformation properties of their joint wave function by permuting their spatial location is referred to as their \emph{exchange statistics}. For spacetime dimensions $d+1 \geq 4$, the exchange statistics of $n$ point particles is governed by representations of the symmetric group $S_{n}$. Experimentally, such systems are observed to transform under the two one-dimensional irreducible representations of $S_{n}$, often referred to as the \emph{trivial} and the \emph{sign} representations, which in turn classify point particles into \emph{bosons} and \emph{fermions}, respectively. However, it is well-know that in (2+1)d the exchange statistics of point particles is not characterised by the symmetric group, but instead by the braid group. Given $n$ particles arranged along a line in the interior of the 2-disk $\mathbb D^{2}$, the braid group is defined via $n$$-$1 generators $\{\sigma_{i}\}_{i=1,\ldots,n-1}$ that satisfy the relation:
\begin{align}
	\sigma_{i}\sigma_{i+1}\sigma_{i}=\sigma_{i+1}\sigma_{i}\sigma_{i+1}  \; ,  \q \forall \; i \in \{1,\ldots,n-2\} \; .
\end{align}
Enumerating the $n$ particles from left to right by the integers $1,2,3,\ldots, n$, we can interpret each $\sigma_{i}$ as corresponding to the clockwise exchange of the particles $i$ and $i+1$.

Building on the tensor structure defined by the comultiplication map $\Delta$, it was shown that the representations of the twisted quantum double algebra admit a representation of the braid group. This representation is interpreted as the exchange statistics of point-like particles in the (2+1)d Dijkgraaf-Witten model. In order to define such braid statistics, we first need to introduce an invertible element of ${\rm Tube}^G_\alpha(\mathbb S^1_\triangle)\otimes {\rm Tube}^G_\alpha(\mathbb S^1_\triangle)$ called the \emph{R-matrix}:
\begin{align}
	R:=\sum_{x,y\in G} |(\looSy{x})\xrightarrow{\mathbb 1_{G}}\rangle \otimes |(\looSy{y})\xrightarrow{x}\rangle \; .
\end{align}
The $R$-matrix is compatible with the comultiplication map via the relation
\begin{align}
	R\Delta \big(|(\looSy{x})\xrightarrow{a}\rangle \big)  R^{-1}=\sigma \circ \Delta \big(|(\looSy{x})\xrightarrow{a}\rangle \big)
\end{align}
for all $|(\looSy{x})\xrightarrow{a}\rangle\in {\rm Tube}^G_\alpha(\mathbb S^1_\triangle)$, where $\sigma$ is the transposition map
\begin{align}
	\sigma:V\otimes W\rightarrow W\otimes V \; ,
\end{align}
which permutes the order of vector spaces in the tensor product. It follows from this relation that, for any two modules $V$ and $W$ of the twisted quantum double, the operator $\hat{R}:=\sigma\circ R$ defines a map
\begin{align}
	\hat{R}:V\otimes W\xrightarrow{\sim}W \otimes V \; ,
\end{align}
which is also a module isomorphism via
\begin{align}
	\hat{R} \circ \Delta \big(|(\looSy{x})\xrightarrow{a}\rangle \big) \triangleright (V\otimes W)=\Delta \big(|(\looSy{x})\xrightarrow{a}\rangle \big)\circ \hat{R}\triangleright (V\otimes W) \; .
\end{align}
Given the three-particle vector space $(V\otimes W)\otimes Z$, it follows that $\hat{R}$ fulfills the \emph{hexagon equations}:
\begin{align}
	\label{eq:hexAlg}
	({\rm id} \otimes \Delta)\hat{R} = \phi^{-1} \, ({\rm id} \otimes \hat{R}) \, \phi \, ( \hat{R} \otimes {\rm id}) \, \phi^{-1} \q , \q 
	(\Delta \otimes {\rm id})\hat{R} = \phi\, ( \hat{R} \otimes {\rm id}) \, \phi^{-1} \, ( {\rm id} \otimes \hat{R}) \, \phi \; ,
\end{align}
which can be equivalently represented in terms of commutative diagrams as in definition \ref{def:braidedmonoidalcat}. These equations in turn ensure the \emph{quasi-Yang-Baxter equation}
\begin{align}
	\label{eq:quasiyangbaxter}
	\begin{tikzpicture}[scale=1,baseline=0em, line width=0.4pt]
	\matrix[matrix of math nodes,row sep =3em,column sep=2em, ampersand replacement=\&] (m) {
		{} \&[-7em] {} \&[-4em] {} \& (V \otimes W )\otimes Z \& {} \&[-4em] {} \&[-7em] {} 
		\\[-2em]
		{} \& {} \& (W \otimes V) \otimes Z \& {}  \& V \otimes (W \otimes Z) \& {} \& {}
		\\[-1em]
		{} \& W \otimes (V \otimes Z) \& {} \& {} \& {} \& V \otimes (Z \otimes W) \& {}
		\\
		W \otimes (Z \otimes V) \& {} \& {} \& {} \& {} \& {} \& (V \otimes Z) \otimes W
		\\
		{} \& (W\otimes Z)\otimes V \& {} \& {} \& {} \& (Z\otimes V)\otimes W \& {}
		\\[-1em]
		{} \& {} \& (Z\otimes W)\otimes V \& {} \& Z\otimes (V\otimes W) \& {} \& {}
		\\[-2em]
		{} \& {} \& {} \& Z \otimes (W \otimes V) \& {} \& {} \& {}
		\\
	};	
	\path
	(m-1-4) edge[->] node[pos=0.4, left, yshift=0.6em] {$\hat{R} \otimes {\rm id}$}  (m-2-3)
	edge[->] node[pos=0.4, right, yshift=0.6em] {$\phi$}  (m-2-5)
	(m-2-3) edge[->] node[pos=0.5, right, yshift=-0.2em] {$\phi$}  (m-3-2)
	(m-3-2) edge[->] node[pos=0.5, right] {${\rm id}\otimes\hat{R}$}  (m-4-1)
	(m-4-1) edge[->] node[pos=0.5, right] {$\phi^{-1}$}  (m-5-2)
	(m-5-2) edge[->] node[pos=0.5, right, yshift=0.2em, xshift=0.1em] {$\hat{R}\otimes{\rm id}$}  (m-6-3)
	(m-6-3) edge[->] node[pos=0.5, left,yshift=-0.6em] {$\phi$}  (m-7-4)
	(m-2-5) edge[->] node[pos=0.5, left, yshift=-0.2em, xshift=-0.1em] {${\rm id}\otimes \hat{R}$}  (m-3-6)
	(m-3-6) edge[->] node[pos=0.5, left] {$\phi^{-1}$}  (m-4-7)
	(m-4-7) edge[->] node[pos=0.5, left] {$\hat{R}\otimes {\rm id}$}  (m-5-6)
	(m-5-6) edge[->] node[pos=0.5, left, yshift=0.2em] {$\phi^{-1}$}  (m-6-5)
	(m-6-5) edge[->] node[pos=0.5, right,yshift=-0.6em] {${\rm id}\otimes \hat{R}$}  (m-7-4)
	;
	\end{tikzpicture}
\end{align}
such that the composition of operators on the dodecagon are equal, and as such $\hat{R}$ defines a \emph{braid group representation} on the modules of ${\rm Tube}^G_\alpha(\mathbb S^1_\triangle)$.

\subsection{Simple representations of the (3+1)d tube algebra}

We derived in sec.~\ref{sec:twistedTriple} the twisted quantum triple algebra ${\rm Tube}^G_\pi(\mathbb T^2_\triangle)$. In the following, we describe its representation theory which we interpret as defining the simple loop-like excitations in the (3+1)d Dijkgraaf-Witten model. We will then describe the statistics of these simple loop-like excitations. Our exposition follows closely the one of sec.~\ref{sec:2+1dsimple} and sec.~\ref{sec:particlefusion}.

Akin to the twisted quantum double algebra, in order to find the simple representations of the twisted quantum triple algebra, we first decompose the algebra into a direct sum of sub-algebras. Letting $G^{2}_{\rm comm.}:=\{(a,b)\in G^{2}\,|\, ab=ba\}$, we define an equivalence relation on $G^{2}_{\rm comm.}$ given by $(a,b)\sim (a',b')$ if there exists a $k\in G$ such that $(a',b')=(a^{k},b^{k})$. The set of such equivalence classes forms a partition of $G^2_{\rm comm.}$ into disjoint subsets. Given any pair of disjoint equivalence classes $D_1,D_2$ and any pair of $G$-coloured graph-states of the form
\begin{equation}
	| (\looDSy{x_1}{y_1} ) \xrightarrow{a} \rangle \, , \, | (\looDSy{x_2}{y_2} ) \xrightarrow{b} \rangle
\end{equation}
such that $(x_1,y_1)\in D_1$ and $(x_2,y_2)\in D_2$ it follows from the ${\rm Tube}^G_\pi(\mathbb T^2_\triangle)$ algebra product defined in \eqref{eq:tqtalgprod} that
\begin{align}
	| (\looDSy{x_1}{y_1} ) \xrightarrow{a} \rangle
	\star
	| (\looDSy{x_2}{y_2} ) \xrightarrow{b} \rangle
	=
	0 \; .
\end{align} 
A consequence of this observation is that each equivalence class $D\subset G^2_{\rm comm.}$ naturally defines a sub-algebra ${\rm Tube}^G_\pi(\mathbb T^2_\triangle)_{D}\subset{\rm Tube}^G_\pi(\mathbb T^2_\triangle)$ given by
\begin{align}
	{\rm Tube}^G_\pi(\mathbb{T}^{2}_{\triangle})_{D}
	:={\rm Span}_{\mathbb{C}}\{  | (\looDSy{x}{y} ) \xrightarrow{a} \rangle  \}
	_{\substack{\forall a \in G \\ \forall (x,y)\in D}} \; .
\end{align}
Noting that the set of equivalence classes forms a partition of $G^2_{\rm comm.}$ it follows that
\begin{align}
	{\rm Tube}^G_\pi(\mathbb{T}^{2}_{\triangle})=\bigoplus_{D}{\rm Tube}^G_\pi(\mathbb{T}^{2}_{\triangle})_{D}
\end{align}
where the direct sum is over all equivalence classes $D\subset G^2_{\rm comm.}$.
Utilising this decomposition of ${\rm Tube}^G_\pi(\mathbb{T}^2_\triangle)$ we can find the simple modules in terms of the simple modules of ${\rm Tube}^G_\pi(\mathbb {T}^2_\triangle)_{D}$ for each $D\subset G^2_{\rm comm.}$.

At this point, let us remark that in limiting cases the twisted quantum triple reduces to either the (2+1)d twisted quantum double or the (1+1)d group algebra. Indeed,  given the equivalence class $D_{\mathbb 1_{G}}:=(\mathbb 1_{G}, \mathbb 1_{G})$, the corresponding sub-algebra is naturally isomorphic to the (1+1)d tube algebra for $G$ and the trivial 2-cocycle. Similarly, given an equivalence class containing an element of the form $(\mathbb 1_{G},a)$ or $(a,\mathbb 1_{G})$ the twisted quantum triple is isomorphic to the untwisted quantum double algebra.

Given an equivalence class $D$, we now describe the simple modules of ${\rm Tube}^G_\pi(\mathbb T^2_\triangle)_{D}$. To this end, let us introduce some notations:
We notate each element in $D$ by $(d^x_{i},d^y_{i})$ for $i\in\{1,\ldots, |D| \}$ and we call $(d^x_{1},d^y_{1})\in D$ the \emph{representative element of $D$}.
We next define the set $Q_{D}:=\{q_{1},\ldots q_{|D|} \}$ such that each $q_{i}\in G$ is defined by a non-canonical choice of element in $G$ satisfying the conditions
$(d^x_{1},d^y_{1})=(q^{-1}_{i}d^x_{i}q_{i},q^{-1}_{i}d^y_{i}q_{i})$
and $q_{1}:=\mathbb 1_{G}$.
Finally, we define the \emph{stabiliser group} of $D$ by
\begin{align}
	Z_{D}:=\{a\in G\, |\, (d^x_{1},d^y_{1})=(a^{-1}d^x_{1}a,a^{-1}d^y_{1}a)  \} \; ,
\end{align}
i.e. the subgroup consisting of elements of $G$ that simultaneously commute with both $d^x_{1}$ and $d^y_{1}$. As in the (2+1)d case, if the group $G$ is taken to be abelian, the previous construction vastly simplifies so that each pair of elements forms an equivalence class and the stabiliser group is given by the group itself.

Utilising the conventions outlined above, it follows from equations (\ref{eq:twisted2Coc2},\ref{eq:twisted2Cocc2}) that $\tau^2_{d^x_{1},d^y_{1}}(\pi)\in Z^{2}(Z_{D},\rU(1))$ defines a normalised 2-cocycle of $Z_{D}$ when $\pi\in Z^{4}(G,\rU(1))$ is a normalised 4-cocycle of $G$. For each simple $\tau^2_{d^x_{1},d^y_{1}}(\pi)$-projective representation $(\mathcal{D}^R,V_R)$ of $Z_{D}$, we can then define a simple representation of the twisted quantum triple algebra by a homomorphism $\mc{D}^{D,R}:{\rm Tube}^G_\pi(\mathbb T^2_\triangle)_{D}\rightarrow {\rm End}(V_{D,R})$ where 
\begin{align}\label{eq:TQTsimplemod}
	V_{D,R}:=
	{\rm Span}_{\mathbb{C}}\{ \ket{d^x_{i},d^y_{i},v_{m}}\;| \;\forall \, i=1,\ldots,|D| \; {\rm and}\; m=1,\ldots,{\rm dim}(V_{R}) \} 
\end{align}
and for $i,j\in\{1,\ldots,|D| \}$, $m,n\in\{1,\ldots,{\rm dim}(V_R) \}$ we define
\begin{align}\label{eq:TQTrepelement}
	\mc{D}^{D,R}_{im,jn}\big(|(\looDSy{x}{y} )\xrightarrow{a} \rangle \big)
	:=
	\delta_{x,d^x_{i}}\delta_{y,d^y_{i}}\delta_{x^{a},d^x_{j}}\delta_{y^{a},d^y_{j}}
	\frac{\tau^2_{d^x_{1},d^y_{1}}\pi(q^{-1}_{i},a)}
	{\tau^2_{d^x_{1},d^y_{1}}\pi(q^{-1}_{i}aq_{j},q^{-1}_{j})}
	D^{R}_{m,n}(|\xrightarrow{q^{-1}_{i}aq_{j}}\rangle)
\end{align}
such that
\begin{align}
	\mc{D}^{D,R}\big(|(\looDSy{x}{y})\xrightarrow{a} \rangle \big)
	:=
	\frac{1}{|G|^{\frac{1}{2}}}
	\sum^{|D|}_{i,j=1}\sum^{{\rm dim}(V_R)}_{m,n=1}
	\mc{D}^{D,R}_{im,jn}\big(|(\looDSy{x}{y})\xrightarrow{a} \rangle \big)
	\ket{d^x_{i},d^y_{i},v_{m}}
	\bra{d^x_{j},d^y_{j},v_{n}} \; .
\end{align}
If follows from the definition and the linearity condition \eqref{eq:2dLinearity} that these matrices indeed define an algebra homomorphism, i.e.
\begin{align}
	\sum_{j,n}
	\mc{D}^{D,R}_{im,jn}\big(|(\looDSy{x_1}{y_1} )\xrightarrow{a} \rangle \big)
	\mc{D}^{D,R}_{jn,ko}\big(|(\looDSy{x_2}{y_2} )\xrightarrow{b}  \rangle \big)
	=
	\delta_{x_2,x_1^{a}}
	\delta_{y_2,y_1^{a}} \, \tau^2_{x_1,y_1}(\pi)(a,b)
	\mc{D}^{D,R}_{am,bn}\big(|(\looDSy{x_1}{y_1})\xrightarrow{ab} \rangle \big)
	 \; .
\end{align}
Furthermore the matrices satisfy the conjugation relation
\begin{align}
	\overline{\mc{D}^{D,R}_{im,jn}\big(|(\looDSy{x}{y})\xrightarrow{a} \rangle \big)}
	=
	\frac{1}{\tau^2_{x,y}(\pi)(a,a^{-1})}
	\mc{D}^{D,R}_{jn,im} \big(|(\looDSy{x^{a}}{y^{a}})\xrightarrow{a^{-1}} \rangle \big) \; ,
\end{align}
as well as the following orthogonality and completeness conditions
\begin{align}
	&\frac{1}{|G|}\sum_{x,y,a\in G}
	\mc{D}^{D,R}_{im,jn}\big(|(\looDSy{x}{y})\xrightarrow{a} \rangle \big)
	\overline{\mc{D}^{D',R'}_{i'm',j'n'}\big(|(\looDSy{x}{y})\xrightarrow{a} \rangle \big)}
	=
	\frac{\delta_{D,D'}\delta_{\mc{D}^R,\mc{D}^{R'}}}{|D|{\rm dim}(V_R)}
	\delta_{i,i'}\delta_{j,j'}\delta_{m,m'}\delta_{n,n'}
	\label{eq:tqtgroupoidorthog2}
	\\
&\frac{1}{|G|}\sum_{\{D,\mc{D}^R\}}
\sum_{\substack{
		i,m\\
		j,n
	}}
	|D|{\rm dim}(V_R)\mc{D}^{D,R}_{im,jn}\big(|(\looDSy{x}{y})\xrightarrow{a} \rangle \big)
	\overline{\mc{D}^{D,R}_{im,jn}(|(\looDSy{x'}{y'})\xrightarrow{a'} \rangle)}
	=\delta_{x,x'}\delta_{y,y'}\delta_{a,a'} \; .
	\label{eq:tqtgroupoidorthog1}
	\end{align}
These conditions can be utilised to verify that the set of simple representations indexed by all pairs $(D,R)$ indeed forms the set of all simple modules.

\subsection{Physical interpretation of the quantum triple algebra simple representations}
We now turn our attention to the interpretation of the simple representations of the twisted quantum triple algebra. Given a simple representation labeled by a pair $(D,R)$, analogously to the (2+1)d example, the equivalence class $D$ corresponds to the set of possible $G$-colourings of the $\mathbb T^{2}$ boundary that are in the same orbit of the tube algebra, whereas the vector space $V_{R}$ corresponds to the decomposition of the symmetries of a boundary colouring under the action of the tube algebra. Thus, we similarly call $D$ a \emph{magnetic flux quantum number} and $R$ an \emph{electric charge quantum number}. 

Let us now refine our description of the $G$-colourings of the torus boundary in order to give a more physical interpretation to the tube algebra. As discussed in sec.~\ref{sec:DW}, a $G$-colouring of the torus boundary is a local description of a flat $G$-connection which is itself a group homomorphism in $ {\rm Hom}(\pi_1(\mathbb T^2),G)$. Noting that $\pi_{1}(\mathbb{T}^{2})=\mathbb{Z}\times\mathbb{Z}$, we know that such a homomorphism defines a commuting pair of elements of $G$. This can be visualised by drawing two directed closed lines on the torus, each corresponding to one of its non-contractible cycles, or in other words to the image of the homomorphism for one of the generators of $\pi_{1}(\mathbb{T}^{2})$, and such that they are labeled by two commuting group variables:
\begin{align}
	\label{eq:torusholon}
	\niceTorus{1}  \; .
\end{align}
In this description, there is no obvious distinction between the two non-contractible cycles of the torus. However, within the physical setting of interest, the torus boundary is always embedded in the interior of a spatial three-manifold, and in this case we can distinguish the two non-contractible cycles. 

Let us first consider a single loop-like excitation inside a three-disk $\mathbb{D}^{3}$. This situation can occur by removing a solid torus from a three-disk so as to obtain the manifold $\mathbb{D}^{3}\backslash (\mathbb{D}^{2}\times \mathbb{S}^{1})$. We then choose a basepoint on the torus and define a $G$-colouring via a group homomorphism in ${\rm Hom}(\pi_{1}(\mathbb{D}^{3}\backslash (\mathbb{D}^{2}\times \mathbb{S}^{1})),G)$. Since $\pi_{1}(\mathbb{D}^{3}\backslash (\mathbb{D}^{2}\times \mathbb{S}^{1}))=\mathbb{Z}$, the manifold $\mathbb{D}^{3}\backslash (\mathbb{D}^{2}\times \mathbb{S}^{1})$ possesses only one non-contractible cycle, unlike the torus that possesses two, so that the $G$-colouring is simply the labeling of the cycle by a group variable. Let us now consider the diagram in equation \eqref{eq:torusholon} as being embedded inside the three-disk. The path coloured by $y$ is no longer non-contractible as it can be lifted from the surface and contracted to the basepoint. It immediately follows that $y= \mathbb 1_{G}$.
In this limit the twisted quantum triple algebra reduces to the untwisted quantum double algebra so that loop-like excitations are in one-to-one corespondence with point-like particles of the (2+1)d model with trivial input 3-cocycle. Additionally, in the limiting case that both holonomies are given by the group identity we can interpret the excitation as a point-like particle that carries a trivial flux quantum number and a charge given by a representation of $G$.

In order for $y$ to take a non-trivial value, we need to enforce that the corresponding path is non-contractible. This can be done by removing a solid cylinder $\mathbb D^{2}\times \mathbb{I}$ from $\mathbb{D}^{3}\backslash (\mathbb{D}^{2}\times \mathbb{S}^{1})$ such that the solid cylinder threads through the hole of the torus, and such that $\mathbb D^{2}\times \{0\}$ and $\mathbb D^{2}\times \{1\}$ are incident with the boundary of $\mathbb D^{3}$. This situation can be depicted as
\begin{equation}
	\label{eq:torusholonthreaded}
	\hspace{-3.5em} \niceTorusThreaded{1} \!\!\!\!\!\! .
\end{equation}
Since the solid cylinder introduces a new non-contractible cycle, it is labeled by a non-trivial group element which can be interpreted as a magnetic flux threading through the loop excitation.
The presence of such a non-trivial threading flux constrains the possible magnetic flux and electric charge quantum numbers of the loop excitation by the requirement that the two fluxes must commute. Interestingly the tube algebra demonstrates that the properties of the loop are not affected by any possible charges threading through the loop.

\subsection{Twisted quantum triple comultiplication and fusion of loop-like excitations}\label{sec:loopfusion}

In this section, we describe the generalisation of the comultiplication map of the twisted quantum double introduced in sec.~\ref{sec:particlefusion} to the case of the twisted quantum triple, which in turn generalises the fusion of point-like particles in the (2+1)d model to the case of loop-like excitations in the (3+1)d model. Unlike its twisted quantum double counterpart, we will see that the comultiplication map of the twisted quantum triple only has a non-zero action on pairs of loops which share the same threading flux.

Similarly to the twisted quantum double case, given a pair of loop-like excitations with internal Hilbert spaces $V_{D_1,R_1}$ and $V_{D_2,R_2}$, respectively, we can consider their joint Hilbert space in the absence of any external constraints as the space given by the tensor product $V_{D_1,R_1}\otimes V_{D_2,R_2}$. In order to understand how the twisted quantum triple acts on the corresponding two-loop Hilbert space $V_{D_1,R_1}\otimes V_{D_2,R_2}$, we introduce the \emph{comultiplication map} $\Delta:{\rm Tube}^G_\pi(\mathbb T^2_\triangle)\rightarrow {\rm Tube}^G_\pi(\mathbb T^2_\triangle)\otimes {\rm Tube}^G_\pi(\mathbb T^2_\triangle)$ defined as
\begin{align}
	\Delta\big(|(\looDSy{x}{y})\xrightarrow{a} \rangle \big)
	:=
	\sum_{x_{1},x_{2}\in Z_{y}}\delta_{x_{1}x_{2},x} \,
	\gamma^2_{a,y}(\pi)(x_{1},x_{2})
	|(\looDSy{x_{1}}{y})\xrightarrow{a} \rangle
	\otimes
	|(\looDSy{x_{2}}{y})\xrightarrow{a} \rangle \; ,
\end{align}
where
\begin{align}
	\gamma^2_{a,y}(\pi)(x_{1},x_{2}):=
	\frac{\tau_{y}(\pi)(x_{1},x_{2},a) \, \tau_{y}(\pi)(a,x^{a}_{1},x^{a}_{2})}
	{\tau_{y}(\pi)(x_{1},a,x^{a}_{2})} \; .
\end{align}
Using the 4-cocycle condition, we can verify the following properties of $\Delta$ akin to the twisted quantum double example:
$\Delta$ is an algebra homomorphism, i.e.
\begin{align}
	\Delta \big(|(\looDSy{x_1}{y_1})\xrightarrow{a} \rangle
	\star
	|(\looDSy{x_2}{y_2})\xrightarrow{b} \rangle \big)
	=
	\Delta \big(|(\looDSy{x_1}{y_1})\xrightarrow{a} \rangle \big)
	\star
	\Delta \big(|(\looDSy{x_2}{y_2})\xrightarrow{b} \rangle \big)
\end{align}
which follows from 
\begin{align}
	\gamma^2_{ab,y}(\pi)(x_{1},x_{2}) \, \tau^2_{x_{1}x_{2},y}(\pi)(a,b)
	=
	\tau^2_{x_{1},y}(\pi)(a,b) \, \tau^2_{x_{2},y}(\pi)(a,b)
	\, \gamma^2_{a,y}(\pi)(x_{1},x_{2}) \, \gamma^2_{b,y^{a}}(\pi)(x^{a}_{1},x^{a}_{2}) \; .
\end{align}
$\Delta$ is quasi-coassociative, i.e.
\begin{align}
	(\Delta\otimes {\rm id}) \circ
	\Delta\big(|(\looDSy{x}{y})\xrightarrow{a} \rangle \big)
	=
	\phi({\rm id}\otimes \Delta) \circ
	\Delta\big(|(\looDSy{x}{y})\xrightarrow{a} \rangle \big)
	\phi^{-1}
\end{align}
where we introduced the \emph{twist} $\phi$ defined as
\begin{align}
	\phi:=
	\sum_{\substack{y\in G\\x_{1},x_{2},x_{3}\in Z_{y}}}
	\tau^{-1}_{y}(\pi)(x_{1},x_{2},x_{3})
	|(\looDSy{x_{1}}{y})\xrightarrow{\mathbb 1_{G}} \rangle
	\otimes
	|(\looDSy{x_{2}}{y})\xrightarrow{\mathbb 1_{G}} \rangle
	\otimes
	|(\looDSy{x_{3}}{y})\xrightarrow{\mathbb 1_{G}} \rangle \; .
\end{align}
Given a three-particle vector space $(V \otimes W) \otimes Z$, the twist $\phi$ induces the module isomorphism
\begin{equation}
	\phi: (V \otimes W) \otimes Z \xrightarrow{
		\sim} V \otimes (W \otimes Z) \; .
\end{equation}
The quasi-coassociativity is ensured by the cocycle data relation
\begin{align}
	\gamma^2_{a,y}(\pi)(x_{1},x_{2}) \, \gamma^2_{a,y}(\pi)(x_{1}x_{2},x_{3}) \,
	\tau^2_{y^{a}}(\pi)(x^{a}_{1},x^{a}_{2},x^{a}_{3})
	=
	\gamma^2_{a,y}(\pi)(x_{2},x_{3}) \, \gamma^2_{a,y}(\pi)(x_{1},x_{2}x_{3})
	\tau^2_{y}(\pi)(x_{1},x_{2},x_{3})  \; .
\end{align}
At this point, it is useful to provide an illustration of this comultiplication map. Let $x,x_{1},x_{2}$ correspond to the flux carried by three loop-like excitation, and $y$ the threading flux, the corresponding multiplication map looks like
\begin{align}
	\picDelta{2}{y}{x}{} \;\; \xrightarrow{\, \Delta \,} \;\; \picDelta{1}{y}{x_1}{x_2}
\end{align}
which makes clear that the comultiplication map preserves the threading flux while splitting the holonomy associated with the loop-like excitation on the l.h.s between the two loop-like excitations on the r.h.s.. 

\bigskip \noindent
Given a pair of representations $(\mc{D}^{D_1,R_1},V_{D_1,R_1})$ and $(\mc{D}^{D_2,R_2},V_{D_2,R_2})$ of the twisted quantum triple, the comultiplication $\Delta$ allows us to define the tensor product representation $\mc{D}^{D_1,R_1}\otimes \mc{D}^{D_2,R_2}(\Delta):{\rm Tube}^G_\pi(\mathbb T^2_\triangle)\otimes {\rm Tube}^G_\pi(\mathbb T^2_\triangle)\rightarrow V_{D_1,R_1}\otimes V_{D_2,R_2}$ where
\begin{align}
	\nn
	\mc{D}^{D_1,R_1}\otimes \mc{D}^{D_2,R_2}\big(\Delta \big(|(\looDSy{x}{y})\xrightarrow{a}\rangle \big) \! \big)
	=\!\!\!\!\!\! \sum_{x_{1},x_{2} \in Z_y} \!\!\!\!\delta_{x_1x_2,x}\, \gamma_{a,y}(\pi)(x_{1},x_{2}) \,
	\mc{D}^{D_1,R_1}\big(|(\looDSy{x_{1}}{y})\xrightarrow{a}\rangle \big)
	\otimes
	\mc{D}^{D_2,R_2}\big(|(\looDSy{x_{2}}{y})\xrightarrow{a}\rangle\big)
\end{align}
which is compatible with the algebra product by the requirement that $\Delta$ defines an algebra homomorphism. The semi-simplicity of ${\rm Tube}^G_\pi(\mathbb T^2_\triangle)$ implies that such tensor product representations are generically not simple and as such admit a decomposition into the direct sum of irreducible representations, i.e.
\begin{align}
	V_{D_1,R_1} \otimes V_{D_2,R_2}\simeq \bigoplus_{\{ (D_3,R_3) \}}N^{(D_3,R_3)}_{(D_1,R_1),(D_2,R_2)}V_{D_3,R_3} \; .
\end{align}
Here $N^{(D_3,R_3)}_{(D_1,R_1),(D_2,R_2)}\in \mathbb{Z}^{+}_{0}$ is a non-negative integer, called the \emph{fusion multiplicity}, that defines how many times each irreducible representation $(\mathcal{D}^{D_3,R_3},V_{C_3,R_3})$ occurs in the decomposition of ${(\mc{D}^{D_1,R_1} \otimes \mc{D}^{D_2,R_2}(\Delta),V_{D_1,R_1}\otimes V_{D_2,R_2})}$. Using the orthogonality relations of the representations, we find the following expression for the fusion multiplicities:
\begin{align}
	N^{(D_3,R_3)}_{(D_1,R_1),(D_2,R_2)}
	:=
	\frac{1}{|G|}\sum_{y,a\in G \atop x \in Z_y}
	\text{tr}
	\Big(
	\mc{D}^{D_1,R_1}\otimes \mc{D}^{D_2,R_2}\big(\Delta\big(|(\looDSy{x}{y})\xrightarrow{a}\rangle \big) \big)
	\overline{\mc{D}^{D_3,R_3}\big(|(\looDSy{x}{y})\xrightarrow{a}\rangle \big)}
	\Big) \; .
\end{align}
We visualise the fusion process of two loop-like excitations with the same threading flux as follows:
\begin{align}
	\picDelta{1}{}{D_1,R_1}{D_2,R_2}
	\;\; \xrightarrow{\text{fusion}} \;\;
	\bigoplus_{\{(D_3,R_3)\}}N_{(D_1,R_1),(D_2,R_2)}^{(D_3,R_3)}
	\;\;  \picDelta{2}{}{D_3,R_3}{} \; .
\end{align}
The definition of the fusion multiplicities shows in particular that it only makes sense to fuse two loops which share the same threading flux in sharp contrast to the (2+1)d example where no such constraint exists.

\subsection{Twisted quantum triple $R$-matrix and braiding of loop-like excitations\label{sec:loopBraiding}}

In the twisted quantum double discussion, we introduced the so-called $R$-matrix and defined the exchange statistics of point-like particles in the 2-disk. Although the exchange statistics of point-like particles in (3+1)d are characterised by representations of the symmetric group, it has been shown that the exchange statistics of loop-like excitations in the 3-disk can be characterised by representations of the so-called \emph{loop-braid group} \cite{lin2005motion,Baez:2006un} or \emph{necklace group} \cite{bellingeri2016braid,Bullivant:2018djw}. In the following, we restrict to the linear necklace group.

As before, let us consider the 3-disk with a solid cylinder removed. The exchange statistics of loop-like excitations threaded by the solid cylinder correspond to the so called linear necklace group which is isomorphic to the braid group. Labelling $n$ loops by $1,2,\ldots,n$ along the positive horizontal axis as in the following diagram, the generator $\sigma_{i}$ of the braid group correspond to exchanging loops enumerated by $i$ and $i+1$ by passing loop $i$ horizontally through loop $i+1$:
\begin{equation*}
	\picBraid{1} \; \to \; \picBraid{2} \; \to \; \picBraid{3} \; .
\end{equation*}
As for the twisted quantum double, we can build on the tensor structure defined by the comultiplication map $\Delta$ in order to show that the representations of the twisted quantum triple algebra admit a representation of the braid group which is interpreted as the exchange statistics of loop excitations in the (3+1)d Dijkgraaf-Witten model. In order to define the braid statistics, we introduce an invertible element of ${\rm Tube}^G_\pi(\mathbb T^2_\triangle)\otimes {\rm Tube}^G_\pi(\mathbb T^2_\triangle)$ which we also call the $R$-matrix:
\begin{align}
	R:=\sum_{\substack{y,a\in G\\x,x'\in Z_{y}}} |(\looDSy{x}{y})\xrightarrow{\mathbb 1_{G}}\rangle \otimes |(\looDSy{x'}{y})\xrightarrow{x}\rangle \; .
\end{align}
The compatibility of this $R$-matrix with the comultiplication map is ensured via
\begin{align}
	R\, \Delta \big(|(\looDSy{x}{y})\xrightarrow{z}\rangle \big)R^{-1}=\sigma \circ \Delta \big(|(\looDSy{x}{y})\xrightarrow{z}\rangle \big)
\end{align}
for all $|(\looDSy{x}{y})\xrightarrow{z}\rangle\in {\rm Tube}^G_\pi(\mathbb T^2_\triangle)$, where $\sigma$ is the transposition map
\begin{align}
	\sigma:V\otimes W\rightarrow W\otimes V \; ,
\end{align}
which permutes the order of vector spaces in the tensor product.
Unlike in the twisted quantum double example the $R$-matrix is not a module isomorphism for the tensor product of any two modules but instead only a module isomorphism on a subspace of the tensor product. Let $V$ and $W$ be a pair of modules of ${\rm Tube}^G_\pi(\mathbb T^2_\triangle)$ we define the projector $\Delta(\mathbb{1}):V\otimes W\rightarrow V\otimes W$ by the comultiplication of the identity element
\begin{align}
\Delta(\mathbb{1}):=\sum_{\substack{y\in G\\x\in Z_{y}}}
\Delta(|(\looDSy{x}{y})\xrightarrow{\mathbb 1_{G}}\rangle)
\end{align}
and define
\begin{align}
V\; \underline{\otimes} \; W:=\Delta(\mathbb{1})\triangleright (V\otimes W)\subset V\otimes W \; .
\end{align}
This subspace corresponds to the subspace of the tensor product of modules which carry the same threading flux. For any two modules $V,W$ of the twisted quantum triple algebra the operator $\hat{R}:=\sigma\circ R$ defines a map
\begin{align}
	\hat{R}:V\; \underline{\otimes} \; W\xrightarrow{\sim}W \; \underline{\otimes} \; V \; ,
\end{align}
which is a module isomorphism:
\begin{align}
	\hat{R} \circ \Delta \big(|(\looDSy{x}{y})\xrightarrow{a}\rangle \big)\triangleright (V\; \underline{\otimes} \; W)=\Delta \big(|(\looDSy{x}{y})\xrightarrow{a}\rangle \big) \circ\hat{R}\triangleright (V\; \underline{\otimes} \; W) \; .
\end{align}
Additionally, given the three-particle vector space $(V\; \underline{\otimes} \; W)\; \underline{\otimes} \; Z$, it follows that $\hat{R}$ satisfies the \emph{hexagon equations}, which in turn implies the \emph{quasi-Yang-Baxter equation} defined in equation \eqref{eq:quasiyangbaxter}, so that $\hat{R}$ defines a braid group representation on the modules of ${\rm Tube}^G_\pi(\mathbb T^2_\triangle)$.

\newpage
\section{Category theoretical aspects\label{sec:cat}}
In this section, we introduce some notions from category theory in order to reformulate and make more precise some of the results derived earlier. In particular, we present the technology of loop groupoids. This is used to make more rigorous the notion of lifted models introduced previously, which is in turn used to rederive tube algebras in any dimensions.

\subsection{Preliminaries\label{sec:catPrelimi}}
In order to introduce the relevant notations, we review below some basic definitions of category theory. More details can be found in \cite{ etingof2016tensor, mac1971category}.

\begin{definition}[\emph{Category}]
	A \emph{category} $\mathcal{C}$ consists of:
	\begin{enumerate}[itemsep=0.4em,parsep=1pt,leftmargin=4em]
		\item[$\bul$] A collection of \emph{objects} denoted by ${\rm Ob}(\mathcal C)$.
		\item[$\bul$] A collection of \emph{morphisms} between objects denoted by ${\rm Hom(\mathcal C)}$ such that each morphism has a source and a target object. Given two objects $a$ and $b$, the collection of morphisms from $a$ to $b$ is denoted by $ {\rm Hom}_{\mathcal C}(a,b) \ni f:a\to b$.
		\item[$\bul$] A composition rule $\circ:{\rm Hom}_{\mathcal C}(a,b)\times {\rm Hom}_{\mathcal C}(b,c) \to {\rm Hom}_{\mathcal C}(a,c)$ of morphisms such that the composition of $f: a \to b$ and $g: b \to c$ is denoted by $f \circ g:a \to c$. This composition rule is \emph{associative}, i.e. for $f:a\to b $, $g:b \to c$ and $h:c \to d$, we have
		\begin{equation*}
			(f \circ g) \circ h = f \circ (g \circ h) \; ,
		\end{equation*} 
		and for every object $x \in {\rm Ob}(\mathcal C)$ there exists an \emph{identity} morphism ${\rm id}_x \in {\rm Hom}_{\mathcal{C}}(x,x)$ such that for every $f:a \to b \in {\rm Hom}_{\mathcal C}(a,b)$, we have $f \circ {\rm id}_a = f = {\rm id}_b \circ  f$.
	\end{enumerate}
\end{definition}
\noindent
We often depict relations between morphisms using commutative diagrams so that points represent objects and arrows represent morphisms between. For instance, the composition of two morphisms $f:a \to b$ and $g: b \to c$ is depicted by
\begin{equation*}
	(a \xrightarrow{\q f \q}b) \circ (b \xrightarrow{\q g \q} c) = a \xrightarrow{\q f \circ g \q} c \; .
\end{equation*}

\begin{definition}[\emph{Monoidal category}]
	A \emph{monoidal category} is a sextuple $(\cC, \otimes, \mathbbm{1}, \alpha, \ell, r)$ that consists of:
	\begin{enumerate}[itemsep=0.4em,parsep=1pt,leftmargin=4em]
		\item[$\bul$] A category $\cC$.
		\item[$\bul$] A binary functor $\otimes: \cC \times \cC \to \cC$ referred to as the \emph{tensor product}.
		\item[$\bul$] A unit object $\mathbbm{1}_\cC \in {\rm Ob}(\cC)$.
		\item[$\bul$] Three natural isomorphisms $\alpha$, $\ell$, $r$ referred to as the \emph{associator}, the \emph{left unitor} and the \emph{right unitor}, respectively, defined as:
		\begin{align*}
			\alpha_{a,b,c} : (a \otimes b ) \otimes c &\xrightarrow{\sim} a \otimes (b \otimes c) \\
			\ell_a : \mathbbm{1} \otimes a &\xrightarrow{\sim} a \\
			r_a : a \otimes \mathbb{1} &\xrightarrow{\sim} a \; .
		\end{align*}
		subject to the coherence relations encoded in the commutative diagrams:
		\begin{equation*}
		\begin{tikzpicture}[scale=1,baseline=0em, line width=0.4pt]
			\matrix[matrix of math nodes,row sep =4em,column sep=-3em, ampersand replacement=\&] (m) {
				{}  \& {} \& {} \&[4em] (a \otimes b) \otimes (c \otimes d) \&[4em] {} \& {} \& {}
				\\
				((a \otimes b) \otimes c) \otimes d \& {} \& {} \& {} \& {} \& {} \& a \otimes (b \otimes (c \otimes d))
				\\
				{} \& (a \otimes (b \otimes c)) \otimes d  \& {} \& {} \& {} \& a \otimes ((b \otimes c) \otimes d) \& {}\\
			};	
			\path
			(m-2-1) edge[->] node[pos=0.6, left, xshift=-0.5em] {$\alpha_{a\otimes b,c,d}$}  (m-1-4)
			edge[->] node[pos=0.6, left, xshift=-0.5em] {$\alpha_{a,b,c} \otimes {\rm id}_d$}  (m-3-2)
			(m-1-4) edge[->] node[pos=0.4, right, xshift=0.5em] {$\alpha_{a,b,c \otimes d}$} (m-2-7)
			(m-3-2) edge[->] node[pos=0.5, below] {$\alpha_{a,b \otimes c,d}$}  (m-3-6)
			(m-3-6) edge[->] node[pos=0.4, right,xshift=0.5em] {${\rm id}_a \otimes \alpha_{b,c,d}$}  (m-2-7)	
			;
		\end{tikzpicture}
		\end{equation*}
		\begin{equation*}
		\begin{tikzpicture}[scale=1,baseline=0em, line width=0.4pt]
			\matrix[matrix of math nodes,row sep =4em,column sep=1em, ampersand replacement=\&] (m) {
				{}  \&  a \otimes b   \& {}
				\\
				(a \otimes \mathbbm{1}) \otimes b) \& {} \& a \otimes (\mathbbm{1} \otimes b) \\
			};	
			\path
			(m-2-1) edge[->] node[pos=0.6, left, xshift=-0.5em] {$r_a \otimes {\rm id}_{b}$}  (m-1-2)	
			edge[->] node[pos=0.5, below] {$\alpha_{a,\mathbbm{1},c,b}$}  (m-2-3)
			(m-2-3) edge[->] node[pos=0.6, right, xshift=0.5em] {${\rm id}_a \otimes \ell_{b}$}  (m-1-2)		
			;
		\end{tikzpicture} .
		\end{equation*}
		These consistency conditions are usually referred to as the \emph{pentagon} and the \emph{triangle} relations, respectively.
	\end{enumerate}
\end{definition}
\noindent
A monoidal category as defined above is sometimes referred to as \emph{weak} monoidal category since the morphisms $\alpha$, $\ell$ and $r$ weakens the associativity and the unit conditions.
\begin{example}[\emph{Category of $G$-graded vector spaces}]
	Let $G$ be a finite (possible non-abelian) group. A $G$-graded vector space is a vector space of the form $V = \bigoplus_{a \in G}V_a$. We consider the category $\mathbb C$--${\rm Vec}^{\alpha}_G$ whose objects are $G$-graded complex-valued vector spaces. The tensor product is defined according to
	\begin{equation*}
		(V \otimes W)_a = \bigoplus_{b,c \in G \atop b \cdot c = a} V_b \otimes W_c \; . 
	\end{equation*}
	This category has $|G|$ simple objects denoted by $\delta_{a \in G}$, i.e. objects satisfying ${\rm End}(\delta_a) = \mathbb C$, provided by the one-dimensional $G$-graded vector spaces. It follows from the definition that the tensor product of simple objects boils down to the group multiplication, i.e. $\delta_a \otimes \delta_b \cong \delta_{a b}$. It is enough to define the associator on the simple objects. Thus, we are looking for an isomorphism characterized by a group 3-cochain $\alpha: G^3 \to \mathbb{C}^\times $ such that
	\begin{equation*}
		\alpha_{\delta_a, \delta_b, \delta_c} = \alpha(a,b,c) \cdot {\rm id}_{\delta_{abc}}: (\delta_a \otimes \delta_b) \otimes \delta_c \xrightarrow{\sim} \delta_a \otimes (\delta_b \otimes \delta_c) \; .
	\end{equation*}
	The pentagon relation above is then satisfied if $\alpha$ is a 3-cocycle in $H^3(G,\mathbb{C}^\times) \simeq H^3(G,\rU(1))$, i.e.
	\begin{equation}
		\alpha(a,b,c) \, \alpha(a,b \cdot c,d) \, \alpha(b,c,d)= \alpha(a \cdot b,c,d) \, \alpha(a,b,c \cdot d)
	\end{equation}
	for every $a,b,c,d \in G$. Furthermore, it follows from the triangle relation above that if the right and left unitors are trivial, then the 3-cocycle $\alpha$ is normalized, i.e. $\alpha(a, \mathbbm{1}, b)= 1$, $\forall \, a,b \in G$. 
\end{example}
\noindent
It turns out that the category of $G$-graded vector spaces is the relevant structure to describe the (2+1)d Hamiltonian realisation of Dijkgraaf-Witten theory. Indeed, we can show that for a model whose input data is $\mathbb C$--${\rm Vec}^{\alpha}_G$, the bulk excitations are provided by the objects of the so-called Drinfel'd center $\mathcal{Z}(\mathbb C$--${\rm Vec}^{\alpha}_G)$ of the category. But the Drinfel'd center $\mathcal{Z}(\mathbb C$--${\rm Vec}^{\alpha}_G)$ is equivalent to the category of modules of the Drinfel'd double $\mathcal{D}^\alpha(G)$, which we showed to be a representative of the (2+1)d tube algebra. By construction $\mathcal{Z}(\mathbb C$--${\rm Vec}^{\alpha}_G)$ is a braided monoidal category:

\begin{definition}[\emph{Braided monoidal category}]\label{def:braidedmonoidalcat}
	Let $\cC \equiv (\cC, \otimes, \mathbbm{1}, \alpha, \ell, r)$ be a monoidal category with tensor product $\otimes$. We define a \emph{braiding} on $\cC$ as a natural isomorphism $R: a\otimes b \xrightarrow{\sim} b \otimes a$ subject to the coherence relations encoded in the commutative diagrams:
	\begin{equation*}
	\begin{tikzpicture}[scale=1,baseline=0em, line width=0.4pt]
		\matrix[matrix of math nodes,row sep =4em,column sep=2em, ampersand replacement=\&] (m) {
			{} \& (c \otimes a )\otimes b \& {} 
			\\
			(a \otimes c) \otimes b \& {} \& c \otimes (a \otimes b)
			\\
			a \otimes (c \otimes b) \& {} \& (a \otimes b) \otimes c
			\\
			{} \& a \otimes (b \otimes c) \& {}
			\\
		};	
		\path
		(m-1-2) edge[->] node[pos=0.4, left, xshift=-0.3em] {$R_{c,a} \otimes {\rm id}_b$}  (m-2-1)
				edge[->] node[pos=0.4, right, xshift=0.3em] {$\alpha_{c,a,b}$}  (m-2-3)
		(m-2-1) edge[->] node[pos=0.5, right] {$\alpha_{a,c,b}$}  (m-3-1)
		(m-2-3) edge[->] node[pos=0.5, left] {$R_{c,a \otimes b}$}  (m-3-3)
		(m-3-1) edge[->] node[pos=0.6, left, xshift=-0.3em] {${\rm id}_a \otimes R_{c,b}$}  (m-4-2)
		(m-3-3) edge[->] node[pos=0.6, right, xshift=0.3em] {$\alpha_{a,b,c}$}  (m-4-2)
		;
	\end{tikzpicture} , 
	\begin{tikzpicture}[scale=1,baseline=0em, line width=0.4pt]
		\matrix[matrix of math nodes,row sep =4em,column sep=2em, ampersand replacement=\&] (m) {
			{} \& b \otimes (c \otimes a) \& {} 
			\\
			(b \otimes c) \otimes a \& {} \& b \otimes (a \otimes c)
			\\
			a \otimes ( b \otimes c ) \& {} \& (b \otimes a) \otimes c
			\\
			{} \& (a \otimes b) \otimes c  \& {}
			\\
		};	
		\path
		(m-1-2) edge[->] node[pos=0.4, left, xshift=-0.4em] {$\alpha_{b,c,a}^{-1}$}  (m-2-1)
		edge[->] node[pos=0.4, right, xshift=0.3em] {${\rm id}_a \otimes R_{c,a}$}  (m-2-3)
		(m-2-1) edge[->] node[pos=0.5, right] {$R_{b \otimes c,a}$}  (m-3-1)
		(m-2-3) edge[->] node[pos=0.5, left] {$\alpha_{b,a,c}^{-1}$}  (m-3-3)
		(m-3-1) edge[->] node[pos=0.6, left, xshift=-0.3em] {$\alpha_{a,b,c}^{-1}$}  (m-4-2)
		(m-3-3) edge[->] node[pos=0.6, right, xshift=0.3em] {$R_{b,a}\otimes {\rm id}_c$}  (m-4-2)
		;
	\end{tikzpicture} .
	\end{equation*}
	The consistency conditions above are usually referred to as the \emph{hexagon} relations. We then define a \emph{braided monoidal category} as a pair $(\cC, R)$.
\end{definition}
\noindent
Note that given $\mathcal{Z}(\mathbb C$--${\rm Vec}^{\alpha}_G)$ the hexagon relations above are the analogues of \eqref{eq:hexAlg}. It turns out that for an abelian group $A$, the category of $A$-graded vector spaces can be turned into a braided monoidal category $\mathbb C$--${\rm Vec}^{\alpha,R}_A$ by introducing a family of isomorphisms characterized by a group 2-cochain $R : A^2 \to \mathbb{C}^\times$ such that
\begin{equation*}
	R_{\delta_a,\delta_b} = R(a,b)\cdot {\rm id}_{\delta_{ab}} : \delta_a \otimes \delta_b \xrightarrow{\sim} \delta_b \otimes \delta_a \; .
\end{equation*}
The hexagon equations above are then satisfied if
\begin{align}
	\label{eq:hex1}
	\alpha(c,a,b)\,  R(c,a + b)\, \alpha(a,b,c) &= R(c,a) \, \alpha(a,c,b) \, R(c,b) \\
	\label{eq:hex2}
	R(b + c,a)\, \alpha(b,a,c) &= R(c,a) \, \alpha(b,c,a) \, R(b,a) \; \alpha(a,b,c) \; .
\end{align}
Note however that the (2+1)d Dijkgraaf-Witten model only requires its monoidal version as input. Furthermore it is not possible to define such braided category when the group is non-abelian.

\subsection{Monoidal 2-category of $G$-graded 2-vector spaces}

As mentioned above, the input category of the (2+1)d Hamiltonian realisation of Dijkgraaf-Witten model is the category of $G$-graded vector spaces such that its Drinfel'd center describes the bulk excitations of the model. Analogously, the relevant input category in (3+1)d is believed to be the 2-category of $G$-graded 2-vector spaces \cite{mackaay2000finite, douglas2018fusion}.

So far, we have only considered 1-categories, i.e. categories that contain objects and (1-)morphisms between objects. It is possible to generalize these constructions by providing additional structures. For instance, we can define 2-categories which, on top of objects and 1-morphisms between objects, contain 2-morphisms between 1-morphisms. In the following, we are only interested in a specific 2-category, namely the 2-category of $G$-graded 2-vector spaces \cite{kapranov19942, 1998math......5030M, mackaay2000finite}. But first, let us define 2-vector spaces:

\begin{definition}[\emph{2-vector spaces}]
	We call a \emph{2-matrix} $\mathcal{M}$ a matrix whose entries denoted by $\mathcal{M}_{mn}$
	are finite-dimensional complex valued (1-)vector spaces. We then define a 2-vector space as a formal symbol $\mathcal{V}[p]$ with $p = 0, 1, 2, \ldots$, such that the collection ${\rm Hom}(\mathcal{V}[p], \mathcal{V}[q])$ of 1-morphisms from $\mathcal{V}[p]$ to
	$\mathcal{V}[q]$ corresponds to the set of all $p \times q$ 2-matrices $\mathcal{M}$. The
	collection ${\rm Hom}(\mathcal{M},\mathcal{N})$ of 2-morphisms from $\mathcal{M}$ to $\mathcal{N}$ such
	that $\mathcal{M}, \mathcal{N} : \mathcal{V}[p] \to \mathcal{V}[q]$ then corresponds to the set of all matrices $\mathcal{T}$ of linear operators $\mathcal{T}_{mn} :\mathcal{M}_{mn} \to \mathcal{N}_{mn}$. 
\end{definition}
\noindent
We define the category $\mathbb C$--$2{\rm Vec}_G$ as the 2-category whose objects are $G$-graded complex 2-vector spaces as defined above. Similarly to the category of $G$-graded vector spaces, simple objects $\delta_a$ are labeled by group elements in $G$ and the tensor product between simple objects is provided by the group multiplication. We choose the underlying 1-category of $\mathbb C$--$2{\rm Vec}_G$ to be \emph{strict}, i.e. the associator $\alpha^0$, the right unitor and the left unitors are chosen to be trivial. Although the associator is trivial, it is possible to weaken the corresponding pentagon equation by introducing a so-called \emph{pentagonator} 2-morphism defined as:
\begin{equation*}
	\alpha^0_{\delta_a \otimes \delta_b ,\delta_c,\delta_d} \circ \alpha^0_{\delta_a,\delta_b,\delta_c \otimes \delta_d} \; \xRightarrow{\pi_{\delta_a,\delta_b,\delta_c,\delta_d}} \;
	(\alpha^0_{\delta_a,\delta_b,\delta_c} \otimes {\rm id}_{\delta_d}) \circ \alpha^0_{\delta_a,\delta_b\otimes \delta_c,\delta_d} \circ ({\rm id}_{\delta_a} \otimes \alpha^0_{\delta_b,\delta_c,\delta_d}) \; .
\end{equation*} 
This pentagonator 2-morphism is determined by a group 4-cochain $\pi: G^4 \to \mathbb C^\times$. The pentagonator must satisfy some coherence relations \cite{kapranov19942, delcamp2018gauge} that enforces the 4-cochain $\pi$ to be a 4-cocycle in $H^4(G,\mathbb C^\times) \simeq H^4(G,\rU(1))$. We denote the corresponding 2-category by $\mathbb C$--$2{\rm Vec}^\pi_G$. Similarly, we could define a 2-morphism that weakens the triangle relation. However, we choose it to be trivial, which on turns implies that $\pi$ is a normalized 4-cocycle. In the following, we often identify simple objects and the group variables labeling them so that the pentagonator 2-morphism will be written $\pi_{a,b,c,d}$, $a,b,c,d \in G$, instead.

So the input data of $\mathbb C$--$2{\rm Vec}^\pi_G$ is the same as the one of the (3+1)d Hamiltonian model introduced earlier, namely a finite group $G$ and a normalised representative of a cohomology class in $H^4(G,\rU(1))$, as expected. Furthermore, it is possible to relate the definition of $\pi$ to the $2 \rightleftharpoons 3 $ and $1 \rightleftharpoons 4$ Pachner moves, the same way the 4-cocycle appears as the amplitude of the corresponding Pachner operators \eqref{eq:P23} and \eqref{eq:P41}. In the following, we explain how to recover the quantum triple algebra starting from this data using the technology of loop groupoids. 

While finishing this manuscript, Kong et al. published a beautiful work where they define a generalization of the center construction for the case of \emph{monoidal bicategories} \cite{Kong:2019brm}. Starting from $\mathbb C$--$2{\rm Vec}^\pi_G$, they find a 2-category whose objects and 1-morphisms can be mapped to the irreducible modules of the twisted quantum triple. Although the two approaches seem to yield the same result, we believe our approach makes more transparent the physics of the defect excitations.

\subsection{Delooping and classifying space}
In this part, we review the concepts of delooping groupoid and classifying space of groupoid. More details can be found in \cite{eilenberg1953groups, eilenberg1954groups, may1992simplicial,  hatcher2002algebraic, willerton2008twisted}.

\begin{definition}[\emph{Groupoid}]
	A (finite) groupoid $\mathcal{G}$ is a category such that there is a finite number of objects, and a finite number morphisms which are all invertible.
\end{definition}
\noindent
Given a finite groupoid $\mathcal{G}$, we can construct a so-called simplicial set $\mathcal{B} \mathcal{G}$ referred to as the classifying space of the groupoid:

\begin{definition}[\emph{Classifying space of a groupoid}\label{def:class1}]
	Let $\mathcal{G}$ be a finite groupoid. We define the classifying space $\mathcal{B}\mathcal{G}$ of $\mathcal{G}$ as the simplicial set obtained by gluing together abstract $n$-simplices identified with strings of $n$ composable morphisms $x_0 \xrightarrow{g_1}  x_1 \xrightarrow{g_2} \cdots \xrightarrow{g_n} x_n $ in $\mathcal{G}$. A given $n$-simplex can be represented as a standard $n$-simplex whose vertices are labeled by $x_0, x_1, \ldots, x_n$ and whose oriented edges are labeled by morphisms and composition of morphisms.\footnote{By identifying abstract simplices with labeled standard simplices, we are building the topological space referred to as the geometrical realisation of $\mathcal{BG}$. But we loosely identify simplicial sets and their geometrical realisations in this paper.} As a simplicial set, $\mathcal{B}\mathcal{G}$ comes equipped with \emph{face} homomorphisms such that the $i$-th face of a given $n$-simplex is obtained by removing the object $x_i$ from the string of morphisms and compose the corresponding adjacent morphisms if $i \neq 0,n$. For instance the three (1-)faces of the 2-simplex $x_0 \xrightarrow{g_1} x_1 \xrightarrow{g_2} x_2$ reads $x_0 \xrightarrow{g_1} x_1$, $x_1 \xrightarrow{g_2} x_2$ and $x_0 \xrightarrow{g_1 \circ g_2} x_2$. 
\end{definition}
\noindent
In the following, we are particularly interested in the classifying space of the one-object groupoid associated with any finite group:
\begin{definition}[\emph{Delooping of a group}]
	Let $G$ be a (finite) group. The delooping of $G$ is the one-object groupoid $\overline{G}$ that consists of a single object denoted by $\bul$ and ${\rm Hom}(\bul,\bul) = G$ such that the composition of morphisms is given by the group multiplication. Informally, we think of $\overline{G}$ as
	\begin{equation}
		\overline{G} = \{ \, \looSy{g}   \, \equiv \, \bul \xrightarrow{g} \bul \, | \, g \in G \, \} \; .
	\end{equation}
\end{definition}
\noindent 
The classifying space of the one-object groupoid $\overline{G}$ coincides as a simplicial set with the classifying space $BG$ of the finite group $G$ as usually defined in algebraic topology \cite{eilenberg1953groups, hatcher2002algebraic}: 
\begin{definition}[\emph{Classifying space of a finite group}\label{def:class2}]
	Let $G$ be a finite group. We denote by $EG$ the simplicial set whose $n$-simplices are identified with ordered ($n$+1)-tuples $(g_0,g_1,\ldots,g_n)$ in $G^{n+1}$. We define the $i$-th face homomorphism $\partial_i = \partial_i^{(n)} : G^{n+1} \to G^n$ on the set of $n$-simplices as $\partial_i (g_0, \ldots, g_n) := (g_0,\ldots,g_{i-1},g_{i+1}, \ldots, g_n)$ such that the boundary of an $n$-simplex is given by the homomorphism $\partial^{(n)} =  \partial_0 - \partial_1 + \cdots + (-1)^n\partial_n$. The group $G$ has a left action $\triangleright$ on $EG$ by left multiplication, i.e. $g \triangleright (g_0, \ldots, g_n) = (gg_0, \ldots ,gg_n)$, and the classifying space $BG$ of the finite group $G$ is obtained as the quotient space $BG=EG/G$. Since $(g_0, \ldots, g_n) \sim (gg_0, \ldots, gg_n)$, $BG$ can be equivalently obtained in terms of $n$-tuples $[g_1|\ldots|g_n]$ via the map $[g_1|\ldots|g_n] \to (\mathbbm{1},g_1 ,g_1 g_2,\dots,g_1 \cdots g_n)$ so that to each $n$-simplex $(g_0, \ldots, g_n)$, we now assign the $n$-tuple $(g_0, \ldots, g_n) \to [g_{0}^{-1} g_1| g_1^{-1} g_2| \ldots | g_{n-1}^{-1}g_n]$. As in definition \ref{def:class1}, 1-simplices of the simplicial set are now labeled by product of group variables, and the boundary map $\partial$ reads
	\begin{equation*}
		\partial ^{(n)}[g_1|\ldots|g_n] = [g_2| \ldots| g_n]+\sum_{i=1}^{n-1}(-1)^i[g_1|\ldots|g_{i-1}|g_ig_{i+1}|g_{i+2}| \ldots| g_n]+(-1)^n[g_1|\ldots|g_{n-1}]
	\end{equation*}
	which agrees with the earlier definition via the identification $[g_1| \ldots| g_n] \equiv \bul \xrightarrow{g_1} \cdots \xrightarrow{g_n} \bul$.
\end{definition}
\noindent
 Since the boundary homomorphism satisfies $\partial \circ \partial = 0$, the simplicial set $BG$ forms a chain complex whose chain groups are provided by the $n$-simplices. Thus, we identify $n$-tuples $[g_1| \ldots| g_n]$ with $n$-chains which are chosen to be valued in $\rU(1)$. This defines the set of $n$-chains $C_n(BG,\rU(1))$. Similarly, we define $n$-cochains as functions which assign to any $n$-simplex of $BG$ an element of $\rU(1)$. By dualising, the boundary operator $\partial$, we can finally define the cohomology $H^n(BG, \rU(1))$. The algebraic cohomology $H^n(G,\rU(1))$ of the group $G$ is then defined as the simplicial cohomology $H^n(BG, \rU(1))$ of its classifying space. More generally, we define the cohomology of a groupoid $\mathcal{G}$ as the simplicial cohomology of its classifying space $\mathcal{BG}$. 

Finally, given a groupoid $\mathcal{G}$, a specific point in the classifying space $\mathcal{BG}$ is identified by a string of morphisms $x_0 \xrightarrow{g_1} \cdots \xrightarrow{g_n} x_n$ and a set of coordinates $0 \leq t_1 \leq t_2 \leq \ldots \leq t_n$. In the following, we denote such a point by $x_0 \xrightarrow{g_1,t_1} x_1 \xrightarrow{g_2,t_2} \cdots \xrightarrow{g_n,t_n}x_n$.

\medskip \noindent
This delooping procedure can be generalized to any (pointed) category: Given a category $\mathcal{C}$, we can define its delooping $\overline{\mathcal{C}}$ where $\overline{\mathcal{C}}$ has a single object $\bul$ and ${\rm Hom}(\bul,\bul) = \mathcal{C}$. The composition of the 1-morphisms is then provided by the monoidal structure of $\mathcal{C}$, i.e. the composite of $\bul \xrightarrow{a} \bul \xrightarrow{b} \bul$ is  $\bul \xrightarrow{a \otimes b} \bul$. Applying the delooping procedure to the monoidal category $\mathbb C$--${\rm Vec}^{\alpha}_G$ of $G$-graded vector spaces, we obtain a 2-category which consists of a single object $\bul$, finitely many simple 1-morphisms labeled by group variables in $G$ and a 1-associator 2-morphism $\alpha$ whose definition in terms of commutative diagrams reads
\begin{equation}
	\PrePTwoTwoCat{}
\end{equation}
where we identify simple objects and the corresponding group variables for convenience. Using a slightly abusive notation, this can be equivalently depicted as
\begin{equation}
	\label{eq:assocCommut}
	\PTwoTwoCat{}
\end{equation}
As simplicial complexes in the corresponding classifying space, the diagrams above defines a $2 \leftrightharpoons 2$ move as expected.
Similarly, the delooping of the 2-category $\mathbb C$--$2{\rm Vec}^{\pi}_G$ of $G$-graded 2-vector spaces, we obtain a 3-category that consists of a single object $\bul$, simple 1-morphisms labeled by group variables in $G$ and a trivial 1-associator 2-morphisms $\alpha^0$ whose pentagon equation is weakened by the presence of a pentagonator 3-morphism $\pi$ determined by a group 4-cocycle. Using the notation above, we have
\begin{equation}
	\label{eq:TwoThreeCat}
	\PTwoThreeCat{2}
	\xRrightarrow{\pi_{a,b,c,d}} 
	\PTwoThreeCat{1} \; ,
\end{equation}
which heuristically confirms that the category underlying the (3+1)d lattice model for which the amplitude associated with the $2 \leftrightharpoons 3$ move is given by a 4-cocycle, is indeed the 2-category $\mathbb C$--$2{\rm Vec}^{\pi}_G$.

\subsection{Loop groupoid}
This part is the heart of our construction. We review the definition of loop groupoid and explain how it is related to the notion of transgression maps of groupoid cocycles. To do so, we follow closely the work of Willerton \cite{willerton2008twisted}. A similar approach can be found in \cite{bartlett2009unitary}. We then explain how this construction can be used to recover the quantum double and quantum triple algebra, as well as defining lifted topological models.

\bigskip \noindent
So the next step of our construction requires to define the so-called \emph{loop groupoid} $\Lambda \mathcal{G}$ of a finite groupoid $\mathcal{G}$. It can be succinctly defined as the functor category ${\rm Fun}(\overline{\mathbb{Z}}, \mathcal{G})$  where $\overline{\mathbb{Z}}$  is the delooping of $\mathbb{Z}$. Since $\pi_1(\mathbb{S}^1) = \mathbb{Z}$, or in other words $\mathcal{B} \overline{\mathbb Z} \simeq \mathbb{S}^1$, it is possible to think of the objects in $\Lambda \mathcal{G}$ as being `loops' in $\mathcal{G}$, hence the name. In practice, we use the following equivalent definition:
\begin{definition}[\emph{Loop groupoid}]
	Let $\mathcal{G}$ be a finite groupoid. We define the loop groupoid $\Lambda \mathcal{G}$ as the finite groupoid whose objects are given by endomorphisms $g \in {\rm End}_\mathcal{G}(x)$, for each $x \in {\rm Ob(\mathcal{G})}$, and whose morphisms read $g \xrightarrow{h} h^{-1} \circ g \circ h$ where $g \in {\rm End}_\mathcal{G}(x)$ and $h \in {\rm Hom}_\mathcal{G}(x,y)$. The composition of the 1-morphisms is inherited from the one in $\mathcal{G}$. 
\end{definition} 
\noindent 
In the following, we are interested in the loop groupoid $\Lambda \overline{G}$ of the one-object groupoid $\overline G$. In this case, the objects of $\Lambda \overline G$ can be identified with group elements in $G$ so that the loop groupoid effectively describes the action of the group onto itself by conjugation. This action becomes of course trivial in the case of an abelian group.

Let $G$ be a finite (possibly \emph{non-abelian}) group. Let us consider the classifying space $\mathcal{B}\Lambda\overline{G}$ of its loop groupoid such that vertices are identified with endomorphisms $x \in {\rm End}_{\overline G}(\bul)$ and edges with morphisms $x \xrightarrow{g}g^{-1}xg$. In the following, we make use of the shorthand notation $x^g \equiv g^{-1}xg$. The $n$-simplices are identified with the following strings of morphisms
\begin{equation}
	\label{eq:SHnot}
	(\looSy{x})  \xrightarrow{g_1} 	( \looSy{x^{g_1}} )  \xrightarrow{g_2} \cdots \xrightarrow{g_n}(\!\!\! \looSy{x^{g_1 \cdots g_n}} \!\!\!) \equiv 
	(\looSy{x})  \xrightarrow{g_1}  \cdots \xrightarrow{g_n} \; ,
\end{equation}
where we used an explicit notation for the objects in order to make the construction more transparent. A point of the classifying space is then identified with
\begin{equation}
	(\looSy{x})  \xrightarrow{g_1,t_1} \cdots \xrightarrow{g_n,t_n} \; .
\end{equation}
Let us also consider a different space, namely the \emph{free loop space} on $\mathcal{LB}\overline G$. Given a topological space $X$, its free loop space $\mathcal{L}X$ is the topological space of all the loops in $X$, namely $\mathcal{L}X = {\rm Maps}(\mathbb{S}^1,X)$.\footnote{The \emph{free} loop space is to differentiate with the \emph{based} loop space which is the space of all loops fixed at a certain base point.} We can define a map $	\textsf{\small LIF}: \mathcal{B}\Lambda \overline G \to 
\mathcal{LB} \overline G \equiv \mathcal{L}BG$ which maps a specific point of the topological space $\mathcal{B}\Lambda \overline G$ to a loop in $\mathcal{LB} \overline G$ parameterized by $t$ such that: 
\begin{equation*}
	\label{eq:LIF}
	\textsf{\small LIF}(t) \, : \, 
	(\looSy{x})  \xrightarrow{g_1,t_1} \cdots \xrightarrow{g_n,t_n} 
	\;\;\;\;\; \longmapsto \;\;\; 
	\bul \xrightarrow{g_1,t_1} \cdots \xrightarrow{g_i,t_i} \xrightarrow{x^{g_1 \cdots g_i},t} \xrightarrow{g_{i+1},t_{i+1}}\cdots \xrightarrow{g_n,t_n} \;\; , \; {\rm for} \; t_i \leq t \leq t_{i+1} \; .
\end{equation*}
Let us explain in detail what this map does in the case of a 2-simplex in $\mathcal{B}\Lambda \overline G$: Firstly, it associates to this 2-simplex a prism whose top and bottom faces are identified. In this case, a 3d triangulation for this prism is obtained using a convention adapted from conv.~\ref{conv:dlifting}. Secondly, a specific point of the original 2-simplex is associated to a loop in $\mathcal{LB}\overline G$ parameterized by $t$. The map then returns a specific point of this loop. The 3-simplex to which the point belongs depends on the value of $t$. In the following, we refer to this map as the \emph{lifting} map in analogy with the lifting procedure alluded in sec.~\ref{sec:Lifting1}.\footnote{In Willerton's work \cite{willerton2008twisted}, the same map is referred to as the `Parmesan map'.} Most importantly, Willerton showed in \cite{willerton2008twisted} that the lifting map $\textsf{\small LIF}$ is a \emph{homotopy equivalence.} 

\medskip \noindent
The lifting map $\textsf{\small LIF} : \mathcal{B}\Lambda \overline G \to \mathcal{L}BG$ naturally induces another map $\mathcal{B} \Lambda \overline G \times \mathbb{S}^1 \to BG$. Via the identification between $n$-simplices and simplicial $n$-chains, the latter map induces yet another map at the level of chains $C_n(\mathcal{B}\Lambda \overline G ,\rU(1)) \to C_{n+1}(BG, \rU(1))$ defined as
\begin{equation}
	(\looSy{x})  \xrightarrow{g_1} \cdots \xrightarrow{g_n} 
	\;\;\;\;\;\; \longmapsto \;\;\;\; 
	\sum_{i=0}^n(-1)^{n-i-1}
	\bul \xrightarrow{g_1} \cdots \xrightarrow{g_i} \xrightarrow{x^{g_1 \cdots g_i}} \xrightarrow{g_{i+1}}\cdots \xrightarrow{g_n} 
\end{equation}
which can be rewritten as follows in terms of the notation introduced in def.~\ref{def:class2}
\begin{equation}
	[g_1| g_2 |\ldots |g_n]_x
	\;\;\;\; \longmapsto \;\;\;\; \sum_{i=0}^n (-1)^{n-i-1}[g_1|\ldots|g_i|(g_1 \cdots g_i)^{-1}x(g_1 \cdots g_i)|g_{i+1}|\ldots|g_{n}] \; .
\end{equation}  
The ($n$+1)-chains appearing on the r.h.s are identified with the ($n$+1)-simplices obtained after triangulation of the topological space $\mathcal{L}BG$ according to conv.\ref{conv:dlifting}, which is obtained by lifting $\mathcal{B}\Lambda \overline G$. Finally, using the fact that the algebraic cohomology of a groupoid is defined as the simplicial cohomology of its classifying space, and dualising the map above, we obtain a map at the level of algebraic cocycles:\footnote{In the case where we work with an abelian group $A$, the cohomology of $\Lambda \overline A$ is equal to the group cohomology so that we actually define a map $\tau:Z^{n+1}(A,\rU(1)) \to Z^n(A, \rU(1))$.}
\begin{equation}
\begin{array}{ccccl}
		\tau : & Z^{n+1}(G, \rU(1)) & \xrightarrow{\q} & Z^{n}(\Lambda \overline G,\rU(1)) & \\
		\label{eq:S1transForm}
		& \omega & \xmapsto{\q} &\tau_x(\omega)(g_1,\ldots,g_{n}) & \equiv \tau(\omega)\big([g_1| \ldots | g_{n}]_x \big) \\
		& & & & := \prod_{i=0}^{n}\omega(g_1, \ldots, g_i,\, x^{g_1 \cdots g_i} , g_{i+1}, \ldots, g_{n})^{(-1)^{n-i}} \; .
\end{array}
\end{equation}
Henceforth, we refer to this map as the $\mathbb{S}^1$-transgression map. Let us illustrate this definition with an example. The $\mathbb{S}^1$-transgression map acts on a 4-cocycle $\pi \in Z^4(G, \rU(1))$ as
\begin{equation}
	\label{eq:transS14Coc}
	\tau_x(\pi)(a,b,c) = \frac{\pi(a,a^{-1}xa,b,c) \, \pi(a,b,c,(abc)^{-1}xabc)}{\pi(x,a,b,c) \, \pi(a,b,(ab)^{-1}xab,c)} 
\end{equation}
which is precisely the formula \eqref{eq:S1trans4} obtained in the previous section. So the $\mathbb{S}^1$-transgression of a group cocycle does not result in another group cocycle but a groupoid cocycle, namely a cocycle of the loop groupoid of its delooping. This explains why it does not satisfy the usual cocycle condition, but instead a `twisted' version. It is now easy to derive these conditions using the boundary map of the simplicial set $\mathcal{B} \Lambda \overline G$. For convenience we reproduce below the one satisfied by $\tau(\pi)$:
\begin{equation}
	\label{eq:twisted3CocBis}
	d^{(3)}\tau_x(\pi)(a,b,c,d) = \frac{\tau_{a^{-1}xa}(\pi)(b,c,d) \, \tau_{x}(\pi)(a,bc,d) \, \tau_x(\pi)(a,b,c)}{\tau_{x}(\pi)(ab,c,d) \, \tau_{x}(\pi)(a,b,cd)} = 1 \; .
\end{equation}
Interestingly, this whole procedure can be iterated. Indeed, it is possible to define a lifting map $\textsf{\small LIF}^{2} : \mathcal{B}\Lambda^2 \overline G \to \mathcal{L}^2BG$ which ultimately leads to the $\mathbb{T}^2$-transgression map $\tau^{2} : Z^{n+2}(G,\rU(1)) \to Z^n(\Lambda^2 \overline G,\rU(1))$ defined according to
\begin{equation*}
\tau_{x,y}^{2}(\omega)(g_1,\ldots,g_{n}) = \prod_{i=0}^{n}\tau_{y}(\omega)(g_1, \ldots, g_i,\, x^{g_1 \cdots g_i} , g_{i+1}, \ldots, g_{n})^{(-1)^{n+1-i}} \; .
\end{equation*}
Applying this definition to $\pi \in Z^4(G,\rU(1))$ results in
\begin{align}
	\tau_{x,y}^{2}(\pi)(a,b) := \frac{\tau_y(\pi)(x,a,b) \, \tau_y(\pi)(a,b,(ab)^{-1}xab)}{\tau_y(\pi)(a,a^{-1}xa,b)} 
\end{align}
which is precisely the 2-cochain appearing in the multiplication of the twisted quantum triple algebra.

\bigskip \noindent
With a little stretch of formalism, the lifting map $\textsf{\small LIF}$ induces a transgression map on the level of morphisms determined by group cocycles. Indeed, let us consider the delooping $\mathbb C$--$2\overline{{\rm Vec}}^{\pi}_G$ of the 2-category of $G$-graded 2-vector spaces so that $\pi$ is now a pentagonator 3-morphisms which weakens the pentagon coherence relation of the trivial 1-associator 2-morphism. There are finitely many simple morphisms in this category and they are labeled by group variables in $G$. Since in many ways it is enough to consider simple morphisms only, many properties can be directly inferred from the study of $\overline G$ carried out above. The homotopy equivalence between $\mathcal{B}\Lambda \overline G$ and $\mathcal{L}BG$ implies 
\begin{equation}
	(\looSy{x})  \xrightarrow{a} \;\;\;  \simeq
	\begin{tikzpicture}[scale=1,baseline=-2.25em, line width=0.4pt]
		\matrix[matrix of math nodes,row sep =3em,column sep=3em, ampersand replacement=\&] (m) {
			{\bul} \& 	{\bul}
			\\
			{\bul}  \&  {\bul}
			\\
			};	
			\path
			(m-2-1) edge[->] node[pos=0.5, left] {${\sss x}$}  (m-1-1)
			(m-2-2) edge[->] node[pos=0.5, right] {${\sss a^{-1}xa}$}  (m-1-2)	
			(m-1-1) edge[->] node[pos=0.5, above] {${\sss a}$} (m-1-2)
			(m-2-1) edge[->] node[pos=0.5, above] {${\sss a}$} (m-2-2)
			(m-2-1) edge[->] node[pos=0.5, above, sloped] {${\sss x \cdot a}$} (m-1-2)	
			; 
	\end{tikzpicture} \; ,
\end{equation}
where we are still using  the shorthand notation defined in \eqref{eq:SHnot}.
Moreover, recall that the composition of the morphisms on the l.h.s in $\Lambda \overline G$ reads
\begin{equation}
	(\looSy{x})  \xrightarrow{a} \, \circ  \;\, (\looSy{x^a})  \xrightarrow{b} \;\,  = \, (\looSy{x})  \xrightarrow{ab} \; .
\end{equation}
Let us now suppose that the associativity with respect to this composition rule is weakened by an 1-associator 2-morphism denoted by $\tau(\pi)$. Analogously to \eqref{eq:assocCommut}, this can be represented in terms of commutative diagrams in $\Lambda \overline G$: 
\begin{equation}
	\assocLoop{1}
	\xRightarrow{\tau_x(\pi)_{a,b,c}} 
	\assocLoop{2} \; .
\end{equation}
This representation can be conveniently used to check \eqref{eq:twisted3CocBis} explicitly.
But by virtue of the lifting map $\textsf{\small LIF}$, the diagrams depicted above, thought as simplices in $\mathcal{B}\Lambda \overline G$, are equivalent to the simplicial complexes depicted below in $\mathcal{L}BG$
\begin{equation}
	\begin{tikzpicture}[scale=1,baseline=-0.25 em, line width=0.4pt]
		\matrix[matrix of math nodes,row sep =2em,column sep=2em, ampersand replacement=\&] (m) {
			{\bul} \&[-0.5em] {} \& {\bul} \& {} \\[-0.5em]
			{} \& {\bul} \& {} \& {\bul} \\
			{\bul} \& {} \& {\bul} \& {} \\
			{} \& {\bul} \& {} \& {\bul}
			\\
		};	
		\path	
		(m-1-1) edge[->] node[pos=0.5, above] {${\sss b}$} (m-1-3)
		(m-2-2) edge[->] node[pos=0.5, above] {${\sss a \cdot b \cdot c}$} (m-2-4)
		(m-1-1) edge[<-] node[pos=0.6, left] {${\sss a}$} (m-2-2)
		(m-1-3) edge[->] node[pos=0.4, right] {${\sss c}$} (m-2-4)
		(m-3-1) edge[sha] node[pos=0.5, above] {${\sss b}$} (intersection of m-3-1--m-3-3 and m-2-2--m-4-2)
		(intersection of m-3-1--m-3-3 and m-2-2--m-4-2) edge[->, shb]  (m-3-3)
		(m-4-2) edge[->] node[pos=0.5, below] {${\sss a \cdot b \cdot c}$} (m-4-4)
		(m-3-1) edge[<-] node[pos=0.6, left] {${\sss a}$} (m-4-2)
		(m-3-3) edge[->] node[pos=0.4, right] {${\sss c}$} (m-4-4)
		(m-1-1) edge[<-] node[pos=0.5, left] {${\sss x^a}$} (m-3-1)
		(m-2-2) edge[<-] node[pos=0.2, right] {${\sss x}$} (m-4-2)
		(m-1-3) edge[<-,shorten >= 1.2em] (intersection of m-1-3--m-3-3 and m-2-2--m-2-4)
		(intersection of m-1-3--m-3-3 and m-2-2--m-2-4) edge[shb] node[pos=0.5, right] {${\sss x^{ab}}$} (m-3-3)
		(m-2-4) edge[<-] node[pos=0.5, right] {${\sss x^{abc}}$} (m-4-4)	
		(m-2-2) edge[->] node[pos=0.4, above, sloped] {${\sss a \cdot b}$} (m-1-3)
		(m-4-2) edge[->] node[pos=0.5, below, sloped] {${\sss a \cdot b}$}(m-3-3)
		;
	\end{tikzpicture}
	\xRrightarrow{\tau_x(\pi)_{a,b,c}} 
	\begin{tikzpicture}[scale=1,baseline=-0.25 em, line width=0.4pt]
		\matrix[matrix of math nodes,row sep =2em,column sep=2em, ampersand replacement=\&] (m) {
		{\bul} \&[-0.5em] {} \& {\bul} \& {} \\[-0.5em]
		{} \& {\bul} \& {} \& {\bul} \\
		{\bul} \& {} \& {\bul} \& {} \\
		{} \& {\bul} \& {} \& {\bul}
		\\
		};	
		\path	
		(m-1-1) edge[->] node[pos=0.5, above] {${\sss b}$} (m-1-3)
		(m-2-2) edge[->] node[pos=0.2, above] {${\sss a \cdot b \cdot c}$} (m-2-4)
		(m-1-1) edge[<-] node[pos=0.6, left] {${\sss a}$} (m-2-2)
		(m-1-3) edge[->] node[pos=0.4, right] {${\sss c}$} (m-2-4)
		(m-3-1) edge[sha] node[pos=0.5, above] {${\sss b}$} (intersection of m-3-1--m-3-3 and m-2-2--m-4-2)
		(intersection of m-3-1--m-3-3 and m-2-2--m-4-2) edge[->, shb]  (m-3-3)
		(m-4-2) edge[->] node[pos=0.5, below] {${\sss a \cdot b \cdot c}$} (m-4-4)
		(m-3-1) edge[<-] node[pos=0.6, left] {${\sss a}$} (m-4-2)
		(m-3-3) edge[->] node[pos=0.4, right] {${\sss c}$} (m-4-4)
		(m-1-1) edge[<-] node[pos=0.5, left] {${\sss x^a}$} (m-3-1)
		(m-2-2) edge[<-] node[pos=0.2, right] {${\sss x^{}}$} (m-4-2)
		(m-1-3) edge[<-, shorten >= 1.2em] (intersection of m-1-3--m-3-3 and m-2-2--m-2-4)
		(intersection of m-1-3--m-3-3 and m-2-2--m-2-4) edge[shb] node[pos=0.5, right] {${\sss x^{ab}}$} (m-3-3)
		(m-2-4) edge[<-] node[pos=0.5, right] {${\sss x^{abc}}$} (m-4-4)	
		(m-1-1) edge[->, sha] node[pos=0.55, above, sloped] {${\sss b \cdot c}$} (m-2-4)
		(m-3-1) edge[sha] (intersection of m-3-1--m-4-4 and m-2-2--m-4-2)
		(intersection of m-3-1--m-4-4 and m-2-2--m-4-2) edge[->, shab] node[pos=0.5, above, sloped] {${\sss b \cdot c}$} (m-4-4)
		;
	\end{tikzpicture}
\end{equation}
where we kept the triangulation of the prisms implicit. Keeping in mind that the prisms should be triangulated, these diagrams correspond exactly to the lifted $2 \leftrightharpoons 2$ Pachner move $(2 \leftrightharpoons 2) \times \mathbb{S}^1$ studied in sec.~\ref{sec:Lifting1}. And we showed in sec.~\ref{sec:Lifting1} that such a move could be decomposed into four (non-trivial) $2 \leftrightharpoons 3$ moves. But to every such $2 \leftrightharpoons 3$ move can be associated a pentagonator 3-morphism in $\mathbb C$--$2\overline{{\rm Vec}}^{\pi}_G$ as illustrated in \eqref{eq:TwoThreeCat}, so that we can write
\begin{equation}
	\tau_x(\pi)_{a,b,c} = \pi_{x,a,b,c}^{-1} \circ \pi_{a,a^{-1}xa,b,c}  \circ  \pi_{a,b,(ab)^{-1}xab,c}^{-1} \circ \pi_{a,b,c,(abc)^{-1}xabc}
\end{equation}
which is the obvious analogue of \eqref{eq:transS14Coc}. So in the same way the $\mathbb{S}^1$-transgression map defines a loop groupoid 3-cocycle from a group 4-cocycle, we argue that it maps the pentagonator 3-morphism $\pi$ in $\mathbb C$--$2\overline{{\rm Vec}}^{\pi}_G$ to a 1-associator 2-morphism $\tau(\pi)$ in its loop groupoid. Therefore, similarly to the relation between the input 3-cocycle of the (2+1)d Dijkgraaf-Witten model and the associator of $\mathbb C$--${\rm Vec}^{\alpha}_G$, the $\mathbb S^1$-transgression of the input 4-cocycle of the (3+1)d model corresponds to the 1-associator of the loop groupoid of $\mathbb C$--$2\overline{{\rm Vec}}^{\pi}_G$.

\subsection{Groupoid algebra and quantum triple}

In this part, we use the loop groupoid technology in order to redefine the twisted quantum double and the twisted quantum triple algebras derived in sec.~\ref{sec:tube} as the tube algebras associated with the excitations of the (2+1)d and (3+1)d models, respectively. Given a groupoid, it is always possible to define an algebra as follows:
\begin{definition}[\emph{Groupoid algebra}]
	Let $\mathcal G$ be a groupoid and $\mathbb k$ a field. We define the groupoid algebra $\mathbb{k} [\mathcal G]$ as the algebra over $\mathbb k$ determined by the vector space spanned by the morphisms in $\mathcal{G}$ and whose multiplication rule is provided by the composition rule of morphisms in $\mathcal{G}$ whenever it is defined, else is zero. 
\end{definition}
\noindent
A primary example of this construction is the quantum double as the \emph{twisted} groupoid algebra of the loop groupoid of the delooping of a group \cite{willerton2008twisted}:
\begin{example}[\emph{Drinfel'd double algebra}]
	Let us apply the definition of the groupoid algebra to the loop groupoid $\Lambda \overline G$. As basis for the vector space $\mathbb{C}[\Lambda \overline G]$ is provided by
	\begin{equation}
		| (\looSy{x})  \xrightarrow{a} \ra \q {\rm for} \; x,a \in G
	\end{equation} 
	which are in one-to-one correspondence with the morphisms in $\Lambda \overline G$. Furthermore, the multiplication rule between two basis elements reads
	\begin{equation}
		| (\looSy{x})  \xrightarrow{a} \ra \star
		| (\looSy{y})  \xrightarrow{b} \ra = \delta_{a^{-1}xa,y} 
		| (\looSy{x})  \xrightarrow{a \cdot b} \ra
	\end{equation}
	where the delta function ensures that the composition in $\Lambda \overline G$ is defined. This reproduces the quantum double multiplication rule. Let $\alpha \in Z^3(G,\rU(1))$, we know that $\tau(\alpha) \in Z^2(\Lambda \overline G, \rU(1))$ by definition.\footnote{Here we can think of $\alpha$ as determining the associator of $\mathbb C$--${\rm Vec}^{\alpha}_G$.} This loop groupoid 2-cocycle can be used in order to twist this multiplication rule as follows:
	\begin{equation}
		| (\looSy{x})  \xrightarrow{a} \ra \star
		| (\looSy{y})  \xrightarrow{b} \ra = \delta_{a^{-1}xa,y} \, \tau_x(\alpha)(a,b) \,
		| (\looSy{x})  \xrightarrow{a \cdot b} \ra \; .
	\end{equation}
	Hence, as an algebra, the twisted quantum double is isomorphic to the twisted groupoid algebra $\mathbb{C}^{\tau(\alpha)}[\Lambda \overline G]$. 
\end{example}
\noindent

\medskip \noindent
Let us now adapt this construction so as to define the quantum triple algebra as the twisted groupoid algebra $\mathbb{C}^{\tau^2(\pi)}[\Lambda^2 \overline G]$ of $\Lambda^2 \overline G$ obtained by iterating twice the loop groupoid construction over the one-object groupoid $\overline G$. Objects in $\Lambda^2 \overline G$ are provided by endomorphisms $x \in {\rm End}_{\Lambda \overline G}(y)$, which by definition of the morphisms in $\Lambda \overline G$ must be labeled by group variables in the centralizer $Z_y$, and for every $x \in {\rm End}_{\Lambda \overline G}(y)$ we have morphisms $x \xrightarrow{g} g^{-1} \circ x \circ g$ such that $g \in {\rm Hom}_{\Lambda \overline G}(y, y^g)$. By analogy with the loop groupoid, we represent morphisms in $\Lambda^2 \overline G$ as 
\begin{equation}
	(\looDSy{x}{y}) \! \xrightarrow{g} (\looDSy{x^g}{y^g})
	\equiv (\looDSy{x}{y}) \! \xrightarrow{g} 
\end{equation}
such that $xyx^{-1}y^{-1}= \mathbb{1}$. We then define the twisted groupoid algebra as before. A basis for the vector space $\mathbb{C}[\Lambda^2 \overline G]$ is provided by
\begin{equation}
	| (\looDSy{x}{y}) \! \xrightarrow{a} \ra  \q {\rm for} \; y,a \in G, \; x \in Z_y \; ,
\end{equation}
which are in one-to-one correspondence with the quantum triple algebra basis element.
Let $\pi \in Z^4(G, \rU(1))$, we know that $\tau^{2}(\pi) \in Z^2(\Lambda^2 \overline G, \rU(1))$ by definition.\footnote{Here we can think of $\pi$ as determining the pentagonator of $\mathbb C$--${2\rm Vec}^{\pi}_G$ so that $\tau(\pi)$ determines the 1-associator in the loop groupoid of its delooping.} As before, this 2-cocycle can be used to define the following twisted multiplication rule  
\begin{equation}
	| (\looDSy{x_1}{y_1}) \xrightarrow{a} \ra \star | (\looDSy{x_2}{y_2}) \xrightarrow{b} \ra = \delta_{a^{-1}x_1a,x_2}\, \delta_{a^{-1}y_1a,y_2}  \, \tau_{x_1,y_1}^{2}(\pi)(a,b) \, 	| (\looDSy{x_1}{y_1})  \xrightarrow{a \cdot b} \ra
\end{equation}
which is exactly \eqref{eq:tqtalgprod}. Hence, as an algebra, the quantum triple is isomorphic to the twisted groupoid algebra $\mathbb{C}^{\tau^2(\pi)}[\Lambda^2 \overline G]$ of the loop groupoid $\Lambda^2 \overline G$ of the loop groupoid $\Lambda \overline G$ of the one-object groupoid $\overline G$ of the group $G$. 

\bigskip \noindent 
So we showed how Willerton's derivation of the twisted quantum double algebra could be easily generalized so as to recover the twisted quantum triple algebra from the groupoid algebra construction of loop groupoids. However, the way the twisting is added to the multiplication rule is somewhat \emph{ad hoc}. In the following, we use the technology of loop groupoids to define more rigorously the lifting procedure presented in sec.~\ref{sec:Lifting1}. This lifting procedure will in turn be used to rederive the (2+1)d and (3+1)d tube algebras in terms of the (1+1)d one, making the introduction of the twist naturally descending from the definition of the topological model.

\subsection{Loop groupoid colouring and lifted models\label{sec:Lifting2}}

Recall that given a finite group $G$ and a closed oriented ($d$+1)-manifold $\cM$, the partition function of the Dijkgraaf-Witten model is performed over homotopy classes of maps $[\gamma]:\cM \to BG$, while the topological action is provided by the canonical pairing $\langle  \gamma^\star \omega, [\cM]\rangle$ between the pull-back of the cocycle $\omega \in Z^{d+1}(BG,\mathbb R / \mathbb Z)$ onto $\cM$ and the fundamental class $[\cM] \in H_{d+1}(\cM, \mathbb Z)$ of $\cM$. Let us now suppose that we have a closed ($d$+1)-manifold of the form $\mathcal{N} \times \mathbb S^1$. We explained earlier that given a finite group $G$, the lifting map $\textsf{\small LIF}: \mathcal{B}\Lambda \overline G \to 
\mathcal{LB} \overline G \equiv \mathcal{L} BG$ is a homotopy equivalence. It implies the equivalence $\mathcal{B}\Lambda \overline G \times \mathbb S^1 \simeq BG$ which in turn induces the $\mathbb S^1$-transgression map sending a group ($n$+1)-cocycle to a loop groupoid $n$-cocycle. It follows that given a ($d$+1)-manifold of the form $\mathcal{N} \times \mathbb S^1$, we can define an equivalent model in one lower-dimensional. This model is obtained by summing over homotopy classes $[\tilde \gamma] : \mathcal{N} \to \mathcal B \Lambda \overline G$ while the topological action is provided by the pairing $\la \tilde \gamma ^\star \tau (\omega), [\mathcal{N}]\ra $ between the pull-back of the loop groupoid cocycle $\tau(\omega) \in Z^d(\mathcal{B}\Lambda \overline G, \mathbb R / \mathbb Z)$ and the fundamental class $[\mathcal{N}]$. 

The remark above suggests that given a triangulated cobordism $\mathcal{C}_\triangle \times \mathbb S^1$, the partition function $\mathcal{Z}^G_\omega[\mathcal{C}_\triangle \times \mathbb{S}^1]$ should be equal to the lower dimensional partition function $\mathcal{Z}^{\Lambda \overline G}_{\tau(\omega)}[\mathcal{C}_\triangle]$ obtained by summing over loop groupoid colourings. This is indeed the right language to formalise the lifting mechanism presented in sec.~\ref{sec:Lifting1}. In order to check this, let us first introduce the notion of groupoid colouring:

\begin{definition}[\emph{Finite groupoid colouring}\label{eq:GroupoidCol}]
	Let $\mc{G}$ be a finite groupoid and $\mc{M}$ a $d$-manifold endowed with a triangulation $\mc{M}_\triangle$. We define a $\mc{G}$-colouring on $\mc{M}_\triangle$ in a way reminiscent of the definition of the classifying space $\mathcal{BG}$ as follows: To every 0-simplex $(v_{0})\subset \mc{M}_{\triangle}$, we associate an object $x_{v_0}\in {\rm Ob}(\mc{G})$. To every 1-simplex $(v_{0}v_{1})\subset \mc{M}_{\triangle}$, we assign a morphism $x_{v_0}\xrightarrow{g_{v_0v_1}} x_{v_1} \in {\rm Hom}_{\mc{G}}(x_{v_0},x_{v_1})$ whose source and target objects agree with the objects assigned to the 0-simplices $(v_{0})$ and $(v_{1})$, respectively. Furthermore, for every 2-simplex $(v_{0}v_{1}v_{2}) \subset \mc{M}_{\triangle}$, we enforce that a given morphism is obtained as the composition of the other two, i.e. $g_{v_0v_2} = g_{v_0v_1} \circ g_{v_1v_2}$. We notate the set of groupoid colourings of $\mc{M}_{\triangle}$ as ${\rm Col}(\mc{M}_{\triangle},\mc{G})$.\footnote{It follows immediately from the definition that we have the identification ${\rm Col}(\mc{C}_{\triangle}, \overline G)={\rm Col}(\mc{C}_{\triangle},{G})$.} 
\end{definition}
\noindent
Let us apply this definition to the loop groupoid $\Lambda \overline G$. Recall that the set of objects in  $\Lambda\overline{G}$  is $G$ and that the set of morphisms reads $\{x\xrightarrow{g}x^{g}\}_{\forall x,g\in G}$. A $\Lambda\overline{G}$-colouring $\mathfrak{g}$ of $\mc{C}_{\triangle}$ assigns group variables to every 0- and 1-simplices of $\mc{C}_{\triangle}$ such that given a 1-simplex $(v_0v_1) \subset \mathcal{C}_\triangle$, we have $\mathfrak{g}[v_0]=x_{v_0} \in G$, $\mathfrak{g}[v_0v_1] = g_{v_0v_1}$, and $\mathfrak{g}[v_1]=x_{v_1} := x_{v_0}^{g_{v_0v_1}}$. Furthermore, for every 2-simplex $(v_{0}v_{1}v_{2})\in\mc{C}_{\triangle}$, the constraint on the boundary morphisms $x_{v_0} \xrightarrow{g_{v_0v_1}}x_{v_1}$, $x_{v_1}\xrightarrow{g_{v_1v_2}}x_{v_2}$ and $x_{v_0}\xrightarrow{g_{v_0v_2}}x_{v_2}$ enforces that $g_{v_0v_1} \cdot g_{v_1v_2}=g_{v_0v_2}$.
This agrees exactly with the prescription provided in sec.~\ref{sec:Lifting1} for a colouring of $\mathcal{C}_\triangle$ given a $G$-colouring of $\mathcal{C}_\triangle \times \mathbb S^1$. Therefore, we have the following identification
\begin{align}
{\rm Col}(\mc{C}_{\triangle}\times\mathbb{S}^{1},G)
=
{\rm Col}(\mc{C}_{\triangle},\Lambda\overline{G}) 
\end{align}
between $G$-colourings on $\mc{C}_\triangle \times \mathbb S^1$ and loop groupoid $\Lambda \overline G$-colourings on $\mathcal{C}_\triangle$.

Let us now consider the colouring $\mathfrak{g} \in {\rm Col}( \mathcal{C}_\triangle, \Lambda \overline G)$ compatible with a given $G$-colouring of $\mathcal{C}_\triangle \times \mathbb S^1$. The amplitude associated with a $G$-coloured lifted $d$-simplex $(v_0\ldots v_d) \times \mathbb S^1$ is equal to the amplitude of the $\Lambda \overline G$-coloured $d$-simplex $(v_0 \ldots v_d)$:
\begin{align}
	\tau(\omega)(\mathfrak{g}[v_{0}\ldots v_{d}]):=\tau_{\mathfrak{g}[v_0]}(\omega)(\mathfrak{g}[v_0v_1],\mathfrak{g}[v_1v_2],\ldots,\mathfrak{g}[v_{d-1}v_d]) 
\end{align}
such that $\tau(\omega)$ is now interpreted as a loop groupoid $d$-cocycle.

Using the above conventions we can write the ($d$+1)-dimensional Dijkgraaf-Witten partition for $\mc{C}_{\triangle}\times\mathbb{S}^{1}$, where $\mc{C}_{\triangle}$ is an oriented, triangulated cobordism with boundary $\partial\mc{C}_{\triangle}=\overline{\mc{C}_{\triangle,0}}\sqcup \mc{C}_{\triangle,1}$ as follows:
\begin{align*}
	\mc{Z}^{\Lambda\overline{G}}_{\tau(\omega)}[\mc{C}_{\triangle}]
	=
	\frac{1}{|G|^{|\mc{C}_{\triangle}^{0}|-\frac{1}{2}| \partial \mc{C}_{\triangle}^{0} |  } }
	\sum_{\mathfrak g\in{\rm Col}(\mc{C}_{\triangle},\Lambda \overline G)}\prod_{\triangle^{(d)}\subset \mc{C}_{\triangle}}
	\!\!\! \tau(\omega)(\mathfrak g[\triangle^{(d)}])^{\epsilon(\triangle^{(d)})}
	\!\!\! \bigotimes_{\triangle^{(1)}\subset\mc{C}_{\triangle,1} }
	\!\!\!
	\ket{\mathfrak g[\triangle^{(1)}]}
	\!\!\! \bigotimes_{\triangle^{(1)}\subset\mc{C}_{\triangle,0} }
	\!\!\!
	\bra{\mathfrak g[\triangle^{(1)}]} \; ,
\end{align*}
which is equal to $	\mc{Z}^{G}_{\omega}[\mc{C}_{\triangle}\times\mathbb{S}^{1}]$ as provided by \eqref{eq:DW_Zclosed}. It follows from the construction that we can identify physical states in $\mathcal{V}^{\Lambda \overline G}_{\tau(\omega)}[\mc{C}_{\triangle,0/1}]$ with physical states in the original Hilbert spaces $\mathcal{V}^G_\omega[\mc{C}^{\mathbb{S}^{1}}_{\triangle,0/1}]$.

\bigskip \noindent
We  explained earlier that the lifting map $\mathsf{LIF}$ can be iterated so as to define the $\mathbb{S}^1 \times \cdots \times \mathbb{S}^1$-transgression map. Similarly, the lifting procedure we have just defined can be iterated. Letting $\Z$ be the ($d$+1)-dimensional Dijkgraaf-Witten partition function, it is straightforward to construct the $n$-fold lifted Dijkgraaf-Witten model of an $m$-dimensional oriented triangulated cobordism such that $m+n=d+1$. Let $\mc{C}_{\triangle_{m}}$ be an $m$-dimensional triangulated manifold. Previously we showed that a $G$-colouring of $\mc{C}_{\triangle_{m}}\times\mathbb{S}^{1}$ could be defined as a $\Lambda\overline{G}$-colouring of $\mc{C}_{\triangle_{m}}$. We now argue that a $G$-colouring of $\mc{C}_{\triangle_{m}}\times^{n}_{i=1}\mathbb{S}^{1}$ can be defined in terms of a $\Lambda^{n}\overline{G}$-colouring of $\mc{C}_{\triangle_{m}}$, where $\Lambda^{n}\overline{G}$ is the $n$-th loop groupoid of $\overline{G}$ defined by $\Lambda^{n}\overline{G}:=\Lambda(\Lambda^{n-1}\overline{G})$. More specifically, for $n>0$ we can define the groupoid $\Lambda^{n}\overline{G}$ as the groupoid with object set $\{(x_n,\ldots,x_{1})\in G^{n}\, |\, x_{i}x_{j}=x_{j}x_{i},  \, \forall \, i,j\in\{1,\ldots,n \} \}$ and morphism set $\{(x_{n},\ldots,x_{1})\xrightarrow{g}(x^{g}_{n},\ldots,x^{g}_{1})\}_{\forall g\in G,(h_{n},\ldots,h_{1})\in {\rm Ob}(\Lambda^n\overline{G})}$.

Furthermore, it follows naturally that the cocycle data the partition function assigns to the $n$-fold lifting of an $m$-simplex $(v_{0}\ldots v_{m})$ with positive orientation is provided by the evaluation of the $\times_{i=1}^n \mathbb S^1$-transgression $\tau^{n}(\omega)\in Z^{m}(\Lambda^{n}\overline{G},\rU(1) )$ of $\omega$ on the $\Lambda^n \overline G$-coloured $m$-simplex $(v_0, \ldots, v_m)$. Putting everything together, we can write the ($d$+1)-dimensional Dijkgraaf-Witten partition function for $\mc{C}_{\triangle_{m}}\times^{n}_{i=1}\mathbb{S}^{1}$, where $\mc{C}_{\triangle_{m}}$ is an $m$-dimensional, oriented triangulated cobordism with boundary $\partial\mc{C}_{\triangle}=\overline{\mc{C}_{\triangle,0}}\sqcup \mc{C}_{\triangle,1}$ as follows:
\begin{align*}
	\mc{Z}^{\Lambda^n\overline{G}}_{\tau^n(\omega)}[\mc{C}_{\triangle}]
	=
	\frac{1}{|G|^{|\mc{C}_{\triangle}^{0}|-\frac{1}{2}| \partial \mc{C}_{\triangle}^{0} |  } }
	\sum_{\mathfrak g\in{\rm Col}(\mc{C}_{\triangle},\Lambda^n \overline G)}\prod_{\triangle^{(d)}\subset \mc{C}_{\triangle}}
	\!\!\! \tau^{n}(\omega)(\mathfrak g[\triangle^{(d)}])^{\epsilon(\triangle^{(d)})}
	\!\!\! \bigotimes_{\triangle^{(1)}\subset\mc{C}_{\triangle,1} }
	\!\!\!
	\ket{\mathfrak g[\triangle^{(1)}]}
	\!\!\! \bigotimes_{\triangle^{(1)}\subset\mc{C}_{\triangle,0} }
	\!\!\!
	\bra{\mathfrak g[\triangle^{(1)}]} 
\end{align*}
which is equal to $	\mc{Z}^{G}_{\omega}[\mc{C}_{\triangle}\times_{i=1}^n\mathbb{S}^{1}]$ as provided by \eqref{eq:DW_Zclosed}. Using the bijection between ${\rm Col}(\mc{C}_{\triangle_{m}},\Lambda^{n}\overline{G})$ and ${\rm Col}(\mc{C}_{\triangle_{m}}\times^{n}_{i}\mathbb{S}^{1},G)$, we can further identify physical states in $\mathcal{V}^{\Lambda^n \overline G}_{\tau^n(\omega)}[\mc{C}_{\triangle,0/1}]$ with physical states in the original Hilbert spaces  $\mathcal{V}^{ G}_{\omega}[\mc{C}^{\times_{i=1}^n\mathbb{S}^{1}}_{\triangle,0/1}]$. Therefore, we can rewrite the lattice Hamiltonian realisation of Dijkgraaf-Witten theory on a $n$-times compactified $d$-dimensional surface in terms of a ($d$$-$$n$)-dimensional model with the group $G$ replaced by the loop groupoid $\Lambda^{n}\overline{G}$ and the ($d$+1)-cocycle $\omega\in Z^{d+1}(G,\rU(1))$ with a $m$-cocycle $\tau^{n}(\omega)\in Z^{m}(\Lambda^n\overline{G},\rU(1))$.

\subsection{Twisted quantum double as the lifted (1+1)d tube algebra}

Let us now rederive the (2+1)d tube algebra within this new context of lifted models. Let $\Sigma_{\rm 2d}$ be a 2d surface of the form $\Sigma_{\rm 1d} \times \mathbb S^1$ equipped with a triangulation $\Sigma_{{\rm 1d}, \triangle} \times \mathbb S^1$. The input for the model is given by a pair $(G,\alpha)$, where $G$ is a finite group and $\alpha$ is a representative normalised 3-cocycle in a cohomology class $[\alpha]\in H^{3}(G,\rU(1))$. Since one of the spatial directions is compactified, we can apply our lifting procedure and define an equivalent model on $\Sigma_{{\rm 1d}, \triangle}$ whose input data is $(\Lambda \overline G, \tau(\alpha))$. This lifted model assigns to every edge $(v_0v_1) \equiv \triangle^{(1)} \subset \Sigma_{{\rm 1d },\triangle}$ a loop groupoid element $\mathfrak{g}[v_0v_1]$.

Recall that in (2+1)d there is a unique choice for the boundary manifold, namely the circle $\mathbb S^{1}$. Although there is no canonical way to triangulate the circle, we know from the previous discussions that any two choices will define Morita equivalent algebras. Here we choose to triangulate the circle by applying conv.~\ref{conv:dlifting} to the point $\mathbb o$, i.e. with a single edge and identified vertices. By doing so, we will use the lifting procedure to define the tube algebra in terms of the (1+1)d tube algebra example. Graphically, we depict this 2d tube as $\mathbb{S}^{1}\times \I$ as follows:
\begin{equation}
	\tubem{\mathbb{S}^{1}}:=\tubem{\mathbb o}\times \mathbb{S}^{1}
	= 	\GrCyl{0.8}{0}{1}{}{}
	\times\mathbb{S}^{1} \; .
\end{equation} 
The Hilbert space $\mc{V}^G_\alpha[\mathfrak{T}[\mathbb S^1]]=\mc{V}^{\Lambda\overline{G}}_{\tau(\alpha)}[\tubem{\mathbb o}]$ is spanned by $\Lambda \overline G$-coloured graph-states of the form:
\begin{align*}
	\mc{V}^{\Lambda\overline{G}}_{\tau(\alpha)}[\tubem{\mathbb{o}}]
	={\rm Span}_{\mathbb{C}}
	\big\{
	\, \big| \mathfrak{g}[\GrCyl{0.8}{0}{1}{}{}] \big\rangle \, 
	\big\}_{\forall \mathfrak{g}\in {\rm Col}(\mathfrak{T}[\mathbb{o}],\Lambda \overline G)} 
	\equiv 
	{\rm Span}_{\mathbb{C}}
	\big\{
	\, \big| \GrCyl{0.8}{0}{1}{x}{a} \big\rangle \, 
	\big\}_{\forall a,x\in G}\; ,
\end{align*}
which is equipped with the canonical inner product
\begin{align}
	\big\langle \, \GrCyl{0.8}{0}{1}{x_1}{a} \, \big| \, \GrCyl{0.8}{0}{1}{x_2}{b} \, \big\rangle
	=\delta_{a,b} \, \delta_{x_1,x_2} \; .
\end{align}
Repeating the computation of sec.~\ref{sec:1+1dtube}, the algebra product on $\mathcal{V}^{\Lambda \overline G}_{\tau(\alpha)}[\mathfrak{T}[\mathbb o]]$ reads
\begin{align}
	\nn\big|
	\GrCyl{0.8}{0}{1}{x_1}{a}
	\big\rangle
	\star 
	\big|
	\GrCyl{0.8}{1}{2}{x_2}{b}
	\big\rangle
	&= \mathbb{P}_{\tubem{\mathbb{o}} \cup_{\mathbb{o}} \tubem{\mathbb{o}}} \circ \mathfrak{G} \triangleright
	\big( \, \big| \GrCyl{0.8}{0}{1}{x_1}{a} \big\rangle
	\otimes \big| \GrCyl{0.8}{1}{2}{x_2}{b} \big\rangle \, \big) \\
	&=
	\delta_{x_2,x_1^a} \, \mathbb{P}_{\tubem{\mathbb{o}} \cup_{\mathbb{o}} \tubem{\mathbb{o}}} \triangleright \big( \, \big|
	\GrCylDouble{0.8}{0}{1}{2}{x_1}{a}{b}
	\big\rangle \, \big) \; .
\end{align}
Applying definition \eqref{eq:opA}, the action of the operator $\mathbb{P}$ is expressed in terms of the partition function $\mathcal{Z}^{\Lambda \overline G}_{\tau(\alpha)}$ as follows:
\begin{align}
	\nn
	&\mathbb{P}_{\tubem{\mathbb{o}} \cup_{\mathbb{o}} \tubem{\mathbb{o}}} \triangleright \big( \, \big|
	\GrCylDouble{0.8}{0}{1}{2}{x_1}{a}{b}
	\big\rangle \, \big) \\[-0.5em]
	& \q = \mathcal{Z}^{\Lambda \overline G}_{\tau(\alpha)} \Bigg[\GrCylZ{0.8}{0}{1}{2}{1'}{a}{b}{x_1} \! \Bigg]
	\big|
	\GrCylDouble{0.8}{0}{1}{2}{x_1}{a}{b}
	\big\rangle
	\\
	& \q =
	\nn
	\frac{1}{|G|}\sum_{k}\frac{\tau_x(\alpha)(a,k)}{\tau_{x^a}(\alpha)(k,k^{-1}b)}
	\big|
	\GrCylDoubleSpe{0.8}{0}{1'}{2}{x_1}{ak}{k^{-1}b}{x_1^{ab}}
	\big\rangle \; .
\end{align}
It now remains to apply the triangulation changing isomorphism between ground states subspaces so as to recover the initial triangulation. Following sec.~\ref{sec:DWopen} and \ref{sec:fixedpoint}, this isomorphism is expressed as the 2d partition function for the pinched interval cobordism given by the 2-simplex $\snum{(012)}$. Explicitly, the triangulation changing operator reads
\begin{align}
	\mathcal{Z}^{\Lambda \overline G}_{\tau(\alpha)} \Bigg[
	\GrCylPinched{0.8}{0}{2}{1'}{}{}
	\Bigg]
	=\frac{1}{|G|^\frac{1}{2}}\sum_{y,c,d\in G}\tau_y(\alpha)(c,d)
	\big| \, \GrCyl{0.8}{0}{2}{y}{cd} \, \big\rangle \big\langle \, \GrCylDouble{0.8}{0}{1'}{2}{y}{c}{d} \, \big|
\end{align}
so that
\begin{equation}
	\big| \, \GrCylDoubleSpe{0.8}{0}{1'}{2}{x_1}{ak}{k^{-1}b}{x_1^{ab}} \, \big\rangle
	\simeq \frac{1}{|G|^\frac{1}{2}}\, \tau_x(\alpha)(ak,k^{-1}b) 
	\big| \,  \GrCyl{0.8}{0}{2}{x_1}{ab} \, \big\rangle \; .
\end{equation}
Putting everything together, the algebra product of ${\rm Tube}^{\Lambda \overline G}_{\tau(\alpha)}(\mathbb o)$ is given by
\begin{equation}
	\nn\big|
	\GrCyl{0.8}{0}{1}{x_1}{a}
	\big\rangle
	\star 
	\big|
	\GrCyl{0.8}{1}{2}{x_2}{b}
	\big\rangle
	= \delta_{x_2,x_1^a} \,
	\frac{1}{|G|^\frac{1}{2}}\tau_{x_1}(\alpha)(a,b)
	\big|
	\GrCyl{0.8}{0}{2}{x_1}{ab}
	\big\rangle \; ,
\end{equation}
where we made use of the 2-cocycle condition $d^{(2)}\tau_{x_1}(\alpha)(a,k,k^{-1}b) =1$. As expected, this reproduces exactly \eqref{eq:2dalgprod}.

\subsection{Twisted quantum triple as the twice lifted (1+1)d tube algebra}

Let us now rederive the (3+1)d tube algebra for torus-boundaries. Let $\Sigma_{\rm 3d}$ be a 3d surface of the form $\Sigma_{\rm 1d} \times \mathbb S^1 \times \mathbb S^1$ equipped with a triangulation $\Sigma_{{\rm 1d}, \triangle} \times \mathbb T^2$. The input for the model is given by a pair $(G,\pi)$, where $G$ is a finite group and $\pi$ is a representative normalised 4-cocycle in a cohomology class $[\pi]\in H^{4}(G,\rU(1))$. Since two of the spatial directions are compactified, we can apply our lifting procedure twice so as to define an equivalent model on $\Sigma_{{\rm 1d}, \triangle}$ whose input data is $(\Lambda^2 \overline G, \tau^2(\pi))$. This twice lifted model assigns to every edge $(v_0v_1) \equiv \triangle^{(1)} \subset \Sigma_{{\rm 1d },\triangle}$ a loop groupoid element $\mathfrak{g}[v_0v_1]$ which in turn assigns one group variable to the bulk of the edge and two group variables to its boundary 0-simplices according to conv.~\ref{eq:GroupoidCol}.

Recall that in (3+1)d, there are several possible choices of boundary manifold but we focus on the case of the torus $\mathbb T^2$. Although there is no canonical way to triangulate the torus, we know from the previous discussions that any two choices will define Morita equivalent algebras. Here we choose to triangulate the torus by applying twice conv.~\ref{conv:dlifting} to the point $\mathbb o$. By doing so, we will use twice the lifting procedure to define the tube algebra in terms of the (1+1)d tube algebra example. Graphically, we depict this 3d tube as $\mathbb{T}^{2}\times \I$ as follows:
\begin{equation}
	\tubem{\mathbb{T}^{2}}:=\tubem{\mathbb o}\times \mathbb{S}^{1} \times \mathbb S^1
	= 	\GrCyl{0.8}{0}{1}{}{}
	\times\mathbb{S}^{1} \times \mathbb S^1 \; .
\end{equation} 
The Hilbert space $\mc{V}^{G}_{\pi}[\tubem{\mathbb T^2}]=\mc{V}^{\Lambda^2\overline{G}}_{\tau^2(\pi)}[\tubem{\mathbb o}]$ is spanned by $\Lambda^2 \overline G$-coloured graph-states of the form:
\begin{align*}
	\mc{V}^{\Lambda^2\overline{G}}_{\tau^2(\pi)}[\tubem{\mathbb{o}}]
	={\rm Span}_{\mathbb{C}}
	\big\{
	\, \big| \mathfrak{g}[\GrCyl{0.8}{0}{1}{}{}] \big\rangle \, 
	\big\}_{\forall \mathfrak{g}\in {\rm Col}(\mathfrak{T}[\mathbb{o}],\Lambda^2 \overline G)} 
	\equiv 
	{\rm Span}_{\mathbb{C}}
	\big\{
	\, \big| \GrCylT{0.8}{0}{1}{x}{y}{a} \big\rangle \, 
	\big\}_{\substack{\forall a,y \in G \\ \forall x \in Z_y}}\; ,
\end{align*}
which is equipped with the canonical inner product
\begin{align}
\big\langle \, \GrCylT{0.8}{0}{1}{x_1}{y_1}{a} \, \big| \, \GrCylT{0.8}{0}{1}{x_2}{y_2}{b} \, \big\rangle
=\delta_{a,b} \, \delta_{x_1,x_2} \, \delta_{y_1,y_2} \; .
\end{align}
Repeating the computation of sec.~\ref{sec:1+1dtube}, the algebra product on $\mathcal{V}^{\Lambda^2 \overline G}_{\tau^2(\pi)}[\mathfrak{T}[\mathbb o]]$ reads
\begin{align*}
	\nn\big|
	\GrCylT{0.8}{0}{1}{x_1}{y_1}{a}
	\big\rangle
	\star 
	\big|
	\GrCylT{0.8}{1}{2}{x_2}{y_2}{b}
	\big\rangle
	&= \mathbb{P}_{\tubem{\mathbb{o}} \cup_{\mathbb{o}} \tubem{\mathbb{o}}} \circ \mathfrak{G} \triangleright
	\big( \, \big| \GrCylT{0.8}{0}{1}{x_1}{y_1}{a} \big\rangle
	\otimes \big| \GrCylT{0.8}{1}{2}{x_2}{y_2}{b} \big\rangle \, \big) \\
	&=
	\delta_{x_2,x_1^a}\, \delta_{y_2,y_1^a} \, \mathbb{P}_{\tubem{\mathbb{o}} \cup_{\mathbb{o}} \tubem{\mathbb{o}}} \triangleright \big( \, \big|
	\GrCylDoubleT{0.8}{0}{1}{2}{x_1}{y_1}{a}{b}
	\big\rangle \, \big) \; .
\end{align*}
Applying definition \eqref{eq:opA}, the action of the operator $\mathbb{P}$ is expressed in terms of the partition function $\mathcal{Z}^{\Lambda^2 \overline G}_{\tau^2(\pi)}$ as follows:
\begin{align}
	\nn
	&\mathbb{P}_{\tubem{\mathbb{o}} \cup_{\mathbb{o}} \tubem{\mathbb{o}}} \triangleright \big( \, \big|
	\GrCylDoubleT{0.8}{0}{1}{2}{x_1}{y_1}{a}{b}
	\big\rangle \, \big) \\[-0.5em]
	& \q = \mathcal{Z}^{\Lambda^2 \overline G}_{\tau^2(\pi)} \Bigg[\GrCylZT{0.8}{0}{1}{2}{1'}{a}{b}{x_1}{y_1} \! \Bigg]
	\big|
	\GrCylDoubleT{0.8}{0}{1}{2}{x_1}{y_1}{a}{b}
	\big\rangle
	\\
	& \q =
	\nn
	\frac{1}{|G|}\sum_{k}\frac{\tau^2_{x_1,y_1}(\pi)(a,k)}{\tau_{x_1^a, y_1^a}^2(\pi)(k,k^{-1}b)}
	\big|
	\GrCylDoubleSpeT{0.8}{0}{1'}{2}{x_1}{y_1}{ak}{k^{-1}b}{x_1^{ab},y_1^{ab}}
	\big\rangle \; .
\end{align}
It now remains to apply the triangulation changing isomorphism between ground states subspaces so as to recover the initial triangulation. Following sec.~\ref{sec:DWopen} and \ref{sec:fixedpoint}, this isomorphism is expressed as the 2d partition function for the pinched interval cobordism given by the 2-simplex $\snum{(012)}$. Explicitly, the triangulation changing operator reads
\begin{align}
	\mathcal{Z}^{\Lambda^2 \overline G}_{\tau^2(\pi)} \Bigg[
	\GrCylPinched{0.8}{0}{2}{1'}{}{}
	\Bigg]
	=\frac{1}{|G|^\frac{1}{2}}\sum_{z,\tilde z,c,d\in G}\tau^2_{z,\tilde z}(\pi)(c,d)
	\big| \, \GrCylT{0.8}{0}{2}{z}{\tilde z}{cd} \, \big\rangle \big\langle \, \GrCylDoubleT{0.8}{0}{1'}{2}{z}{\tilde z}{c}{d} \, \big|
\end{align}
so that
\begin{equation}
	\big| \,\GrCylDoubleSpeT{0.8}{0}{1'}{2}{x_1}{y_1}{ak}{k^{-1}b}{x_1^{ab},y_1^{ab}} \, \big\rangle
	\simeq \frac{1}{|G|^\frac{1}{2}}\, \tau^2_{x_1,y_1}(\pi)(ak,k^{-1}b) 
	\big| \,  \GrCylT{0.8}{0}{2}{x_1}{y_1}{ab} \, \big\rangle \; .
\end{equation}
Putting everything together, the algebra product of ${\rm Tube}^{\Lambda^2 \overline G}_{\tau^2(\pi)}(\mathbb o)$ is given by
\begin{equation}
	\nn\big|
	\GrCylT{0.8}{0}{1}{x_1}{y_1}{a}
	\big\rangle
	\star 
	\big|
	\GrCylT{0.8}{1}{2}{x_2}{y_2}{b}
	\big\rangle
	= \delta_{x_2,x_1^a} \, \delta_{y_2,y_1^a} \,
	\frac{1}{|G|^\frac{1}{2}}\tau^2_{x_1,y_1}(\pi)(a,b)
	\big|
	\GrCylT{0.8}{0}{2}{x_1}{y_1}{ab}
	\big\rangle \; ,
\end{equation}
where we made use of the 2-cocycle condition $d^{(2)}\tau^2_{x_1,y_1}(\pi)(a,k,k^{-1}b) =1$. As expected, this reproduces exactly \eqref{eq:tqtalgprod}.

\newpage
\section{Discussion}

Gauge models of topological phases have been under intense scrutiny in recent years. These models are especially relevant in (3+1)d where they describe a large class of systems displaying non-trivial topological order \cite{lan2017classification}. In this paper, we studied in detail the Hamiltonian realisation of Dijkgraaf-Witten theory for general spacetime dimensions. The goal of this paper was two-fold: Introduce tools that can be used in order to classify excitations and study their statistics, and provide a rigorous treatment of the dimensional reduction arguments typically used in the condensed matter literature.

Firstly, we presented a systematic way of constructing lattice Hamiltonian realisations of Dijgraaf-Witten theory in terms of pinched interval cobordisms. We then exposed a general program to study the excitations yielded by these Hamiltonians in terms of tube algebras. The tube algebras in (1+1)d, (2+1)d and (3+1)d for the case of loop-like excitations were derived explicitly. We then presented in detail their representation theory together with their quasi-Hopf-like algebraic structure. In particular, in (3+1)d we defined the compatible comultiplication rule and $R$-matrices that encode the fusion and the braiding of loop-like excitations, respectively.

Secondly, we described in detail the situation when one of the spatial directions is compactified using the technology of loop-groupoids. More specifically, we explained that given a ($d$+1)-dimensional model whose input data is a finite group $G$ and a normalised group ($d$+1)-cocycle $\omega\in Z^{d+1}(G,{\rm U}(1))$, upon compactification the ground state subspace of the model can be expressed  as a \emph{lifted} $d$-dimensional model whose input cocycle is now a $d$-dimensional loop-groupoid cocycle and that such states can be defined in terms of loop-groupoid coloured graph-states. As an application we then showed that the lifted models can be utilised to express higher-dimensional tube algebras in terms of lifted lower dimensional models.

The tools introduced in this paper admit several direct generalisations. For instance, they can straightforwardly be adapted in order to study tube algebras in (3+1)d with different boundary conditions, hence classifying excitations beyond the loop-like ones. Each possible oriented boundary condition is determined by a genus-$\mathsf{g}$ surface. The case $\mathsf{g}=0$ corresponds to the sphere $\mathbb S^2$ and gives rise to a classification of point-like particles in terms of irreducible representations of the input group $G$ that are independent of the choice of input 4-cocycle. The case $\mathsf{g}=1$ was the one studied in detail in the present manuscript. For an arbitrary $\mathsf{g}$, the corresponding algebra can be defined utilising the constructions presented in sec.~\ref{sec:tube}. Indeed, given an oriented boundary surface $\Sigma_{\mathsf{g}}$ with genus $ \mathsf{g} \geq 1$, the corresponding tube algebra can be specified, up to Morita equivalence, by a groupoid $\Gamma_{\Sigma_{\mathsf{g}}}$ with objects the set of all $\{h_{1},k_{1},\ldots, h_{\mathsf{g}},k_{\mathsf{g}} \}\in G^{2\mathsf{g}}$ such that $\prod^{\mathsf{g}}_{i=1}[h_{i},k_{i}]=\mathbb{1}_{G}$ which correspond to group homomorphisms $\gamma:\pi_{1}(\Sigma_{\mathsf{g}})\rightarrow G$, and morphisms $(h_{1},k_{1},\ldots,h_{\mathsf{g}},k_{\mathsf{g}})\xrightarrow{g}(h^{g}_{1},k^{g}_{1},\ldots,h^{g}_{\mathsf{g}},k^{g}_{\mathsf{g}})$ for all $g\in G$ \cite{AB}. Such morphisms define the ground state basis states for a triangulation of $\Sigma_{\mathsf{g}}\times \mathbb I$ where $\Sigma_{\mathsf{g}}$ is realised by a triangulation of the $2\mathsf{g}$-gon with one independent vertex and edges appropriately identified. The source and target objects of a morphism then correspond to $G$-colorings of $\Sigma_{\mathsf{g}}\times \{0\}$ and $\Sigma_{\mathsf{g}}\times \{1\}$, respectively, while the morphisms specify the total $G$-coloring of $\Sigma_{\mathsf{g}}\times \mathbb I$. The tube algebra is then given by the twisted groupoid algebra:
\begin{align*}
	&\big| (h_{1},k_{1},\ldots,h_{\mathsf{g}},k_{\mathsf{g}})\xrightarrow{g}(h^{g}_{1},k^{g}_{1},\ldots,h^{g}_{\mathsf{g}},k^{g}_{\mathsf{g}})
	\big\rangle
	\star
	\big|
	(h'_{1},k'_{1},\ldots,h'_{\mathsf{g}},k'_{\mathsf{g}})\xrightarrow{g'}(h'^{g'}_{1},k'^{g'}_{1},\ldots,h'^{g'}_{\mathsf{g}},k'^{g'}_{\mathsf{g}})
	\big\rangle
	\\
	&\q =
	\Big(\prod^{\mathsf{g}}_{i=1}
	\delta_{h'_{i},h^{g}_{i}}
	\delta_{k'_{i},k^{g}_{i}}
	\Big)
	\frac{\beta_{(h_{1},k_{1},\ldots,h_{\mathsf{g}},k_{\mathsf{g}})}(g,g')}{\sqrt{|G|}}
	\big|
	(h_{1},k_{1},\ldots,h_{\mathsf{g}},k_{\mathsf{g}})\xrightarrow{gg'}(h^{gg'}_{1},k^{gg'}_{1},\ldots,h^{gg'}_{\mathsf{g}},k^{gg'}_{\mathsf{g}})
	\big\rangle
\end{align*}
where $\beta_{(h_{1},k_{1},\ldots,h_{\mathsf{g}},k_{\mathsf{g}})}(g,g')$ is defined by an appropriate combination of the input 4-cocycle $\pi\in H^{4}(G,U(1))$ and can be expressed explicitly from the pinched cobordism defining the gluing of the states. It is a normalised groupoid 2-cocycle in $H^{2}(\Gamma_{\Sigma_{\mathsf{g}}},\rU(1))$ that satisfies
\begin{align*}
	\frac{
		\beta_{(h^{a}_{1},k^{a}_{1},\ldots,h^{a}_{\mathsf{g}},k^{a}_{\mathsf{g}})}(b,c) \,
		\beta_{(h_{1},k_{1},\ldots,h_{\mathsf{g}},k_{\mathsf{g}})}(a,bc)}{\beta_{(h_{1},k_{1},\ldots,h_{\mathsf{g}},k_{\mathsf{g}})}(a,b) \, \beta_{(h_{1},k_{1},\ldots,h_{\mathsf{g}},k_{\mathsf{g}})}(ab,c)
	}
	=1 \; .
\end{align*}
The simple modules can then be found, analogously to the case of the twisted quantum double and twisted quantum triple, by first reducing the algebra to subalgebras given by objects related by conjugation and then resolving each such algebra by the irreducible representations of the stabiliser group of a representative object \cite{willerton2008twisted}.

Furthermore, it is possible to enrich the present constructions to accommodate lattice models that have a higher gauge theory interpretation \cite{ Kapustin:2013uxa, Bullivant:2016clk,Bullivant:2017sjz,delcamp2018gauge, Delcamp:2019fdp, zhu2018topological, Wen:2018zux, Wan:2018bns, Wan:2018djl, Wan:2019oyr}. This generalization was formally stated in \cite{AB} in the strict 2-group setting using the language of groupoids for general spacetime dimensions and choices of boundary manifold. The explicit derivation of the tube algebras and their simple modules within this context will be presented in a forthcoming paper.
 
The tube algebra program can also be adapted to study excitations of gapped boundaries in topological phases of matter \cite{Barkeshli:2014cna, cong2016topological, Bullivant:2017qrv, wang2018gapped}. In this scenario, the `tube' is generalised to have two forms of boundary, a physical gapped boundary corresponding to the boundary of the spatial manifold, and a boundary introduced by removing a local neighbourhood of an excitation incident on the boundary of the spatial manifold. Analogously to the bulk excitations, the boundary tube algebra can be directly applied to understanding the fusion and braiding structures of point-particle excitations constrained to the boundary of a three-dimensional spatial manifold. Additionally, the boundary tube algebra theory can be applied to give an algebraic approach to classifying domain walls of arbitrary codimension between different topological phases and provides a canonical method to define exactly solvable Hamiltonian models in such contexts.
Interestingly, due to the topology of the problem, a generalisation of the notion of lifted models also plays an important role when considering point-like boundary excitations of the (3+1)d Dijkgraaf-Witten model in comparison to the boundary excitations of the (2+1)d model. This approach will be applied to the Dijkgraaf-Witten and higher gauge theory models of topological phases of matter in a forthcoming work.

\begin{center}
	\textbf{Acknowledgments}
\end{center}
\noindent
CD would like to thank Apoorv Tiwari for several discussions on related topics.
This project has received funding from the European Research Council (ERC) under the European Union’s Horizon 2020 research and innovation programme through the ERC Starting Grant WASCOSYS (No. 636201). CD is funded by the Deutsche Forschungsgemeinschaft (DFG, German Research Foundation) under Germany’s Excellence Strategy – EXC-2111 – 390814868. AB would like to thank Yidun Wan and Jacqueline Lawrence for related discussions. AB is funded by the EPSRC doctoral prize fellowship.

\newpage
\titleformat{name=\section}[display]
{\normalfont}
{\footnotesize\centering {APPENDIX \thesection}}
{0pt}
{\large\bfseries\centering}
\appendix

\section{Membrane-net picture\label{sec:membrane}}

Given a three-dimensional surface $\Sigma_{\rm 3d}$ endowed with a triangulation $\Sigma_{{\rm 3d}, \triangle}$, we defined in sec.~\ref{sec:DWHam} the lattice Hamiltonian realisation of Dijkgraaf-Witten theory. In (3+1)d, the input data of this model is a finite group $G$ and a cohomology class $[\pi] \in H^4(G,\rU(1))$. This Hamiltonian yields point-like charge and string-like flux excitations, which are classified by irreducible representations of the twisted quantum triple algebra. Since the underlying graph is a triangulation, it is easy to define the Hamiltonian projector directly in terms of the corresponding partition function. Furthermore, the local unitary transformations, with respect to which the ground states of the Hamiltonian are fixed point wave functions, can be expressed directly in terms of Pachner operators.

In this appendix we would like to propose a different formulation of the same model in terms of so-called \emph{membrane-nets}. The goal of this reformulation is two-fold. Firstly, it provides a definition of the model that is analogous to string-net models. Secondly, we will argue that it can be used to shed light on the fusion and the braiding statistics of loop-like excitations \cite{Baez:2006un, Wang:2014xba, Jiang:2014ksa, PhysRevB.91.035134, Cheng:2017ftw, Bullivant:2018pju, Bullivant:2018djw, Wang:2019diz}.  We assume in this appendix that the group is \emph{abelian}. We make this restriction because we are mainly interested in the specificity of dealing with loop-like objects instead of point-like ones. Furthermore, we want to exploit the fact that there is a graphical correspondence between the definition of the lattice Hamiltonian and the statistics of abelian loop-like excitations. More precisely, we will explain how the membrane-nets graphical calculus can be used in order to provide \emph{spacetime} diagrams for the fusion and braiding processes of these excitations.

\bigskip \noindent
\emph{Since the point of this appendix is merely to provide some intuition regarding the statistics of loop-like excitations, the following exposition is looser than in the main text. However, the model we are going to present and the one defined previously are strictly equivalent, it is thus possible to refer to the previous sections which provides a more rigorous treatment of the more technical points.}

\subsection{String-net models\label{sec:string}}

\bigskip \noindent
Levin and Wen introduced in \cite{Levin:2004mi} \emph{string-nets} as a systematic way to construct exactly solvable models displaying topological order in two dimensions. A string-net is, as the name suggests, a network of oriented strings. These strings are labeled by super-selection sectors that must satisfy compatibility conditions at every node of the network referred to as \emph{branching rules}. In the case where all the nodes are chosen to be three-valent, we can think of the underlying graph as being the one-skeleton of the polyhedral decomposition dual to a 2d triangulation. Linear superpositions of string-nets configurations form the Hilbert space of the system. In general, a string-net model corresponds to the lattice Hamiltonian realisation of a \emph{Turaev-Viro} topological quantum field theory \cite{Turaev:1992hq, Barrett:1993ab, Turaev:1994xb}, whose input data is a \emph{spherical/fusion category}. Such lattice Hamiltonian yields bulk point-like excitations that come in two types, namely magnetic fluxes and electric charges. 

As for the gauge model introduced in sec.~\ref{sec:DWHam}, local unitary transformations can be defined at the level of the network so as to implement a wave function renormalisation group flow. Fixed point wave functions with respect to this renormalisation flow are then found to be ground states of parent Hamiltonians, the local transformations specifying uniquely the fixed point wave functions. We distinguish several local unitary transformations, one of them is the so-called \emph{F-move}, which is nothing else than the Poincar\'e dual of the $2\leftrightharpoons 2$ Pachner move:
\begin{equation}
	\label{eq:strF}
	\hexagonMoveONE{0.23}{2} \xrightarrow{\alpha(a,b,c)} \hexagonMoveFIVE{0.23}{2} \; .
\end{equation}
Let us now suppose that the input data of the string-net model is the category $\mathbb C$--${\rm Vec}^{\alpha}_A$ of $A$-graded vector spaces so that the set of super-selection sectors is taken to be a finite abelian group $A$ and the branching rules are provided by the group multiplication.\footnote{We explained earlier that the category of $A$-graded vector spaces is indeed the relevant structure to describe the lattice Hamiltonian realisation of (2+1)d Dijkgraaf-Witten theory, which is equivalent to the string-net model under consideration.} In this case, the amplitude of the $F$-move depicted above is provided by the group 3-cocycle $\alpha$. The fact that the amplitude associated with such a move must be a 3-cocycle follows from the  \emph{pentagon} coherence relation that $\alpha$ must satisfy for the process to be self-consistent. 

Bulk excitations of this string-net model can be studied using the tube algebra approach as explained in sec.~\ref{sec:tube}, where the relevant tube is the cylinder $\mathbb{S}^1 \times \mathbb{I}$. Within this context, bulk excitations are found to be the objects of the so-called \emph{Drinfel'd center} category $\mathcal{Z}(\mathbb C$--${\rm Vec}^{\alpha}_A)$, which is a braided fusion category. But under \emph{Tannaka duality}, these objects correspond to the modules of the twisted quantum double quasi-Hopf algebra.

\medskip \noindent
Let us now consider the system of \emph{abelian} anyons described by $\mathbb C$--${\rm Vec}^{\alpha}_A$.\footnote{Given an appropriate choice of $\alpha$, these abelian anyons labeled by $A$ can be thought of as the pure flux excitations of the corresponding string-net model.} In this context, we can think of a string in the diagrams above as being a worldline for a (point-like) abelian anyon labeled by the corresponding super-selection sector. In which case, the three-valent nodes are interpreted as the fusion of two particles. Taking the time direction to be downwards, the string-nets above represent the worldlines of three particles labeled by $a$, $b$ and $c$ that are fusing with each other, so that the l.h.s and the r.h.s only differ in the fusion pattern. Within this spacetime interpretation, the $F$-move depicted above corresponds to a change of ordering in which the particles are fused, and this process is accompanied with a $\rU(1)$ phase expressed in terms of the group 3-cocycle. The pentagon coherence relation can be reformulated as follows: Given four particles fusing according to a specific pattern, two different sequences of $F$-moves can be performed so as to obtain the same alternative fusing pattern. Self-consistency imposes that the collective phases associated with these two sequences must be equal, which in turn implies the 3-cocycle condition. 

So equation \eqref{eq:strF} can be interpreted either as the defining $F$-move local transformation of the string-net model, or as the associativity of the fusion process of abelian anyons. Henceforth, upon describing the fusion of anyonic excitations, we will refer to equation \eqref{eq:strF} as being the \emph{string diagram} representation of the associator isomorphism that is determined by the 3-cocycle $\alpha$. One goal of this appendix is to reproduce this somewhat trivial statement in three dimensions in terms of membrane-nets. More precisely, we will present to which extent local unitary transformations in terms of membrane-nets can be interpreted as surface diagrams associated with the fusion of loop-like objects upon compactification of one of the spatial directions. The same graphical calculus will then be used in order to provide spacetime diagrams for the corresponding braiding process.

\subsection{Membrane-nets and lattice Hamiltonian}

In light of the correspondence between one of the defining local transformations of string-net models and the string diagram representation of the associator isomorphism, we would like to derive a higher-dimensional version of the string-net formalism for the model \eqref{eq:Ham}. We refer to this generalization as \emph{membrane-nets}, and their construction follows closely the two-dimensional one.

A \emph{membrane-net} is a three-dimensional network of oriented two-dimensional membranes. These membranes are labeled by super-selection sectors that must satisfy compatibility conditions at every edge of the network so that only certain combination of super-selection sectors are allowed. More precisely, let  $\Sigma$ be a \emph{closed} 3d surface endowed with a triangulation $\Sigma_\triangle$. We consider the polyhedral decomposition $\Sigma_\Upsilon$ dual to the triangulation $\Sigma_\triangle$ such that $i$-simplices $\triangle^{(i)}$ are dual (3$-$$i$)-cells $\Upsilon^{(3-i)}$. Since we are interested in the membrane-net model equivalent to the model we introduced in sec.~\ref{sec:DWHam}, we label each such 2-cell with a group variable $g \in A$, where $A$ is a finite \emph{abelian} group, so that each labeling defines a different graph-state. Branching rules are enforced at every link (or 1-cell), namely the oriented product of the group variables labeling the membranes meeting at a 1-cell must vanish. In order to express these branching rules, we need to introduce a convention regarding the orientation of the membranes:
\begin{convention}[\emph{Orientation of the membranes}\label{conv:orMemb}] 
	Given a membrane net, to every 1-cell where three membranes meet, we can assign a dual 2-simplex, whose boundary 1-simplices are oriented either inwards or outwards from the corresponding dual membrane with respect to the paper plane. In the case where the dual 1-simplex is oriented outwards, we decorate the corresponding group labeling with a bar, i.e. $\bar{\sss *}$. For instance, consider the  membrane-net
	\begin{equation*}
		\dualHamB{0.23}{1} \; \equiv \; \dualHamB{0.23}{3} \; ,
	\end{equation*}
	it follows from the convention that the oriented product of the group variables is indeed the identity. Note that since the group $A$ is abelian, we write the product rule additively. The orientation chosen is such that it reproduces the conventions of the model we defined previously on the dual triangulation. For notational convenience, we will often write a minimal labeling only so that the remaining labels can be deduced from the branching rules, e.g.
	\begin{equation*}
		\convLabels{0.23}{1} \; \equiv \; \convLabels{0.23}{2}\; .
	\end{equation*}
	Note that the branching structure is such that, given two membranes meeting at a 1-cell, the labeling of the remaining one is always the sum of the other two.
\end{convention}
\noindent
Recall that the local unitary transformations are the defining feature of topological order. We expressed in sec.~\ref{sec:DWHam} these local transformations in terms of Pachner operators. Since the current model is strictly equivalent to the one described there, these Pachner operators are still relevant. However they must now be defined with respect to the dual polyhedral decomposition. In particular, the Pachner $2 \leftrightharpoons 3$ operator yields for instance the isomorphism
\begin{equation}
	\label{eq:P23d} 
	\Bigg|\;\; \dualPTwoThree{0.23}{1} \;\; \Bigg\rangle
	\, \simeq \,  \pi(a,b,c,d)^{-1} \,
	\Bigg| \;\; \dualPTwoThree{0.23}{2} \;\; \Bigg\rangle \; ,
\end{equation}
where as before $\pi$ is chosen to be \emph{normalized} 4-cocycle, and such that the remaining labels can be deduced from the branching rules imposed at every 1-cell. In the following, we notate this move as $(2 \leftrightharpoons 3)^\star$. Similarly, the Pachner $1 \leftrightharpoons 4$ yields for instance
\begin{equation}
	\label{eq:P41d}
	\Bigg| \;\; \dualPFourOne{0.23}{1} \;\; \Bigg\rangle
	\, \simeq \, \pi(a,b,c,d)^{+1} \,
	\Bigg| \;\; \dualPFourOne{0.23}{2} \;\; \Bigg\rangle \; .
\end{equation}
In the following, we notate this move as $(1 \leftrightharpoons 4)^\star$. It is interesting to consider a special case of eq.~\eqref{eq:P23d} that is obtained by setting the group variables $b$ and $c$ to the identity, namely
\begin{equation}
	\Bigg| \;\; \specialPTwoThree{0.23}{1} \;\; \Bigg\rangle 
	\, \simeq \, \Bigg| \;\; \specialPTwoThree{0.23}{2} \;\; \Bigg\rangle \; ,
\end{equation}
where we made use of the fact that the 4-cocycle $\pi$ is normalized, i.e. $\pi(a,\zo,\zo,d) = 1$. It follows from this last expression that a closed membrane (homeomorphic to a two-sphere) can be fused to a neighboring membrane as follows:
\begin{equation}
	\label{eq:spec}
	\Bigg| \;\; \specialSphere{0.23}{1} \;\Bigg\rangle \simeq \Bigg| \;\; \specialSphere{0.23}{2} \; \Bigg\rangle \; .
\end{equation}
Recall that the membrane-net model under consideration is strictly equivalent to the model introduced in sec.~\ref{sec:DWHam} at the difference that the degrees of freedom now live on the 2-cells of the dual polyhedral decomposition. Nevertheless, within this formalism, it is not very natural to define the lattice Hamiltonian in terms of the corresponding partition function. Instead, it is defined directly in terms of the local transformations presented above, in a fashion akin to two-dimensional string-net models.

\bigskip \noindent
So let us define the parent Hamiltonian whose ground states are the fixed point wave functions satisfying equations (\ref{eq:P23d}, \ref{eq:P41d}). To every 1-cell $\Upsilon^{(1)}$ of the polyhedral decomposition, we assign an operator $\mathbb B_{\Upsilon^{(1)}}$ which enforces the branching rules, i.e. penalizes non-flat $A$-connections. For instance, one has
\begin{equation}
	\mathbb B_{\Upsilon^{(1)}} \triangleright 
	\Bigg| \;\; \dualHamB{0.23}{2} \;\; \Bigg\rangle  = \delta_{a+b,c} \, \Bigg| \;\; \dualHamB{0.23}{2} \;\; \Bigg\rangle \; .
\end{equation}
To every 3-cell $\Upsilon^{(3)}$, we assign an operator $\mathbb A_{\Upsilon^{(3)}}$ which modifies the gauge field configuration of the 2-cells $\Upsilon^{(2)} \subset \Upsilon^{(3)}$ adjacent to $\Upsilon^{(3)}$ by `fusing' a closed membrane of defect into the boundary of $\Upsilon^{(3)}$. We decompose such $\mathbb A$-operator as $\mathbb A_{\Upsilon^{(3)}} = 1/|A|\sum_{k \in A}\mathbb A^k_{\Upsilon^{(3)}}$, where the action of the operator $\mathbb{A}^k_{\Upsilon^{(3)}}$ is defined graphically in terms of the local transformations presented above by inserting a closed membrane of defect labeled by $k$ as follows:\footnote{The reader familiar with string-net models should recognize that this definition is a natural generalization of the two-dimensional case. Indeed, for string-net models, the analogue of the $\mathbb A$-operator is defined by fusing a closed loop of defect into the boundary of a plaquette so as to change the gauge field configuration. See \cite{Levin:2004mi} for further details.}
\begin{equation}
	\mathbb{A}^k_{\Upsilon^{(3)}}\; \triangleright \; 
	\Bigg|\;\; \dualHamA{0.22}{1} \;\; \Bigg\rangle 
	= \Bigg|\;\; \dualHamA{0.22}{2} \;\; \Bigg\rangle \; .
\end{equation}
Using four times the move \eqref{eq:spec}, we fuse the membrane of defect into the boundary, and finally we use four times the $(1 \leftrightharpoons 4)^\star$ move \eqref{eq:P41d} in order to recover the original polyhedral decomposition:
\begin{equation*}
	 \Bigg|\;\; \dualHamA{0.22}{3} \;\; \Bigg\rangle 
	  \simeq \frac{\pi(a+b,c,d,k) \, \pi(a,b,c+d,k)}{\pi(b,c,d,k) \, \pi(a,b+c,d,k) } \,
	\Bigg| \;\; \dualHamA{0.22}{4} \;\; \Bigg\rangle \; .
\end{equation*}
Remember that the amplitude of the move \eqref{eq:P41d} depends on the orientation of the corresponding dual 4-simplex according to conv.~\ref{conv:orFour}. Since we choose the initial state to be the same one as in sec.~\ref{sec:DWHam}, but on the dual membrane-net, we recover the same amplitude as in sec.~\ref{sec:DWHam} as expected.

The operators defined above are strictly equivalent to the ones defined previously, but on the dual polyhedral decomposition, thus they satisfy exactly the same properties. In particular, all the operators commute with each other so that the lattice Hamiltonian projector reads
\begin{equation}
	\label{eq:Hamd}
	\mathbb H = - \sum_{\Upsilon^{(1)} \subset \Sigma_\Upsilon}\mathbb B_{\Upsilon^{(1)}} - \sum_{\Upsilon^{(3)} \subset \Sigma_\Upsilon}\mathbb A_{\Upsilon^{(3)}} \; ,
\end{equation}
where the sums run over the 1-cells and the 3-cells of the polyhedral decomposition $\Sigma_\Upsilon$, respectively. It is possible to check explicitly that ground states of this Hamiltonian do satisfy equations \eqref{eq:P23d} and \eqref{eq:P41d}.

\subsection{Lifted models and surface diagrams}

We explained in sec.~\ref{sec:Lifting1} that when one of the spatial directions is compactified, it is possible to express the Hamiltonian model in terms of another model in one-lower dimensional. This process was made rigorous in sec.~\ref{sec:Lifting2} using the language of loop groupoids. We do not intend to repeat this analysis here but merely to represent the so-called lifted Pachner operators in terms of membrane nets. 

Let us consider a 3d surface $\Sigma_{\rm 3d}$ of the form $\Sigma_{\rm 2d} \times \mathbb S_1$, with $\Sigma_{\rm 2d}$ a Riemann surface. Let us endow $\Sigma_{\rm 2d}$ with a two-dimensional polyhedral decomposition $\Sigma_{{\rm 2d}, \Upsilon}$ dual to a triangulation  $\Sigma_{{\rm 2d}, \triangle}$, so that every 0-cell is adjacent to three 1-cells. We then lift $\Upsilon_{\rm 2d}$ to a three-dimensional polyhedral decomposition $\Sigma_{{\rm 3d},\Upsilon}$ such that $\Sigma_{{\rm 3d},\Upsilon}$ is dual to the three-dimensional triangulation $\Sigma_{{\rm 3d},\triangle}$ we would obtain according to conv.~\ref{conv:dlifting}. We know from the analysis carried out in sec.~\ref{sec:Lifting2} that is it possible to define a $\Lambda \overline A$-coloured model on $\Sigma_{{\rm 2d}, \Upsilon}$ that is equivalent to the $A$-coloured three-dimensional one. Upon this process, the input group 4-cocycle is replaced by its $\mathbb S^1$-transgression $\tau(\pi)$, which satisfies the usual group 3-cocycle condition since the group $A$ is abelian. 

It follows that the $\mathbb S^1$-transgression appears as the amplitude of the so-called lifted Pachner operators. This could be confirmed explicitly by repeating the computations performed in sec.~\ref{sec:Lifting1}, but this time within the membrane-net formalism. For instance, this would amount to lifting the $F$-move \eqref{eq:strF} analogously to the way we lifted the $2 \rightleftharpoons 2$ move, and show that it can be decomposed into four $(2 \rightleftharpoons 3)^\star$ moves. Doing this computation, we would recover the same amplitude as before. Nevertheless, it is enough to consider the membrane-nets dual to the initial and final $A$-coloured three-dimensional triangulations appearing in the definition of the lifted $2 \leftrightharpoons 2$ move. Although there is unique way to define the membrane-net dual to a given coloured triangulation, there are several ways to display this lifted $F$-move graphically. The choice we make is motivated by the fact that we want to be able to interpret the final result in terms  of spacetime diagrams for the associativity relation of the fusion of loop-like objects. So the membrane-net representation of the lifted $F$-move reads
\begin{equation}
	\label{eq:assocLoop}
	\q \slantTwoTwoA{1} \xrightarrow{\tau_x(\pi)(a,b,c)} \q\;\;
	\slantTwoTwoA{2}
\end{equation}
where each dot represents a 0-cell dual to one of the 4-simplices appearing in \eqref{eq:lifted22move}. By definition, the membrane-nets above are merely obtained by considering the Poincar\'e dual of the triangulations appearing in \eqref{eq:lifted22move}. But, its interpretation as a lifted $F$-move is also very clear. Indeed, we can think of the membranes labeled by $a,b,c \in A$ as being obtained by `sweeping' the string-nets appearing in \eqref{eq:strF} along a circle $\mathbb{S}^1$ living in a plane orthogonal to the paper one. This membrane-net representation can then be used to check that the $\mathbb{S}^1$-transgression $\tau(\pi)$ must be a group 3-cocycle by considering a lifted version of the pentagon coherence relation. 

\bigskip \noindent
We emphasized in sec.~\ref{sec:string} how the $F$-move operator, whose amplitude is given by a group 3-cocycle, could be interpreted as the string diagram representation of the associator for the fusion of abelian anyons labeled by $A$. Since the amplitude of the lifted $F$-move operator \eqref{eq:assocLoop} is also given in terms of a group 3-cocycle, and since point-like objects are mapped to loop-like objects under such lifting, it is natural to expect a similar version of this interpretation to persist.

Indeed, it is tempting to interpret the membrane-net representation of the lifted $F$-move in terms of so-called \emph{surface diagrams for} the associativity of the fusion of loop-like abelian excitations. Within this spacetime interpretation, the time direction is taken to be downwards so that the diagrams on both side represent the worldsheets of three loop-like excitations labeled by $a,b,c \in A$, linked by a background string labeled by $x \in A$, fusing with each other, the fusion being provided by the group product rule. The two sides only differ in the order in which the loop-like excitations are fused. Therefore, we interpret \eqref{eq:assocLoop} as the \emph{surface diagram} representation of the associator isomorphism determined by the group 3-cocycle $\tau(\pi )$, with respect to the fusion product of the loop-like excitations. The coherence relation the associator isomorphism must satisfy can then be represented in terms of surface diagrams. 
\subsection{Loop braiding}
Given a 3d topological model displaying abelian loop-like excitations, the statistical phase acquired by a loop upon braiding with another loop when they are threaded by the same string, may be computed by solving the \emph{hexagon relations} with respect to $\mathbb S^1$-transgression of the input 4-cocycle \cite{Wang:2014xba}. In this part, we wish to provide some intuition for this statement using the arguments presented previously.

Let us consider \eqref{eq:assocLoop} and interpret it as a defining relation in terms of surface diagrams for the associativity of the fusion of the abelian loop-like excitations. It follows from the discussion above that by embedding the loops in a compactified 3d manifold, it is possible to identify this associativity relation with the lifted $F$-move for the lifted membrane-net model whose input data is $(\Lambda \overline A, \tau(\pi))$ so that the associator is determined by the $\mathbb S^1$-transgression $\tau(\pi)$ of the 4-cocycle $\pi$. We now would like to define graphically the corresponding braiding isomorphism.

Let us first briefly review the two-dimensional case. We mentioned earlier how string diagrams provide a graphical calculus for the fusion statistics of point-like excitations. These excitations can also be braided, in which case the statistical phase that determines this \emph{braiding} isomorphism is provided by a group 2-cochain $R$ valued in $\rU(1)$.\footnote{As explained in detail in sec.~\ref{sec:cat}, this braiding isomorphism can be used to turn the monoidal category of $A$-graded vector spaces into a braided one.} There is also a string-diagram representation of the braiding move that corresponds to resolving the crossing of two lines:
\begin{equation}
	\label{eq:strR}
	\Rmove{0.23}{2} \xrightarrow{R(a,b)} \Rmove{0.23}{1} \; ,
\end{equation}
where as before, the time direction is taken to be downwards so that the string-diagram on the l.h.s corresponds to the worldlines of two point-like  excitations labeled by $a,b \in A$ that exchange position before fusing. As explained in sec.~\ref{sec:catPrelimi}, the braiding isomorphism must satisfy the so-called \emph{hexagon relations} \eqref{eq:hex1} and \eqref{eq:hex2}. These algebraic equations can be represented graphically in terms of string diagrams. For instance, the first one reads
\begin{equation}
	\label{eq:hex}
	\begin{tikzpicture}[scale=1,baseline=0em]
	\matrix[matrix of math nodes,row sep =-1.5em,column sep=2em, ampersand replacement=\&] (m) {
		{} \& \hexagonMoveTHREE{0.23}{2}  \& \hexagonMoveONE{0.23}{2} \& {} \\
		\hexagonMoveFOUR{0.23}{2}  \& {} \& {} \& \hexagonMoveFIVE{0.23}{2} \\
		{} \& \hexagonMoveTWO{0.23}{2}  \& \hexagonMoveSIX{0.23}{2} \& {} \\
	};	
	\path
	(m-2-1) edge[->, shorten <= 1em, shorten >= -2em, line width=0.4pt] node[pos=1.1, above, sloped] {${\sss \alpha(c,a,b)}$} (m-1-2)
	edge[->, shorten <= -2em, shorten >= 1em, line width=0.4pt] node[pos=0, above, sloped] {${\sss R(c,a)}$} (m-3-2)
	(m-1-2) edge[->, shorten <= -0.5em, shorten >= -0.5em, line width=0.4pt] node[pos=0.5, above] {${\sss R(c,a+b)}$} (m-1-3)
	(m-3-2) edge[->, shorten <= -0.5em, shorten >= -0.5em, line width=0.4pt] node[pos=0.5, above] {${\sss \alpha(a,c,b)}$} (m-3-3)
	(m-2-4) edge[<-, shorten <= 1em, shorten >= -2em, line width=0.4pt] node[pos=1.1, above, sloped] {${\sss \alpha(a,b,c)}$} (m-1-3)
	edge[<-, shorten <= -2em, shorten >= 1em, line width=0.4pt] node[pos=0, above, sloped] {${\sss R(c,b)}$} (m-3-3)
	;
	\end{tikzpicture} \!\! .
\end{equation}
In light of these two-dimensional results, we are looking for an isomorphism determined by a 3-cochain $R_*( * , *)$ which corresponds to a lifted version of the braiding move \eqref{eq:strR}, the same way \eqref{eq:assocLoop} is a lifted version of the associativity \eqref{eq:strF}. We propose the following definition in terms of surface diagrams:
\begin{equation}
	\label{eq:braidLoop}
	\loopMoveA{1} \q \xrightarrow{R_x(a,b)} \q\loopMoveA{2}
\end{equation}
where as before each black dot represents a 0-cell which can be thought as dual to a  4-simplex. This definition satisfies several criteria: Firstly, it does correspond to a lifted version of \eqref{eq:strR}. Indeed, if we omit the background membrane labeled by the group variable $x \in A$, we are left with the surface diagrams obtained by sweeping the string diagrams appearing in \eqref{eq:strR} along a circle $\mathbb S^1$. Secondly, we can check explicitly that the isomorphisms determined by the cochains $\tau( \pi)$ and $R_*(*, *)$ defined according to \eqref{eq:assocLoop} and \eqref{eq:braidLoop} do satisfy the lifted version of the hexagon relations, which are obtained by lifting each string diagram appearing in the original hexagon relation the same way \eqref{eq:braidLoop} is obtained from \eqref{eq:strR}. The algebraic equation corresponding to the lifted version of \eqref{eq:hex} reads
\begin{equation}
	\tau_x(\pi)(c,a,b)\,  R_x(c,a+b)\, \tau_x(\pi)(a,b,c) = R_x(c,a) \, \tau_x(\pi)(a,c,b) \, R_x(c,b) \; .
\end{equation}
We leave it to the reader to draw the corresponding sequence of surface diagrams. Interestingly, given a 3-cochain $\alpha$, it is always possible to construct a `trivial' solution to the algebraic equation above via
\begin{equation}
	\tau^0_x(\pi)(a,b,c) = \frac{\tau_x(\alpha)(b,c) \, \tau_x(\alpha)(a,b+c)}{\tau_x(\alpha)(a+b,c) \, \tau_x(\alpha)(a,b)}
	\q , \q
	R^0_x(a,b) = \frac{\tau_x (\alpha)(a,b)}{\tau_x (\alpha)(b,a)}
\end{equation} 
such that 
\begin{equation}
	\tau_x (\alpha)(a,b) = \frac{\alpha(x,a,b) \, \alpha(a,b,x)}{\alpha(a,x,b)} \; .
\end{equation}
It then turns out that the 0-cells depicted by a black dot in \eqref{eq:braidLoop} can be thought as dual to the six 4-simplices paired with $\alpha$ appearing in the definition of $R^0_x(a,b)$. Similarly, we can identify the $0$-cells marked with a black dot in \eqref{eq:assocLoop} with the 3-simplices appearing in the definition of $\tau^0_x(\pi)(a,b,c)$.

We can further analyse the surface diagrams obtained above from a spacetime point of view, i.e. by interpreting the different membranes as worldsheets of loop-like objects. Doing so, we realise that the surface diagrams provide a representation of the loop braiding. Indeed, taking the time direction to be downwards, the surface diagram on the l.h.s of \eqref{eq:braidLoop} corresponds to the worldsheets of abelian loop-like excitations threaded by a string exchanging position before fusing with each other. This can be made more explicit by drawing the \emph{movie} of the surface diagram: 
\begin{equation}
	\loopMoveAred{} \q\q  \longleftrightarrow \q\q \loopMovie{}
\end{equation}
which can be used to confirm a posteriori the definition \eqref{eq:braidLoop}.

Since we defined graphically an associator and a braiding isomorphisms with respect to the fusion of abelian loop-like excitations, we should be able to construct a corresponding braided monoidal category whose axioms are recalled in sec.~\ref{sec:catPrelimi}. Objects are labeled by pairs of group elements such that one represents the threading flux and the other one the flux of the loop itself. Borrowing the notation of the twisted quantum triple elements, we notate these objects
\begin{equation}
		(\looDSy{x}{a})  \q {\rm for} \; x,a \in A  \; .
\end{equation}
However, $x$ does not necessarily label a loop-like object but merely a linked flux tube. Given such a linked flux labeled by $x \in A$, we define the tensor product as
\begin{equation}
	(\looDSy{x}{a}) \otimes (\looDSy{x}{b}) = (\looDSy{x}{a+b}) \; ,
\end{equation}
which we should think of as the fusion of two abelian loop-like excitations linked to the same third flux. Graphically, this can be represented as
\begin{equation*}
	\hspace{-1.2em}\Qalgebra{1} \hspace{-1.2em} \otimes \hspace{-1.2em} \Qalgebra{2} \hspace{-1.2em} = \Qalgebra{3} =  \hspace{-1.2em}  \Qalgebra{4}  \; .
\end{equation*}
We then choose an associator isomorphism that is determined by $\tau(\pi)$ and then define a compatible braiding isomorphism. Together with the graphical representation above, the spacetime diagrams corresponding to these morphisms are the ones displayed earlier.

\bibliographystyle{JHEP}
\bibliography{ref_cat}

\end{document}